\documentclass[aps,twocolumn,nofootinbib,showpacs,prd,10pt,superscriptaddress]{revtex4-2}
\usepackage{graphicx}
\usepackage[english]{babel}
\usepackage[T1]{fontenc}
\usepackage{mathrsfs}
\usepackage[tbtags]{amsmath}
\usepackage{amssymb}
\usepackage{amsxtra}
\usepackage{amsopn}
\usepackage{latexsym}
\usepackage[mathcal]{eucal}
\usepackage{mathtools}
\usepackage{slashed}
\usepackage{xcolor}
\usepackage[hyperfootnotes=false,bookmarks=false]{hyperref}

\usepackage{braket}
\usepackage{dsfont}
\usepackage{enumerate}
\usepackage{lipsum}

\newcommand{\BE}{\begin{equation}}
\newcommand{\EE}{\end{equation}}
\newcommand{\BA}{\begin{align}}
\newcommand{\EA}{\end{align}}

\newcommand{\mc}{\mathcal}
\newcommand{\psibar}{\overline{\psi}}
\newcommand{\cbar}{\overline{c}}
\newcommand{\gammabar}{\overline{\gamma}}
\newcommand{\avg}[1]{\left\langle#1\right\rangle}
\newcommand{\txtin}{\text{in}}
\newcommand{\txtout}{\text{out}}

\begin{document}

\title{Dynamical gluon mass generation, the Schwinger mechanism, spontaneous symmetry\\breaking, and color confinement in the linear covariant gauges of QCD}

\author{Giorgio Comitini}
\email{giorgio.comitini@dfa.unict.it}
\affiliation{Dipartimento di Fisica e Astronomia ``E. Majorana'', Universit\`a di Catania, Via S. Sofia 64, I-95123 Catania (CT), Italy}

\date{\today}

\begin{abstract}
   We critically re-examine the Kugo-Ojima (KO) scenario for color confinement and identify several inconsistencies and a crucial error in its derivation. Within the KO framework, under customary assumptions, we show that the BRST quartet structure incorporates a unique mechanism for dynamical gluon mass generation -- namely, the Schwinger mechanism, in the formulation developed by Aguilar, Binosi, Ferreira, and Papavassiliou (ABFP). We show that the Schwinger mechanism is a Higgs mechanism, in that it occurs together with the spontaneous breaking of the global gauge charge, as it should. The ABFP bound state is the associated (unphysical) composite Goldstone boson. The zero-momentum value of the gluon propagator evaluated in the Pinch Technique--Background-Field Method (PT-BFM) naturally emerges in the derivation, setting a Renormalization-Group invariant scale for mass generation and suggesting a new nonperturbative interpretation of the PT-BFM. BRST symmetry takes over the role of the broken symmetry, while the color symmetry of the spectrum remains unbroken. We obtain a unique, well-defined, unbroken, BRST-exact asymptotic color charge operator that realizes the confinement of color in the sense of the absence of colored asymptotic states from the BRST cohomology. Still within the KO framework, we take the first fundamental step towards a proof of the nonperturbative unitarity of the linear covariant gauges of QCD, and we discuss the experimental evidence in favor of the Schwinger mechanism. We lay down the foundations for a unified theory of gauge-boson mass generation in the linear covariant gauges of Yang-Mills theories.
\end{abstract}

\maketitle

\section{Introduction}\label{sec:intro}

In a series of landmark papers published between 1978 and 1979 \cite{KO78b,Oji78,KO78a,KO79b,KO79c,KO79d,KO79a,Kug78}, T. Kugo and I. Ojima (KO) formulated the canonical quantization of Yang-Mills theories in linear covariant gauges (LCG) and introduced the renowned ``Kugo-Ojima criterion for color confinement''. In what follows, we begin by giving an account of the criterion and its flaws, then we move on to describe the purpose of the present paper and its main findings.

The intuitive observation that underlies the KO criterion for confinement is straightforward: Quantum Chromodynamics (QCD) is a Maxwell theory and, like any Maxwell theory, it possesses a Gauss law. In QCD, the (integral) Gauss law equates the color charge $Q^{a}$ contained in all of space to the flux of the chromoelectric field $\vec{\mathcal{E}}^{a}$ at infinity,
\begin{equation}
   Q^{a}=-g^{-1}\oint_{R\to{\infty}}dS\ \hat{n}\cdot\vec{\mathcal{E}}^{a}\ ,
\end{equation}
$g$ being the strong coupling constant. If the carriers of the strong force -- namely, gluons -- acquire a mass, then the chromoelectric field at infinity dies off faster than $1/R^{2}$; the vanishing of the chromoelectric flux at infinity then forces the vanishing of the color charge of every possible state of the theory, yielding $Q^{a}=0$.

In the canonical quantization of the LCGs of Yang-Mills theory, the Gauss law becomes an operatorial equation, and it gets corrected by a term that can be expressed as an anticommutator \cite{Oji78,KO79a},
\begin{equation}\label{YMcharge}
   Q^{a}=-g^{-1}\oint_{R\to{\infty}}dS\ \hat{n}\cdot\vec{\mathcal{E}}^{a}+\left\{Q_{B},g^{-1}\int d^{3}{\bf x}\ D_{0}\cbar^{a}\right\}\ .
\end{equation}
In the above equation, $Q^{a}$ is the Noether charge associated with global gauge (color) transformations, $\overline{c}^{a}$ is the Faddeev-Popov antighost field \cite{FP67a,FP67b}, and $D_{\mu}^{ab}$ is the (adjoint) gauge-covariant derivative. The operator $Q_{B}$ is known as the ``Becchi-Rouet-Stora-Tyutin (BRST) charge''. $Q_{B}$ is nilpotent -- $Q_{B}^{2}=0$ -- and it plays a central role in the KO quantization framework, as it allows one to distinguish between the physical and the unphysical states of the theory \cite{KO78b,KO78a}: a state is physical if and only if it belongs to the so-called ``cohomology'' $\mc{H}_{\text{phys}}$ of the BRST charge,
\begin{equation}
   \mc{H}_{\text{phys}}=\text{ker}\{Q_{B}\}/\text{im}\{Q_{B}\}\ ,
\end{equation}
where $\text{ker}\{Q_{B}\}$ (resp. $\text{im}\{Q_{B}\}$) is the kernel (resp. image) of $Q_{B}$, and the division sign denotes a vector space quotient. In the indefinite-metric Hilbert space on which the theory is quantized, physical states are represented by BRST-invariant states $\ket{\text{phys}}$,
\begin{equation}
   Q_{B}\ket{\text{phys}}=0\ .
\end{equation}

Since the second term on the right-hand side (RHS) of Eq.~\eqref{YMcharge} is ``BRST-exact'' -- i.e. it is the (anti)commutator between the BRST charge and some other operator --, the Gauss law holds in its original form when evaluated between two physical states:
\begin{equation}
   \bra{\text{phys}}gQ^{a}+\oint_{R\to{\infty}}dS\ \hat{n}\cdot\vec{\mathcal{E}}^{a}\ket{\text{phys}^{\prime}}=0\ .
\end{equation}
The argument that relates color confinement to the gluon mass is thus still meaningful in the quantum formulation of LCGs, with one qualification: it is the color of \textit{physical states} that can be expected to vanish if gluons acquire a mass\footnote{Here and throughout, by gluons ``acquiring a mass'' we mean that the dressed transverse form factor $\Delta(p^{2})$ of the gluon propagator has no pole at zero momentum squared, $p^{2}=0$, implying the absence of massless transverse gluons from both the physical and the unphysical sectors of the spectrum.}. Colored states can (and will) be present in the unphysical sector of the theory regardless of the gluon mass.

In early 1979, immediately after formulating the framework of canonical quantization, Kugo and Ojima discovered that a massless particle $\chi^{a}$ belonging to the unphysical sector of the spectrum of the linear covariant gauges spontaneously breaks the global gauge charge $Q^{a}$ whenever the gauge bosons of Yang-Mills theory acquire a mass \cite{KO79d,KO79a},
\begin{equation}\label{ugen}
   \bra{\Omega}gQ^{a}\ket{\chi^{b}}\propto \delta^{ab}(1+u)\ .
\end{equation}
In the above equation, $u$ is a coefficient that we will define in due time. If the RHS of the equation is different from zero, then $Q^{a}\ket{\Omega}\neq 0$, implying that $Q^{a}$ is spontaneously broken, and that $\chi^{a}$ is the Goldstone boson associated with its breaking \cite{GSW62}.

The particle $\chi^{a}$ appearing in the Goldstone equation Eq.~\eqref{ugen} belongs to a quartet of stable massless unphysical particles -- termed the ``elementary BRST quartet'' \cite{Kug78,KO79a} -- whose existence in the unphysical sector of the spectrum of the linear covariant gauges is enforced by the Slavnov-Taylor identities of the theory \cite{Tay71,Sla72,BRS76} regardless of mass generation. If the gauge boson remains massless, $\chi^{a}$ is the unphysical longitudinal polarization of the gauge boson \cite{KO79c,KO79a}. When the gauge boson becomes massive, the identity of $\chi^{a}$ depends on the theory. In modern literature \cite{AA10,AA11}, within the elementary BRST quartet, $\chi^{a}$ has come to be known as its ``first parent''\footnote{\cite{AA10,AA11} mistakenly identify the first parent of the elementary BRST quartet of QCD with the massless longitudinal gluon polarization. When gluons dynamically acquire a mass, that polarization either becomes massive, or it disappears from the spectrum, hence it cannot be a member of the massless elementary quartet.}.

The discovery of the spontaneous breaking of $Q^{a}$ by the first parent of the elementary BRST quartet came as both a blessing and curse. It was a blessing because it provided a clear-cut criterion to set QCD apart from Higgs theories \cite{KO79d,KO79a}: if color symmetry is broken, as is the case in Higgs theories \cite{EB64,Higgs64,GHK64,Higgs66,Kib67,Wei67,Sal68,tHoo71}, then the generator of color transformations is ill-defined, the spectrum of the theory does not organize itself into representations of the color charge, and the relation between the gauge boson mass, the color charge, and color confinement encoded in the Gauss law is lost, making it possible for Higgs theories not to display color confinement. Under the hypothesis that the exact pole structure of the propagators of Higgs theories is the same as that obtained in perturbation theory, Kugo and Ojima verified \cite{KO78b,KO79b,KO79d,KO79a} that, in these theories, the unphysical Goldstone bosons associated with the spontaneous breaking in Eq.~\eqref{ugen} are indeed the unphysical degrees of freedom of the Higgs field, that the massive gauge bosons and the Higgs boson belong to the BRST cohomology, hence to the physical spectrum, and that the physical $S$-matrix of Higgs theories is unitary.

That discovery, however, also came as a curse, because it demanded an explanation of how, in QCD, color symmetry can be preserved when a gluon mass is generated. Without a theory of dynamical gluon mass generation at their disposal, Kugo and Ojima proposed a sufficient criterion for color confinement \cite{KO79d,KO79a}: since the right-hand side of Eq.~\eqref{ugen} is proportional to a coefficient $1+u$, color confinement would be realized in QCD if gluons acquire a mass, \textit{and} if the coefficient $1+u=0$.\\

The logic that underlies the KO criterion for confinement, insofar as it requires $1+u=0$, is flawed. To appreciate this, it suffices to make a few observations. First of all, $u=0$ in Abelian theories. If the KO criterion were meaningful in any general sense, Abelian Yang-Mills theories would be prevented from showing confinement. Yet various examples of confining Abelian theories are known -- the oldest of which being the Schwinger model\footnote{Two-dimensional QED with massless fermions.} \cite{Sch62a,Sch62b,LS71,KS74,CKS74,Ito75,Nak75,Nak76a,Nak77a,Nak77b,Nak77c}.

Second, as we will review in the present paper, in the linear covariant gauges of Quantum Electrodynamics (QED), the Noether charge associated with global gauge transformations is well known to be spontaneously broken in the KO framework; moreover, it is known \textit{not} to coincide with the electric charge operator. These facts signal that the spontaneous breaking of the Noether charge associated with global gauge transformations is a \textit{structural property} of linear covariant gauges -- independent of whether a gauge-boson mass is generated or not -- and that the actual charge operator, if it exists, must be obtained from the Noether charge mostly on a per-theory basis. Since Eq.~\eqref{ugen} features the Noether charge, that equation on its own \textit{cannot} provide a criterion for confinement based on the preservation of color symmetry.

Third, in non-Abelian theories, $1+u$ is equal both to a ratio between Lehmann-Symanzik-Zimmerman (LSZ) coefficients, and to the zero-momentum limit of a certain \textit{unrenormalized} Green's function. As such, its value is ill-defined\footnote{Indeed, historically, the only way to make sense of the KO criterion was to conjecture that, at least in the Landau gauge, $1+u=0$ is realized as $1+u=1/\infty$ \cite{Kug95,HvSA96,vSHA98,WA01,AvS01,Zwa02,FA02,FAR02,LvS02,FA03,ADFM04,PLNvS04,Fis06,FP06a,FP06b,DSVV09,Kon09,FMP09,VZ12}. This conjecture has long been proven false \cite{SS96,Cuc97,LSWP98a,LSWP98b,BBLW00,BBLW01,BHLP04,SIMS05,BBMM06,CM07,CM08a,CM08b,BIMS09,ISI09,BMM10,ABBC12,BLLM12,OS12,BBCO15,CDMO18,AN04,AP06,ABP08,NF00,FN03,FN06a,Ste06,ABP09,ABFP24}.} and, if renormalized, not Renormalization-Group invariant (RGI) \cite{Ste06,ABP09,ABFP24}. A criterion for confinement relying on an ill-defined quantity which does not satisfy the basic property of RG invariance is extremely questionable, if not plainly meaningless.

Another reason why the lack of RG invariance of $1+u$ is concerning is that said coefficient appears on the RHS of an equation whose left-hand side (LHS) is supposed to be RGI. If $Q^{a}$ were the actual color charge operator, as is implied by the KO criterion, $Q^{a}Q^{a}$ would be the Casimir operator, and a maximally commuting subalgebra of $\{Q^{a}\}_{a=1}^{8}$ would act as a Cartan subalgebra. In both cases, the eigenvalues of these operators would be fixed by the representation theory of SU(3), leaving no room for a RHS which is not RGI.

And indeed, while preparing the present paper, we made an unexpected discovery: despite having been introduced almost fifty years ago, Eq.~\eqref{ugen} has never been normalized correctly. Specifically, in \cite{KO79d,KO79a,NO90}, Kugo and Ojima systematically overlook a textbook step in the calculation of matrix elements of asymptotic states -- namely, dividing the matrix element by the appropriate LSZ coefficient\footnote{\label{fn:koerror}The error is introduced in Eqs. (2.24), (6.19), (4.4.1 -- 21) of \cite{KO79d,KO79a,NO90}, respectively. There the asymptotic field $\chi$ is defined without a multiplicative LSZ coefficient -- see Eqs.~(2.16a), (3.21a), (4.1.3 -- 22) of \cite{KO79d,KO79a,NO90}, respectively. In order to obtain a meaningful and finite matrix element, the equations as a whole must be divided manually by the LSZ coefficient of $\chi$. Those definitions should be contrasted with, say, (4.17) or (4.41) of \cite{KO79a}, in which the LSZ coefficients are introduced explicitly. Had the latter definitions been employed, the correct result would have been obtained automatically, without requiring manual normalization.}. Reinstating the missing coefficient on the RHS of Eq.~\eqref{ugen} solves the puzzle of the lack of RG invariance caused by $1+u$.

This error was never corrected in the subsequent literature. For instance, in a later, very famous revival of the KO criterion \cite{Kug95}, the omitted LSZ coefficient is still missing\footnote{See Eq.~(2.8), which uses the same conventions as \cite{KO79d,KO79a,NO90}.}, and the original Goldstone equation is not even reported. There, instead, in support of $1+u=0$, Kugo derives relations between the LSZ coefficients of local current operators with the aim of obtaining a well-defined color charge operator, a procedure which is quite different from the original arguments presented in \cite{KO79d,KO79a,NO90}. The latter maintained that, in the presence of a gluon mass, the condition $1+u=0$ would be \textit{sufficient} to obtain a well-defined color charge \textit{without} redefining the global gauge charge.

Ref.~\cite{Kug95} is problematic in its own right, and deserves further discussion. One of the LSZ coefficients in the main relation therein derived by Kugo -- namely, the coefficient $w$ in Eq.~(2.12) -- is identically equal to zero whenever gluons acquire a mass. This is obvious: as Kugo and Ojima themselves had been correctly maintaining all throughout \cite{KO79d,KO79a,NO90}, if the transverse gluon propagator has no massless pole, the gluon field-strength tensor $F_{\mu\nu}^{a}$ has no massless spectrum\footnote{As we will discuss later, non-Abelian Yang-Mills theories can technically escape this conclusion, but the ways in which they can do so in QCD are unappealing. In fact, KO never mention these explicitly in \cite{KO79d,KO79a,NO90}, and they do not affect the argument being made here.}. On the RHS of the second of Eq.~(2.12) is a massless asymptotic field $\beta^{a}$; therefore, in the absence of a massless transverse gluon pole, regardless of whether $1+u=0$ or not, the coefficient $w$ is \textit{zero}. A few lines below, Kugo proceeds to divide by $w$, thus obtaining a result which is invalid in the presence of a gluon mass. That the result is useless, in addition to being invalid, is also obvious. As maintained by Kugo and Ojima all throughout \cite{KO79d,KO79a,NO90}, the spatial integral of the spatial divergence of an operator with no massless spectrum is zero: when gluons are massive, regardless of whether $1+u=0$ or not, the term $\partial^{\mu}F_{\mu\nu}^{a}$ in Eq.~(2.14) contributes \textit{zero} to the equation. This -- namely, the vanishing of the chromoelectric flux at infinity -- is, in fact, the whole point of the KO criterion for confinement, insofar as it \textit{does not} require $1+u=0$. In the presence of a gluon mass, the ``well-defined'' color charge operator defined by Kugo in \cite{Kug95} is thus still very much ill-defined: it is the very same operator from which he started -- namely, the spontaneously broken Noether charge. No further arguments are provided in \cite{Kug95} as to why the criterion $1+u=0$ should hold, leaving us with nothing but a missing coefficient, and the incorrectly normalized Goldstone equation of \cite{KO79d,KO79a,NO90}.\\

Once all LSZ coefficients are accounted for properly, one finds that the correct form of the Goldstone equation originally derived by Kugo and Ojima reads
\begin{align}\label{symbreak0prel}
   &\bra{\Omega}Q^{a}\ket{\chi^{b}(p)}=i\,g^{-1}\,M\,(1+u)\ \hat{p}^{0}\,\delta^{ab}\,(2\pi)^{3}\delta^{(3)}({\bf p})\ ,
\end{align}
where $M=[-i\Delta(0)]^{-1/2}$ is the gluon mass, as defined from the zero-momentum limit of the transverse form factor $\Delta(p^{2})$ of its \textit{unrenormalized} propagator, $g$ is the bare strong coupling constant, and $\hat{p}=(|{\bf p}|,{\bf p})$ is a massless on-shell momentum\footnote{Following Peskin and Schroeder \cite{PS95}, in our conventions, the one-particle momentum-space integration measure is $\frac{1}{2\hat{p}^{0}}\,\frac{d^{3}{\bf p}}{(2\pi)^{3}}$, so the distribution $\hat{p}^{0}\,(2\pi)^{3}\delta^{(3)}({\bf p})$ does not vanish.}. As we anticipated earlier, and as we will see in the present paper, the combination of constants on the RHS of Eq.~\eqref{symbreak0prel} is finite and RGI, as it should. This combination is also \textit{not} equal to zero, as it is setting the RGI scale for gluon mass generation.

In the corrected form reported in Eq.~\eqref{symbreak0prel}, the Goldstone equation of Kugo and Ojima is telling a story which is very different from the one the KO scenario for confinement has historically seemed to support \cite{Kug95,HvSA96,vSHA98,WA01,AvS01,Zwa02,FA02,FAR02,LvS02,FA03,ADFM04,PLNvS04,Fis06,FP06a,FP06b,DSVV09,Kon09,FMP09,VZ12}. The story Eq.~\eqref{symbreak0prel} is telling is that of the Schwinger mechanism \cite{Sch62a,JJ73,EF74,Smi74,Cor81}, in the formulation which, in modern times, has been put forth by A. C. Aguilar, D. Binosi, M. N. Ferreira and J. Papavassiliou (ABFP) \cite{AP06,ABP08,AP09,AIMP12,IP13,ABP16,ABFP16,ABP17,BP18,Pap22,FP23,FP24,FP25}.

\newpage

The purpose of the present paper is to show that, within the KO framework \cite{KO78b,Oji78,KO78a,KO79b,KO79c,KO79d,KO79a,Kug78}, under customary assumptions, the quartet structure of the unphysical degrees of freedom of QCD in linear covariant gauges incorporates the ABFP Schwinger mechanism (SM) \cite{AP06,ABP08,AP09,AIMP12,IP13,ABP16,ABFP16,ABP17,BP18,Pap22,FP23,FP24,FP25} as the unique mechanism of dynamical gluon mass generation. When a gluon mass is generated, the massless colored scalar that triggers the ABFP Schwinger mechanism is the nonperturbative first parent of the elementary BRST quartet of the linear covariant gauges of QCD.

In this paper we will demonstrate that the Schwinger mechanism occurs with the spontaneous breaking of the global gauge charge $Q^{a}$, hence it is a Higgs mechanism. We must, however, be clear about what we mean by this statement. As we said before, and as we will show explicitly, the spontaneous breaking of the global gauge charge is a structural property of linear covariant gauges, occurring regardless of mass generation. When gauge bosons do not acquire a mass, the associated unphysical Goldstone boson is the gauge boson's massless longitudinal polarization \cite{KO79c}. When they acquire a mass, the unphysical Goldstone boson is a massless scalar particle \cite{Ito75,Nak75,Nak76a,KO79b}. In QCD, the SM is a Higgs mechanism in the sense that the gluon mass is generated via the ``eating'' of the scalar Goldstone boson associated with the spontaneous breaking of $Q^{a}$, without any colored field acquiring a vacuum expectation value, and without the color symmetry of the spectrum being violated. In the present paper we are going to show concretely how this happens.

The Goldstone bosons that trigger the SM in QCD are (obviously) not present in its Lagrangian. As we will see, they can be interpreted as massless colored scalar bound states, dynamically generated in the unphysical (gauge) sector of the spectrum. As such, they are not observable in the physical world, exactly like the Goldstone bosons of Higgs models are not. In a sense, they are artifacts of the linear covariant gauges, existing for the purpose of maintaining the internal consistency and symmetries of the theory in the presence of a gluon mass. Within the formalism of LCGs, they behave just like ordinary particles, except that they do not belong to the BRST cohomology, and that they are associated with states whose norm is not positive definite\footnote{Additionally, except for the Landau gauge, their asymptotic fields are dipole fields, as is customary for Goldstone bosons in linear covariant gauges \cite{Nak75}.}.

In the present paper we are not going to address the question of how the bound states are formed. The topic of bound-state formation in the SM lies beyond the author's area of expertise and is currently an active field of research. We will, however, mention that -- like ordinary bound states -- the formation of the ABFP bound states is governed by Bethe-Salpeter equations (BSE) \cite{AIMP12,IP13,ABP17,BP18,Pap22,FP24} which must be solved numerically. While we will not derive these BSEs explicitly, since both the KO framework and the ABFP SM are set in the linear covariant gauges of QCD, and satisfy the same Schwinger-Dyson equations (SDE), once we show that the first parent of the elementary BRST quartet possesses the same quantum numbers and Bethe-Salpeter amplitudes as the ABFP bound state, and that it produces massless longitudinal poles -- so-called ``Schwinger poles'' -- in the same amputated vertices and with the same residues as the ABFP bound states, the results of Papavassiliou and collaborators can be imported into the KO framework.

What we are going to show is that \textit{if} a gluon mass is dynamically generated, \textit{then}, in the KO framework, the ABFP bound state exists, and it triggers the Schwinger mechanism. In this paper, rather than providing a preliminary account of the SM, we will show how -- equation by equation -- the entire mechanism emerges naturally from the quartet structure and dynamics of the unphysical degrees of freedom of QCD.

In the process of deriving the mechanism, we will clarify the actual meaning of the Kugo-Ojima constant $1+u$. As it turns out, $1+u$ \textit{must} be ill-defined and non-RGI, because its role in the theory is that of turning the ill-defined and non-RGI ratio $M/g$ in Eq.~\eqref{symbreak0prel} into the finite and RGI quantity $\widehat{M}/g=M(1+u)/g$. This quantity enters the Goldstone equation
\begin{equation}\label{curreq0}
   \bra{\Omega}J_{\mu}^{a}(x)\ket{\chi^{b}(p)}=i\widehat{M}/g\,p_{\mu}\,\delta^{ab}\,e^{-ipx}
\end{equation}
which is presented here for the first time. In Eq.~\eqref{curreq0}, $J_{\mu}^{a}(x)$ is the (Noether) color current, $\chi^{a}$ is the ABFP bound state and $\widehat{M}=M(1+u)$ is the (square root of the inverse of the) zero-momentum limit of the \textit{unrenormalized} background-background gluon propagator evaluated in the Pinch Technique--Background-Field Method (PT-BFM) scheme \cite{Abb80,Cor81,CP89,Pil97,GHS99,GHS01,BP02b,BP02a,BP04,BP09,CPB11,BQ13}.

The historical development of the PT-BFM goes hand in hand with the study of dynamical gluon mass generation, having been invented by J. M. Cornwall for the specific purpose of describing the gluon mass \cite{Cor81}. At no point in our study were we aiming to compute quantities defined in the PT-BFM. That in Eq.~\eqref{curreq0} is quite simply the result one gets from a direct calculation. The way by which the result is obtained will allow us to advance a new interpretation of the PT-BFM.

In QCD, as is evident from Eq.~\eqref{curreq0}, $\widehat{M}/g$ plays the same role as the pion decay constant in low-energy nuclear interactions, and as the Higgs field's vacuum expectation value in weak interactions. If, in addition to being RGI, $\widehat{M}/g$ were also shown to be gauge invariant\footnote{Currently, we have no indication that it is, but also no proof to the contrary.}, then that constant would be the obvious candidate for the fundamental nonperturbative scale of the strong interactions, emerging from the quantum dynamics itself of the theory.

In this paper we are going to show that, when a gluon mass is generated in the KO framework, the global gauge (Noether) charge $Q^{a}$ does not merely generate color transformations. Asymptotically\footnote{I.e., when it acts on the asymptotic (in/out) states of the theory.}, it generates the constant-parameter (``global'') limit of residual gauge transformations. The generators of these transformations separate into two independent asymptotic charges. The first of these is spontaneously broken for structural reasons which are easy to understand, and which we will discuss; this breaking causes the behavior displayed by Eqs.~\eqref{symbreak0prel} and \eqref{curreq0}, and it has the ABFP bound state as its associated Goldstone boson. The broken symmetry is protected by BRST symmetry, which -- as we will see -- quite literally replaces it. The second term is the generator of asymptotic color transformations. The latter is well-defined, unbroken, and BRST-exact. Color confinement in the original sense of Kugo and Ojima -- namely, in the sense of the absence of colored asymptotic states from the physical spectrum -- then follows, \textit{without} the condition $1+u=0$ being realized.

In the Kugo-Ojima framework, the mechanism of color confinement is the dynamical mechanism which generates a gluon mass while preserving the color symmetry of the spectrum \cite{KO79a}. From what we said, it follows that this mechanism is the ABFP Schwinger mechanism.

With the elementary BRST quartet of the linear covariant gauges of QCD having found its nonperturbative completion in the Goldstone boson of the Schwinger mechanism, and with color confined in the original sense of Kugo-Ojima, all colored asymptotic states arrange themselves into unphysical BRST quartets decoupled from the physical BRST cohomology. The nonperturbative unitarity of the physical $S$-matrix of the theory $S_{\text{phys}}$ then follows immediately by virtue of BRST symmetry \cite{KO78b,KO78a,Nak79,NO90}. The Schwinger mechanism formulated in the KO framework thus constitutes the first known example of a nonperturbative, confining, unitary, infrared completion of QCD. As we said, within the framework, in the presence of dynamical gluon mass generation, and under customary assumptions, this completion is also unique.

Above, the unitarity of the physical $S$-matrix must be understood in the sense that $S_{\text{phys}}^{\dagger}S_{\text{phys}}=S_{\text{phys}}S_{\text{phys}}^{\dagger}=1$. By ``unitarity'', however, one usually also means that the inner product of the physical Hilbert space $\mc{H}_{\text{phys}}$ is positive definite. In perturbation theory, one can check the positive definiteness of physical states field by field; in an intrinsically nonperturbative theory like QCD, with stable bound states formed in composite channels, this cannot be done in a systematic way \cite{Nak79}. While proving unitarity in the sense of the positive definiteness of the inner product of the BRST cohomology $\mc{H}_{\text{phys}}$ lies beyond the scope of the present paper, we will discuss the additional assumptions under which this property holds.

The techniques introduced and employed in this work can be readily generalized to the nonperturbative study of gauge-boson mass generation in linear covariant gauges of Yang-Mills theories beyond QCD. Our main finding, in this respect, is that, in the KO framework, the ABFP Schwinger mechanism, the (Abelian) Stueckelberg mechanism, the Schwinger mechanism of the Schwinger model, and the Higgs mechanism of Higgs models, can all be described using the same language. Specifically, in each of these theories, the first parent of the elementary BRST quartet produces massless longitudinal ``Schwinger poles'' in the amputated vertices and/or in the gauge-boson polarization tensor, which qualifies them all as Schwinger mechanisms. Since all of them occur through the ``eating'' of the scalar Goldstone boson associated with the spontaneous breaking of the global gauge charge, they all qualify as Higgs mechanisms as well. Our results thus pave the way for a unified theory of gauge-boson mass generation in the linear covariant gauges of Yang-Mills theory.

The present paper is organized as follows. In Sec.~\ref{sec:review} we review the quantization of Yang-Mills theories in LCGs and the Kugo-Ojima framework. In Sec.~\ref{sec:ssblcg} we discuss the flaws in the analysis of spontaneous symmetry breaking presented by Kugo and Ojima in \cite{KO79d,KO79a}, we provide our preliminary analysis, and we introduce our first assumptions on the behavior of QCD. In Sec.~\ref{sec:mlessspect} we study the massless component of the gluon spectrum in dynamically massive QCD, and we show that the ABFP Schwinger mechanism is active under customary KO assumptions. In Sec.~\ref{sec:symbreak}, building on the results of the previous section, we complete our analysis of spontaneous symmetry breaking and address color confinement. In Sec.~\ref{sec:discussion} we discuss our results and present our conclusions. Throughout this paper, we focus on the gauge sector of QCD, introducing quarks only when necessary.

\section{Yang-Mills theory in linear covariant gauges}\label{sec:review}

In this section, we set the notation and we review the quantization of Yang-Mills theories in LCGs and the Kugo-Ojima framework. In Sec.~\ref{sec:pymt} we review the canonical quantization of LCGs, their symmetries, and the Schwinger-Dyson equations. In Sec.~\ref{sec:framework} we review indefinite-metric Hilbert spaces, the BRST cohomology, the postulate of asymptotic completeness, and the structure of the elementary BRST quartet. In Sec.~\ref{sec:spectr} we discuss spectral representations and we review the quartet mechanism and the proof of the unitarity of the physical $S$-matrix in the KO framework.

\newpage

\subsection{Canonical quantization of Yang-Mills theory in LCGs, symmetries, and the Schwinger-Dyson equations}\label{sec:pymt}

The linear covariant gauges of Yang-Mills theory are defined by the Faddeev-Popov (FP) action \cite{FP67a,FP67b,NO90}
\begin{align}
   \mc{S}_{\text{FP}}&=\int d^{4}x\ \Big\{-\frac{1}{4}\,F_{\mu\nu}^{a}F^{a\mu\nu}-i\partial^{\mu}\cbar^{a}D_{\mu}c^{a}+\\
   \notag&\qquad\qquad\qquad\ \ +\frac{\xi}{2}\,B^{a}B^{a}+B^{a}\partial^{\mu}A_{\mu}^{a}\Big\}\ ,
\end{align}
where $A_{\mu}^{a}$ is the gluon field, $B^{a}$ is the Nakanishi-Lautrup (NL) field \cite{Nak66,Lau67,Nak73}, $c^{a}$ and $\cbar^{a}$ are the anticommuting Faddeev-Popov ghost and antighost fields \cite{FP67a,FP67b}, with Hermiticity assignment \cite{KO78b}
\begin{equation}
   (c^{a})^{\dagger}=c^{a}\ ,\qquad(\cbar^{a})^{\dagger}=\cbar^{a}\ , 
\end{equation}
$D_{\mu}^{ab}=\delta^{ab}\partial_{\mu}-gf^{abc}A_{\mu}^{c}$ is the covariant derivative in the adjoint representation, $F_{\mu\nu}^{a}=\partial_{\mu}A_{\nu}^{a}-\partial_{\nu}A_{\mu}^{a}+gf^{abc}A_{\mu}^{b}A_{\nu}^{c}$ is the gluon field-strength tensor, $\xi$ is the gauge parameter, $g$ is the strong coupling constant, and $f^{abc}$ are the structure constants of the gauge group. The fields, gauge parameter, and coupling are bare (unrenormalized) quantities.

The theory is quantized canonically \cite{KO78a,KO79c,NO90} by assigning (anti)commutation relations $[\Pi,\mc{F}]_{\mp}=-i$ to the fields $\mc{F}$ and their conjugate momenta $\Pi$. The conjugate momentum $\Pi^{a\mu}$ of the gluon field is
\begin{equation}\label{conjmom}
   \Pi^{a\mu}=F^{a\mu 0}+\eta^{\mu0}B^{a}\ ,
\end{equation}
where $\eta=\text{diag}(+1,-1,-1,-1)$ is the Minkowski metric, whereas the conjugate momenta $\Pi^{a}$ and $\overline{\Pi}^{a}$ of the ghost and antighost fields are, respectively \cite{KO78a,NO90},
\begin{equation}
   \Pi^{a}=i\partial_{0}\cbar^{a}\ ,\qquad \overline{\Pi}^{a}=-iD_{0}c^{a}\ .
\end{equation}
The conjugate momentum $\Pi^{a}_{B}$ of the NL field is zero,
\begin{equation}
   \Pi^{a}_{B}=0\ .
\end{equation}

The equations
\begin{equation}
   \Pi^{a0}=B^{a}\ ,\qquad \Pi^{a}_{B}=0
\end{equation}
form a system of second-class constraints \cite{Dir64} which is solved by ignoring $\Pi^{a}_{B}$ and by treating $B^{a}$ as the conjugate momentum of the temporal component $A^{a}_{0}$ of the gluon field. The equal-time canonical (anti)commutation relations assigned to the elementary fields of Yang-Mills theory then read
\begin{align}\label{cancomm}
   \notag[B^{a}({\bf x},t),A^{b}_{0}({\bf y},t)]&=-i\delta^{ab}\delta^{(3)}({\bf x}-{\bf y})\ ,\\
   [F^{ai0}({\bf x},t),A^{b}_{j}({\bf y},t)]&=-i\delta^{ab}\delta^{i}_{j}\delta^{(3)}({\bf x}-{\bf y})\ ,\\
   \notag\{\partial_{0}\cbar^{a}({\bf x},t),c^{b}({\bf y},t)\}&=-\delta^{ab}\delta^{(3)}({\bf x}-{\bf y})\ ,\\
   \notag\{D_{0}c^{a}({\bf x},t),\cbar^{b}({\bf y},t)\}&=\delta^{ab}\delta^{(3)}({\bf x}-{\bf y})\ ,
\end{align}
where $i$ and $j$ are spatial indices.

The Hamiltonian operator $H$ derived from the action $\mc{S}_{\text{FP}}$ is
\begin{align}
   \notag H&=\int d^{3}{\bf x}\ \bigg\{-\frac{1}{2}\,\Pi^{ai}\Pi^{ai}-D_{i}\Pi^{ai}A^{a}_{0}+\frac{1}{4}\,F^{aij}F^{a}_{ij}+\\
   \notag&\qquad\qquad\ +i\overline{\Pi}^{a}\Pi^{a}+gf^{abc}\,\Pi^{a}A_{0}^{b}c^{c}+i\partial^{i}\cbar^{a}D_{i}c^{a}+\\
   &\qquad\qquad\ -\frac{\xi}{2}\,B^{a}B^{a}-B^{a}\partial^{i}A_{i}^{a}\bigg\}
\end{align}
and it is Hermitian, $H^{\dagger}=H$. The field equations obtained from $H$ are
\begin{align}\label{feqs}
   \notag\partial^{\nu}F_{\nu\mu}^{a}+gf^{abc}A^{b\,\nu}F^{c}_{\nu\mu}+igf^{abc}\partial_{\mu}\cbar^{b}c^{c}-\partial_{\mu}B^{a}=0\ ,\\
   \partial^{\mu}D_{\mu}c^{a}=0\ ,\quad D^{\mu}\partial_{\mu}\cbar^{a}=0\ ,\quad\partial^{\mu}A_{\mu}^{a}+\xi B^{a}=0\ .
\end{align}

The first of Eq.~\eqref{feqs} are the Maxwell-Yang-Mills equations (MYME). In the absence of gauge fixing, their temporal component $\mu=0$ is a constraint equation\footnote{I.e. an equation enforcing a relation between the fields and their conjugate momenta, rather than a dynamical equation describing their evolution in time.} known as the ``Gauss constraint'' \cite{HT92}, or Gauss law. Thanks to the term $\partial_{0}B^{a}=\partial_{0}\Pi^{a0}$, the linear covariant gauges promote the Gauss constraint to a dynamical equation.

The trade-off for having a dynamical Gauss law is the introduction of unphysical dynamical degrees of freedom into Yang-Mills theory. For each color index, three of these -- the FP ghosts and NL field -- are introduced explicitly in the FP Lagrangian. A fourth one is introduced indirectly. In the absence of gauge fixing, the MYME are invariant under (local) gauge transformations, causing one gluonic degree of freedom to decouple from the equations. Once the gauge is fixed, gauge invariance is lost, and that degree of freedom becomes an active participant in the dynamics of the theory.

The second and third of Eq.~\eqref{feqs} govern the dynamics of the FP ghosts, whereas the dynamical equation of the NL field is given by the divergence of the MYME,
\begin{equation}
   \partial^{2}B^{a}-gf^{abc}\partial^{\mu}(A^{b\,\nu}F^{c}_{\nu\mu}+i\partial_{\mu}\cbar^{b}c^{c})=0\ ,
\end{equation}
where $\partial^{2}=\partial^{\mu}\partial_{\mu}$. The last of Eq.~\eqref{feqs} -- the ``gauge-fixing condition'' -- is a constraint relating the divergence of the gluon field to the NL field.

In addition to Poincar\'{e} transformations generated by the momentum operator $P_{\mu}$ and by the Lorentz generators $M_{\mu\nu}$ \cite{NO90}, the Faddeev-Popov action and the field equations of LCGs are covariant under three internal symmetries \cite{NO90}: global gauge (color) transformations, ghost scale transformations, and BRST transformations. Global gauge transformations are generated by the global gauge (color) charge $Q^{a}=\int d^{3}{\bf x}\ J_{0}^{a}$, where $J_{\mu}^{a}$ is the conserved global gauge (color) Noether current
\begin{equation}\label{colorcurr}
   J^{a}_{\mu}=f^{abc}\left(A^{b\nu}F_{\nu\mu}^{c}+A^{b}_{\mu}B^{c}+i\partial_{\mu}\cbar^{b}c^{c}-i\cbar^{b}D_{\mu}c^{c}\right)\ .
\end{equation}
The corresponding (infinitesimal) transformations of the elementary Yang-Mills fields read\footnote{These commutators can be rigorously evaluated by smearing, i.e. by first computing the equal-time commutator of their operand with the current $J_{0}^{a}$, and then integrating in three-space.}
\begin{align}\label{color}
   [iQ^{a},A_{\mu}^{b}]&=-f^{abc}A_{\mu}^{c}\ ,\quad [iQ^{a},B^{b}]=-f^{abc}B^{c}\ ,\\
   \notag [iQ^{a},c^{b}]&=-f^{abc}c^{c}\ ,\quad\ \ [iQ^{a},\cbar^{b}]=-f^{abc}\cbar^{c}\ . 
\end{align}
Ghost scale transformations are generated by the ghost charge $Q_{C}=\int d^{3}{\bf x}\ J_{C}^{0}$, where $J_{C}^{\mu}$ is the conserved Noether ghost current
\begin{equation}
   J_{C}^{\mu}=i\cbar^{a}D^{\mu}c^{a}-i\partial^{\mu}\cbar^{a}c^{a}\ .
\end{equation}
These only act on the ghost fields:
\begin{equation}\label{ghost}
   [iQ_{C},c^{a}]=c^{a}\ ,\qquad [iQ_{C},\cbar^{a}]=-\cbar^{a}\ .
\end{equation}

BRST transformations are generated by the BRST charge $Q_{B}=\int d^{3}{\bf x}\ J_{B}^{0}$, where $J_{B}^{\mu}$ is the conserved Noether BRST current
\begin{equation}\label{brstcurr}
   J_{B}^{\mu}=B^{a}D^{\mu}c^{a}-\partial^{\mu}B^{a}c^{a}+\frac{ig}{2}\,f^{abc}\,\partial^{\mu}\cbar^{a}c^{b}c^{c}\ .
\end{equation}
These act on the elementary fields as
\begin{align}\label{brst}
   \notag [i Q_{B},A_{\mu}^{a}]&=D_{\mu}c^{a}\ ,\qquad\{iQ_{B},c^{a}\}=-\frac{g}{2}f^{abc}c^{b}c^{c}\ ,\\
   \{iQ_{B},\cbar^{a}\}&=iB^{a}\ ,\qquad\ \ [iQ_{B},B^{a}]=0\ .
\end{align}
Recalling that, at the classical level, infinitesimal gauge transformations act on the gluon field as
\begin{equation}
   \delta A_{\mu}^{a}=g^{-1}D_{\mu}\Lambda^{a}=g^{-1}\partial_{\mu}\Lambda^{a}+f^{abc}A_{\mu}^{b}\Lambda^{c}\ ,
\end{equation}
where $\Lambda^{a}$ is an arbitrary function, we see that the first of Eq.~\eqref{brst} has the form of an infinitesimal gauge transformation parametrized by the ghost field $c^{a}$. Since the ghost field satisfies the Faddeev-Popov-Gribov equation \cite{Gri78} $\partial^{\mu}D_{\mu}c^{a}=0$\footnote{Ghosts are, quite literally, quantized zero modes of the FP operator $\partial^{\mu}D_{\mu}$, whereas antighosts are quantized zero modes of its adjoint $D_{\mu}\partial^{\mu}$.}, BRST transformations preserve the gauge-fixing condition\footnote{In an operator formalism, quantum fields are always on the shell of the field equations: at the operatorial level, there is no difference between ``on-shell'' and ``off-shell'' BRST transformations.},
\begin{equation}
   [iQ_{B},\partial^{\mu}A_{\mu}^{a}+\xi B^{a}]=\partial^{\mu}D_{\mu}c^{a}=0\ .
\end{equation}
Therefore, BRST transformations are \textit{residual} gauge transformations parametrized by ghosts, acting on the FP ghosts and NL field in addition to the gluon field.

$Q_{B}$ and $Q_{C}$ are Hermitian operators that commute with the generators $P_{\mu},M_{\mu\nu}$ of the Poincar\'{e} group \cite{NO90}. As charges, they are assumed to annihilate the vacuum,
\begin{equation}\label{brstvac}
   Q_{B}\ket{\Omega}=0\ ,\qquad Q_{C}\ket{\Omega}=0\ ,
\end{equation}
and they generate the so-called ``BRST algebra'' \cite{NO90},
\begin{align}
   \notag\{Q_{B},Q_{B}\}&=2Q_{B}^{2}=0\ ,\quad [iQ_{C},Q_{B}]=Q_{B}\ ,\\
   &\qquad\ [Q_{C},Q_{C}]=0\ .
\end{align}
The first of these expresses the nilpotency of the BRST charge.

The global gauge charge $Q^{a}$ is also Hermitian, and it commutes with the Poincar\'{e} generators and with the BRST algebra\footnote{Here, again, commutators should be evaluated by smearing.}:
\begin{equation}\label{chargealgebra}
   [Q^{a},Q_{B}]=0\ ,\qquad [Q^{a},Q_{C}]=0\ .
\end{equation}
While Eq.~\eqref{color} appears to imply that $Q^{a}$ generates the color algebra,
\begin{equation}\label{coloralgebra}
   [Q^{a},Q^{b}]=if^{abc}Q^{c}\ ,
\end{equation}
in the KO framework -- as we will discuss at length in Secs.~\ref{sec:ssblcg} and \ref{sec:symbreak} -- $Q^{a}$, as well as $[Q^{a},Q^{b}]$, are ill-defined as operators. It follows, first of all, that $Q^{a}$ does not annihilate the vacuum, $Q^{a}\ket{\Omega}\neq 0$, and second, that Eq.~\eqref{coloralgebra} does not hold.

Finally, using Eq.~\eqref{brst}, it is easy to put the MYME in the form \cite{Oji78}
\begin{equation}\label{feqojima}
   \partial^{\nu}F^{a}_{\nu\mu}+gJ_{\mu}^{a}-\{Q_{B},D_{\mu}\cbar^{a}\}=0\ .
\end{equation}
By integrating the temporal component of this equation in three-dimensional space, and by defining $\mc{E}^{ai}=F^{ai0}$, one obtains Eq.~\eqref{YMcharge}.
\\

From Eqs.~\eqref{cancomm} and \eqref{feqs} one derives the Schwinger-Dyson equations of the theory. Let us recall how this is done using the gluon propagator as an example.

Let $\ket{\Omega}$ be the vacuum state of the theory, let $T\{\cdots\}$ denote a time-ordered product, and let $\Delta^{ab}_{\mu\nu}(p)$ denote the (unrenormalized) momentum-space dressed gluon propagator,
\begin{equation}
   \Delta_{\mu\nu}^{ab}(p)=\int d^{4}x\ e^{ipx}\ \bra{\Omega}T\{A_{\mu}^{a}(x)A_{\nu}^{b}(0)\}\ket{\Omega}\ .
\end{equation}
Moreover, let $K_{\mu}^{a}$ be the kernel operator defined as
\begin{equation}\label{kernelop}
   K_{\mu}^{a}=gf^{abc}\left[\partial^{\nu}(A_{\nu}^{b}A_{\mu}^{c})+A^{b\,\nu}F^{c}_{\nu\mu}+i\partial_{\mu}\cbar^{b}c^{c}\right]\ ,
\end{equation}
so that the MYME read
\begin{equation}
   \partial^{\sigma}(\partial_{\sigma}A_{\mu}^{a}-\partial_{\mu}A_{\sigma}^{a})-\partial_{\mu}B^{a}+K_{\mu}^{a}=0\ .
\end{equation}
Using $\langle \cdots\rangle$ as a shorthand for $\bra{\Omega}\cdots\ket{\Omega}$, we evaluate the vacuum expectation value of the time-ordered product between the MYME and the gluon field,
\begin{equation}
   \langle T\{[\partial^{\sigma}(\partial_{\sigma}A_{\mu}^{a}-\partial_{\mu}A_{\sigma}^{a})-\partial_{\mu}B^{a}+K_{\mu}^{a}]A_{\nu}^{b}\}\rangle = 0\ .
\end{equation}
By extracting the derivatives from the time-ordered product, using $B^{a}=-\xi^{-1}\,\partial^{\mu}A_{\mu}^{a}$, and switching to momentum space, we obtain the equation
\begin{equation}\label{deq664}
   i\Delta_{(0)}^{-1}(p)\cdot \Delta(p)+\langle T\{K A\}\rangle_{\text{F.T.}}(p)=i\ ,
\end{equation}
where Lorentz and color indices are suppressed, ``F.T.'' denotes a Fourier transform, and $\Delta_{(0)}^{-1}(p)^{ab}_{\mu\nu}$ is the tree-level inverse gluon propagator,
\begin{equation}
   \Delta_{(0)}^{-1}(p)^{ab}_{\mu\nu}=i\delta^{ab}\,\left[\eta_{\mu\nu}\,p^{2}-(1-\xi^{-1})\,p_{\mu}p_{\nu}\right]\ .
\end{equation}
The $i$ on the RHS of Eq.~\eqref{deq664} originates from equal-time commutators which appear in the equation while extracting the derivatives from the time-ordered product. Denoting the gluon polarization tensor with $\Pi_{\mu\nu}^{ab}(p)$,
\begin{equation}\label{glupoldef}
   \Pi_{\mu\nu}^{ab}(p)=\Delta_{(0)}^{-1}(p)^{ab}_{\mu\nu}-\Delta^{-1}(p)^{ab}_{\mu\nu}\ ,
\end{equation}
we finally obtain the gluon propagator's SDE in the form
\begin{equation}\label{gpsde}
   \Pi_{\mu\nu}^{ab}(p)=i\,\left[\langle T\{K A\}\rangle_{\text{F.T.}}(p)\cdot \Delta^{-1}(p)\right]_{\mu\nu}^{ab}\ ,
\end{equation}
where, on the RHS, the (tensorial) inverse propagator is contracted with the indices of the gluon field $A$ and amputates the external gluon leg of the Green's function $\langle T\{K A\}\rangle$.

\begin{figure}[h]
    \includegraphics[width=0.35\textwidth]{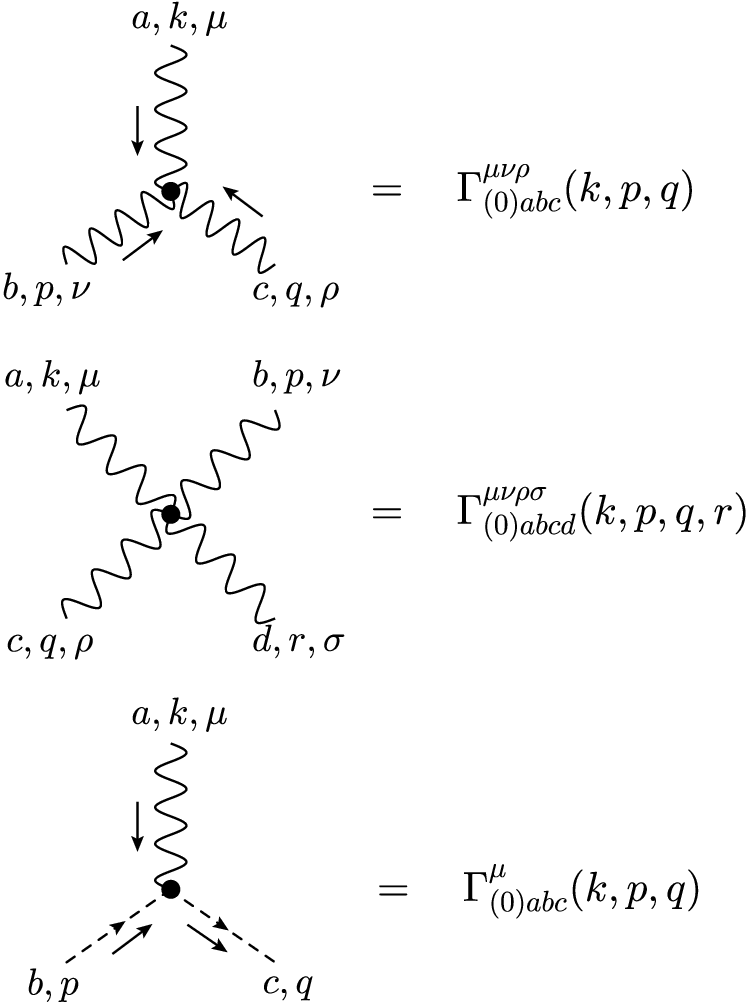}
    \caption{Tree-level vertices of Yang-Mills theory in LCGs. Arrows indicate the direction of momentum flow.}
    \label{QCDVertices}
\end{figure}

In Sec.~\ref{sec:goldbound} we will report both the explicit analytic form and the ordinary diagrammatic representation of the SDE in Eq.~\eqref{gpsde}. In this section, we limit ourselves to listing the tree-level vertices that figure in it, which can be read off directly from the kernel operator $K_{\mu}^{a}$ after symmetrization. These are the three-gluon vertex $\Gamma^{\mu\nu\rho}_{(0)abc}$, four-gluon vertex $\Gamma^{\mu\nu\rho\sigma}_{(0)abcd}$, and ghost-antighost-gluon vertex $\Gamma_{(0)abc}^{\mu}$, respectively,
\begin{align}\label{emvertices}
   &\Gamma^{\mu\nu\rho}_{(0)abc}(k,p,q)=\\
   \notag&\quad=f^{abc}[\eta^{\mu\nu}(k-p)^{\rho}+\eta^{\nu\rho}(p-q)^{\mu}+\eta^{\rho\mu}(q-k)^{\nu}]\ ,\\
   \notag&\Gamma^{\mu\nu\rho\sigma}_{(0)abcd}(k,p,q,r)=-i[f_{abe}f_{cde}(\eta^{\mu\rho}\eta^{\nu\sigma}-\eta^{\mu\sigma}\eta^{\nu\rho})+\\
   \notag&\qquad\qquad\qquad\qquad\ \ +f_{ace}f_{bde}\left(\eta^{\mu\nu}\eta^{\rho\sigma}-\eta^{\mu\sigma}\eta^{\nu\rho}\right)+\\
   \notag&\qquad\qquad\qquad\qquad\ \ +f_{ade}f_{bce}\left(\eta^{\mu\nu}\eta^{\rho\sigma}-\eta^{\mu\rho}\eta^{\nu\sigma}\right)]\ ,\\
   \notag&\Gamma_{(0)abc}^{\mu}(k,p,q)=if_{abc}\,q^{\mu}\ ,
\end{align}
see Fig.~\ref{QCDVertices}. Here and in what follows, we adopt the convention that vertices carry no prefactor of the coupling constant $g$.

\subsection{Indefinite-metric Hilbert spaces, the BRST cohomology, asymptotic completeness, and the elementary BRST quartet}\label{sec:framework}

As is well known \cite{Bleu50,Gup50,Nak72,NO90}, the linear covariant gauges of Yang-Mills theory are quantized on indefinite-metric Hilbert spaces. In this section, we begin with a concise overview of some of the key features of these spaces, and then we move on to discuss the KO quantization framework. The reader is referred to \cite{Nak72,NO90} for a comprehensive treatment of these topics.\\

Indefinite-metric Hilbert spaces are inner-product vector spaces that satisfy all the defining properties of Hilbert spaces, except for the positive definiteness of the inner product. The latter, denoted by $\langle \cdot | \cdot \rangle$, is only required to be non-degenerate, so that
\begin{equation}
   \braket{\phi|\psi}=0\ \quad \forall\, \ket{\phi}
\end{equation}
implies $\ket{\psi}=0$. It follows that, for each state $\ket{\psi}\neq 0$, there must exist at least one state $\ket{\phi}$ with nonzero inner product with $\ket{\psi}$, $\braket{\phi|\psi}\neq 0$.

Given a state $\ket{\psi}$, with an abuse of language, the number $\braket{\psi|\psi}$ is called its norm. States $\ket{\psi}$ such that $\braket{\psi|\psi}=0$ (resp. $\braket{\psi|\psi}<0$) are called zero-norm (resp. negative-norm) states.

Given an operator $A$, an operator $A^{\dagger}$ such that
\begin{equation}
   \bra{\phi}A^{\dagger}\ket{\psi}=\bra{\psi}A\ket{\phi}^{\star}\ \quad \forall\,\ket{\phi},\ket{\psi}\ ,
\end{equation}
where the star denotes complex conjugation, is called the adjoint of $A$. An operator such that $A=A^{\dagger}$ is said to be Hermitian.

On indefinite-metric Hilbert spaces, Hermitian operators can have both real and complex-conjugate eigenvalues. To see this, let $\ket{a_{1}}$ (resp. $\ket{a_{2}}$) be an eigenvector of a Hermitian operator $A$ with eigenvalue $a_{1}$ (resp. $a_{2}$). Then
\begin{align}
   \bra{a_{1}}A\ket{a_{2}}=a_{2}\braket{a_{1}|a_{2}}\ ,\ \bra{a_{1}}A\ket{a_{2}}=a_{1}^{\star}\braket{a_{1}|a_{2}}\ ,
\end{align}
hence
\begin{equation}
   (a_{1}^{\star}-a_{2})\braket{a_{1}|a_{2}}=0\ .
\end{equation}
It follows that either $\ket{a_{1}}$ and $\ket{a_{2}}$ are orthogonal to each other, $\braket{a_{1}|a_{2}}=0$, or $\ket{a_{1}}$ and $\ket{a_{2}}$ have complex-conjugate eigenvalues, $a_{2}=a_{1}^{\star}$. Setting $\ket{a_{1}}=\ket{a_{2}}=\ket{a}$,
\begin{equation}
   (a^{\star}-a)\braket{a|a}=0\ ,
\end{equation}
we see that non-real eigenvalues correspond to zero-norm eigenstates.

In addition to complex-conjugate eigenvalues, on indefinite-metric Hilbert spaces, Hermitian operators can have so-called ``generalized eigenstates'' (or ``generalized eigenvectors'') \cite{Nak72}. Given an operator $A$, a state $\ket{a}$ is said to be a generalized eigenstate of $A$ of order $n$ with ``generalized eigenvalue'' $a$ if
\begin{equation}
   (A-a)^{n}\ket{a}=0\ ,\qquad (A-a)^{n-1}\ket{a}\neq0
\end{equation}
for some integer $n\geq 1$. Generalized eigenstates of order $n=1$ are ordinary eigenstates. Generalized eigenstates of order $n=2$, i.e. states $\ket{a}$ such that
\begin{equation}
   (A-a)^{2}\ket{a}=0\ ,\qquad (A-a)\ket{a}\neq0\ ,
\end{equation}
are called ``dipole states'' or ``dipoles'' (with respect to the operator $A$). Generalized eigenstates of order $n\geq 2$ are called ``multipole states''.

If $\ket{a}$ is a generalized eigenstate of order $n$ of an operator $A$ with generalized eigenvalue $a$, then all states of the form $(A-a)^{k}\ket{a}$ for $k=0,\dots,n-1$ are generalized eigenstates of $A$ of order $n-k$ with the same generalized eigenvalue. If $A$ is Hermitian and $a$ is real, then these states have zero norm for every $k$ such that $2k\geq n$,
\begin{equation}
   \bra{a}(A-a)^{2k}\ket{a}=0\ .
\end{equation}
Regardless of their order, it can be shown \cite{Nak72} that if $\ket{a_{1}}$ and $\ket{a_{2}}$ are generalized eigenstates of a Hermitian operator $A$ with generalized eigenvalues $a_{1}$ and $a_{2}$, and $a_{1}^{\star}\neq a_{2}$, then $\ket{a_{1}}$ and $\ket{a_{2}}$ are orthogonal to each other, $\braket{a_{1}|a_{2}}=0$. In particular, non-real generalized eigenvalues correspond to zero-norm generalized eigenstates.

In what follows, with an abuse of language, we shall refer to indefinite-metric Hilbert spaces simply as ``Hilbert spaces''.\\

In Yang-Mills theories, the ghost charge $Q_{C}$ and the mass (squared) operator $P^{2}$ provide well-known examples of Hermitian operators with, respectively, non-real, complex-conjugate eigenvalues, and generalized eigenstates.

Consider the state $\ket{c^{a}(x)}=c^{a}(x)\ket{\Omega}$, where $c^{a}(x)$ is the ghost field and $\ket{\Omega}$ is the vacuum state of the theory. Using Eqs.~\eqref{ghost} and \eqref{brstvac}, it is easy to see that $\ket{c^{a}(x)}$ is an eigenvector of $Q_{C}$ with eigenvalue $-i$:
\begin{equation}
   Q_{C}\,c^{a}(x)\ket{\Omega}=[Q_{C},c^{a}(x)]\ket{\Omega}=-ic^{a}(x)\ket{\Omega}\ .
\end{equation}
Similarly, $\ket{\cbar^{a}(x)}=\cbar^{a}(x)\ket{\Omega}$, with $\cbar^{a}(x)$ the antighost field, is an eigenvector of $Q_{C}$ with eigenvalue $+i$. Since $[c^{a}(x)]^{2}=0$, $[\cbar^{a}(x)]^{2}=0$ by the anticommuting nature of the FP ghosts, and since $c^{a}$ and $\cbar^{a}$ are Hermitian operators, these eigenvectors have zero norm, e.g.
\begin{equation}
   \braket{c^{a}(x)|c^{a}(x)}=\bra{\Omega}[c^{a}(x)]^{2}\ket{\Omega}=0\ ,
\end{equation}
as expected. In order to work with an operator with real eigenvalues, it is customary to define an anti-Hermitian ghost number operator $N_{C}=iQ_{C}$, $N_{C}^{\dagger}=-N_{C}$, so that
\begin{equation}
   N_{C}\ket{c^{a}(x)}=\ket{c^{a}(x)}\ ,\qquad N_{C}\ket{\cbar^{a}(x)}=-\ket{\cbar^{a}(x)}\ .
\end{equation}
Eigenstates of $N_{C}$ with real eigenvalues are necessarily zero-norm states.

In the KO framework, it is customary to assume that states with nonzero ghost number are generated by acting on the vacuum state with the ghost or the antighost field operators \cite{KO78b,Oji78,KO78a,KO79b,KO79c,KO79d,KO79a,Kug78}. Under this assumption, $N_{C}$ has integer eigenvalues. In terms of $N_{C}$, the BRST algebra reads
\begin{align}
   \notag\{Q_{B},Q_{B}\}&=2Q_{B}^{2}=0\ ,\quad [N_{C},Q_{B}]=Q_{B}\ ,\\
   &\qquad\ [N_{C},N_{C}]=0\ .
\end{align}
From the second of these, it follows that the BRST charge increases the ghost number of states by one unit.

Consider now the MYME of the free photon field in LCGs, together with their gauge-fixing condition,
\begin{equation}
   \partial^{\nu}F_{\nu\mu}-\partial_{\mu}B=0\ ,\qquad \partial^{\mu}A_{\mu}+\xi B=0\ .
\end{equation}
The dynamical equation of the NL field, in this case, reads $\partial^{2}B=0$, whence it follows that $\partial^{2}(\partial^{\mu}A_{\mu})=0$. Plugging the gauge-fixing condition into the MYME, we see that
\begin{equation}
   \partial^{2}A_{\mu}=(1-\xi^{-1})\,\partial_{\mu}(\partial^{\nu}A_{\nu})\ .
\end{equation}
From the above equation, it follows that, in the Feynman gauge ($\xi=1$), and in the Feynman gauge only, the photon field satisfies the free wave equation $\partial^{2}A_{\mu}=0$. Beyond the Feynman gauge, on the other hand, operating on both sides of the equation with $\partial^{2}$, we obtain
\begin{equation}\label{photondipole}
   (\partial^{2})^{2}A_{\mu}=0\ ,\qquad \partial^{2}A_{\mu}\neq 0\qquad(\xi\neq 1).
\end{equation}
Eq.~\eqref{photondipole} is called a ``dipole equation'', and the photon field that satisfies it is called a ``dipole field''. In the free photon field's Fock space, the dipole corresponds to order-2 generalized eigenstates of the mass operator $P^{2}$ with generalized eigenvalue zero \cite{Nak72,NO90}:
\begin{equation}
   P^{2}\ket{A_{L}(p)}=-i(1-\xi)|{\bf p}|\ket{A_{S}(p)}\ ,\quad P^{2}\ket{A_{S}(p)}=0\ ,
\end{equation}
where $\ket{A_{L,S}(p)}$ are the (unphysical) longitudinal and scalar photon polarizations of momentum $p$. Accordingly, for $\xi \neq 1$, the free-photon propagator $\Delta_{\mu\nu}(p)$,
\begin{equation}\label{photdipoles}
   \Delta_{\mu\nu}(p)=\frac{-i}{p^{2}+i\epsilon}\,\left\{\eta_{\mu\nu}-(1-\xi)\,\frac{p_{\mu}p_{\nu}}{p^{2}+i\epsilon}\right\}\ ,
\end{equation}
contains a double pole (dipole) at $p^{2}=0$.

Finally, consider the field equations of the Stueckelberg model \cite{Stu38a,Stu38b,DTT88,RRA04} in LCGs -- reviewed in Appendix~\ref{app:stschw} --,
\begin{align}\label{stfeq}
   \partial^{\nu}F_{\nu\mu}-\partial_{\mu}B-m\,\partial_{\mu}\phi+m^{2}A_{\mu}=0\ ,\\
   \notag\partial^{2}\phi-m\,\partial^{\mu}A_{\mu}=0\ ,\quad\partial^{\mu}A_{\mu}+\xi B=0\ .
\end{align}
Here $\phi$ -- the ``Stueckelberg field'' -- is a scalar field, and $m$ is a mass parameter. The second and third of Eq.~\eqref{stfeq}, together, yield
\begin{equation}
   \partial^{2}\phi+\xi m B=0\ .
\end{equation}
It follows that, in the Landau gauge ($\xi=0$), $\partial^{2}\phi=0$: the Stueckelberg field satisfies the free wave equation. Beyond the Landau gauge, on the other hand, $\partial^{2}B=0$ implies that the Stueckelberg field is a dipole field,
\begin{equation}
   (\partial^{2})^{2}\phi=0\ ,\qquad\partial^{2}\phi\neq 0\qquad(\xi\neq 0)\ .
\end{equation}
As we review in the Appendix, the photon field $A_{\mu}$ that solves the Stueckelberg model,
\begin{equation}
   A_{\mu}=U_{\mu}+m^{-1}\partial_{\mu}\phi+m^{-2}\partial_{\mu}B\ ,
\end{equation}
contains the Stueckelberg field, the NL field, and the operator $U_{\mu}$ associated with the physical massive photon modes of the model. Correspondingly, for $\xi\neq 0$, both the Stueckelberg field's propagator $\Delta_{\phi\phi}(p^2)=\int d^{4}x\,e^{ipx}\,\avg{T\{\phi(x)\phi(0)\}}$ and the (massive) photon propagator $\Delta_{\mu\nu}(p)$ contain double poles (dipoles) at $p^{2}=0$:
\begin{align}\label{stueckdipoles}
   \Delta_{\phi\phi}(p^{2})&=\frac{i}{p^{2}+i\epsilon}-\frac{i\xi m^{2}}{(p^{2}+i\epsilon)^{2}}\ ,\\
   \notag\Delta_{\mu\nu}(p)&=\frac{-i}{p^{2}-m^{2}+i\epsilon}\,\left\{\eta_{\mu\nu}-\frac{p_{\mu}p_{\nu}}{p^{2}+i\epsilon}\right\}+\frac{-i\xi p_{\mu}p_{\nu}}{(p^{2}+i\epsilon)^{2}}\ .
\end{align}

In Appendix~\ref{app:stschw} and \ref{app:higgsmod} we review, respectively, the Schwinger model and Higgs models in linear covariant gauges. In these theories as well, in all LCGs except for the Landau gauge, massless dipoles are found in the propagators of the Goldstone bosons and of the massive gauge bosons \cite{Nak75}.\\

Generalized eigenstates and states with non-positive norm spoil the ordinary probabilistic interpretation of the quantum formalism, hence they must represent unphysical states. And indeed, in the electroweak sector of the Standard Model, they are well-known to be associated with the unphysical modes of the theory -- namely, the longitudinal and the scalar photon polarizations, the scalar polarization of the $W$ and $Z$ fields, the unphysical components of the Higgs field, and the FP ghosts \cite{KO78b,KO79b,KO79c}.

Understanding why unphysical states exist in the Hilbert space $\mc{H}$ of the theory is straightforward. As we saw in the previous section, after gauge fixing, the BRST charge $Q_{B}$ still generates a residual gauge symmetry parametrized by ghosts. Since $Q_{B}$ is not the zero operator, there exist states in $\mc{H}$ which are not invariant under these transformations, and there exist states which are their images. These are called, respectively, parent and daughter states \cite{NO90,AA10,AA11}:
\begin{equation}
   Q_{B}\ket{\text{parent}}=\ket{\text{daughter}}\neq 0\ .
\end{equation}
Since $Q_{B}$ is nilpotent, $Q_{B}^{2}=0$, acting on a daughter state with $Q_{B}$ yields zero. Moreover, parent states are not in the image of $Q_{B}$, for otherwise they would be annihilated by $Q_{B}$. Parents and daughters thus form ``BRST doublets''.

Daughter states have zero norm,
\begin{equation}
   \braket{\text{daughter}|\text{daughter}}=\braket{\text{parent}|Q_{B}^{2}|\text{parent}}=0\ .
\end{equation}
By the non-degeneracy of the inner product, there must then exist a state $\ket{\text{parent}^{\prime}}$ in $\mc{H}$ that has nonzero inner product with $\ket{\text{daughter}}$,
\begin{equation}\label{deq682}
   0\neq \braket{\text{parent}^{\prime}|\text{daughter}}=\braket{\text{parent}^{\prime}|Q_{B}|\text{parent}}\ .
\end{equation}
$\ket{\text{parent}^{\prime}}$ is distinct from $\ket{\text{parent}}$: if $\ket{\text{parent}}$ has ghost number $n_{C}$, then $\ket{\text{daughter}}$ has ghost number $n_{C}+1$ and $\ket{\text{parent}^{\prime}}$ has ghost number $-n_{C}-1\neq n_{C}$. From Eq.~\eqref{deq682}, it follows that $Q_{B}\ket{\text{parent}^{\prime}}\neq 0$. Therefore $\ket{\text{parent}^{\prime}}$ is the parent state of a second BRST doublet,
\begin{equation}
   Q_{B}\ket{\text{parent}^{\prime}}=\ket{\text{daughter}^{\prime}}\neq 0\ 
\end{equation}
with $\ket{\text{daughter}^{\prime}}$ having ghost number $-n_{C}\neq n_{C}+1$, so that $\ket{\text{daughter}^{\prime}}\neq \ket{\text{daughter}}$. This second doublet is coupled to the first by the inner product, hence it forms with it a ``metric pair'' of BRST doublets called a ``BRST quartet''.

BRST quartets organize the structure of the unphysical states of $\mc{H}$. Given that $Q_{B}$ commutes with the mass operator $P^{2}$, $[Q_{B},P^{2}]=0$, members of quartets which are (generalized) eigenstates of the mass operator form multiplets of equal mass\footnote{The mass operator $P^{2}$ is customarily assumed to have real positive (generalized) eigenvalues \cite{NO90}.}. Similarly, members of quartets carrying the quantum numbers of a charge $Q$ that commutes with the BRST charge, $[Q_{B},Q]=0$, all have those same quantum numbers.

A state which is annihilated by the BRST charge, $Q_{B}\ket{\psi}=0$, equivalently, $\ket{\psi}\in\text{ker}\{Q_{B}\}$, is called ``BRST-closed''; a state which is in the image of the BRST charge, $\ket{\psi}\in\text{im}\{Q_{B}\}$, equivalently, $\ket{\psi}=Q_{B}\ket{\phi}$ for some $\ket{\phi}$, is called ``BRST-exact''. States $\ket{\text{phys}}$ which are BRST-closed but not BRST-exact,
\begin{equation}
   Q_{B}\ket{\text{phys}}=0\ ,\qquad \ket{\text{phys}}\notin\text{im}\{Q_{B}\}\ ,
\end{equation}
are called ``BRST singlets''.

Given a BRST-closed state $\ket{\psi}$ and any state $\ket{\phi}$, the state
\begin{equation}
   \ket{\psi^{\prime}}=\ket{\psi}+Q_{B}\ket{\phi} 
\end{equation}
is BRST-closed as well, and it has the same inner product as $\ket{\psi}$ with any other BRST-closed state. With respect to the inner product, BRST-closed states thus form equivalence classes modulo the addition of BRST-exact states. The quotient space of this equivalence relation is called the ``BRST cohomology'', and it is denoted by $\mc{H}_{\text{phys}}$,
\begin{equation}
   \mc{H}_{\text{phys}}=\text{ker}\{Q_{B}\}/\text{im}\{Q_{B}\}\ .
\end{equation}
Members of BRST quartets either do not belong to $\mc{H}_{\text{phys}}$, or they correspond to its zero element.

BRST singlets are invariant under the residual gauge transformations generated by $Q_{B}$, and they are not in its image, hence they represent physical states. Since they correspond to unique elements of the BRST cohomology $\mc{H}_{\text{phys}}$, the latter is identified with the physical Hilbert space of the theory. This identification is at the heart of the Kugo-Ojima framework for the quantization of Yang-Mills theories \cite{KO78b,Oji78,KO78a,KO79b,KO79c,KO79d,KO79a,Kug78}.

In the full Hilbert space $\mc{H}$ of the theory, the elements of $\mc{H}_{\text{phys}}$ are represented by BRST-closed states. For this reason, with an abuse of notation, BRST-closed states are often denoted by $\ket{\text{phys}}$ even if they are BRST-exact.\\

In the KO quantization framework, the Hilbert space $\mc{H}$ of the theory is assumed to be ``asymptotically complete'' \cite{KO78b,Oji78,KO78a,KO79b,KO79c,KO79d,KO79a,Kug78}. Following \cite{NO90}, we introduce the definitions and results needed for our purposes. We refer the reader to Sec.~1.4 and Appendices~2 and 4 therein for a detailed exposition of the underlying axiomatic formulation.

Let $\phi$ be a scalar Heisenberg field, and suppose that $\phi$ satisfies the equation
\begin{equation}
   (\partial^{2}+m_{R}^{2})\phi(x)=J(x)\ ,
\end{equation}
where $m_{R}$ is a finite (``renormalized'') real number and $J(x)$ is an operator with no discrete spectrum. Let $D_{R/A}(x;m^{2})$ be the retarded and advanced (free) propagators
\begin{align}\label{ffretadv}
   D_{R}(x;m^{2})&=-\int\frac{d^{4}p}{(2\pi)^{4}}\ \frac{e^{-ipx}}{p^{2}-m^{2}+i\epsilon\,\text{sgn}(p^{0})}\ ,\\
   \notag D_{A}(x;m^{2})&=-\int\frac{d^{4}p}{(2\pi)^{4}}\ \frac{e^{-ipx}}{p^{2}-m^{2}-i\epsilon\,\text{sgn}(p^{0})}\ ,
\end{align}
where $\text{sgn}(x)$ is the sign function. Define an ``in'' and an ``out'' field $\phi^{\text{in}/\text{out}}(x)$ as\\
\begin{equation}\label{yangfeldman}
   Z_{\phi}^{1/2}\,\phi^{\text{in}/\text{out}}(x)=\phi(x)-\int d^{4}y\ D_{R/A}(x-y;m_{R}^{2})\ J(y)\ .
\end{equation}
In the above ``Yang-Feldman equation'' \cite{YF50}, $Z_{\phi}^{1/2}$ is a coefficient -- customarily assumed to be real and positive -- known as the ``Lehmann-Symanzik-Zimmerman (LSZ) coefficient'' of $\phi$\footnote{As we will see in a moment, at variance with ordinary field-strength renormalization constants, which are defined by introducing a renormalization scale $\mu$, LSZ coefficients are specifically defined on the mass shell of \textit{unrenormalized} Green's functions.}.

By taking the $t=x^{0}\to\pm\infty$ limits of Eq.~\eqref{yangfeldman}, we see that
\begin{align}
   \phi(x)&\to Z_{\phi}^{1/2}\,\phi^{\text{in}}(x)\quad\ \,\text{as}\quad t\to -\infty\ ,\\
   \notag\phi(x)&\to Z_{\phi}^{1/2}\,\phi^{\text{out}}(x)\quad\text{as}\quad t\to +\infty\ .
\end{align}
Moreover,
\begin{equation}
   (\partial^{2}+m_{R}^{2})\,\phi^{\text{in}/\text{out}}(x)=0\ ,
\end{equation}
that is, $\phi^{\text{in}/\text{out}}(x)$ are free fields of mass $m_{R}^{2}$. In and out fields are collectively called ``asymptotic fields'', and they are denoted by the superscript ``as.'', $\phi^{\text{as.}}(x)=\phi^{\text{in}/\text{out}}(x)$. Because of the integral in the second term of Eq.~\eqref{yangfeldman}, asymptotic fields are not local fields.

The LSZ coefficient $Z_{\phi}^{1/2}$ of a field $\phi(x)$ is fixed so that the commutation relations of asymptotic fields are normalized canonically -- i.e., like the commutator of ordinary free fields,
\begin{equation}\label{lszcomm}
   [\phi^{\text{as.}}(x),\phi^{\text{as.}}(y)]=iD(x-y;m_{R}^{2})\ ,
\end{equation}
where $D(x;m^{2})$ is the function
\begin{equation}\label{ffcommut}
   D(x;m^2)=-i\int\frac{d^{4}p}{(2\pi)^{3}}\ e^{-ipx}\ \text{sgn}(p^{0})\ \delta(p^{2}-m^2)\ .
\end{equation}
On indefinite-metric Hilbert spaces, the RHS of Eq.~\eqref{lszcomm} can have a minus sign, extra factors of $\pm i$, or be zero, if $\phi^{\text{as.}}(x)$ creates states with non-positive norm.

If a two-point function $\bra{\Omega}\phi_{1}(x)\phi_{2}(y)\ket{\Omega}$ of local fields $\phi_{1,2}(x)$ has a discrete spectrum, then, in general, there exist two asymptotic fields $\phi_{1,2}^{\text{as.}}(x)$ such that
\begin{align}
   &\bra{\Omega}\phi_{1}(x)\phi_{2}(y)\ket{\Omega}\big|_{\text{discr. spectr.}}=\\
   \notag&\qquad=(Z_{\phi_{1}}Z_{\phi_{2}})^{1/2}\bra{\Omega}\phi_{1}^{\text{as.}}(x)\phi_{2}^{\text{as.}}(y)\ket{\Omega}\ .
\end{align}
The full set of the asymptotic fields of the theory can thus be determined by analyzing the discrete spectrum of its two-point functions. The corresponding LSZ coefficients can be computed from the residues of time-ordered correlators at their momentum space poles
\begin{align}
   &\bra{\Omega}T\left\{\phi_{1}(x)\phi_{2}(y)\right\}\ket{\Omega}_{\text{F.T.}}\big|_{\text{pole}}=\\
   \notag&\qquad =(Z_{\phi_{1}}Z_{\phi_{2}})^{1/2}\bra{\Omega}T\{\phi_{1}^{\text{as.}}(x)\phi_{2}^{\text{as.}}(y)\}\ket{\Omega}\ .
\end{align}

For non-scalar fields, the definition of asymptotic field uses the appropriate free-field differential operator and Green's functions in place of $\partial^{2}+m_{R}^{2}$ and $D_{R,A}(x;m^{2}_{R})$, $D(x;m^{2}_{R})$ \cite{YF50}. The definition of asymptotic field for a multipole field is given in \cite{NO78}. In theories like QCD, whose spectrum contains stable bound states, one needs to define asymptotic fields for composite operators as well, as done in \cite{Zim58}. Thanks to the phenomenon of the ``eating'' of the Goldstone boson, to be discussed in Sec.~\ref{sec:resgtgb}, we will never need to make reference to this definition.

With the asymptotic fields $\{\phi^{\text{as.}}(x)\}$ normalized canonically, one can define corresponding creation and annihilation operators as is customarily done in free-field theories. The creation operators thus obtained act on the unique vacuum state of the theory $\ket{\Omega}$ producing ``asymptotic (in/out) states'' $\ket{\text{in}/\text{out}}$, while the annihilation operators are assumed to annihilate $\ket{\Omega}$. Asymptotic states represent (one-particle and multiparticle) scattering states: they describe sets of particles which are free in the $t\to\pm \infty$ limit.

Unstable bound states and, more generally, particles that decay in finite time, do not survive the $t\to\pm\infty$ limit, hence they have no associated asymptotic field, nor asymptotic states. In the formalism, they figure as resonances of finite width in the scattering amplitudes of stable states.

The $S$-matrix can be defined either as the collection of inner products
\begin{equation}\label{Smat}
   S_{\alpha\beta}=\braket{\alpha,\text{out}|\beta,\text{in}}\ ,
\end{equation}
or as the operator that transforms in fields to out fields and vice-versa,
\begin{equation}\label{Sop}
   \phi^{\text{out}}(x)=S^{-1}\phi^{\text{in}}(x)S\ .
\end{equation}
As a consequence of the Hermiticity of the Hamiltonian, the $S$-matrix is unitary, $S^{-1}=S^{\dagger}$; moreover, it stabilizes the vacuum, $S\ket{\Omega}=\ket{\Omega}$. Note that the $S$-matrix defined by Eqs.~\eqref{Smat} and \eqref{Sop} is the $S$-matrix of the full indefinite-metric Hilbert space $\mc{H}$: it describes the scattering of both physical and unphysical states. In physical applications, one is interested in the unitarity of the \textit{physical} S-matrix $S_{\text{phys}}$, which is defined on the physical Hilbert space $\mc{H}_{\text{phys}}$. This topic is postponed until the next section.

By acting on the vacuum state $\ket{\Omega}$ with the in (resp. out) creation operators of every asymptotic field $\phi^{\text{in}}(x)$ (resp. $\phi^{\text{out}}(x)$) of the theory, one generates a Fock space denoted by $\mc{H}_{\text{in}}$ (resp. $\mc{H}_{\text{out}}$). The postulate of ``asymptotic completeness'' consists in requiring that the Hilbert space of the theory $\mc{H}$ is equal to both these spaces:
\begin{equation}
   \mc{H}_{\text{in}}=\mc{H}_{\text{out}}=\mc{H}\ .
\end{equation}\\

In the present paper, following Kugo and Ojima \cite{KO78a,KO79b,KO79c,KO79a}, we will make use of three key properties of asymptotic fields. The first is a trivial consequence of their definition. Since on the RHS of Eq.~\eqref{yangfeldman} $J(x)$ has no discrete spectrum, one has
\begin{equation}\label{1psme}
   \bra{\Omega}\phi(x)\ket{\Psi(p^{2}=m^{2})}=Z_{\phi}^{1/2}\bra{\Omega}\phi^{\text{as.}}(x)\ket{\Psi(p^{2}=m^{2})}
\end{equation}
for every one-particle state $\ket{\Psi(p^{2}=m^{2})}$ of mass $m^{2}$: the matrix element of a local operator between the vacuum and any one-particle state is equal to the matrix element of the corresponding asymptotic field multiplied by its LSZ coefficient.

The second property relies on asymptotic completeness, and follows from the ``Greenberg-Robinson theorem'' \cite{Gre62,Rob62,NO78,NO90}. The latter states that the (anti)commutator of two asymptotic fields is a function ($c$-number):
\begin{equation}
   [\phi_{1}^{\text{as.}}(x),\phi_{2}^{\text{as.}}(y)]_{\mp}=c\text{-number}\ .
\end{equation}
The Greenberg-Robinson theorem is valid in spacetime dimensions $d\geq 3$, and it has known counterexamples in two-dimensional non-Abelian theories \cite{DF73,RB77,NO78}. It follows that our analysis\footnote{As well as the original analysis \cite{KO78a,KO79b,KO79c,KO79a} of Kugo and Ojima.} \textit{is not valid in two spacetime dimensions}.

The third property concerns the action of symmetries on asymptotic fields. Under the assumption of asymptotic completeness, if $Q$ is the generator of an \textit{unbroken} symmetry acting on local fields $\phi(x),\psi(x)$ as
\begin{equation}
   [iQ,\phi(x)]=\psi(x)\ ,
\end{equation}
then it can be shown \cite{KO79a,NO90} that the symmetry is preserved after taking the asymptotic limit $t\to\pm \infty$:
\begin{equation}\label{asymsymm}
   [iQ,Z_{\phi}^{1/2}\phi^{\text{as.}}(x)]=Z_{\psi}^{1/2}\psi^{\text{as.}}(x)\ ,
\end{equation}
where $\phi(x)\to Z_{\phi}^{1/2}\,\phi^{\text{as.}}(x)$, and similarly for $\psi(x)$. For the case of a broken symmetry generator, we postpone our discussion until Sec.~\ref{sec:resgtgb}.\\

To conclude this section, let us introduce the main protagonist of this paper: the ``elementary BRST quartet'' of Yang-Mills theory. Consider the two-point functions
\begin{equation}
   \avg{T\{A_{\mu}^{a}(x)B^{b}(y)\}}\ ,\qquad \avg{T\{D_{\mu}c^{a}(x)\cbar^{b}(y)\}}\ .
\end{equation}
By taking the divergence of both Green's functions, using the antighost equal-time canonical anticommutation relations Eq.~\eqref{cancomm}, the ghost field equations Eq.~\eqref{feqs}, the BRST transformations Eq.~\eqref{brst}, and appealing to Lorentz covariance, one can show that \cite{KO79a}
\begin{align}\label{quarteteq}
   \avg{T\{A_{\mu}^{a}(x)B^{b}(0)\}}_{\text{F.T.}}&=-\frac{\delta^{ab}\,p_{\mu}}{p^{2}+i\epsilon}\ ,\\
   \notag\avg{T\{D_{\mu}c^{a}(x)\cbar^{b}(0)\}}_{\text{F.T.}}&=\frac{i\delta^{ab}\,p_{\mu}}{p^{2}+i\epsilon}\ .
\end{align}
Since these Fourier transforms have poles at $p^{2}=0$, the fields $A_{\mu}^{a}$, $B^{a}$, $D_{\mu}c^{a}$, and $\cbar^{a}$ \textit{always} possess a discrete spectrum at $p^{2}=0$, regardless of the dynamics of the theory. The corresponding massless asymptotic fields are customarily denoted by $\chi^{a}$, $\beta^{a}$, $\gamma^{a}$ and $\gammabar^{a}$,
\begin{align}\label{quartetasym}
   \notag A_{\mu}^{a}&\to Z_{\chi}^{1/2}\,\partial_{\mu}\chi^{a}+\cdots\ ,\\
   B^{a}&\to Z_{\chi}^{-1/2}\,\beta^{a}+\cdots\ ,\\
   \notag D_{\mu}c^{a}&\to \widetilde{Z}_{3}^{-1/2}\,\partial_{\mu}\gamma^{a}+\cdots\ ,\\
   \notag \cbar^{a}&\to \widetilde{Z}_{3}^{1/2}\,\gammabar^{a}+\cdots\ .
\end{align}
Here the arrows denote the limits $t \to\pm \infty$, the ellipsis denotes other potential asymptotic contributions, the superscript ``as.'' has been suppressed, and $Z_{\chi}^{1/2}$ and $\widetilde{Z}_{3}^{1/2}$ are LSZ coefficients. The latter are fixed in pairs by the residues of the poles in Eq.~\eqref{quarteteq}.

Absent massless bound states in the channels $f^{abc}A_{\mu}^{b}c^{c}$ and $f^{abc}c^{b}c^{b}$\footnote{See the derivation in Sec.~\ref{sec:asympt} for dynamically massive QCD, \cite{KO79b} for Higgs models, and \cite{KO79c} for Yang-Mills theories without dynamical gauge-boson mass generation.}, from Eqs.~\eqref{brst} and \eqref{asymsymm}, it follows that $\chi^{a}$, $\beta^{a}$, $\gamma^{a}$ and $\gammabar^{a}$ form a quartet of asymptotic fields related to one another by BRST transformations,
\begin{align}\label{emquartet0}
   [iQ_{B,R},\chi^{a}]=\gamma^{a}\ ,\qquad \{iQ_{B,R},\gamma^{a}\}=0\ ,\\
   \notag\{iQ_{B,R},\gammabar^{a}\}=i\beta^{a}\ ,\qquad [iQ_{B,R},\beta^{a}]=0\ ,
\end{align}
where $Q_{B,R}=(Z_{\chi}\widetilde{Z}_{3})^{1/2}\,Q_{B}$ is a redefined (``renormalized'') BRST charge. The canonically normalized mixed inner products between their one-particle states are fixed by Eq.~\eqref{quarteteq} to be \cite{KO78b,KO79b,KO79c,KO79a}
\begin{align}\label{emprods0}
   \braket{\chi^{a}(p)|\beta^{b}(q)}&=-(2\hat{p}^{0})(2\pi)^{3}\delta^{ab}\delta^{(3)}({\bf p}-{\bf q})\ ,\\
   \notag\braket{\gamma^{a}(p)|\gammabar^{b}(q)}&=i(2\hat{p}^{0})(2\pi)^{3}\delta^{ab}\delta^{(3)}({\bf p}-{\bf q})\ ,
\end{align}
where $\hat{p}=(|{\bf p}|,{\bf p})$. Being associated with the elementary fields of Yang-Mills theory, the quartet formed by $\chi^{a}$, $\beta^{a}$, $\gamma^{a}$ and $\gammabar^{a}$ is known as the ``elementary BRST quartet''.

Within the elementary BRST quartet, $\chi^{a}$ and $\gammabar^{a}$ are not BRST invariant, hence they are parents, whereas $\gamma^{a}$ and $\beta^{a}$ are the image of the former, hence they are daughters. $\chi^{a}$ and $\gamma^{a}$ are customarily called the ``first'' parent and daughter, whereas $\gammabar^{a}$ and $\beta^{a}$ are called the ``second'' parent and daughter \cite{AA10,AA11}.

The identity of $\beta^{a}$, $\gamma^{a}$ and $\gammabar^{a}$ is known: they are massless excitations associated, respectively, to the NL field $B^{a}$ \footnote{Equivalently, because of the gauge-fixing condition $\partial^{\mu}A_{\mu}+\xi B^{a}=0$, to the scalar polarization of the gauge boson.}, to the ghost field $c^{a}$ and to the antighost field $\cbar^{a}$. These are the fields which are explicitly added to the Yang-Mills Lagrangian in the process of gauge fixing.

The identity of $\chi^{a}$, on the other hand, is theory-dependent. In theories without gauge-boson mass generation, $\chi^{a}$ is the longitudinal gauge-boson polarization \cite{KO79c}. In theories in which the gauge bosons acquire a mass, it is a massless scalar: modulo mixing with $\beta^{a}$, it is the Stueckelberg field in the Stueckelberg model, a fermion-antifermion state in the Schwinger model \cite{Ito75,Nak75,Nak76a,Nak77a,Nak77b,Nak77c}, and the unphysical components of the Higgs field in Higgs models \cite{KO79b}. These theories are reviewed in Appendices \ref{app:photon}, \ref{app:stschw} and \ref{app:higgsmod}.

The identity of $\chi^{a}$ in QCD with dynamical gluon mass generation has never been established. While in some of the more recent literature \cite{AA10,AA11} $\chi^{a}$ is identified with a massless longitudinal gluon polarization, this identification is inconsistent: if a massless longitudinal polarization existed in the presence of a gauge boson mass, a Lorentz transformation would mix it with a massive transverse polarization, thus contradicting the fact that Lorentz transformations preserve the mass of states\footnote{The scalar polarization $\partial^{\mu}A_{\mu}$, on the other hand, is Lorentz-invariant, hence it can exist and remain massless.}. When gauge bosons acquire a mass, their longitudinal polarization either becomes massive, or it ceases to exist as an asymptotic state.

In the older literature, instead, Kugo noted that scalar bound states could play the role of $\chi^{a}$ \cite{Kug78}; this possibility appears not to have been explored subsequently. Interestingly, one of the works he cites in this context, a 1974 paper by J. Smit \cite{Smi74}, is among the precursors of modern studies of the Schwinger mechanism in QCD. In the present paper, we will show that Kugo's observation was, in fact, correct.

Finally, let us discuss normalization. In our conventions, the LSZ coefficients $Z_{\chi}^{1/2}$, $\widetilde{Z}_{3}^{1/2}$ of the elementary BRST quartet are tracked explicitly -- see Eq.~\eqref{quartetasym}. This is at variance with the generic definitions employed by Kugo and Ojima while discussing their confinement criterion -- see Eqs.~(2.16), (3.21), (4.1.3 -- 22) -- (4.1.3 -- 25), (2.8) of \cite{KO79d,KO79a,NO90,Kug95}, respectively --, and in agreement with the definitions they employ when concretely analyzing the asymptotic spectrum of Yang-Mills theories -- see e.g. Eqs. (4.17) or (4.41) of \cite{KO79a}. With KO's generic conventions, the asymptotic fields create states which are not normalized. As a consequence, when using those, in order to correctly evaluate the matrix element of an operator, one has to divide the result by the square root of (the modulus of) the norm of the asymptotic states -- namely, by one factor of $Z^{1/2}$ for each of the non-normalized asymptotic states in the equation. With our conventions, once the LSZ coefficients are fixed to yield the canonical commutation relations in Eq.~\eqref{lszcomm}, the asymptotic fields create and annihilate states which are already canonically normalized, and one does not need to normalize the matrix elements manually.

\subsection{Spectral representations, the quartet mechanism, and the unitarity of the physical $S$-matrix}\label{sec:spectr}

In much of the literature on the analytic structure of QCD Green's functions, it is stated that the gluon propagator does not admit a K\"{a}ll\'{e}n-Lehmann (KL) spectral representation \cite{Kal52,Leh54}. This statement is based on two distinct arguments: the purported existence of complex-conjugate singularities in the propagator's complex-energy plane, and the ``violation'' of the positivity of its spectral density. Given the extensive use we will make of spectral representations in the present paper, we deem it necessary to spend a few words to discuss these arguments.

The existence of complex-conjugate singularities -- specifically, of complex-conjugate poles -- in the analytic continuation of the transverse form factor of the gluon propagator from Euclidean space to the complex-energy plane is predicted by a number of approaches to infrared QCD, ranging from Gribov-Zwanziger formulations \cite{Gri78,Zwa89,DGSVV08,dBP24}, to reconstructions from lattice data \cite{BT20,FOS20,DOS18}, to perturbative methods \cite{SIR16b,SC18,HK19,HK20}\footnote{The latter include works by the present author, who has since abandoned the idea that these singularities play any role in the infrared dynamics of QCD \cite{Com26b}.}, to -- to some extent -- functional methods \cite{FH20}. Following Stingl \cite{Sti85,Sti94}, these complex-conjugate poles are often interpreted as short-lived gluonic excitations, realizing gluon confinement through the absence of transverse gluon states from the asymptotic space $\mc{H}_{\text{in}/\text{out}}$ of the theory.

Stingl's interpretation, unfortunately, is incorrect, as it disregards the fact that the complex-conjugate poles in question are located on the \textit{first (principal) Riemann sheet} of the propagator\footnote{Additionally, in \cite{Sti85,Sti94} Stingl employs an analytic continuation from Euclidean to Minkowski space -- namely, the Wick rotation $p^{2}_{E}\to -p^{2}$, $p_{E}$ being the Euclidean momentum -- which is not valid in the presence of these complex-conjugate poles \cite{HK21b}.}. Short-lived excitations of the quantum fields, however, are associated with poles on the \textit{second} Riemann sheets of two-point functions, reached by analytically continuing the energy variable beyond the branch cut on the Minkowski axis $p^{2}>0$ \cite{JS61,ELOP66,PDGResonances2026}.

Complex-conjugate poles on the first Riemann sheet of two-point functions, on the other hand, are known to correspond to propagator terms which -- far from being damped in time as those of short-lived excitations -- \textit{exponentially blow up} as $t\to\pm\infty$ \cite{LW69,Nak72,HK21b}. As a consequence, in Minkowski space, propagators with these poles do not even admit Fourier transforms \cite{HK21b}, let alone spectral representations\footnote{Ways to define Fourier-like transforms by means of deformations of the energy integration contour are still known in this case, but -- as can be easily expected -- they lead to theories which break Lorentz symmetry \cite{Nak72,ABCM25}.}.

To comply with Fourier-transformability, Lorentz invariance, causality, and locality, theories with complex-conjugate poles in the first Riemann sheets require a reformulation of quantum field theory which, despite decades of reasearch \cite{LW69,Nak72,ABCM25}, has failed to materialize. Since the Kugo-Ojima framework is built upon standard quantum-field-theoretical postulates adapted to indefinite-metric Hilbert spaces \cite{NO90}, these singularities have no place in the present work, and will not be considered further.

As for the gluon propagator lacking a K\"{a}ll\'{e}n-Lehmann representation due to its spectral density not being positive definite, this claim finds no support in the literature. In the first paper ever published on the KL representation \cite{Kal52}, G. K\"{a}ll\'{e}n derives a photon spectral density\footnote{See Eq.~(45) therein and the discussion surrounding that equation.} which contains \textit{both} positive \textit{and} negative contributions. Then, \textit{a posteriori}, he argues that, for the \textit{specific} case of the photon propagator, the spectral density is non-negative because it can be computed from the correlator of \textit{gauge-invariant} currents. His argument relies on the fact that the scalar and the longitudinal photon polarizations contribute to the spectral density with equal weight and opposite sign, hence they cancel each other out, whereas the remaining contributions are provided by the transverse photon polarizations, and they are non-negative. The framework in which K\"{a}ll\'{e}n derives this result is the Gupta-Bleuler quantization of QED \cite{Gup50,Bleu50}, which is the precursor of the Kugo-Ojima framework for the quantization of Yang-Mills theories \cite{KO78b}. Within the latter, the cancellation mechanism K\"{a}ll\'{e}n describes in \cite{Kal52} is known as the ``quartet mechanism'' \cite{KO78b,KO78a,KO79a,Kug78}.

From its very inception \cite{Kal52}, when used in gauge theories, the KL construction has only ever placed positivity constraints on spectral densities of \textit{gauge-invariant} correlators. In LCGs, the gauge boson fields $A_{\mu}^{a}$ of non-Abelian Yang-Mills theories are not gauge-invariant operators, nor their correlators can be computed from those of gauge-invariant currents. Therefore, one should not expect their spectral densities to be positive. The gluon propagator does not ``violate'' the positivity of the KL representation because, in gauge theories, the KL representation places no positivity constraint on spectral densities of gauge-dependent correlators.

Related to this matter is the common misconception that the positivity ``violation'' of the gluon spectral density, \textit{taken on its own}, provides information on the presence or absence of gluons from the physical spectrum of theory. This belief is easily disproven. On the one hand, it is well known that the spectral densities of the gauge-boson propagators in Higgs models -- whose fields create physical particles, e.g. the $W/Z$ bosons -- violate positivity \cite{DvEGP21}. On the other hand, in the Landau gauge, the spectral density $\rho_{\phi}(p^{2})$ of the Stueckelberg field's propagator in Eq.~\eqref{stueckdipoles} is a positive Dirac delta, $\rho_{\phi}(p^{2})=\delta(p^{2})$, yet the quanta of the Stueckelberg field do not belong to the physical spectrum -- see Appendix~\ref{app:stschw}.\\

This being said, let us recall how KL representations are formulated in the KO framework \cite{KO78b,KO79b,KO79c,KO79a,NO90}. Spectral representations, in general, are defined by resolutions of the identity with respect to some eigenbasis. In a KL representation, one chooses an eigenbasis of the momentum operator $P_{\mu}$. On indefinite-metric Hilbert spaces, such an eigenbasis will be a generalized eigenbasis, in the sense that it will contain generalized eigenstates of the mass operator $P^{2}$. The technology to deal with these states is well-known, having been introduced in \cite{KO79b,KO79c,KO79a}; we postpone its introduction until Sec.~\ref{sec:asympt}, where we will concretely define dipole states of the first parent of the elementary BRST quartet of dynamically massive QCD.

In the KO framework, the additional structure provided by the BRST algebra and by asymptotic completeness is leveraged to organize the KL representation into physical and unphysical Fock sectors. An asymptotic field $\phi^{\text{as.}}$ creates physical (in or out) particles if $[Q_{B},\phi^{\text{as.}}]_{\mp}=0$ and if there exists no asymptotic field $\psi^{\text{as.}}$ such that $[Q_{B},\psi^{\text{as.}}]_{\pm}\propto\phi^{\text{as.}}$. The asymptotic fields of unphysical particles, on the other hand, form BRST quartets organized as in Eq.~\eqref{emquartet0}.

As we saw in the previous section, by acting on the unique vacuum state of the theory $\ket{\Omega}$ with the creation operators associated with the asymptotic (in/out) fields, one obtains the Fock spaces $\mc{H}_{\text{in}/\text{out}}$ of scattering states. By the postulate of asymptotic completeness, $\mc{H}=\mc{H}_{\text{in}/\text{out}}$, every state in $\mc{H}$ can be expanded in either in or out states. It follows that the identity operator $\mathds{1}$ of the Hilbert space admits a resolution in terms of these states.

Denote the subspace spanned by asymptotic (in/out) states containing $n$ unphysical particles with $\mc{H}_{\text{in}/\text{out}}^{(n)}$, so that
\begin{equation}
   \mc{H}=\bigoplus_{n=0}^{+\infty}\ \mc{H}_{\text{in}/\text{out}}^{(n)}\ .
\end{equation}
Suppressing the ``in/out'' label, each of these spaces has an associated projector $\mathds{P}_{n}$, so that
\begin{equation}\label{idres}
   \mathds{1}=\mathds{P}_{0}+\sum_{n=1}^{+\infty}\ \mathds{P}_{n}\ .
\end{equation}
Since $\mc{H}_{\text{in}/\text{out}}^{(0)}$ contains no unphysical particles, one has $[Q_{B},\mathds{P}_{0}]=0$.

In \cite{KO78a,KO79a}, Kugo and Ojima proved that, for every $n\geq 1$, the projector $\mathds{P}_{n}$ can be expressed recursively as
\begin{equation}\label{quartetmaster}
   \mathds{P}_{n}=\{Q_{B},\mathds{R}_{n}\}\ ,
\end{equation}
where $\mathds{R}_{n}$ is an operator defined as
\begin{align}
   \notag\mathds{R}_{n}&=-\frac{1}{n}\,\sum_{k}\Big\{\gammabar_{k}^{\dagger}\,\mathds{P}_{n-1}\chi_{k}+\chi_{k}^{\dagger}\,\mathds{P}_{n-1}\gammabar_{k}+\\
   &\qquad\qquad\qquad\quad+\sum_{l}\omega_{kl}\,\beta_{k}^{\dagger}\,\mathds{P}_{n-1}\gammabar_{l}\Big\}\ .
\end{align}
In the above expression, $\{\chi_{k}^{\dagger},\beta_{k}^{\dagger},\gamma_{k}^{\dagger},\gammabar_{k}^{\dagger}\}$ form a complete set of creation operators for the unphysical particles -- satisfying the BRST transformations in Eq.~\eqref{emquartet0} --, operators without a dagger are the corresponding annihilation operators, the index $k$ collects momentum, spin, internal and quartet indices, and $\omega_{kl}$ is a theory-dependent matrix. Explicitly, $\mathds{P}_{n}$ reads
\begin{align}\label{pnproj}
   \mathds{P}_{n}&=\frac{1}{n}\,\sum_{k}\Big\{-\beta_{k}^{\dagger}\,\mathds{P}_{n-1}\chi_{k}-\chi_{k}^{\dagger}\,\mathds{P}_{n-1}\beta_{k}+\\
   \notag&-\sum_{l}\omega_{kl}\,\beta_{k}^{\dagger}\,\mathds{P}_{n-1}\beta_{l}+i\gamma_{k}^{\dagger}\,\mathds{P}_{n-1}\gammabar_{k}-i\gammabar_{k}^{\dagger}\,\mathds{P}_{n-1}\gamma_{k}\Big\}\ .
\end{align}

The resolution of the identity in Eq.~\eqref{idres} can then be used to formulate the KL representation of field correlators. Given two local operators $\mc{O}_{1}(x)$, $\mc{O}_{2}(y)$, one has
\begin{equation}
   \bra{\Omega}\mc{O}_{1}(x)\mc{O}_{2}(y)\ket{\Omega}=\sum_{n=0}^{+\infty}\ \bra{\Omega}\mc{O}_{1}(x)\mathds{P}_{n}\mc{O}_{2}(y)\ket{\Omega}\ .
\end{equation}
Once each of the projectors in the equation is written down explicitly, it yields a sum of terms of the form
\begin{equation}\label{deq602}
   \bra{\Omega}\mc{O}_{1}(x)\ket{\{k\}}\bra{\{l\}}\mc{O}_{2}(y)\ket{\Omega}\ ,
\end{equation}
where $\ket{\{k\}}$ and $\ket{\{l\}}$ are multiparticle asymptotic (in/out) states, and $k,l$ denote species, momentum, spin, internal indices and, if applicable, quartet indices. Because of the presence of zero-norm states, the projectors in Eq.~\eqref{deq602} are, in general, not diagonal, $\ket{\{k\}}\neq\ket{\{l\}}$. In the sum, non-diagonal and negative-norm-state projectors contribute with terms which, in general, are not positive\footnote{This lack of positivity in no way implies the confinement of particles created and annihilated by the operators $\mc{O}_{1,2}$, if any.}.

Finally, due to the existence of multipole asymptotic fields, the Fourier transform of correlators will, in general, contain multipole contributions. Well-known examples are provided by the dipoles in Eqs.~\eqref{photdipoles} and \eqref{stueckdipoles}. In Sec.~\ref{sec:asympt} and in Appendix~\ref{app:schwpoles}, we will show how these dipoles emerge concretely from the asymptotic states of the theory.\\

Consider now two BRST-invariant operators $\mc{O}_{1,2}$,
\begin{equation}
   [Q_{B},\mc{O}_{1,2}]_{\mp}=0\ .
\end{equation}
and two BRST-closed states $\ket{\text{phys}}$, $\ket{\text{phys}^{\prime}}$,
\begin{equation}\label{deq866}
   Q_{B}\ket{\text{phys}}=Q_{B}\ket{\text{phys}^{\prime}}=0\ .
\end{equation}
Using Eq.~\eqref{quartetmaster}, we find that
\begin{align}
   \notag&\bra{\text{phys}}\mc{O}_{1}\mc{O}_{2}\ket{\text{phys}^{\prime}}=\sum_{n=0}^{+\infty}\ \bra{\text{phys}}\mc{O}_{1}\mathds{P}_{n}\mc{O}_{2}\ket{\text{phys}^{\prime}}=\\
   \notag&=\bra{\text{phys}}\mc{O}_{1}\mathds{P}_{0}\mc{O}_{2}\ket{\text{phys}^{\prime}}+\\
   \notag&\quad\quad+\sum_{n=1}^{+\infty}\ \bra{\text{phys}}\mc{O}_{1}\{Q_{B},\mathds{R}_{n}\}\mc{O}_{2}\ket{\text{phys}^{\prime}}=\\
   &=\bra{\text{phys}}\mc{O}_{1}\mathds{P}_{0}\mc{O}_{2}\ket{\text{phys}^{\prime}}\ ,
\end{align}
where to obtain the last line we made $Q_{B}$ (anti)commute past $\mc{O}_{1,2}$ and then used Eq.~\eqref{deq866}. It follows that states with $n\geq 1$ unphysical particles do not contribute to correlators of BRST-invariant operators evaluated between physical states \cite{KO79a}.

This property arises from the rigid constraints imposed by BRST symmetry on the structure of unphysical states: their organization into quartets, encoded in the form of $\mathds{P}_{n\geq 1}$, makes it so that unphysical contributions cancel each other out in physical (BRST-invariant) matrix elements. This cancellation mechanism is called the ``quartet mechanism''. It was first exploited by K\"{a}ll\'{e}n to obtain a positive-definite photon spectral density in the framework of the Gupta-Bleuler quantization of QED\footnote{He could avoid the photon dipole because the Gupta-Bleuler framework is formulated in the Feynman gauge.} \cite{Kal52}, and later systematized by Kugo and Ojima in their framework of BRST quantization \cite{KO78b,KO78a,KO79a,Kug78}.

Under the nontrivial assumption that the inner product in $\mc{H}_{\text{in}/\text{out}}^{(0)}$ is positive-definite, the projector $\mathds{P}_{0}$ can be expressed as \cite{KO78b,KO78a,KO79a,Kug78}
\begin{equation}
   \mathds{P}_{0}=\sum_{\{k\}}\ket{\{k\}}\bra{\{k\}}\ ,
\end{equation}
where $\ket{\{k\}}$ are multiparticle states, and the $k$'s, again, denote species, momentum, spin and internal indices. When inserted into the vacuum correlator of a BRST-invariant operator $\mc{O}(x)$, $[Q_{B},\mc{O}(x)]_{\mp}=0$, with its adjoint $\mc{O}^{\dagger}(x)$, after Fourier transforming, $\mathds{P}_{0}$ yields a spectral representation whose density is positive definite:
\begin{align}
   &\bra{\Omega}\mc{O}(x)\mc{O}^{\dagger}(y)\ket{\Omega}=\sum_{n=0}^{+\infty}\ \bra{\Omega}\mc{O}(x)\mathds{P}_{n}\mc{O}^{\dagger}(y)\ket{\Omega}=\\
   \notag&\quad\quad=\bra{\Omega}\mc{O}(x)\mathds{P}_{0}\mc{O}^{\dagger}(y)\ket{\Omega}=\\
   \notag&\quad\quad=\sum_{\{k\}}\,|\bra{\Omega}\mc{O}(0)\ket{\{k\}}|^{2}\,e^{-ip_{\{k\}}(x-y)}=\cdots\ ,
\end{align}
where $p_{\{k\}}$ is the total momentum of the state $\ket{\{k\}}$, and the derivation of the positive-definite density follows textbook steps \cite{PS95}.\\

To conclude this section, let us review the proof of the unitarity of the physical $S$-matrix in the KO framework \cite{KO78b,KO78a}.

First of all, observe that, since the BRST charge commutes with the Hamiltonian, $[Q_{B},H]=0$, and annihilates the vacuum, $Q_{B}\ket{\Omega}=0$, $Q_{B}$ commutes with the $S$-matrix $S$ defined on the full Hilbert space $\mc{H}$,
\begin{equation}
   [Q_{B},S]=0\ .
\end{equation}
It follows that $S$ maps BRST-closed states to BRST-closed states,
\begin{equation}
   Q_{B}\ket{\text{phys}}=0\quad\Longrightarrow\quad Q_{B}S\ket{\text{phys}}=0 ,
\end{equation}
hence that it descends to a well-defined operator $S_{\text{phys}}$ on the BRST cohomology $\mc{H}_{\text{phys}}$,
\begin{equation}
   S_{\text{phys}}:\ \mc{H}_{\text{phys}}\to \mc{H}_{\text{phys}}\ .
\end{equation}
$S_{\text{phys}}$ is the physical $S$-matrix.

The unitarity of $S_{\text{phys}}$ in the sense of $S_{\text{phys}}^{\dagger}S_{\text{phys}}=S_{\text{phys}}S_{\text{phys}}^{\dagger}=1$ follows immediately from the unitarity of $S$. Any pair of elements in $\mc{H}_{\text{phys}}$ is represented in $\mc{H}$ by BRST-closed states $\ket{\text{phys}}$, $\ket{\text{phys}^{\prime}}$, and their image under $S_{\text{phys}}$ is represented in $\mc{H}$ by the BRST-closed states $\ket{S(\text{phys})}=S\ket{\text{phys}}$, $\ket{S(\text{phys}^{\prime})}=S\ket{\text{phys}^{\prime}}$. Since the $S$-matrix is unitary on $\mc{H}$,
\begin{equation}
   \braket{S(\text{phys}^{\prime})|S(\text{phys})}=\braket{\text{phys}^{\prime}|S^{\dagger}S|\text{phys}}=\braket{\text{phys}^{\prime}|\text{phys}}\ ,
\end{equation}
one has $S_{\text{phys}}^{\dagger}S_{\text{phys}}=S_{\text{phys}}S_{\text{phys}}^{\dagger}=1$.

A different matter is that of the positivity of the inner product defined on $\mc{H}_{\text{phys}}$, often referred to as well as the ``unitarity of the $S$-matrix''\footnote{The reason being that operators $A$ on indefinite-metric Hilbert spaces such that $A^{\dagger}A=AA^{\dagger}=1$ are often referred to as ``pseudo-unitary'', whereas the term ``unitary'' is often reserved to analogous operators on positive-metric Hilbert spaces.} \cite{KO78b,KO78a}. In this regard, one observes the following. Since every state in $\mc{H}_{\text{phys}}$ is represented by a BRST-closed state $\ket{\text{phys}}$ in $\mc{H}$, one has
\begin{align}
   \ket{\text{phys}}&=\sum_{n=0}^{+\infty}\,\mathds{P}_{n}\ket{\text{phys}}=\\
   \notag&=\mathds{P}_{0}\ket{\text{phys}}+Q_{B}\sum_{n=1}^{+\infty}\,\mathds{R}_{n}\ket{\text{phys}}\ .
\end{align}
It follows that every state in $\mc{H}_{\text{phys}}$ is represented by a state $\mathds{P}_{0}\ket{\text{phys}}$ in $\mc{H}_{\text{in}/\text{out}}^{(0)}$. Given that the $S$-matrix preserves the inner product of BRST-closed states, the character of the inner products of $\mc{H}_{\text{in}}^{(0)}$ and $\mc{H}_{\text{out}}^{(0)}$ is the same. In particular, the character of the inner product of the BRST cohomology $\mc{H}_{\text{phys}}$ is the same as that of $\mc{H}_{\text{in}/\text{out}}^{(0)}$.

From this observation, one deduces that the inner product of the BRST cohomology $\mc{H}_{\text{phys}}$ is positive definite if and only if every physical asymptotic field $\phi^{\text{as.}}(x)$ -- i.e. every asymptotic field $\phi^{\text{as.}}(x)$ such that $[Q_{B},\phi^{\text{as.}}(x)]_{\mp}=0$, and such that $\phi^{\text{as.}}(x)\not\propto[Q_{B},\psi^{\text{as.}}(x)]_{\pm}$ for some $\psi^{\text{as.}}(x)$ -- creates and annihilates positive-norm states.

The unitarity of the electroweak sector of the Standard Model, minus quarks, quantized within the KO framework, was proven by Kugo and Ojima in \cite{KO78a,KO79b,KO79c} under the customary assumption that the nonperturbative structure of its asymptotic fields is the same as the one provided by perturbation theory\footnote{This assumption lies at the heart of the perturbative formulation of the electroweak sector itself.}. Under this assumption, one can check explicitly that the asymptotic fields of the two massless transverse photon polarizations, of the three massive transverse polarizations of the $W/Z$ bosons, of leptons, and of the Higgs boson, all belong to the BRST cohomology, and all create positive-norm states.

In QCD, by contrast, the perturbative assumption is well known to break down. The reason is twofold. First of all, the spectrum of QCD contains stable bound states whose asymptotic fields are associated with non-elementary, composite operators. Second, KO proved that -- again in their framework and under customary assumptions -- if gluons remain massless as predicted by perturbative QCD, the asymptotic fields associated with the two massless transverse gluon polarizations belong to the BRST cohomology \cite{KO79c}. This would imply the existence of massless gluons among the \textit{physical asymptotic states} of the theory, in contradiction with gluon confinement. Obviously, no such particles are observed experimentally.

In QCD, the binary alternative to the perturbative scenario is that of dynamical gluon mass generation: the infrared dynamics of the strong interactions removes the massless transverse pole present in the tree-level gluon propagator, thus eliminating massless transverse gluons from the asymptotic states of QCD. This idea dates back to the work of J. Schwinger \cite{Sch62a}, and it was considered extremely promising by Kugo and Ojima \cite{KO79d,KO79a,Kug78}: as discussed in the Introduction, dynamical gluon mass generation provides an argument not only for the confinement of gluons, but -- via the Gauss law -- for the confinement of \textit{all color}, in the sense of the absence of colored states from the asymptotic space $\mc{H}_{\text{in}/\text{out}}$ of the theory.

Yet, as soon as KO tried to implement this idea in their framework, they were met with the difficulty that the global gauge (color) charge $Q^{a}$ is spontaneously broken by the first parent $\chi^{a}$ of the elementary BRST quartet \cite{KO79d,KO79a}. From the get-go, this difficulty halted any progress toward establishing the unitarity of QCD and, more generally, towards a complete formulation of dynamically massive, nonperturbative QCD, within the KO framework\footnote{As we mentioned in the Introduction, the criterion $1+u=0$ that would have purportedly avoided spontaneous symmetry breaking, asides from having been derived from flawed arguments and an incorrect equation, is also plainly not verified \cite{SS96,Cuc97,LSWP98a,LSWP98b,BBLW00,BBLW01,BHLP04,SIMS05,CM07,CM08a,CM08b,BIMS09,BBMM06,ISI09,BMM10,ABBC12,BLLM12,OS12,BBCO15,CDMO18,AN04,AP06,ABP08,NF00,FN03,FN06a,Ste06,ABP09,ABFP24}.}.

\section{Spontaneous symmetry breaking in the linear covariant gauges of Yang-Mills theory}\label{sec:ssblcg}

In this section, we discuss the topic of spontaneous symmetry breaking in the linear covariant gauges of Yang-Mills theory. In Sec.~\ref{sec:koanal} we review the classic analysis by Kugo and Ojima and highlight its flaws. In Sec.~\ref{sec:resgtgb} we present our preliminary analysis and we introduce our first assumption on the behavior of QCD under color symmetry.

\subsection{The analysis of Kugo and Ojima and its flaws}\label{sec:koanal}

In \cite{KO79d,KO79a}, Kugo and Ojima tried to circumvent the difficulties arising from the spontaneous breaking of the global gauge charge $Q^{a}$ in dynamically massive QCD by formulating their renowned criterion $1+u=0$. In the present section, we rederive the criterion and we show that it is based on false premises, a flawed line of reasoning, and one trivial but crucial error.\\

Consider the Maxwell-Yang-Mills equations in the form given by Eq.~\eqref{feqojima}:
\begin{equation}
   \partial^{\nu}F^{a}_{\nu\mu}+gJ^{a}_{\mu}-\{Q_{B},D_{\mu}\cbar^{a}\}=0\ .
\end{equation}
This equation remains valid in the presence of matter fields $\phi$ if the latter are minimally coupled to the gauge field \cite{Oji78,NO90},
\begin{equation}
   \partial_{\mu}\phi\to D_{\mu}\phi=\partial_{\mu}\phi-ig A_{\mu}^{a}\tau_{a}\phi
\end{equation}
in the Lagrangian, where the $\tau_{a}$'s are matrices realizing a representation of the gauge group, $[\tau_{a},\tau_{b}]=if_{abc}\tau_{c}$; in this case, the global gauge current $J_{\mu}^{a}$ is the sum of the gauge sector's contribution, Eq.~\eqref{colorcurr}, and of the matter contribution,
\begin{equation}
   J_{\mu}^{a}=(J_{\mu}^{a})_{\text{gauge}}+(J_{\mu}^{a})_{\text{matter}}\ .
\end{equation}
The discussion that follows is therefore quite general\footnote{It also applies to the Stueckelberg model, despite its coupling not being minimal. There $g J_{\mu}$ is given by $m^{2}A_{\mu}-m\partial_{\mu}\phi$ -- see Appendix~\ref{app:stschw}.}.

For the purposes of the present discussion, let us define dynamical gauge-boson mass generation by the following criterion: we say that gauge bosons acquire a mass if the field-strength operator $F_{\mu\nu}^{a}$ has no massless spectrum. In Abelian Yang-Mills theories, one has $F_{\mu\nu}=\partial_{\mu}A_{\nu}-\partial_{\nu}A_{\mu}$, and this criterion is equivalent to the condition that the operator $A_{\mu}$ have no massless \textit{transverse} spectrum. In non-Abelian Yang-Mills theories, the equivalence is reasonable but not immediate. Kugo and Ojima never considered the case in which this equivalence does not hold, so neither will we in the present section. We also ignore the case in which some gauge bosons acquire a mass and some others do not, for it is irrelevant to our discussion.

Let $\ket{\Psi(p^{2}=0)}$ be a massless one-particle state. In view of asymptotic completeness, $\ket{\Psi(p^{2}=0)}$ can be identified with an asymptotic state. If gauge bosons acquire a mass, by definition
\begin{equation}
   \bra{\Omega}F_{\mu\nu}^{a}\ket{\Psi(p^{2}=0)}=0\ ,
\end{equation}
whence it follows that the matrix element of the global gauge current $J_{\mu}^{a}$ between the vacuum and $\ket{\Psi(p^{2}=0)}$ is given by
\begin{align}\label{deq603}
   \notag&\bra{\Omega}gJ_{\mu}^{a}\ket{\Psi(p^{2}=0)}=\bra{\Omega}\{Q_{B},D_{\mu}\cbar^{a}\}\ket{\Psi(p^{2}=0)}=\\
   &\qquad=\bra{\Omega}D_{\mu}\cbar^{a}\,Q_{B}\ket{\Psi(p^{2}=0)}\ .
\end{align}
Since $Q_{B}\ket{\Psi(p^{2}=0)}$ is a one-particle asymptotic state, using Eq.~\eqref{1psme}, we can compute the above matrix element as
\begin{align}\label{deq723}
   &\bra{\Omega}g J_{\mu}^{a}\ket{\Psi(p^{2}=0)}=\\
   \notag&\qquad=Z^{1/2}_{D\cbar}\bra{\Omega}(D_{\mu}\cbar^{a})^{\text{as.}}\,Q_{B}\ket{\Psi(p^{2}=0)}\ .
\end{align}

Next, we identify $Z^{1/2}_{D\cbar}\,(D_{\mu}\cbar^{a})^{\text{as.}}$. The operator $D_{\mu}\cbar^{a}$ reads
\begin{equation}
   D_{\mu}\cbar^{a}=\partial_{\mu}\cbar^{a}+gf^{abc}A_{\mu}^{b}\cbar^{c}\ .
\end{equation}
Ignoring other potential asymptotic contributions, like KO did in their analysis, the asymptotic limit $t\to\pm\infty$ of the first term in the above equation is given by Eq.~\eqref{quartetasym},
\begin{equation}\label{deq943}
   \partial_{\mu}\cbar^{a}\to \widetilde{Z}_{3}^{1/2}\,\partial_{\mu}\gammabar^{a}\ .
\end{equation}
As for the second term, since the asymptotic limit of operator $gf^{abc}A_{\mu}^{b}\cbar^{c}$ may receive a contribution from the asymptotic field $\partial_{\mu}\gammabar^{a}$, we \textit{define} a coefficient $u$ such that
\begin{equation}\label{deq429}
   gf^{abc}A_{\mu}^{b}\cbar^{c}\to \widetilde{Z}_{3}^{1/2}\,u\,\partial_{\mu}\gammabar^{a}\ .
\end{equation}
Observe that, by definition, $u$ is a ratio between LSZ coefficients,
\begin{equation}
   u=\frac{Z_{gA\times \cbar}^{1/2}}{\widetilde{Z}_{3}^{1/2}}\ ,
\end{equation}
where $(A_{\mu}\times \cbar)^{a}=f^{abc}A_{\mu}^{b}\cbar^{c}$; therefore, except for Abelian theories, in which $u=0$ by virtue of $f^{abc}=0$\footnote{In the Abelian case, ghosts are free fields, so one also has $\widetilde{Z}_{3}^{1/2}=1$.}, $u$ is an ill-defined quantity. Putting together Eqs.~\eqref{deq943} and \eqref{deq429}, we obtain
\begin{equation}\label{deq777}
   Z^{1/2}_{D\cbar}\,(D_{\mu}\cbar^{a})^{\text{as.}}=\widetilde{Z}_{3}^{1/2}\,(1+u)\,\partial_{\mu}\gammabar^{a}\ .
\end{equation}
In KO's original derivation, the LSZ coefficients of the elementary BRST quartet are left implicit: the quantity that appears in their equations is
\begin{equation}
   (\partial_{\mu}\gammabar^{a})_{\text{KO}}=\widetilde{Z}_{3}^{1/2}\,\partial_{\mu}\gammabar^{a}\ .
\end{equation}
Using Eq.~\eqref{deq777}, we rewrite Eq.~\eqref{deq723} as
\begin{align}
   &\bra{\Omega}gJ_{\mu}^{a}(x)\ket{\Psi(p^{2}=0)}=\\
   \notag&\qquad=\widetilde{Z}_{3}^{1/2}(1+u)\bra{\Omega}\partial_{\mu}\gammabar^{a}(x)\,Q_{B}\ket{\Psi(p^{2}=0)}\ .
\end{align}

In \cite{KO79d,KO79a}, Kugo and Ojima observed that the spectrum of Yang-Mills theory contains a state that renders the RHS of the above equation different from zero: the one-particle state of the first parent $\chi^{a}$ of the elementary BRST quartet. To show this, they chose
\begin{equation}
   \ket{\Psi(p^{2}=0)}=(\chi^{b}(y))_{\text{KO}}\ket{\Omega}\ ,
\end{equation}
where
\begin{equation}
   (\chi^{b}(y))_{\text{KO}}=Z_{\chi}^{1/2}\,\chi^{b}(y)\ .
\end{equation}
Indeed, using Eq.~\eqref{emquartet0}, we see that
\begin{equation}
   Q_{B}(\chi^{b}(y))_{\text{KO}}\ket{\Omega}=-i\widetilde{Z}_{3}^{-1/2}\,\gamma^{b}(y)\ket{\Omega}\ ,
\end{equation}
from which
\begin{align}
   &\bra{\Omega}gJ_{\mu}^{a}(x)(\chi^{b}(y))_{\text{KO}}\ket{\Omega}=\\
   \notag&\qquad=-i(1+u)\,\partial_{\mu}^{(x)}\bra{\Omega}\gammabar^{a}(x)\,\gamma^{b}(y)\ket{\Omega}
\end{align}
follows. By evaluating the vacuum expectation value of a product of free ghost fields $\gammabar^{a}$ and $\gamma^{b}$, with the positive-frequency Green's function $D^{(+)}(x)$ defined as
\begin{equation}
   D^{(+)}(x)=\int\frac{d^{4}p}{(2\pi)^{3}}\ \theta(p_{0})\ \delta(p^{2})\ e^{-ipx}\ ,
\end{equation}
$\theta(x)$ being a Heaviside theta, we obtain the renowned KO Goldstone equation \cite{KO79d,KO79a,NO90} in the form
\begin{equation}\label{KOGold}
   \bra{\Omega}gJ_{\mu}^{a}(x)(\chi^{b}(y))_{\text{KO}}\ket{\Omega}=-(1+u)\,\delta^{ab}\,\partial_{\mu}D^{(+)}(x-y)\ .
\end{equation}
The spatial integral of the time component of the LHS of Eq.~\eqref{KOGold} reads
\begin{equation}
   \bra{\Omega}gQ^{a}\,(\chi^{b}(y))_{\text{KO}}\ket{\Omega}\ .
\end{equation}
Therefore, by Eq.~\eqref{KOGold}, if the RHS is different from zero, the global gauge charge $Q^{a}$ is spontaneously broken, $Q^{a}\ket{\Omega}\neq 0$, and the first parent $\chi^{a}$ of the elementary BRST quartet is the (unphysical) Goldstone boson associated with the broken symmetry. In the literature, this result is known as the ``converse of the Higgs theorem'' \cite{KO79d,KO79a}.

In \cite{KO79d,KO79a}, Kugo and Ojima put forward the following argument for the criterion $1+u=0$\footnote{See the discussion below Eq.~(6.39) in \cite{KO79a}.}. Since the spontaneous breaking of color symmetry would have severe consequences on the structure of the physical spectrum of the strong interactions, the RHS of Eq.~\eqref{KOGold} must vanish in the presence of dynamical gluon mass generation. Given that $1+u$ is the only undetermined quantity on the RHS of the equation, in dynamically massive QCD, one must have $1+u=0$.\\

The premises of the criterion $1+u=0$ are in open contradiction with the known behavior of ordinary QED \cite{KO79d,KO79a} and of its massive counterpart, the Schwinger model \cite{Ito75,Nak75,Nak76a,Nak77a,Nak77b,Nak77c}, formulated within the KO framework. Indeed, in both these theories, 1. the global gauge (electric) charge $Q$ is spontaneously broken, 2. the electric charge operator does not coincide with the global gauge charge $Q$, and 3. the spectrum does not violate U(1) symmetry. Let us review these results starting from ordinary QED, as analyzed by Kugo and Ojima themselves in \cite{KO79d,KO79a}.

In \cite{KO79d,KO79a}, KO never say explicitly that the global gauge charge $Q$ of ordinary QED is spontaneously broken, nor do they write down a corresponding Goldstone equation. On the contrary, they write\footnote{See the discussion around and below Eq.~(6.28a) in \cite{KO79a}.}: ``It may happen, however, that this equation ($6\cdot28$) is not satisfied in spite of the absence of the spontaneous breakdown, as is seen explicitly in the case of QED.$^{3),8)}$ This is due to the massless contribution from $\alpha_{a}^{\,A}\partial_{\mu}\beta^{a}$ remaining in $g J_{\mu}^{A}$ with some weight $\zeta$''. The equation they make reference to is
\begin{equation}
   \bra{\Omega}g J_{\mu}^{a}(x)\ket{\Psi(p^{2}=0)}=0\ ,
\end{equation}
and it not being satisfied in QED means that, with $\psi$ the electron field, $J_{\mu}=\psibar\gamma_{\mu}\psi$ the electric current, and $e$ QED's coupling constant,
\begin{equation}
   \bra{\Omega}e J_{\mu}(x)\ket{\Psi(p^{2}=0)}\neq0\ .
\end{equation}
The spatial integral of the time component of this equation reads
\begin{equation}
   \bra{\Omega}e Q\ket{\Psi(p^{2}=0)}\neq0\ .
\end{equation}
Therefore, in contrast with the wording they use, KO actually report that, in their framework, the global gauge charge of QED is spontaneously broken, $Q\ket{\Omega}\neq 0$. This phenomenon was first discovered by J. A. Swieca within the Gupta-Bleuler quantization of QED \cite{Swi67}; as we recalled in the previous section, that framework is the precursor of the KO framework.

In \cite{KO79d,KO79a}, the Goldstone equation that expresses the spontaneous breaking of $Q$ in ordinary QED is never reported explicitly, nor its existence is openly acknowledged. Let us reconstruct it from the description provided therein\footnote{In Sec.~\ref{sec:resgtgb} we will rederive the equation using the known asymptotic limit of the operators $\partial^{\nu}F_{\nu\mu}$ and $\partial_{\mu}B$.}.

KO say that a massless contribution of the form $\partial_{\mu}\beta$ remains in $eJ_{\mu}$ with weight $\zeta$. In other words, asymptotically,
\begin{equation}\label{deq622}
   eJ_{\mu}\to \zeta (\partial_{\mu}\beta)_{\text{KO}}\ ,
\end{equation}
where the subscript ``KO'', again, indicates that the asymptotic field is defined with an implicit LSZ coefficient. Shortly after\footnote{See the discussion below Eq.~(6.29b) in \cite{KO79a}.}, they report that $\zeta=1-Z_{3}$, where, in QED \cite{KO79c},
\begin{equation}
   Z_{3}=\frac{1}{1+i\widehat{\Pi}(0)}\ ,
\end{equation}
$\widehat{\Pi}(p^{2})$ being the adimensional photon polarization -- defined from the transverse photon propagator $\Delta(p^{2})$ as
\begin{equation}\label{photprop}
   \Delta(p^{2})=\frac{-i}{p^{2}[1+i\widehat{\Pi}(p^{2})]}\ .
\end{equation}
Therefore, according to KO,
\begin{align}\label{deq963}
   &\bra{\Omega}e J_{\mu}(x)\ket{\Psi(p^{2}=0)}=\\
   \notag&\qquad=\frac{i\widehat{\Pi}(0)}{1+i\widehat{\Pi}(0)}\,\braket{\Omega|(\partial_{\mu}\beta(x))_{\text{KO}}|\Psi(p^{2}=0)}\ .
\end{align}
The coefficient on the RHS of this equation also appears in Swieca's paper \cite{Swi67}.

What is the state that makes the RHS of Eq.~\eqref{deq963} different from zero? Naturally, it is the one-particle state created by the only asymptotic field that, in QED, has non vanishing commutation relations with $\beta$. Namely, the longitudinal photon polarization -- once again, the first parent of the elementary BRST quartet \cite{KO79c}. Therefore, just like in the massive case, we must choose
\begin{equation}
   \ket{\Psi(p^{2}=0)}=(\chi(y))_{\text{KO}}\ket{\Omega}\ .
\end{equation}

Given that $\beta$ and $\chi$ are massless free fields with reciprocal LSZ coefficients, whose associated one-particle states -- see Eq.~\eqref{emprods0} -- have negative mixed inner product,
\begin{equation}
   \braket{\Omega|(\beta(x))_{\text{KO}}(\chi(y))_{\text{KO}}|\Omega}=-D^{(+)}(x-y)\ .
\end{equation}
It follows that, in QED, the Goldstone equation expressing the spontaneous breaking of the global gauge charge $Q$ reads
\begin{align}\label{qedgold}
   &\bra{\Omega}e J_{\mu}(x)(\chi(y))_{\text{KO}}\ket{\Omega}=-\frac{i\widehat{\Pi}(0)}{1+i\widehat{\Pi}(0)}\,\partial_{\mu}D^{(+)}(x-y)\ .
\end{align}

Now, since $u=0$ in Abelian theories, Eq.~\eqref{qedgold} is the same equation as Eq.~\eqref{KOGold}, but for the nonzero prefactor
\begin{equation}\label{zetafac}
   \zeta=\frac{i\widehat{\Pi}(0)}{1+i\widehat{\Pi}(0)}\ .
\end{equation}
Moreover, since in massive theories, using Eqs.~\eqref{emquartet0}, \eqref{deq603}, and \eqref{deq777},
\begin{equation}\label{deq661}
   gJ_{\mu}^{a}\to (1+u)\,(\partial_{\mu}\beta^{a})_{\text{KO}}\ ,
\end{equation}
the same holds for Eq.~\eqref{deq622}, in reference to the last equation. What KO report in \cite{KO79d,KO79a}, therefore, is that \textit{the global gauge charge of ordinary QED, in which no photon mass generation occurs, displays the exact same spontaneous breaking as Yang-Mills theories with gauge-boson mass generation, except for the prefactor $\zeta$ in Eq.~\eqref{zetafac}}. Note that the equations of the massive Abelian case, Eq.~\eqref{KOGold} and \eqref{deq661} with $u=0$, can be formally obtained from Eqs.~\eqref{qedgold} and \eqref{deq622} by setting $\zeta=1$. And in fact, if the photon acquires a mass, $\Delta(0)$ in Eq.~\eqref{photprop} is finite, the corresponding adimensional polarization is infinite, $i\widehat{\Pi}(0)=\infty$, and, formally $\zeta=1$.

The case of the massive photon is realized in the Schwinger model \cite{Sch62a,Sch62b,LS71,KS74,CKS74,Ito75,Nak75,Nak76a,Nak77a,Nak77b,Nak77c}, well known to be confining. Within the KO framework \cite{Ito75,Nak75,Nak76a,Nak77a,Nak77b,Nak77c}, the global gauge charge $Q$ of the Schwinger model was first reported to be spontaneously broken by K. R. Ito and N. Nakanishi \cite{Ito75,Nak75}\footnote{These works predate the formulation of the KO framework, but they nonetheless satisfy the postulates of the framework -- see Appendix~\ref{app:stschw}.}. As we said before, the first parent of the elementary BRST quartet, in this case, is a massless fermion-antifermion state -- see Appendix~\ref{app:stschw}.

The spontaneous breaking of the global gauge charge $Q$ does not lead to a violation of the U(1) symmetry of the spectrum in either ordinary QED, or the Schwinger model. This is obvious for ordinary QED. As for the Schwinger model, within the KO framework, its spectrum contains the massive physical photon and the unphysical elementary BRST quartet -- see Appendix~\ref{app:stschw}. The photon and the quanta of the NL, ghost, and antighost fields are chargeless like in any Abelian theory; the fermion-antifermion first parent state lives in a neutral channel, hence its correct charge assignment is zero. When formulated in the KO framework, all states in the Schwinger model are therefore arranged into neutral electric charge singlets. This should be contrasted with models like the Abelian Higgs model, in which the U(1) symmetry of the spectrum is truly violated: there the Goldstone and Higgs bosons, originally part of the charged Higgs field, acquire different masses as a consequence of the condensation of the Higgs field.

Both in ordinary QED and in the Schwinger model, because of the Goldstone equations Eqs.~\eqref{KOGold} and \eqref{qedgold}, the global gauge charge $Q$ acts nontrivially on the first parent of the elementary BRST quartet, despite them being chargeless. Evidently, in neither theory the global gauge charge $Q$ acts as a mere electric charge operator. This was acknowledged by KO in \cite{KO79d,KO79a} when discussing massless theories, where for the latter they defined a new charge operator.\\

The logical conclusion to draw from the spontaneous breaking of the global gauge charge in both massless and massive QED is clear: in the LCGs of Yang-Mills theory, one cannot infer anything about either gauge-boson mass generation or the violation of electric/color symmetry of the spectrum from the spontaneous breaking of the global gauge charge alone. The latter occurs even when gauge bosons remain massless, and even when the underlying symmetry of the theory -- in the case of QED, global U(1) symmetry -- is preserved by its spectrum. These considerations invalidate the premises on which the criterion $1+u=0$ is formulated and, partly, the analysis of mass generation and symmetry breaking presented by Kugo and Ojima in \cite{KO79d,KO79a}. Let us illustrate the second point with two concrete examples.

First, in Remark $ii)$ below Eq.~(6.26) of \cite{KO79a}, KO state that ``[if $1+u\neq 0$] the spontaneous breakdown of a charge $Q^{A}$ is \textit{in one-to-one correspondence} with the appearance of a \textit{massive} gauge boson $U_{\mu}^{A}$'' (emphasis theirs). This statement is \textit{false}: in ordinary QED, $Q$ is spontaneously broken, $1+u=1$, yet no photon mass generation occurs\footnote{\label{fn:Uexist}En passing, we should also note that, in the context of that quote, KO did not \textit{prove} the existence of the massive gauge boson: throughout Sec.~6.1 of \cite{KO79a}, they \textit{assumed} it exists, see Eq.~(6.11) therein. As far as we know, no nonperturbative proof of the existence of massive gauge-boson asymptotic states in Higgs theories is available to date.}.

Second, below Eq.~(6.29b) of \cite{KO79a}, they state that ``[if $1+u\neq0$] the case $\zeta=1$ corresponds to the spontaneous breakdown'', the implication being that $\zeta\neq1$ corresponds to the absence of spontaneous symmetry breaking. This implication is made explicit in the quote we reported before, in which KO say ``in spite of the absence of spontaneous breakdown, as is seen explicitly in the case of QED''. This statement is thus also \textit{false}: in ordinary QED, $1+u=1$, $\zeta\neq 1$, and the global gauge charge is nevertheless spontaneously broken.

On what basis did Kugo and Ojima make those statements in the first place? The answer is threefold. First of all, throughout their work, Kugo and Ojima conflate the spontaneous breaking of the global gauge charge with the spontaneous breaking of the underlying symmetry of the theory -- e.g., U(1) symmetry in QED or color symmetry in QCD. The known behavior of QED unequivocally shows there is a clear distinction between these two notions.

Second, they assume there is no symmetry breaking if gauge bosons remain massless, despite the clear evidence to the contrary. This assumption is essentially an identification: Ref.~\cite{KO79c}, in which they analyze the case of Yang-Mills theory with no gauge-boson mass generation, is titled ``\textit{Pure Yang-Mills theories without spontaneous symmetry breaking}''.

Third, in Secs.~6.1 and 6.2 of \cite{KO79a}, they pursue a flawed line of reasoning. The latter can be summarized as follows. Starting from the (wrong) implicit assumption of no symmetry breaking in the absence of mass generation, KO study the case in which a mass is generated, and they find that the global gauge charge is spontaneously broken. (This is unsurprising: it is broken even in the absence of mass generation.) Their non-Abelian Goldstone equation has a factor of $1+u$ on its RHS, so they cave in exceptions for the case in which $1+u=0$, in an attempt to identify a criterion by which color symmetry is not broken in dynamically massive QCD. Then, when confronted with the fact that the global gauge charge is spontaneously broken even in massless theories like QED\footnote{K. Aoki made them aware of this fact, presumably through the work of Swieca -- see the Acknowledgments section of \cite{KO79b}.}, they refuse to acknowledge symmetry breaking for what it is. Instead, they use the periphrasis ``It may happen, however, that this equation ($6\cdot28$) is not satisfied in spite of the absence of the spontaneous breakdown [...] due to the massless contribution from $\alpha_{a}^{\,A}\partial_{\mu}\beta^{a}$''. In order to show that a well-defined charge operator exists in massless theories to preserve the symmetry, they introduce for these a new, well-defined symmetry generator, which however is only meaningful if $\zeta\neq 1$\footnote{For context, other authors provide different solutions to the problems posed by the ``Swieca phenomenon'', see e.g. G. Morchio and F. Strocchi in \cite{MS03}.}. Having formulated the criterion $1+u=0$ for the non-Abelian massive case, and having drawn a distinction between $\zeta=1$ and $\zeta\neq 1$ which solely rests on the redefinition they themselves gave of the charge, they do not consider the possibility that the spectrum of dynamically massive theories might respect the symmetry regardless of the spontaneous breaking of the global gauge charge (as it happens in both massless and massive QED). Finally, they declare that their redefinition of the charge operator ``preserves the whole arguments made previously'' by replacing one charge with another in the arguments and in modified equations\footnote{See the discussion below Eq.~(6.29b).}.

The problem with the last statement is that the ``arguments made previously'', which in the meantime led to the introduction of the criterion $1+u=0$, specifically applied to the global gauge charge $Q^{a}$, and specifically moved from the premise that \textit{the global gauge charge generates color transformations, therefore it must not be spontaneously broken to preserve the symmetry of the spectrum}. If 1. said charge is spontaneously broken regardless of gauge-boson mass generation, 2. it is broken regardless of whether the symmetry of the spectrum is preserved or not, and 3. the actual charge operator, in any case, does not coincide with the global gauge charge, then it is not at all clear why the condition $1+u=0$ should be satisfied. On the contrary, the criterion appears out of place, given that, within the KO framework, $1+u=0$ does not hold in any known theory.

``Confinement'' is not an argument in favor of the criterion either: $1+u=0$ does not hold in the Schwinger model, whose exact solutions in the KO framework are known \cite{Ito75,Nak75,Nak76a,Nak77a,Nak77b,Nak77c}, which confines electric charge, yet has $1+u=1$ like every Abelian theory. In \cite{KO79a}, prompted by Nakanishi\footnote{See Remark $ii)$ below Eq.~(6.26) of \cite{KO79a}.}, who had formulated those solutions, Kugo and Ojima explicitly exclude the Schwinger mechanism as a valid mechanism of dynamical gluon mass generation on the grounds that the global gauge charge of the Schwinger model is spontaneously broken \cite{Ito75,Nak75,Nak76a,Nak77a,Nak77b,Nak77c}. Yet they do not provide a single example of a Yang-Mills theory in which the global gauge charge is not spontaneously broken, and they do not explain why this form of symmetry breaking is supposed to conflict with confinement. If anything, its occurrence in the Schwinger model shows that it does not.\\

Finally, in addition to resting on flawed arguments, the KO criterion is also derived from an incorrect equation. Indeed, as we mentioned earlier, the asymptotic field $(\chi^{a})_{\text{KO}}$ used in \cite{KO79d,KO79a} to evaluate the matrix element of the global gauge current is defined with an implicit LSZ coefficient, hence it creates asymptotic states which are not normalized -- see the discussion at the end of Sec.~\ref{sec:framework}.

The correct equation is obtained using, instead, the normalized asymptotic field $\chi^{a}=Z_{\chi}^{-1/2}(\chi^{a})_{\text{KO}}$. In the massive case, it reads
\begin{align}\label{KOGoldCorr}
   &\bra{\Omega}J_{\mu}^{a}(x)\,\chi^{b}(y)\ket{\Omega}=\\
   \notag&\qquad=-g^{-1}Z_{\chi}^{-1/2}(1+u)\,\delta^{ab}\,\partial_{\mu}D^{(+)}(x-y)\ .
\end{align}
To establish whether $Q^{a}$ is spontaneously broken or not, one must determine the value of the combination on the RHS of Eq.~\eqref{KOGoldCorr},
\begin{equation}\label{deq915}
   Z_{\chi}^{-1/2}(1+u)/g\ ,
\end{equation}
\textit{as a whole}: $1+u=0$, by itself, is \textit{not} a sufficient condition for the absence of spontaneous symmetry breaking.

As we will see in Sec.~\ref{sec:ssb}, the combination in Eq.~\eqref{deq915} is finite, Renormalization-Group invariant, and not equal to zero. This outcome is fully expected: as we saw above, in Abelian theories, the global gauge charge is spontaneously broken both in the massless and in the massive case, regardless of whether the underlying symmetry is preserved or not. There is no reason to believe non-Abelian theories should behave differently.\\

In conclusion, through false premises, flawed arguments, and an incorrect equation, Kugo and Ojima formulated a criterion -- $1+u=0$ -- that lacks any valid justification. The spontaneous breaking of the global gauge charge, \textit{per se}, poses no issues for the consistency of dynamically massive QCD: as exemplified by QED, it does not automatically entail the violation of the symmetry of the spectrum.

Still, if we wish to address color confinement, we must identify the true color charge operator of dynamically massive QCD. Doing so, however, is impossible without first answering a question that was never asked in the original analysis of Kugo and Ojima: why is the global gauge charge of the linear covariant gauges of Yang-Mills theory always spontaneously broken, in the first place?

\subsection{Residual gauge symmetry, spontaneous symmetry breaking, color symmetry, and Goldstone bosons}\label{sec:resgtgb}

The aim of this section is to provide a preliminary analysis of the spontaneous breaking of the global gauge charge in the Kugo-Ojima framework, postponing a detailed analysis in the context of dynamically massive QCD to Sec.~\ref{sec:symbreak}. Starting from a review of residual gauge transformations in Abelian theories, we clarify the origin of this phenomenon. Then we discuss symmetry preservation in the presence of a spontaneously broken global gauge charge, and we make our first assumptions about the behavior of QCD. Finally, we make some remarks on the cohomological role of Goldstone bosons in Higgs mechanisms.\\

Consider the theory of the free photon field quantized in linear covariant gauges, with field equations
\begin{align}
   &\partial^{\nu}F_{\nu\mu}-\partial_{\mu}B=0\ ,\qquad\partial^{\mu}A_{\mu}+\xi B=0\ ,\\
   \notag&\qquad\qquad\partial^{2}c=0\ ,\qquad\partial^{2}\cbar=0\ .
\end{align}
As in every Abelian theory, the NL field $B$ is a massless free field, $\partial^{2}B=0$. In particular, $B$ coincides with its asymptotic field, and it admits an ordinary expansion in creation and annihilation operators
\begin{equation}
   B(x)=\int\frac{d^{3}{\bf p}}{(2\pi)^{3}}\,\frac{1}{\sqrt{2\hat{p}^{0}}}\, \left\{\beta_{{\bf p}}\,e^{-i\hat{p}x}+\beta_{{\bf p}}^{\dagger}\,e^{i\hat{p}x}\right\}\ ,
\end{equation}
where $\hat{p}=(|{\bf p}|,{\bf p})$.

Let $C[\Lambda]$ be the operator \cite{Nak72,NO90}
\begin{equation}
   C[\Lambda]=e^{-1}\int d^{3}{\bf x}\ \left(\Lambda \dot{B}-\dot{\Lambda}B\right)\ ,
\end{equation}
where the dot denotes a time derivative and $\Lambda$ is a function satisfying $\partial^{2}\Lambda=0$ that decreases sufficiently fast at spatial infinity. For later convenience, we have introduced a fictitious coupling $e$ in the equation. $C[\Lambda]$ is a conserved charge,
\begin{equation}
   \frac{d}{dt}\,C[\Lambda]=0\ ,
\end{equation}
it is BRST-exact,
\begin{equation}
   C[\Lambda]=\left\{Q_{B},e^{-1}\int d^{3}{\bf x}\ \left(\Lambda\dot{\cbar}-\dot{\Lambda}\cbar\right)\right\}\ ,
\end{equation}
and it is obtained from a current $j_{\mu}^{(\Lambda)}$ defined as
\begin{equation}\label{deq451}
   j_{\mu}^{(\Lambda)}=e^{-1}\left(\Lambda\,\partial_{\mu}B-\partial_{\mu}\Lambda\, B\right)\ ,\quad C[\Lambda]=\int d^{3}{\bf x}\ j_{0}^{(\Lambda)}\ .
\end{equation}
As an operator, it is well-defined. Indeed, by introducing a three-dimensional Fourier transform for $\Lambda$,
\begin{equation}
   \Lambda(x)=\int\frac{d^{3}{\bf p}}{(2\pi)^{3}}\,\frac{1}{\sqrt{2\hat{p}^{0}}}\, \left\{\Lambda_{{\bf p}}\,e^{-i\hat{p}x}+\Lambda_{{\bf p}}^{\star}\,e^{i\hat{p}x}\right\}\ ,
\end{equation}
we see that
\begin{equation}
   C[\Lambda]=ie^{-1}\ \int\frac{d^{3}{\bf p}}{(2\pi)^{3}}\ \left(\Lambda_{{\bf p}}\beta_{{\bf p}}^{\dagger}-\Lambda_{{\bf p}}^{\star}\beta_{{\bf p}}\right)\ .
\end{equation}

Using Eq.~\eqref{cancomm} adapted to the Abelian case, and the field equations, it is easy to see that $C[\Lambda]$ generates the transformations
\begin{equation}\label{deq455}
   [iC[\Lambda],A_{\mu}(x)]=-e^{-1}\partial_{\mu}\Lambda(x)\ ,\qquad \partial^{2}\Lambda=0\ ,
\end{equation}
whereas $B$, $c$ and $\cbar$ are left invariant.  Since $C[\Lambda]$ preserves the gauge-fixing condition,
\begin{equation}
   [iC[\Lambda],\partial^{\mu}A_{\mu}+\xi B]=-e^{-1}\partial^{2}\Lambda=0\ ,
\end{equation}
Eq.~\eqref{deq455} tells us that the operator generates \textit{residual} gauge transformations. These transformations are symmetries of the field equations. Yet, they are not symmetries of the vacuum:
\begin{equation}
   C[\Lambda]\ket{\Omega}=ie^{-1}\ \int\frac{d^{3}{\bf p}}{(2\pi)^{3}}\ \Lambda_{{\bf p}}\beta_{{\bf p}}^{\dagger}\,\ket{\Omega}\neq 0\ .
\end{equation}
It follows that $C[\Lambda]$ is spontaneously broken as an operator \cite{Nak72,NO90}.

Does the spontaneous breaking of $C[\Lambda]$ entail the violation of symmetries of the theory? Of course not. Quite the contrary, spontaneous breakdown is a structural property of gauge transformations. In fact, under the latter, the photon field acquires a term which is a function,
\begin{equation}\label{deq888}
   A_{\mu}\to A_{\mu}+e^{-1}\partial_{\mu}\Lambda\ .
\end{equation}
This, in turn, translates to a nontrivial transformation
\begin{equation}
   0=\avg{A_{\mu}}\to \avg{A_{\mu}}+e^{-1}\partial_{\mu}\Lambda=e^{-1}\partial_{\mu}\Lambda\neq 0
\end{equation}
of the vacuum expectation value $\avg{A_{\mu}}$ of $A_{\mu}$. A transformation that modifies the vacuum expectation value of an operator can only be implemented through a charge that does not annihilate the vacuum -- that is, by a spontaneously broken charge. The spontaneous breaking of $C[\Lambda]$ is thus a trivial property of the gauge transformations it generates.

Do the broken transformations generated by $C[\Lambda]$ have an associated Goldstone boson? The answer is yes: the Goldstone boson is the longitudinal photon polarization -- that is, the first parent $\chi$ of the elementary BRST quartet of the free photon theory. To see this, start from the commutation relations
\begin{equation}\label{deq399}
   [\chi(x),B(y)]=-iD(x-y)
\end{equation}
which we derive in Appendix~\ref{app:photon}. In the above equation, $D(x)$ is the commutator of a free massless scalar field,
\begin{equation}
   D(x)=D(x;m^{2}=0)\ ,
\end{equation}
where $D(x;m^{2})$ is defined in Eq.~\eqref{ffcommut}. Using Eq.~\eqref{deq399}, evaluate the equal-time commutators
\begin{align}
   [\chi(x),B(y)]|_{x^{0}=y^{0}}&=0\ ,\\
   \notag[\chi(x),\dot{B}(y)]|_{x^{0}=y^{0}}&=-i\delta^{(3)}({\bf x}-{\bf y})\ ,
\end{align}
and use them to compute
\begin{equation}
   [iC[\Lambda],\chi(x)]=-e^{-1}\Lambda(x)\ .
\end{equation}
The Goldstone equation
\begin{equation}\label{PhotGold}
   \braket{\Omega|[C[\Lambda],\chi(x)]|\Omega}=ie^{-1}\Lambda(x)
\end{equation}
then follows.

The fact that the Goldstone boson associated with $C[\Lambda]$ is a polarization -- i.e. one of four components of a vector field~-- rather than a scalar is a remarkable property of the transformation. This is possible because, although $\chi(x)$ is not invariant under Lorentz transformations, the polarizations it transforms into all commute with $B(x)$ -- see Appendix~\ref{app:photon} -- hence they do not contribute to Eq.~\eqref{PhotGold} following a Lorentz transformation.

Consider now the limit in which $\Lambda$ becomes a constant, $\Lambda\to\lambda=\text{const}$,
\begin{equation}
   \Lambda_{{\bf p}}\to\sqrt{\hat{p}^{0}/2}\ \lambda\ (2\pi)^{3}\,\delta^{(3)}({\bf p})\ .
\end{equation}
In this limit, $C[\Lambda]$ contains the creation and annihilation operators $\beta_{{\bf 0}}^{\dagger},\beta_{{\bf 0}}$, hence it is an ill-defined operator\footnote{\label{fn:symbreakilldef}The ill-definedness of operators like $\beta^{\dagger}_{{\bf 0}}$ is usually motivated by their ability to create an arbitrary number of linearly independent zero-momentum states out of the vacuum $\ket{\Omega}$, thus violating the uniqueness of the Poincar\'{e}-invariant vacuum state. Especially in a gauge theory, we tend to disagree with this framing of the issue. Regardless of the uniqueness of vacuum, there exists no Lorentz transformation boosting a quantum of momentum $(|{\bf p}|,{\bf p})$ to one of momentum $(0,{\bf 0})$, and vice-versa. Since quanta created by $\beta^{\dagger}_{{\bf 0}}$ then do not live in the same irreducible representation of the Poincar\'{e} group as those created by $\beta^{\dagger}_{{\bf p}\neq {\bf 0}}$, $\beta^{\dagger}_{{\bf 0}}$ does not a priori live in the same representation space as that of the field operator $\beta(x)$.},
\begin{equation}
   C[\lambda]=\lambda e^{-1}\int d^{3}{\bf x}\ \dot{B}=\frac{i\lambda e^{-1}}{2}\, [\sqrt{2\hat{p}^{0}}\, \left(\beta_{{\bf p}}^{\dagger}-\beta_{{\bf p}}\right)]_{{\bf p}=0}\ .
\end{equation}
Yet, its commutation relations remain well-defined, and nontrivial:
\begin{equation}\label{deq640}
   [C[\lambda],\chi(x)]=ie^{-1}\lambda\ ,
\end{equation}
while those with $A_{\mu}$, $B$, $c$ and $\cbar$ vanish. The current corresponding to this case is
\begin{equation}
   j_{\mu}^{(\lambda)}=\lambda\,e^{-1}\partial_{\mu}B\ ,
\end{equation}
and it can be used to evaluate Eq.~\eqref{deq640} without relying on the ill-defined operator $C[\lambda]$: we have
\begin{equation}
   [j_{0}^{(\lambda)}({\bf y},t),\chi({\bf x},t)]=i\lambda\,e^{-1}\,\delta^{(3)}({\bf y}-{\bf x})\ ,
\end{equation}
which, when integrated in $d^{3}{\bf y}$, yields Eq.~\eqref{deq640}.

Let us now consider an interacting Abelian theory quantized in linear covariant gauges, with field equations
\begin{align}
   &\partial^{\nu}F_{\nu\mu}-\partial_{\mu}B+eJ_{\mu}=0\ ,\qquad\partial^{\mu}A_{\mu}+\xi B=0\ ,\\
   \notag&\qquad\qquad\partial^{2}c=0\ ,\qquad\partial^{2}\cbar=0
\end{align}
featuring the matter current $J_{\mu}$. As in the previous section, we assume that matter is minimally coupled, so that, canonically, $J_{\mu}$ generates the U(1) transformations of the matter fields.

Since the NL field is still a massless free field, we can again introduce an operator $C[\Lambda]$,
\begin{equation}
   C[\Lambda]=e^{-1}\int d^{3}{\bf x}\ \left(\Lambda \dot{B}-\dot{\Lambda}B\right)\ .
\end{equation}
Given that the gauge sector of the interacting theory has the same canonical commutation relations as those of the free photon field, every statement we made about $C[\Lambda]$ remains valid in the interacting case, up to the point at which we identified the Goldstone boson.

Does $C[\Lambda]$ have an associated Goldstone boson also in the interacting theory? The answer is yes: the associated Goldstone boson is still the first parent of the elementary BRST quartet $\chi$, regardless of its identity. In fact, Eq.~\eqref{deq399} follows from the commutation relations
\begin{equation}
   [\chi_{{\bf p}},\beta_{{\bf k}}^{\dagger}]=-(2\pi)^{3}\,\delta^{(3)}({\bf p}-{\bf k})
\end{equation}
between the creation/annihilation operators of $\chi$ and $\beta$, which in turn follow from the first of Eq.~\eqref{quarteteq} \cite{KO78b,KO78a,KO79b,KO79c,KO79d,KO79a,Kug78}. The latter reflect Slavnov-Taylor identities which hold in the LCGs of every Yang-Mills theory.

From what we said, understanding the role of the first parent of the elementary BRST quartet in Abelian theories is straightforward: Abelian theories have a residual gauge symmetry, Eq.~\eqref{deq888}, and the first parent of the elementary BRST quartet, entering the gauge-boson field as
\begin{equation}
   A_{\mu}= \partial_{\mu}\chi+\cdots\ ,
\end{equation}
and transforming under $C[\Lambda]$ as
\begin{equation}
   \chi(x)\to\chi(x)+e^{-1}\Lambda(x)\ ,
\end{equation}
is the degree of freedom that concretely implements these transformations.

How is the invariance of physical states $\ket{\text{phys}}$ under the residual gauge transformations generated by $C[\Lambda]$ enforced in Abelian theories? The condition $C[\Lambda]\ket{\text{phys}}=0$ cannot be implemented: $C[\Lambda]$ is spontaneously broken, hence it fails to annihilate the vacuum. Physical states are instead required to be invariant under BRST transformations, $Q_{B}\ket{\text{phys}}=0$.

The connection between $Q_{B}$ and $C[\Lambda]$ is manifest: from Eq.~\eqref{brstcurr}, we see that, in the Abelian case,
\begin{equation}
   Q_{B}=-e\, C[c]\ ,
\end{equation}
where $c$ is the ghost field. When acting on the gauge-boson field, the BRST charge generates the transformation
\begin{equation}
   A_{\mu}\to A_{\mu}-e\,\theta\,\partial_{\mu}c\ ,
\end{equation}
where $\theta$ is a Grassmann number. Thanks to $\avg{c}=0$, this transformation now translates to trivial transformations of $A_{\mu}$'s vacuum expectation value
\begin{equation}
   0=\avg{A_{\mu}}\to \avg{A_{\mu}}-e\,\theta\,\partial_{\mu}\avg{c}=0\ ;
\end{equation}
as a consequence, $Q_{B}$ does not need to be -- and, in fact, it is not -- spontaneously broken. In a very concrete sense, $Q_{B}$ protects the broken symmetry generated by $C[\Lambda]$ by playing the role of its unbroken counterpart. By passing to the BRST cohomology, one achieves the goal of removing not only $\chi$, but also $B$, $c$ and $\cbar$, from the physical spectrum of the theory.\\

What is the relation between residual gauge transformations and the spontaneous breaking of the global gauge charge? To answer this question, let us inspect the Maxwell-Yang-Mills equations. After rewriting them as
\begin{equation}
   J_{\mu}=e^{-1}\,\partial^{\nu}F_{\mu\nu}+j_{\mu}^{(1)}\ ,
\end{equation}
where $j_{\mu}^{(\Lambda)}$ is defined in Eq.~\eqref{deq451}, we see that, in the interacting theory, the current $J_{\mu}$ contains a term that generates constant residual gauge transformations of $\chi$. In addition to that term, in the asymptotic limit, as we review in Appendix~\ref{app:photon}, the field-strength tensor $F_{\mu\nu}$ of a massless gauge boson contributes a similar term to the equation:
\begin{equation}
   \partial^{\mu}F_{\mu\nu}\to Z_{3}\,\partial_{\nu}B=e\,Z_{3}\,j_{\nu}^{(1)}\ .
\end{equation}
If the gauge boson acquires a mass, on the other hand, $F_{\mu\nu}$ has no massless spectrum, and it yields no such term. Overall, we find that the discrete massless spectrum of $J_{\mu}$ is entirely given by
\begin{equation}\label{deq222}
   J_{\mu}\to \zeta\,j_{\mu}^{(1)}\ ,
\end{equation}
where $\zeta=1-Z_{3}$ in the massless case, and $\zeta=1$ in the massive one -- see also Eq.~\eqref{zetafac}. This is the result Kugo and Ojima described in \cite{KO79d,KO79a}, albeit mischaracterizing $\zeta\neq 1$ as a case in which there is no symmetry breaking.

We thus see that, in an interacting Abelian theory, regardless of whether the gauge boson acquires a mass or not, the global gauge charge $Q=\int d^{3}{\bf x}\,J_{0}$ receives a contribution from $C[\Lambda=1]$. Since $C[\Lambda]$ generates spontaneously broken transformations, $Q$ does as well. The spontaneous breaking of the global gauge charge is, in fact, in one-to-one correspondence with that of $C[\Lambda]$.

The spontaneous breaking of $C[\Lambda]$ is well understood: as we saw above, it is a structural property of residual gauge transformations, unrelated to whether or not the U(1) symmetry of the theory is preserved, and to whether or not a gauge-boson mass is dynamically generated. The existence of the associated Goldstone boson -- namely, the first parent of the elementary BRST quartet -- is guaranteed by the Slavnov-Taylor identities of the theory, regardless of its underlying symmetry or dynamics.

Correspondingly, in interacting Abelian theories, the global gauge charge is always spontaneously broken in the KO framework. This phenomenon can be understood as a dynamical effect arising from coupling the current $J_{\mu}$ to the other fields: $Q$ is unbroken ($\zeta=0$) in a non-interacting theory, it is broken with coefficient $\zeta\neq0,1$ in an interacting theory with no gauge-boson mass generation, and it is broken with coefficient $\zeta=1$ in an interacting theory with gauge-boson mass generation.

The underlying U(1) symmetry of the theory can then be spontaneously broken \textit{in addition} to residual gauge symmetry. Since $Q$ generates genuine global U(1) transformations, in addition to residual gauge transformations, and since the discrete massless spectrum of $J_{\mu}$ is entirely given by $j_{\mu}^{(1)}$, if the underlying U(1) symmetry is spontaneously broken, the associated Goldstone boson is the same as that of $C[\Lambda]$ -- namely, it is the first parent $\chi$ of the elementary BRST quartet. Since $\chi$ does not belong to the BRST cohomology, the Goldstone boson of a broken U(1) gauge symmetry does not belong to the physical spectrum. By this mechanism, spontaneously broken gauge symmetries do not give rise to physical Goldstone bosons.

Finally, if the gauge boson acquires a mass, the field-strength tensor $F_{\mu\nu}$ has no massless spectrum, and the spatial integral of the spatial divergence of the electric field operator vanishes \cite{KO79d,KO79a,NO90}
\begin{equation}
   \int d^{3}{\bf x}\ \partial^{\mu}F_{0\mu}=\oint d^{2}{\bf x}^{i}\,F_{0i}=0\ .
\end{equation}
In this case, the global gauge charge $Q$ is entirely given by the integral
\begin{equation}
   Q=\int d^{3}{\bf x}\ j_{0}^{(1)}=C[\Lambda=1]\ ,
\end{equation}
and it generates nothing but the constant-parameter limit of residual gauge transformations.\\

To what extent does the above discussion apply to non-Abelian theories? As is well known, non-Abelian theories have no residual gauge invariance at the level of the local fields: the transformation
\begin{equation}
   A_{\mu}^{a}\to A_{\mu}^{a}+g^{-1}\partial_{\mu}\Lambda^{a}+f^{abc}A_{\mu}^{b}\Lambda^{c}
\end{equation}
is not a symmetry of the field equations for any function $\Lambda^{a}$. Nevertheless, in these theories, the structure we described in the Abelian case survives in its entirety at the level of the asymptotic fields.

Indeed, as we saw in Sec.~\ref{sec:framework}, the existence of the elementary BRST quartet $\chi^{a}$, $\beta^{a}$, $\gamma^{a}$, $\gammabar^{a}$ is guaranteed by the Slavnov-Taylor identities of any Yang-Mills theory, with the first parent $\chi^{a}$ entering the asymptotic limit of the gauge-boson field $A_{\mu}^{a}$ as
\begin{equation}
   A_{\mu}^{a}\to Z_{\chi}^{1/2}\,\partial_{\mu}\chi^{a}+\cdots\ .
\end{equation}
Therefore, one can still define a generator $C[\Lambda]$ by
\begin{equation}
   C[\Lambda]\propto g^{-1}\int d^{3}{\bf x}\,\left(\Lambda^{a}\dot{\beta}^{a}-\dot{\Lambda}^{a}\beta^{a}\right)\ ,
\end{equation}
where now $\beta^{a}$ is the massless asymptotic field associated with the NL field $B^{a}$, obeying $\partial^{2}\beta^{a}=0$. By virtue of commutation relations
\begin{equation}
   [\chi^{a}_{{\bf p}},\beta^{b\dagger}_{{\bf k}}]=-(2\pi)^{3}\,\delta^{ab}\,\delta^{(3)}({\bf p}-{\bf k})
\end{equation}
which are still satisfied by the asymptotic creation/annihilation operators, $C[\Lambda]$ generates the transformations
\begin{equation}
   [iC[\Lambda],\chi^{a}(x)]\propto-g^{-1}\Lambda^{a}(x)
\end{equation}
of the asymptotic field $\chi^{a}$. Just like in the Abelian case, these transformations are spontaneously broken, and $\chi^{a}$ is the associated Goldstone boson. In Sec.~\ref{sec:chargestruct}, in the context of dynamically massive QCD, we will show that $C[\Lambda]$ generates symmetries of the \textit{asymptotic} field equations, and we will establish the connection between these symmetries and asymptotic residual gauge symmetry.

Similarly, the MYME read
\begin{equation}\label{deq334}
   J_{\mu}^{a}=g^{-1}\partial^{\nu}F_{\mu\nu}^{a}+g^{-1}\{Q_{B},D_{\mu}\cbar^{a}\}\ .
\end{equation}
In the asymptotic limit, as we saw in the previous section,
\begin{equation}\label{jaslim}
   J_{\mu}^{a}\to g^{-1}\,Z_{\chi}^{-1/2}(1+u)\,\partial_{\mu}\beta^{a}\ .
\end{equation}
This is the term that Kugo and Ojima tried to suppress in \cite{KO79d,KO79a}. If gauge bosons remain massless\footnote{It is unclear whether this can actually happen in non-Abelian theories.}, $\partial^{\nu}F_{\nu\mu}^{a}$ will provide a similar contribution to the RHS of Eq.~\eqref{deq334} \cite{KO79c,KO79d,KO79a}. There is no reason to believe these contributions will cancel.

Just like in the Abelian case, $\partial_{\mu}\beta^{a}$ is proportional to the current that generates $C[\Lambda]$ in the constant-$\Lambda^{a}$ limit. Therefore, in the non-Abelian case too, $J_{\mu}^{a}$ always generates spontaneously broken transformations. Being a general feature of the theory, this does not imply, \textit{per se}, the violation of its underlying symmetry -- in our case, color symmetry. We will have more to say about these transformations in Sec.~\ref{sec:chargestruct}.\\

Given that, in the linear covariant gauges of Yang-Mills theory, the global gauge charge $Q^{a}$ is always spontaneously broken, how can we distinguish a theory in which the underlying symmetry is \textit{actually} broken from one in which symmetry breaking is a harmless structural property? In the strongest possible form, the symmetry is preserved if all the Green's functions of the theory -- i.e., its vacuum correlators -- transform covariantly under it. While an unbroken charge would guarantee this covariance, the converse is clearly not true: although $Q^{a}$ is always broken, evidently, Green's functions do not violate symmetry in the LCGs of \textit{all} Yang-Mills theories.

In Appendix~\ref{app:softtheorem}, using the results of Secs.~\ref{sec:mlessspect} and \ref{sec:ssb}, we will formulate a soft theorem for dynamically massive QCD that provides a necessary and sufficient condition for the covariance of Green's functions under color transformations, regardless of the spontaneous breaking of $Q^{a}$. There we will show that the matrix elements whose vanishing enforces the preservation of color symmetry \textit{do not} participate in dynamical gluon mass generation, hence that dynamical gluon mass generation does not imply the violation of color symmetry.

While the preservation of color symmetry in its strongest form admits a characterization in terms of a soft theorem, the issue of color confinement -- in the sense of the absence of colored asymptotic states from the physical spectrum -- is more easily addressed by invoking symmetry requirements that apply directly to the asymptotic fields of the theory. In this respect, we observe that, as far as symmetry is concerned, color confinement only requires that the spectrum of the theory be organized into representations of SU(3), so that the question of whether colored asymptotic states are unphysical or not can be posed meaningfully.

In view of asymptotic completeness, said condition is equivalent to asking that asymptotic fields be organized into SU(3) representations: color symmetry is preserved at the level of the spectrum if and only if there exist well-defined operators $(\widetilde{Q}^{a})^{\text{as.}}=(\widetilde{Q}^{a})^{\text{in}/\text{out}}$ whose action is known on the in/out states of the theory and such that, for every asymptotic field $\phi_{I}^{\text{as.}}$,
\begin{equation}\label{ccond}
   [(\widetilde{Q}^{a})^{\text{as.}},\phi_{I}^{\text{as.}}]=T_{IJ}^{a}\,\phi_{J}^{\text{as.}}\ ,\qquad S(\widetilde{Q}^{a})^{\txtout}=(\widetilde{Q}^{a})^{\txtin}S\ ,
\end{equation}
where the $T_{IJ}^{a}$'s realize representations of the color algebra. Note that the $(\widetilde{Q}^{a})^{\text{as.}}$'s in Eq.~\eqref{ccond} cannot be arbitrary operators: it is the color (gauge) symmetry of the theory that must be preserved, not an arbitrary SU(3)\footnote{An arbitrary SU(3) could be a custodial symmetry, like custodial SU(2) in the SU(2) Higgs-Kibble model \cite{Kib67,tHoo71,KO79b}. That would not be the same symmetry as the original gauge symmetry.}. Therefore, in Eq.~\eqref{ccond}, $(\widetilde{Q}^{a})^{\text{as.}}$ must be in some unique relation with the global gauge charge $Q^{a}$.

Now, it is hard to see how the first of Eq.~\eqref{ccond} can be satisfied with $(\widetilde{Q}^{a})^{\text{as.}}$ in a unique relation with $Q^{a}$, if the asymptotic fields do not transform faithfully under $Q^{a}$ in the first place. In light of this consideration, a theory with an unbroken color symmetry should satisfy the following properties. Let $\phi_{I}(x)$ be any local operator, composite or not, and let $[Q^{a},\phi_{I}(x)]$ be the commutator defined as
\begin{equation}
   [Q^{a},\phi_{I}(x)]=\lim_{R\to \infty}\,\int_{|{\bf y}|<R}d^{3}{\bf y}\ [J_{0}^{a}({\bf y},t),\phi_{I}({\bf x},t)]\ .
\end{equation}
Regardless of the spontaneous breaking of $Q^{a}$, the commutator induces a representation of the color algebra on the local fields,
\begin{equation}
   [Q^{a},\phi_{I}(x)]=T^{a}_{IJ}\,\phi_{J}(x)\ .
\end{equation}
Choose the $\phi_{I}$'s so that the representation is irreducible. In a theory that preserves the underlying color symmetry, this representation should be preserved by the asymptotic limit of the local fields,
\begin{equation}\label{symass}
   [Q^{a},\phi_{I}^{\text{as.}}(x)]=T^{a}_{IJ}\,\phi_{J}^{\text{as.}}(x)\ ,
\end{equation}
where, if $\phi_{I}(x)$ transforms nontrivially under $Q^{a}$, 
\begin{equation}\label{aslimsym}
   \phi_{I}(x)\to Z_{\phi}^{1/2}\,\phi_{I}^{\text{as.}}(x)
\end{equation}
as $t\to \pm \infty$. If the asymptotic limit of $\phi_{I}$ contains more than one asymptotic field\footnote{E.g., a local field with Lorentz indices can develop distinct asymptotic fields in its individual tensorial channels.}, we allow for these relations to hold for each such field. Moreover, if $\phi_{I}$ transforms trivially under $Q^{a}$, we allow for constants $\phi_{I}^{(0)}$ in Eq.~\eqref{aslimsym},
\begin{equation}
   \phi_{I}(x)\to \phi_{I}^{(0)}+Z_{\phi}^{1/2}\,\phi_{I}^{\text{as.}}(x)\ ,
\end{equation}
to account for the fact that, while the vacuum expectation value of asymptotic fields is zero, color-neutral operators can develop vacuum condensates without breaking symmetry. Note that the lack of constants in Eq.~\eqref{aslimsym} implicitly precludes colored local operators from developing vacuum condensates, which is the correct behavior for theories with an unbroken color symmetry. In Higgs theories, the nonvanishing vacuum expectation value of the Higgs field invalidates these assumptions.

In Sec.~\ref{sec:chargestruct}, when we will analyze the structure of the global gauge charge $Q^{a}$ in the light of the results of Secs.~\ref{sec:mlessspect} and \ref{sec:ssb}, we will assume that the properties enunciated above are satisfied in dynamically massive QCD. Note that this assumption is, in fact, customary: one usually assumes that $Q^{a}$ is unbroken, in which case they hold automatically. In the KO framework, $Q^{a}$ is always spontaneously broken, and these properties must instead be postulated directly. As we will see in Sec.~\ref{sec:chargestruct}, not only is Eq.~\eqref{symass} consistent with a spontaneously broken $Q^{a}$, but it is actually key to understanding spontaneous symmetry breaking in the non-Abelian theory.

Building on this and other assumptions that are customary in the KO framework, in Sec.~\ref{sec:chargestruct} we will show that, if gluons dynamically acquire a mass, then the operators $(\widetilde{Q}^{a})^{\text{as.}}$ of Eq.~\eqref{ccond} exist, they are unique, and they realize color confinement in the sense of the absence of colored asymptotic states from the BRST cohomology.\\

To conclude this section, let us make some remarks on the role of Goldstone bosons in Higgs mechanisms. As in the Introduction, we use the term ``Higgs mechanism'' to refer to any gauge-boson mass generation mechanism that proceeds via the spontaneous breaking of the global gauge charge and the ``eating'' of the associated scalar Goldstone boson.

The textbook understanding of the concept of the ``eating'' of Goldstone bosons \cite{PS95} is misleading, if not outright wrong. Since it stands in the way of a broader understanding of dynamical mass generation in the LCGs of Yang-Mills theories, we must address it explicitly. That understanding can be summarized as follows. Since physical massless vector bosons have two polarizations, gauge bosons need to combine with a Goldstone boson in order to produce a physical massive vector particle, which has three. The third polarization thus originates from the Goldstone boson, which is ``eaten'' by the gauge boson.

The issues with these statements are twofold. First of all, nonperturbatively, there exists no physical nor mathematical law guaranteeing the conservation of the number of physical degrees of freedom of a theory. Quite the contrary, interactions can render particles unstable, thus removing them from the asymptotic states and causing their decrease, or they can produce stable bound states, thus causing their increase. In particular, physical degrees of freedom need not be transferred from one field to another in order for their number to be conserved; there is no such requirement in quantum field theory.

Second, in manifestly Lorentz-covariant formulations of gauge theories -- which include the KO framework and are the modern standard -- the gauge-boson field has four components from the get-go, hence it does not need to ``combine'' with another field to acquire a third polarization: the third polarization is already there. Moreover, in both perturbative and nonperturbative treatments of Higgs models, except for special gauges with no Goldstone bosons like the unitary gauge \cite{Higgs66,Kib67,Wei73}, the third polarization and the Goldstone boson are individual, distinct degrees of freedom, with different masses and a different dynamics\footnote{Except for the high-energy limit, in which their dynamics coincides by virtue of the renowned Goldstone boson equivalence theorem \cite{CLT74,Vay76,LQT77,CG85}.}. The third polarization is therefore \textit{not} the Goldstone boson, and no ``combination'' occurs.

In what sense, then, can gauge bosons be said to acquire a mass by eating a scalar Goldstone boson? To answer this question, it suffices to note that, while nothing guarantees the conservation of the number of physical degrees of freedom, BRST symmetry guarantees the conservation of the number of \textit{unphysical} degrees of freedom. These must be organized into quartets in order to cancel out from gauge-invariant observables. The role of scalar Goldstone bosons in mass generation, in fact, is not that of providing the third massive gauge-boson polarization, but that of replacing its \textit{massless longitudinal polarization} in the elementary BRST quartet. When gauge bosons acquire a mass, the massless longitudinal polarization either becomes massive, or it outright disappears from the spectrum together with the remaining massive polarizations, thus leaving a gap in the elementary quartet that must be filled in by another particle. If no massless scalar is available to replace the longitudinal polarization, gauge-boson mass generation cannot occur in a way that preserves BRST symmetry.

On the contrary, if a massless scalar is available and becomes the first parent of the elementary quartet, then it is truly ``eaten'' by the gauge boson. In fact, when that happens, 1. the Goldstone boson is excluded from the BRST cohomology, hence it disappears from the physical sector of the spectrum, albeit still existing in its unphysical sector, and 2. it becomes part of the gauge boson's one-particle spectrum. To see the latter, recall that the first parent $\chi^{a}$ enters the asymptotic limit of the gauge field as $A_{\mu}^{a}\to Z_{\chi}^{1/2}\,\partial_{\mu}\chi^{a}$. It follows that the longitudinal component of the gauge-boson field creates and annihilates scalar Goldstone bosons, hence that the matrix elements $\braket{\Omega|A_{\mu}^{a}(x)|\chi^{b}}$ and $\braket{\Omega|A_{\mu}^{a}(x)|\beta^{b}}$, in general, are both different from zero. Via the KL spectral representation, the scalar Goldstone boson then contributes to the massless one-particle spectrum of the gauge boson's propagator together with its scalar polarization -- i.e., with the massless quanta $\beta^{a}$ of the NL field $B^{a}\propto \partial^{\mu}A_{\mu}^{a}$.

Finally, since the scalar Goldstone bosons of Higgs mechanisms merely replace the massless longitudinal polarization in the elementary quartet, their existence does not automatically imply that a massive gauge boson will be present in the physical spectrum. The existence of massive gauge-boson asymptotic states depends on the behavior of the spectrum of the propagator at $p^{2}=m^{2}$ for some $m^{2}>0$; their belonging to the physical spectrum depends on whether or not such states are BRST singlets. The mere existence of a scalar Goldstone boson provides no information at all about either of these properties\footnote{See also footnote \ref{fn:Uexist}.}.

\section {The elementary BRST quartet of dynamically massive QCD and the Schwinger mechanism}\label{sec:mlessspect}

In this section, we finally delve into the main matter of the present paper. In Sec.~\ref{sec:asympt} we perform an asymptotic analysis of the massless component of the gluon spectrum in dynamically massive QCD. There we show that, when a gluon mass is generated, an unphysical particle $\chi^{a}$ with the quantum numbers of the color charge -- namely, a massless colored scalar -- appears in the gluon spectrum to complete the elementary BRST quartet of QCD. In Sec.~\ref{sec:goldbound}, by inspection of its dynamical equations, we establish that $\chi^{a}$ can be interpreted as a bound state. We demonstrate that $\chi^{a}$ generates Schwinger poles in the elementary vertices of the theory, and that it triggers the Schwinger mechanism responsible for the dynamical generation of the gluon mass. With the exception of singularities produced by the NL field $B^{a}$ that do not contribute to mass generation, the equations governing the mechanism in the KO framework coincide with those formulated by Aguilar, Binosi, Ferreira, and Papavassiliou \cite{AP06,ABP08,AP09,AIMP12,IP13,ABP16,ABFP16,ABP17,BP18,Pap22,FP23,FP24,FP25}. We thus establish the correspondence between the KO and the ABFP frameworks, and we identify $\chi^{a}$ with the ABFP bound state.

As a corollary of our results, we show that, in Yang-Mills theories that satisfy customary KO assumptions, including the (Abelian) Stueckelberg model and Higgs models, gauge-boson mass generation occurs via the Schwinger mechanism.

\subsection{Asymptotic analysis of the massless component of the gluon spectrum in dynamically massive QCD}\label{sec:asympt}

Let us consider the momentum-space, unrenormalized gluon and ghost dressed propagators $\Delta_{\mu\nu}^{ab}(p)$ and $\mc{G}^{ab}(p^{2})$,
\begin{equation}\label{glupropdef}
   \Delta_{\mu\nu}^{ab}(p)=\int d^{4}x\ e^{ipx}\, \langle T\left\{A_{\mu}^{a}(x)A_{\nu}^{b}(0)\right\}\rangle\ ,
\end{equation}
\begin{equation}\label{ghpropdef}
   \mc{G}^{ab}(p^{2})=\int d^{4}x\ e^{ipx}\, \langle T\left\{c^{a}(x)\,\cbar^{b}(0)\right\}\rangle\ .
\end{equation}
Because of BRST symmetry, from the first of Eq.~\eqref{quarteteq}, using $B^{b}=-\xi^{-1}\partial^{\nu}A_{\nu}^{b}$, it follows that
\begin{equation}
   p^{\nu}\Delta_{\mu\nu}^{ab}(p)=-i\xi\, \frac{p_{\mu}}{p^{2}}\,\delta^{ab}\ .
\end{equation}
$\Delta_{\mu\nu}^{ab}(p)$ can therefore be expressed as
\begin{equation}\label{spectral0}
   \Delta_{\mu\nu}^{ab}(p)=\Delta^{ab}(p^{2})\ t_{\mu\nu}(p)+\frac{-i\xi}{p^{2}}\ \ell_{\mu\nu}(p)\,\delta^{ab}\ ,
\end{equation}
where $t_{\mu\nu}(p)$ and $\ell_{\mu\nu}(p)$ are, respectively, the transverse and the longitudinal projectors
\begin{equation}
   t_{\mu\nu}(p)=\eta_{\mu\nu}-\frac{p_{\mu}p_{\nu}}{p^{2}}\ ,\qquad \ell_{\mu\nu}(p)=\frac{p_{\mu}p_{\nu}}{p^{2}}\ .
\end{equation}
On the other hand, $\mc{G}^{ab}(p^{2})$ can be expressed as
\begin{equation}
   \mc{G}^{ab}(p^{2})=-\frac{J^{ab}(p^{2})}{p^{2}}\ .
\end{equation}
In what follows, as is customary in QCD, and as is confirmed by lattice simulations in Euclidean space \cite{SS96,Cuc97,LSWP98a,LSWP98b,BBLW00,BBLW01,BHLP04,SIMS05,BBMM06,CM07,CM08a,CM08b,BIMS09,ISI09,BMM10,ABBC12,BLLM12,OS12,BBCO15,CDMO18,FN03,FN06a,Ste06}, we will assume that
\begin{equation}
   \Delta^{ab}(p^{2})=\Delta(p^{2})\,\delta^{ab}\ ,\qquad J^{ab}(p^{2})=J(p^{2})\,\delta^{ab}\ ,
\end{equation}
for some scalar functions $\Delta(p^{2})$ and $J(p^{2})$.

From the second of Eq.~\eqref{quarteteq}, we know that the ghost and the antighost field operators $c^{a}$ and $\cbar^{a}$ possess a massless discrete spectrum\footnote{Technically speaking, Eq.~\eqref{quarteteq} only implies that a massless discrete spectrum is possessed by the antighost field $\cbar^{a}$ and by the sum $D_{\mu}c^{a}=\partial_{\mu}c^{a}+gf^{abc}A_{\mu}^{b}c^{c}$. In the Landau gauge ($\xi=0$), the field equations are symmetric under the exchange $c^{a}\leftrightarrow \cbar^{a}$, and it follows that $c^{a}$ has a massless discrete spectrum as well. Beyond the Landau gauge ($\xi\neq 0$), it is hard to see how the quartet structure of the unphysical degrees of freedom of the theory can be preserved if this property does not hold for $c^{a}$ on its own, and there is no reason to believe it does not. Preliminary lattice data \cite{CDMO18} support the absence of deviations from the Landau-gauge behavior.}. In the spectral representation of $\mc{G}^{ab}(p^{2})$, there must then exists a corresponding $1/p^{2n}$ term for some integer $n\geq 1$. $n>1$ would imply the existence of at least $n$ distinct multipole ghost asymptotic fields with the same ghost number, which are not observed in any Standard Model theory \cite{KO78a,KO79b,KO79c,KO79d,KO79a,Kug78}. $n=1$, instead, agrees with the behavior of the second of Eq.~\eqref{quarteteq}, with that of Standard Model theories \cite{KO78a,KO79b,KO79c,KO79d,KO79a,Kug78}, and with the Euclidean lattice data for QCD \cite{SS96,Cuc97,SIMS05,BBMM06,CM08a,BIMS09,ABBC12,BLLM12,CDMO18}. In what follows, we will assume that $n=1$.

Suppressing diagonal color indices and denoting the coefficient of the $1/p^{2}$ term with $J_{0}$, we have
\begin{equation}\label{deq637}
   \mc{G}(p^{2})=-\frac{J_{0}}{p^{2}}-\frac{J(p^{2})-J_{0}}{p^{2}}\ .
\end{equation}
In what follows, as is expected for a simple pole in a ghost propagator, we will assume $J_{0}>0$. If, in agreement with the behavior that ordinarily allows the introduction of asymptotic fields, the massless ghost spectrum is to be truly discrete -- i.e. if it is to be separated from the multiparticle continuum, so that $\lim_{p^{2}\to 0} p^{2}G(p^{2})=-J_{0}$ --, one should also have $J(0)=J_{0}$. In what follows, we will not make this assumption explicitly. However, we note that, in view of asymptotic completeness, for $n=1$, it may be impossible to realize $J(0)\neq J_{0}$ or, more generally, a $J(p^{2})$ with non-saturating behavior (modulo multiplicative normalization), within the KO framework. We remark that Euclidean lattice data \cite{SS96,Cuc97,SIMS05,BBMM06,CM08a,BIMS09,ABBC12,BLLM12,CDMO18} display such behavior, they have $J(0)>0$, hence they are compatible with $J(0)=J_{0}$.

We say that gluons dynamically acquire a mass if the dynamics of the strong interactions removes the massless pole from the transverse component of the dressed gluon propagator, implying the absence of massless transverse gluons from both the physical and the unphysical sectors of the spectrum of the theory. In other words, dynamical gluon mass generation occurs if
\begin{equation}\label{gmass}
   \Delta(0)=\frac{i}{M^{2}}
\end{equation}
for some renormalizable quantity $M^{-2}=Z_{3}\,M_{R}^{-2}$, where $Z_{3}$ is a renormalization factor such that $Z_{3}^{-1}\Delta(p^{2})$ is finite for all $p^{2}$, and $M_{R}^{-2}$ is a constant.

From now on, we will assume that gluons dynamically acquire a mass. Since the zero-momentum limit of the Euclidean form factor $\Delta_{E}(p_{E}^{2})=-i\Delta(-p_{E}^{2})$ is equal to $M^{-2}$, since $\Delta_{E}(p_{E}^{2})$ is usually assumed to be non-negative, and since renormalization constants are usually assumed to be positive, we will assume that $Z_{3}$ is positive and that $M_{R}^{-2}$ is non-negative. Lattice calculations in Euclidean space \cite{Cuc97,LSWP98a,LSWP98b,BBLW00,BBLW01,BHLP04,SIMS05,BBMM06,CM07,CM08a,CM08b,BIMS09,ISI09,BMM10,ABBC12,BLLM12,OS12,BBCO15,FN03,FN06a,Ste06} confirm these assumptions, furthermore showing that $M_{R}^{-2}\neq 0$. In what follows, we will not assume the latter a priori.

While the longstanding focus of research has necessarily been on the determination of the only unknown quantity in Eq.~\eqref{spectral0}~-- namely, the form factor $\Delta(p^{2})$ -- an understanding of the gluon spectrum can only be gained by analyzing how $\Delta(p^{2})$ shapes the full tensorial propagator $\Delta_{\mu\nu}^{ab}(p)$. In this respect, one systematically neglected aspect in studies of the gluon spectrum is the fact that, even in the presence of dynamical gluon mass generation, the full propagator $\Delta_{\mu\nu}^{ab}(p)$ still contains massless poles: dropping diagonal color indices, the propagator in Eq.~\eqref{spectral0} splits as
\begin{equation}\label{glupropsplit}
   \Delta_{\mu\nu}(p)=\Delta_{\mu\nu}^{m\neq 0}(p)+\Delta_{\mu\nu}^{m= 0}(p)\ ,
\end{equation}
where
\begin{equation}
   \Delta_{\mu\nu}^{m\neq 0}(p)=\Delta(p^{2})\ \eta_{\mu\nu}-[\Delta(p^{2})-\Delta(0)]\,\frac{p_{\mu}p_{\nu}}{p^{2}}\ ,
\end{equation}
modulo multiplicative renormalization, is finite as $p^{2}\to 0$, whereas
\begin{equation}
   \Delta_{\mu\nu}^{m=0}(p)=-\frac{p_{\mu}p_{\nu}}{M^{2}}\ \frac{i}{p^{2}}+\xi p_{\mu}p_{\nu}\frac{-i}{(p^{2})^{2}}
\end{equation}
contains a massless simple pole and -- in all but the Landau gauge ($\xi=0$) -- a massless dipole. Both these poles are proportional to $p_{\mu}p_{\nu}$, hence they are longitudinal. Modulo factors of $p_{\mu}p_{\nu}$, $M^{-2}$ is the residue of the massless simple gluon pole.

To understand the split in Eq.~\eqref{glupropsplit}, it is helpful to introduce a spectral density $\sigma(m^{2})$ for the transverse propagator $\Delta(p^{2})$,
\begin{equation}
   \Delta(p^{2})=\int dm^{2}\ \sigma(m^{2})\ \frac{-i}{p^{2}-m^{2}}\ ,
\end{equation}
and to use the identity
\begin{equation}
   \frac{1}{p^{2}-m^{2}}-\frac{1}{-m^{2}}=\frac{p^{2}}{m^{2}(p^{2}-m^{2})}
\end{equation}
to express $\Delta_{\mu\nu}^{m\neq 0}(p)$ as
\begin{align}\label{gpropmassive}
   \Delta_{\mu\nu}^{m\neq 0}(p)&=\int dm^{2}\ \sigma(m^{2})\ \left(-\eta_{\mu\nu}+\frac{p_{\mu}p_{\nu}}{m^{2}}\right)\ \frac{i}{p^{2}-m^{2}}\ .
\end{align}
The presence of the massive vector boson polarization tensor
\begin{align}
   -\eta_{\mu\nu}+\frac{p_{\mu}p_{\nu}}{m^{2}}=\sum_{\sigma=0,\pm1}\ \varepsilon_{\mu}^{\sigma}(p)\,\varepsilon_{\nu}^{\sigma}(p)^{\star}
\end{align}
in the above equation, where $\varepsilon_{\mu}^{\sigma}(p)$ such that $p^{\mu}\varepsilon_{\mu}^{\sigma}(p)=0$ are polarization vectors of mass $m\neq 0$ and spin $\sigma$, indicates that $\Delta_{\mu\nu}^{m\neq 0}(p)$ is a sum over massive spin-1 vector states: $\Delta_{\mu\nu}^{m\neq 0}(p)$ contains the massive component of the gluon spectrum. The precise content of the latter, encoded in the spectral density $\sigma(m^{2})$, remains under active investigation \cite{CMT05,SFK12,DOS14,SIR16b,CPRW18,BT20,DOS20,HK20,HPRT22,HPW25}, and plays no fundamental role in the present work. For our purposes, it suffices to note that -- by Eq.~\eqref{gpropmassive} -- states making up the gluon's massive discrete spectrum, should they exist, are necessarily arranged in transverse polarization triplets.

The massless longitudinal poles in $\Delta^{m=0}_{\mu\nu}(p)$, on the other hand, determine the massless component of the gluon spectrum\footnote{Except for specific kinematic configurations, multiparticle states consisting entirely of massless particles are massive, and their continuum contributes to $\Delta^{(m\neq 0)}_{\mu\nu}(p)$. Since the finiteness of $\Delta(0)$ modulo multiplicative renormalization implies $\sigma(m^{2}=0)=0$, the exceptional massless configurations are suppressed in $\Delta^{(m\neq0)}(p^{2})$, and the multiparticle continuum has no overlap with the gluon's massless discrete spectrum, regardless of the tensorial channel in which it lives. This is in agreement with the behavior that ordinarily allows the introduction of asymptotic fields.}. As we saw, this component is completely fixed by BRST symmetry and by the (finite, modulo multiplicative renormalization) value of $\Delta(0)$. In theories where gauge bosons do not acquire a mass, like ordinary QED and perturbative QCD, massless longitudinal poles in the boson's propagator are a manifestation of its unphysical longitudinal and scalar polarizations \cite{Nak72,KO79c,NO90}. When gauge bosons acquire a mass, on the contrary, there is no massless longitudinal polarization, and the presence of massless longitudinal poles becomes a nontrivial feature of the spectrum. For instance, as discussed in the previous sections and as we review in Appendices~\ref{app:stschw} and \ref{app:higgsmod}, in the Stueckelberg, Schwinger and Higgs models, it is the unphysical scalar Goldstone bosons of the spontaneously broken global gauge charge that, together with the scalar polarization, make up the massless component of the gauge boson's spectrum. We thus ask: what is the structure and content of the massless component of the spectrum of the gluon propagator, when gluons dynamically acquire a mass?\\

To answer this question, we start by repeating the classic asymptotic analysis of \cite{KO79b,KO79c,NO90} for the case of QCD with dynamical gluon mass generation. The analysis will provide us with asymptotic field equations, commutation relations and BRST transformations which we will use throughout this paper, and it will reveal the structure of the massless gluon spectrum. Since the massive component of the spectrum is currently unknown, we limit ourselves to the massless sector\footnote{Massive asymptotic fields commute or anticommute with massless asymptotic fields, so the former decouple from the latter in the asymptotic analysis as soon as commutators are written down.}. We thus introduce massless asymptotic fields $\chi_{\mu}^{a}$ and $\beta^{a}$ for the field operators $A_{\mu}^{a}$ and $B^{a}$ as
\begin{align}\label{asymm0}
   A_{\mu}^{a}&\to Z_{3}^{1/2}\,A_{\mu}^{a\,\text{as.}}=Z_{\chi}^{1/2}\,\chi_{\mu}^{a}+\cdots\ ,\\
   \notag B^{a}&\to Z_{3}^{-1/2}\,B^{a\,\text{as.}}=Z_{\chi}^{-1/2}\,\beta^{a}+\cdots\ ,
\end{align}
where the arrows indicate that either one of the two limits $t\to\pm \infty$ has been taken, $A_{\mu}^{a\,\text{as.}}$ and $B^{a\,\text{as.}}$ are the asymptotic fields associated with $A_{\mu}^{a}$ and $B^{a}$, $Z_{3}$ is the renormalization constant introduced earlier, $Z_{\chi}^{1/2}$ is an LSZ coefficient, and the ellipsis indicates that a massive component may be present in $A_{\mu}^{a\,\text{as.}}$ and $B^{a\,\text{as.}}$\footnote{For scalar fields like $B^{a}$, $c^{a}$, and $\cbar^{a}$, one expects the associated asymptotic field to have a single mass component. Since whether or not fields develop a massive asymptotic component is inessential to the present study, in what follows we will not make this assumption.}. For instance, if the spectrum of $A_{\mu}^{a}$ contains a massive discrete component,
\begin{equation}\label{glumasscomp}
   A_{\mu}^{a}\to Z_{\chi}^{1/2}\,\chi_{\mu}^{a}+Z_{U}^{1/2}\,U_{\mu}^{a}
\end{equation}
for some massive asymptotic field $U_{\mu}^{a}$ creating massive asymptotic states, with LSZ coefficient $Z_{U}^{1/2}$. Because of the transversality of the massive component of the gluon spectrum, if it exists, $U_{\mu}^{a}$ satisfies $\partial^{\mu}U_{\mu}^{a}=0$\footnote{\label{fn:bmassc}Observe that the gauge-fixing condition $\partial^{\mu}A_{\mu}^{a}+\xi B^{a}=0$ constrains a potential massive asymptotic component of $B^{a}$ to be proportional to $\partial^{\mu}U_{\mu}^{a}=0$, hence to vanish. In what follows, we will mostly disregard this observation.}.

As we often did in the previous sections, to keep the notation simple, we will not distinguish between in and out fields. It is, however, understood that with each of the limits $t\to\pm \infty$ is associated one distinct set of asymptotic fields. Observe that the normalization of $A_{\mu}^{a\,\text{as.}}$ and $B^{a\,\text{as.}}$ is largely arbitrary, as the fields that actually create and annihilate asymptotic states are $\chi_{\mu}^{a}$ and $\beta^{a}$. Here and in what follows, we conventionally normalize them using $Z_{3}^{1/2}$. The normalization of $\chi_{\mu}^{a}$ and $\beta^{a}$, on the contrary, is not arbitrary: as discussed in Sec.~\ref{sec:framework}, we want the asymptotic states they create to be normalized canonically. In view of the quartet structure we will derive in what follows, we have already set the LSZ coefficient of $\beta^{a}$ equal to the inverse of that of $\chi_{\mu}^{a}$.

$\chi_{\mu}^{a}$ and $\beta^{a}$ generate the massless pole structure of two-point Green's functions involving $A_{\mu}^{a}$ and $B^{a}$. Other than the gluon propagator, the latter include 
\begin{align}
   \langle T\{A_{\mu}^{a}(x)B^{b}(0)\}\rangle_{\text{F.T.}}&=-\delta^{ab}\,\frac{p_{\mu}}{p^{2}}\ ,\\
   \notag\langle T\{B^{a}(x)B^{b}(0)\}\rangle_{\text{F.T.}}&=0\ ,
\end{align}
the first of which is the first of Eq.~\eqref{quarteteq}, and the second of which is a consequence of the nihilpotency of the BRST charge: $\langle B^{a}(x)B^{b}(0)\rangle=\langle \cbar^{a}(x)Q_{B}^{2}\cbar^{b}(0)\rangle=0$. By isolating the massless discrete spectrum of the two-point Green's functions, suppressing diagonal color indices, we obtain
\begin{align}\label{glumlessasympoles}
   \notag Z_{\chi}\langle T\{\chi_{\mu}(x)\chi_{\nu}(0)\}\rangle_{\text{F.T.}}&=-\frac{ip_{\mu}p_{\nu}}{M^{2}p^{2}}-\,\frac{i\xi p_{\mu}p_{\nu}}{(p^{2})^{2}}\ ,\\
   \langle T\{\chi_{\mu}(x)\beta(0)\}\rangle_{\text{F.T.}}&=-\frac{p_{\mu}}{p^{2}}\ ,\\
   \notag\langle T\{\beta(x)\beta(0)\}\rangle_{\text{F.T.}}&=0\ .
\end{align}
Since, modulo factors of $p_{\mu}p_{\nu}$, $M^{-2}$ is the non-negative residue of the massless simple pole of the gluon propagator, we tentatively set $Z_{\chi}=M^{-2}$. We will confirm this is the correct coefficient in what follows.

After introducing a rescaled and finite (``renormalized'') gauge parameter $\xi_{R}=Z_{3}^{-1}\xi$, Eq.~\eqref{glumlessasympoles} translates to the vacuum expectation values of commutators \cite{KO79b,KO79c,NO90}
\begin{align}\label{vaccomm}
   &\langle[\chi_{\mu}(x),\chi_{\nu}(y)]\rangle=\\
   &\qquad\notag =i\partial_{\mu}\partial_{\nu}D(x-y)-i\xi_{R}M^{2}_{R}\partial_{\mu}\partial_{\nu}E(x-y)\ ,\\
   \notag&\langle[\chi_{\mu}(x),\beta(y)]\rangle=-i\partial_{\mu}D(x-y)\ ,\quad \langle[\beta(x),\beta(y)]\rangle=0\ .
\end{align}
Here the functions $D(x)$ and $E(x)$ are defined as
\begin{align}
   D(x)&=-i\int\frac{d^{4}p}{(2\pi)^{3}}\ e^{-ipx}\ \text{sgn}(p^{0})\ \delta(p^{2})\ ,\\
   \notag E(x)&=-i\int\frac{d^{4}p}{(2\pi)^{3}}\ e^{-ipx}\ \text{sgn}(p^{0})\ \delta^{\prime}(p^{2})\ ,
\end{align}
where $\text{sgn}(x)$ is the sign function and $\delta^{\prime}(x)$ is the first derivative of the Dirac delta. As in Sec.~\ref{sec:framework}, $iD(x-y)$ is the commutator of an ordinary free massless scalar field. The functions $D(x)$ and $E(x)$ satisfy
\begin{equation}
   \partial^{2}D(x)=0\ ,\qquad \partial^{2}E(x)=D(x)\ .
\end{equation}

Thanks to the Greenberg-Robinson theorem \cite{NO78,KO79b,KO79c,NO90}, valid in spacetime dimension $d\geq 3$, we know that the commutator of two asymptotic fields is a function. Therefore, we promote the vacuum expectation values in Eq.~\eqref{vaccomm} to operator identities:
\begin{align}\label{ascomm0}
   \notag[\chi_{\mu}(x),\chi_{\nu}(y)]&=i\partial_{\mu}\partial_{\nu}D(x-y)-i\xi_{R}M^{2}_{R}\partial_{\mu}\partial_{\nu}E(x-y)\ ,\\
   [\chi_{\mu}(x),\beta(y)]&=-i\partial_{\mu}D(x-y)\ ,\quad [\beta(x),\beta(y)]=0\ .
\end{align}

The ghost sector is treated similarly. We introduce massless asymptotic ghost fields $\gamma^{a}$ and $\gammabar^{a}$ as
\begin{align}\label{asymg}
   c^{a}&\to \widetilde{Z}_{3}^{1/2}\,c^{a\,\text{as.}}=\widetilde{Z}_{3}^{1/2}\gamma^{a}+\cdots\ ,\\
   \notag\cbar^{a}&\to \widetilde{Z}_{3}^{1/2}\,\cbar^{a\,\text{as.}}=\widetilde{Z}_{3}^{1/2}\gammabar^{a}+\cdots\ ,
\end{align}
where, again, the normalization of $c^{a\,\text{as.}}$ and $\cbar^{a\,\text{as.}}$ is arbitrary, and the ellipsis denotes potential massive asymptotic components. Observe that the second of Eq.~\eqref{quarteteq} relates the LSZ coefficient of the massless component of $\cbar^{a}$ to that of $D_{\mu}c^{a}$, not to that of $c^{a}$. Therefore, at variance with Eq.~\eqref{asymm0}, the LSZ coefficients that appear in Eq.~\eqref{asymg} are not the inverse of each other. In the ghost sector, it is customary \cite{KO79b,KO79c,NO90} to set the LSZ coefficient of the massless component of $c^{a}$ equal to that of $\cbar^{a}$, and to then fix $\widetilde{Z}_{3}$ as a whole from the discrete spectrum of the only two-point ghost Green's function which does not vanish by ghost number conservation -- namely, from the ghost propagator in Eq.~\eqref{deq637}. Concretely, dropping diagonal color indices, we have
\begin{equation}
   \widetilde{Z}_{3}\langle T\{\gamma(x)\gammabar(0)\}\rangle_{\text{F.T.}} = -\frac{J_{0}}{p^{2}}\ ,
\end{equation}
from which $\widetilde{Z}_{3}=J_{0}$ follows. We reiterate that, if $J(0)\neq J_{0}$, it is unclear how the massless discrete spectrum of the ghost propagator can be isolated to yield this equation, and that no such difficulty arises for the behavior displayed by the lattice data \cite{SS96,Cuc97,SIMS05,BBMM06,CM08a,BIMS09,ABBC12,BLLM12,CDMO18}.

Passing to vacuum expectation values of anticommutators,
\begin{equation}
   \langle \{\gamma(x),\gammabar(y)\}\rangle = - D(x-y)\ ,
\end{equation}
and promoting the latter to operatorial equations by the Greenberg-Robinson theorem \cite{NO78,KO79b,KO79c,NO90}, we obtain
\begin{equation}\label{commgh}
   \{\gamma(x),\gammabar(y)\} = -D(x-y)\ .
\end{equation}\\

At this point in the analysis, \cite{KO79b,KO79c,NO90} postulate the asymptotic completeness of the full Hilbert space of the theory with respect to the asymptotic fields associated with its elementary fields. This postulate is of course invalid in realistic (nonperturbative) QCD, which has stable bound states in composite channels. Moreover, since our goal is not to derive the structure of the full Hilbert space, but merely that of the massless component of the gluon spectrum, the precise form of all existing asymptotic fields is of no interest to us. For these reasons, we will make a weaker assumption than \cite{KO79b,KO79c,NO90}: we will postulate that $(i)$ the Hilbert space of dynamically massive QCD is asymptotically complete with respect to an unspecified set of asymptotic fields that includes $\chi_{\mu}^{a}$, $\beta^{a}$, $\gamma^{a}$, and $\gammabar^{a}$; and $(ii)$ in this set, all asymptotic fields other than $\chi_{\mu}^{a}$, $\beta^{a}$, $\gamma^{a}$, and $\gammabar^{a}$ commute or anticommute with the latter. Observe that asymptotic fields with different masses or quantum numbers always (anti)commute, so this postulate actually concerns the commutation relations of massless asymptotic fields with the same quantum numbers as $\chi_{\mu}^{a}$, $\beta^{a}$, $\gamma^{a}$, and $\gammabar^{a}$. No such fields are known to exist.

Our assumptions, together with Eqs.~\eqref{ascomm0} and~\eqref{commgh}, yield the asymptotic field equations
\begin{align}\label{asfieldeq0}
   &\partial_{\mu}\chi_{\nu}^{a}-\partial_{\nu}\chi_{\mu}^{a}=0\ ,\quad\partial^{\mu}\chi_{\mu}^{a}+\xi_{R}M_{R}^{2}\,\beta^{a}=0\ ,\\
   \notag&\qquad \partial^{2}\beta^{a}=0\ ,\quad\partial^{2}\gamma^{a}=0\ ,\quad \partial^{2}\gammabar^{a}=0\ .
\end{align}
These hold on the basis that 1. their left-hand sides (anti)commute with all of $\chi_{\mu}^{a}$, $\beta^{a}$, $\gamma^{a}$ and $\gammabar^{a}$, hence they commute with all asymptotic fields by assumption $(ii)$ above; 2. by asymptotic completeness $(i)$, the LHSs are then proportional to the identity operator; 3. nonzero proportionality constants would imply non-vanishing vacuum expectation values for the asymptotic fields, hence they must be discarded.

Going ahead, the first equation in Eq.~\eqref{asfieldeq0} has solutions $\chi_{\mu}^{a}=\partial_{\mu}\chi^{a}$. It follows that we can express the asymptotic limit of the elementary fields of dynamically massive QCD as
\begin{align}\label{asymm}
   \notag A_{\mu}^{a}&\to M^{-1}\,\partial_{\mu}\chi^{a}+\cdots\ ,\quad B^{a}\to M\,\beta^{a}+\cdots\ ,\\
   &c^{a}\to \widetilde{Z}_{3}^{1/2}\gamma^{a}+\cdots\ ,\qquad\cbar^{a}\to\widetilde{Z}_{3}^{1/2}\gammabar^{a}+\cdots\ ,
\end{align}
where $\chi^{a}$, $\beta^{a}$, $\gamma^{a}$, and $\gammabar^{a}$ satisfy the asymptotic field equations
\begin{align}\label{asfieldeq}
   &\partial^{2}\chi^{a}+\xi_{R}M_{R}^{2}\,\beta^{a}=0\ ,\quad \partial^{2}\beta^{a}=0\ ,\\
   \notag&\qquad \partial^{2}\gamma^{a}=0\ ,\quad \partial^{2}\gammabar^{a}=0\ .
\end{align}
Observe that, beyond the Landau gauge ($\xi,\xi_{R}\neq0$), $\chi^{a}$ is a dipole field: $(\partial^{2})^{2}\chi^{a}=0$, $\partial^{2}\chi^{a}\neq 0$.

$\chi^{a}$, $\beta^{a}$, $\gamma^{a}$, and $\gammabar^{a}$ are all scalar fields. Indeed, since unbroken transformations of local fields induce corresponding transformations of the asymptotic fields -- see Eq.~\eqref{asymsymm} --, and since Lorentz transformations act within same-mass multiplets, $\partial_{\mu}\chi^{a}$, $\beta^{a}$, $\gamma^{a}$, and $\gammabar^{a}$ in Eq.~\eqref{asymm} all transform into each other under a Lorentz transformation. The scalar nature of $B^{a}$, $c^{a}$, and $\cbar^{a}$ is then inherited by $\beta^{a}$, $\gamma^{a}$, and $\gammabar^{a}$, whereas the derivative in $\partial_{\mu}\chi^{a}$ realizes the vector character of the transformation of $A_{\mu}^{a}$, thus making $\chi^{a}$ a scalar as well.

The commutation relations of $\chi^{a}$ are obtained by integrating those of $\chi_{\mu}^{a}$ while discarding constants forbidden by spacelike commutativity. Overall, we have
\begin{align}\label{ascomm}
   \notag[\chi^{a}(x),\chi^{b}(y)]&=-i\delta^{ab}\left\{D(x-y)-\xi_{R}M^{2}_{R}E(x-y)\right\}\ ,\\
   [\chi^{a}(x),\beta^{b}(y)]&=-i\delta^{ab}\,D(x-y)\ ,\\
   \notag [\beta^{a}(x),\beta^{b}(y)]&=0\ ,\quad \{\gamma^{a}(x),\gammabar^{b}(y)\}=-\delta^{ab}\,D(x-y)\ .
\end{align}
Finally, since the massive components of $A_{\mu}^{a\,\text{as.}}$ and $B^{a\,\text{as.}}$, if they exist, commute with the massless component,
\begin{align}\label{ascommfull}
   &[A_{\mu}^{a\,\text{as.}}(x),\chi^{b}(y)]=\\
   \notag&\qquad=-i\delta^{ab}\,M^{-1}_{R}\left\{\partial_{\mu}D(x-y)-\xi_{R}M^{2}_{R}\partial_{\mu}E(x-y)\right\}\ ,\\
   \notag &[A_{\mu}^{a\,\text{as.}}(x),\beta^{b}(y)]=-i\delta^{ab}M^{-1}_{R}\,\partial_{\mu}D(x-y)\ ,\\
   \notag&[B^{a\,\text{as.}}(x),\chi^{b}(y)]=-i\delta^{ab}M_{R}\,D(x-y)\ ,\\
   \notag &[B^{a\,\text{as.}}(x),\beta^{b}(y)]=0\ .
\end{align}\\

Now that we have field equations and coordinate-space commutation relations for the massless asymptotic fields $\chi^{a}$, $\beta^{a}$, $\gamma^{a}$, and $\gammabar^{a}$, we can define asymptotic creation and annihilation operators, derive their momentum-space commutation relations, and construct asymptotic states. First, however, we must understand how to deal with the dipolar nature of $\chi^{a}$, which only manifests itself beyond the Landau gauge ($\xi\neq 0$).

There are two common ways of solving a dipolar system of equations like $\partial^{2}\chi^{a}+\xi_{R}M_{R}^{2}\,\beta^{a}=0$, $\partial^{2}\beta^{a}=0$. The first is manifestly Lorentz covariant, and employs four-dimensional Fourier transforms instead of the customary three-dimensional ones. This approach is sometimes used to quantize QED beyond the Feynman gauge\footnote{As we noted in Sec.~\ref{sec:framework} for the free photon field, and as we review in Appendix~\ref{app:photon}, in QED with no dynamical photon mass generation, the photon field is a dipole field in all but the Feynman gauge.} and other models, see for instance \cite{Nak66,Nak72,NO90}. The second is to choose a preferred Lorentz frame and to introduce a functional linear combination $\widetilde{\chi}^{a}$ of $\chi^{a}$ and $\beta^{a}$ that behaves like a simple-pole field: one defines
\begin{equation}\label{chidip}
   \widetilde{\chi}^{a}(x)=\chi^{a}(x)+\xi_{R}M_{R}^{2}\,\mathcal{E}_{x}^{(1/2)}\beta^{a}(x)
\end{equation}
where $\mathcal{E}_{x}^{(1/2)}$ \cite{Lau67,Nak72,KO79b,KO79c,NO90} is a noncovariant integro-differential operator such that
\begin{equation}
   \partial^{2}f(x)=0\quad\Longrightarrow\quad\partial^{2}\mathcal{E}_{x}^{(1/2)}f(x)=f(x)\ ,
\end{equation}
whose definition and properties we review in Appendix~\ref{app:dipole}. It is then easy to see that
\begin{equation}
   \partial^{2}\widetilde{\chi}^{a}=0\ ,
\end{equation}
so that both $\widetilde{\chi}^{a}$ and $\beta^{a}$ admit ordinary three-dimensional Fourier expansions in massless modes. This approach is used in \cite{KO79a,KO79b,KO79c,NO90}, and it is the one we will follow in this work. Note that $\widetilde{\chi}^{a}$ reduces to $\chi^{a}$ in the Landau gauge ($\xi=\xi_{R}=0$).

Using the identity \cite{NO90}
\begin{equation}\label{deq855}
   (\mathcal{E}^{(1/2)}_{x}+\mathcal{E}^{(1/2)}_{y})D(x-y)=E(x-y)\ ,
\end{equation}
we see that
\begin{align}\label{ascommdip}
   [\widetilde{\chi}^{a}(x),\widetilde{\chi}^{b}(y)]=[\widetilde{\chi}^{a}(x),\beta^{b}(y)]=-i\delta^{ab}D(x-y)\ .
\end{align}
From this equation, it follows that $\widetilde{\chi}^{a}$ and $\beta^{a}$ are canonically normalized. Our previous identification
\begin{equation}
   Z_{\chi}^{1/2}=M^{-1}
\end{equation}
is thus confirmed\footnote{The same LSZ coefficient is found in Higgs theories \cite{KO79b,KO79a,NO90} -- see also Appendix~\ref{app:higgsmod}. This is expected, since in their ordinary KO treatment those theories satisfy the assumptions of the present section.}.

The creation and annihilation operators of the asymptotic fields $\widetilde{\chi}^{a}$, $\beta^{a}$, $\gamma^{a}$, and $\gammabar^{a}$ are defined as
\begin{align}
   \notag\widetilde{\chi}^{a}(x)&=\int \frac{d^{3}\bf{p}}{(2\pi)^{3}}\,\frac{1}{\sqrt{2\hat{p}^{0}}}\ \left\{e^{-i\hat{p}x}\,\chi^{a}_{{\bf p}}+e^{i\hat{p}x}\,\chi^{a\,\dagger}_{{\bf p}}\right\}\ ,\\
   \beta^{a}(x)&=\int \frac{d^{3}\bf{p}}{(2\pi)^{3}}\,\frac{1}{\sqrt{2\hat{p}^{0}}}\ \left\{e^{-i\hat{p}x}\,\beta^{a}_{{\bf p}}+e^{i\hat{p}x}\,\beta^{a\,\dagger}_{{\bf p}}\right\}\ ,\\
   \notag\gamma^{a}(x)&=\int \frac{d^{3}\bf{p}}{(2\pi)^{3}}\,\frac{1}{\sqrt{2\hat{p}^{0}}}\ \left\{e^{-i\hat{p}x}\,\gamma^{a}_{{\bf p}}+e^{i\hat{p}x}\,\gamma^{a\,\dagger}_{{\bf p}}\right\}\ ,\\
   \notag\gammabar^{a}(x)&=\int \frac{d^{3}\bf{p}}{(2\pi)^{3}}\,\frac{1}{\sqrt{2\hat{p}^{0}}}\ \left\{e^{-i\hat{p}x}\,\gammabar^{a}_{{\bf p}}+e^{i\hat{p}x}\,\gammabar^{a\,\dagger}_{{\bf p}}\right\}\ ,
\end{align}
where $\hat{p}=(|{\bf p}|,{\bf p})$ is a massless on-shell momentum. According to Eqs.~\eqref{ascomm} and \eqref{ascommdip}, we then find that the only non-vanishing (anti)commutation relations between the creation and annihilation operators are
\begin{align}\label{ascommca}
   &[\chi^{a}_{{\bf p}},\chi^{b\,\dagger}_{{\bf q}}]=[\chi^{a}_{{\bf p}},\beta^{b\,\dagger}_{{\bf q}}]=-(2\pi)^{3}\delta^{ab}\delta^{(3)}({\bf p}-{\bf q})\ ,\\
   \notag&\qquad\quad\{\gamma^{a}_{{\bf p}},\gammabar^{b\,\dagger}_{{\bf q}}\}=i\,(2\pi)^{3}\,\delta^{ab}\,\delta^{(3)}({\bf p}-{\bf q})\ .
\end{align}

Using the creation operators, we define one-particle asymptotic states by acting on the unique vacuum state of the theory $\ket{\Omega}$:
\begin{align}\label{asstateinn}
   \ket{\chi^{a}(p)}=\sqrt{2\hat{p}^{0}}\,\chi^{a\,\dagger}_{{\bf p}}\ket{\Omega}\ ,\quad \ket{\beta^{a}(p)}=\sqrt{2\hat{p}^{0}}\,\beta^{a\,\dagger}_{{\bf p}}\ket{\Omega}\ ,\\
   \notag\ket{\gamma^{a}(p)}=\sqrt{2\hat{p}^{0}}\,\gamma^{a\,\dagger}_{{\bf p}}\ket{\Omega}\ ,\quad \ket{\gammabar^{a}(p)}=\sqrt{2\hat{p}^{0}}\,\gammabar^{a\,\dagger}_{{\bf p}}\ket{\Omega}\ .
\end{align}
Since by definition
\begin{equation}
   \chi_{{\bf p}}^{a}\ket{\Omega}=\beta_{{\bf p}}^{a}\ket{\Omega}=\gamma_{{\bf p}}^{a}\ket{\Omega}=\gammabar_{{\bf p}}^{a}\ket{\Omega}=0\ ,
\end{equation}
the commutation relations in Eq.~\eqref{ascommca} imply that the only non-vanishing inner products between these states are
\begin{align}
   \braket{\chi^{a}(p)|\chi^{b}(q)}&=\braket{\chi^{a}(p)|\beta^{b}(q)}=\\
   \notag&=-(2\hat{p}^{0})(2\pi)^{3}\delta^{ab}\delta^{(3)}({\bf p}-{\bf q})\ ,\\
   \notag\braket{\gamma^{a}(p)|\gammabar^{b}(q)}&=i(2\hat{p}^{0})(2\pi)^{3}\delta^{ab}\delta^{(3)}({\bf p}-{\bf q})\ .
\end{align}
The minus sign in the first of Eq.~\eqref{asstateinn} and the vanishing of the $\braket{\beta|\beta}$ inner product imply that $\widetilde{\chi}^{a}$ creates and annihilates negative-norm states, whereas $\beta^{a}$ creates and annihilates zero-norm states -- that is, both field operators are associated with \textit{unphysical} asymptotic states. It is easy to check that the projector $\mathds{P}_{1G}$ onto the linear subspace spanned by these states is
\begin{align}\label{p1g}
   \mathds{P}_{1G}&=\int\frac{d^{3}{\bf p}}{(2\pi)^{3}}\frac{1}{2\hat{p}^{0}}\,\big(\ket{\beta^{a}(p)}\bra{\beta^{a}(p)}+\\
   \notag&\qquad\ -\ket{\chi^{a}(p)}\bra{\beta^{a}(p)}-\ket{\beta^{a}(p)}\bra{\chi^{a}(p)}+\\
   \notag&\qquad\ +i\ket{\gamma^{a}(p)}\bra{\gammabar^{a}(p)}-i\ket{\gammabar^{a}(p)}\bra{\gamma^{a}(p)}\big)\ ,
\end{align}
in agreement with the prediction in Eq.~\eqref{pnproj} for the general form of the terms in $\mathds{P}_{1}$ corresponding to the vacuum projector $\ket{\Omega}\bra{\Omega}$ contained in $\mathds{P}_{0}$.

$\ket{\beta^{a}(p)}$, $\ket{\gamma^{a}(p)}$, and $\ket{\gammabar^{a}(p)}$ are massless eigenstates of the momentum operator $P_{\mu}$. In the Landau gauge ($\xi=\xi_{R}=0$), the same holds for $\ket{\chi^{a}(p)}$. Beyond the Landau gauge, $\ket{\chi^{a}(p)}$ is an eigenstate of the spatial momentum operator $P_{i}$, and a massless dipole state of the Hamiltonian $H=P_{0}$ and of the mass operator $P^{2}$. To prove these statements, observe that, by Eq.~\eqref{asymsymm}, translations of the local fields induce the transformations
\begin{equation}\label{transleq}
   [P_{\mu},\chi^{a}(x)]=-i\partial_{\mu}\chi^{a}(x)
\end{equation}
of $\chi^{a}$, and similarly for $\beta^{a}$, $\gamma^{a}$, and $\gammabar^{a}$. A standard calculation then shows that
\begin{align}\label{deq566}
   \notag P_{\mu}\ket{\beta^{a}(p)}&=\hat{p}_{\mu}\ket{\beta^{a}(p)}\ ,\\
   P_{\mu}\ket{\gamma^{a}(p)}&=\hat{p}_{\mu}\ket{\gamma^{a}(p)}\ ,\\
   \notag P_{\mu}\ket{\gammabar^{a}(p)}&=\hat{p}_{\mu}\ket{\gammabar^{a}(p)}\ .
\end{align}
It follows that these one-particle states are massless eigenstates of $P^{2}$. In the Landau gauge, $\widetilde{\chi}^{a}(x)=\chi^{a}(x)$, and the same holds for $\ket{\chi^{a}(p)}$,
\begin{align}
   \notag P_{\mu}\ket{\chi^{a}(p)}&=\hat{p}_{\mu}\ket{\chi^{a}(p)}\qquad (\xi=\xi_{R}=0)\ .
\end{align}

Beyond the Landau gauge, using the expression provided in Appendix~\ref{app:dipole} for $\mc{E}^{(1/2)}_{x}$ and Eq.~\eqref{transleq}, a straightforward calculation of the commutator $[P_{\mu},\widetilde{\chi}^{a}(x)]$ yields
\begin{equation}\label{deq686}
   [P_{\mu},\widetilde{\chi}^{a}(x)]=-i\partial_{\mu}\widetilde{\chi}^{a}(x)+\frac{i}{2}\,\delta_{\mu}^{0}\,\xi_{R}M_{R}^{2}\ \boldsymbol{\nabla}^{-2}\,\partial_{0}\beta^{a}(x)\ ,
\end{equation}
where $\boldsymbol{\nabla}^{2}$ is the spatial Laplacian, and $\boldsymbol{\nabla}^{-2}$ is its inverse. By a standard calculation, it then follows that
\begin{equation}
   P_{i}\ket{\chi^{a}(p)}=p_{i}\ket{\chi^{a}(p)}\ .
\end{equation}
As for the behavior of $\ket{\chi^{a}(p)}$ with respect to $H=P_{0}$ and $P^{2}$, using Eq.~\eqref{deq686} and
\begin{equation}
   \ket{\chi^{a}(p)}=-i\int d^{3}{\bf x}\ e^{-i\hat{p}x}\,\overleftrightarrow{\partial_{0}}\widetilde{\chi}^{a}(x)\ket{\Omega}\ ,
\end{equation}
where $f\overleftrightarrow{\partial_{0}}g=f(\partial_{0}g)-(\partial_{0}f)g$, one computes
\begin{align}
   P_{0}\ket{\chi^{a}(p)}&=-i\int d^{3}{\bf x}\ e^{-i\hat{p}x}\,\overleftrightarrow{\partial_{0}}\left[P_{0},\widetilde{\chi}^{a}(x)\right]\ket{\Omega}\ .
\end{align}
The above equation yields
\begin{align}\label{chistatedip}
   \notag P_{0}\ket{\chi^{a}(p)}&=\hat{p}_{0}\ket{\chi^{a}(p)}+\frac{\xi_{R}M_{R}^{2}}{2\hat{p}^{0}}\,\ket{\beta^{a}(p)}\ ,\\
   P_{0}^{2}\ket{\chi^{a}(p)}&=\hat{p}_{0}^{2}\ket{\chi^{a}(p)}+\xi_{R}M_{R}^{2}\ket{\beta^{a}(p)}\ ,\\
   \notag P^{2}\ket{\chi^{a}(p)}&=\xi_{R}M_{R}^{2}\,\ket{\beta^{a}(p)}\ ,\\
   \notag(P^{2})^{2}\ket{\chi^{a}(p)}&=0\ .
\end{align}
The first, third and fourth of Eq.~\eqref{chistatedip}, together with $(P_{0}-\hat{p}_{0})^{2}\ket{\chi^{a}(p)}=0$, imply that $\ket{\chi^{a}(p)}$ is a massless dipole of the Hamiltonian $H=P_{0}$ and of the mass operator $P^{2}$.

It can be checked explicitly that the massless discrete spectrum from which we started, including the dipole in the gluon propagator, is recovered by inserting the one-particle projector $\mathds{P}_{1G}$ in Eq.~\eqref{p1g} between the fields $A_{\mu}^{a}$, $B^{a}$, $c^{a}$ and $\cbar^{a}$ in two-point Green's functions, and by using Eq.~\eqref{1psme}, the asymptotic limit
\begin{equation}\label{deq285}
   A_{\mu}^{a}\to M^{-1}\partial_{\mu}\widetilde{\chi}^{a}-\xi M\,\partial_{\mu}(\mathcal{E}^{(1/2)}\beta^{a})\ ,
\end{equation}
and the commutation relations between the asymptotic creation and annihilation operators. The second term in Eq.~\eqref{deq285} provides the dipole terms. For instance, in the two-point function $\langle A_{\mu}^{a}(x)A_{\nu}^{b}(y)\rangle$, that term yields
\begin{align}
   &\langle A_{\mu}^{a}(x)A_{\nu}^{b}(y)\rangle=\langle A_{\mu}^{a}(x)\mathds{P}_{1G}A_{\nu}^{b}(y)\rangle+\cdots=\\
   \notag&=\xi\,\delta^{ab}\,\partial_{\mu}^{(x)}\partial_{\nu}^{(y)}\,\left(\mathcal{E}^{(1/2)}_{x}+\mathcal{E}^{(1/2)}_{y}\right)\,\int\frac{d^{3}{\bf p}}{(2\pi)^{3}}\,\frac{e^{-i\hat{p}(x-y)}}{2\hat{p}^{0}}+\cdots\ ,
\end{align}
which translates to
\begin{align}
   &\langle [A_{\mu}^{a}(x),A_{\nu}^{b}(y)]\rangle=\\
   \notag&=i\xi\,\delta^{ab}\,\partial_{\mu}^{(x)}\partial_{\nu}^{(y)}\,\left(\mathcal{E}^{(1/2)}_{x}+\mathcal{E}^{(1/2)}_{y}\right)\,D(x-y)+\cdots\ .
\end{align}
The dipole in the first of Eq.~\eqref{vaccomm} is then recovered using Eq.~\eqref{deq855} and substituting $\partial^{(y)}_{\nu}$ with $-\partial^{(x)}_{\nu}$. 

For later reference, we report the equal-time (anti)commutation relations between the asymptotic fields and their time derivatives. Using \cite{NO90}
\begin{align}
   \notag D(x-y)|_{x^{0}=y^{0}}&=0\ ,\qquad E(x-y)|_{x^{0}=y^{0}}=0\ ,\\
   \frac{\partial}{\partial y^{0}}\bigg|_{y^{0}=x^{0}}D(x-y)&=\delta^{(3)}({\bf x}-{\bf y})\ ,\\
   \notag \frac{\partial}{\partial y^{0}}\bigg|_{y^{0}=x^{0}}E(x-y)&=0\ ,
\end{align}
we find that the only nonvanishing equal-time (anti)commutation relations are
\begin{align}\label{ascommet}
   &[\chi^{a}({\bf x},t),\dot{\chi}^{b}({\bf y},t)]=[\chi^{a}({\bf x},t),\dot{\beta}^{b}({\bf y},t)]=\\
   \notag&\qquad=[\beta^{a}({\bf x},t),\dot{\chi}^{b}({\bf y},t)]=-i\delta^{ab}\delta^{(3)}({\bf x}-{\bf y})\ ,\\
   \notag &\{\gamma^{a}({\bf x},t),\dot{\gammabar}^{b}({\bf y},t)\}=-\{\gammabar^{a}({\bf x},t),\dot{\gamma}^{b}({\bf y},t)\}=\\
   \notag&\qquad=-\delta^{ab}\,\delta^{(3)}({\bf x}-{\bf y})\ .
\end{align}\\

We now turn to the properties of the asymptotic fields $\chi^{a}$, $\beta^{a}$, $\gamma^{a}$, and $\gammabar^{a}$ under BRST transformations. Recalling that the latter act on the elementary fields of QCD as
\begin{align}\label{brstagain}
   \notag [i Q_{B},A_{\mu}^{a}]&=D_{\mu}c^{a}\ ,\quad\{iQ_{B},c^{a}\}=-\frac{g}{2}f^{abc}c^{b}c^{c}\ ,\\
   \{iQ_{B},\cbar^{a}\}&=iB^{a}\ ,\qquad [iQ_{B},B^{a}]=0\ ,
\end{align}
we follow \cite{KO79b,KO79c,NO90} in postulating there are no massless bound states in the ghost-gluon $f^{abc}A_{\mu}^{b}c^{c}$ and two-ghost $f^{abc}c^{b}c^{c}$ composite channels, so that the asymptotic limit of the former is provided entirely by $\partial_{\mu}\gamma^{a}$, and that of the latter is zero. The standard analysis of \cite{KO79b,KO79c,NO90} then applies unaltered to the massless asymptotic component of the elementary fields of dynamically massive QCD. In particular, by Eq.~\eqref{asymsymm}, using the third of Eq.~\eqref{quartetasym} in addition to the results of this section, we can take the asymptotic limit of Eq.~\eqref{brstagain} and separate the massless from a potential massive component to find
\begin{align}\label{brst0}
   \notag[i Q_{B},Z_{3}^{1/2}M_{R}^{-1}\partial_{\mu}\chi^{a}]&=\widetilde{Z}_{3}^{-1/2}\partial_{\mu}\gamma^{a}\ ,\\
   \{iQ_{B},\widetilde{Z}_{3}^{1/2}\gamma^{a}\}&=0\ ,\\
   \notag \{iQ_{B},\widetilde{Z}_{3}^{1/2}\gammabar^{a}\}&=iZ_{3}^{-1/2}M_{R}\,\beta^{a}\ ,\\
   \notag[iQ_{B},Z_{3}^{-1/2}M_{R}\,\beta^{a}]&=0\ .
\end{align}
By introducing a rescaled (``renormalized'') BRST charge $Q_{B,R}$,
\begin{equation}\label{renqb}
Q_{B,R}=(\widetilde{Z}_{3}Z_{3})^{1/2}\,Q_{B}\ ,
\end{equation}
and by integrating the first of Eq.~\eqref{brst0} while discarding a constant to avoid breaking BRST symmetry, we obtain the following expressions for the BRST transformations of $\chi^{a}$, $\beta^{a}$, $\gamma^{a}$, and $\gammabar^{a}$:
\begin{align}\label{brste}
   [i Q_{B,R},\chi^{a}]&=M_{R}\,\gamma^{a}\ ,\qquad\{iQ_{B,R},\gamma^{a}\}=0\ ,\\
   \notag \{iQ_{B,R},\gammabar^{a}\}&=iM_{R}\,\beta^{a}\ ,\qquad[iQ_{B,R},\beta^{a}]=0\ .
\end{align}
Note that the normalization of $Q_{B,R}$ is largely arbitrary: the operator can be redefined by multiplying it by a finite constant. In Eq.~\eqref{renqb}, we used $Z_{3}$ for the sole purpose of obtaining a recognizable, finite mass scale on the RHSs of the BRST transformations.

Eq.~\eqref{brste} tell us that $\chi^{a}$ and $\beta^{a}$, together with $\gamma^{a}$ and $\gammabar^{a}$, form a BRST quartet. Specifically, they form the elementary, massless, scalar BRST quartet of dynamically massive QCD, within which $\chi^{a}$ is the first parent. In terms of the corresponding one-particle asymptotic states\footnote{Note that, since $[iQ_{B,R},\beta^{a}]=0$, we have $[iQ_{B,R},\widetilde{\chi}^{a}]=[iQ_{B,R},\chi^{a}]=M_{R}\gamma^{a}$.},
\begin{align}
   \notag Q_{B,R}\ket{\chi^{a}(p)}=-iM_{R}\ket{\gamma^{a}(p)}\ ,\quad Q_{B,R}\ket{\gamma^{a}(p)}=0\ ,\\
   Q_{B,R}\ket{\gammabar^{a}(p)}=M_{R}\ket{\beta^{a}(p)}\ ,\quad Q_{B,R}\ket{\beta^{a}(p)}=0\ .
\end{align}

By virtue of asymptotic completeness, the renormalized BRST charge $Q_{B,R}$ admits a unique expression in terms of the asymptotic fields $\chi^{a}$, $\beta^{a}$, $\gamma^{a}$, and $\gammabar^{a}$ \cite{KO78b,KO78a,KO79b,KO79c,NO90}. Indeed, setting
\begin{equation}\label{qbasym}
   Q_{B,R}=M_{R}\int d^{3}{\bf x}\ (\dot{\gamma}^{a}\beta^{a}-\gamma^{a}\dot{\beta}^{a})+\cdots\ ,
\end{equation}
where the ellipsis denotes other potential terms, noting that the explicit terms in Eq.~\eqref{qbasym} yield a time-independent spatial integral, and using the equal-time (anti)commutation relations in Eq.~\eqref{ascommet}, it is easy to see that the explicit terms generate the transformations in Eq.~\eqref{brste}. It follows that any term in the ellipsis of Eq.~\eqref{qbasym} must (anti)commute with all the fields of the elementary BRST quartet, and hence can only contain asymptotic fields not belonging to that quartet. This observation completes the asymptotic analysis of the massless component of the gluon spectrum in dynamically massive QCD.\\

Let us summarize the main results of this section. By inspection of the tensorial structure of the gluon propagator $\Delta_{\mu\nu}^{ab}(p)$ in the presence of dynamical mass generation, we have determined that, in linear covariant gauges, the gluon spectrum splits into an unknown massive component consisting of transverse polarization triplets, and a massless longitudinal component which is entirely fixed by BRST symmetry and by the value of $\Delta(0)$. The asymptotic analysis of the massless component revealed that with the elementary fields of QCD are associated a quartet of colored scalar asymptotic fields $\chi^{a}$, $\beta^{a}$, $\gamma^{a}$, and $\gammabar^{a}$ -- Eq.~\eqref{asymm} -- which satisfy the field equations in Eq.~\eqref{asfieldeq} and the commutation relations in Eq.~\eqref{ascomm}. The corresponding creation and annihilation operators satisfy the commutation relations in Eq.~\eqref{ascommca}; by acting on the vacuum, the former create the one-particle states in Eq.~\eqref{asstateinn}. The one-particle state $\ket{\chi^{a}(p)}$ is a massless dipole state of the Hamiltonian $H=P_{0}$ and of the mass operator $P^{2}$ in all gauges except for the Landau gauge ($\xi=0$), in which it is a simple eigenstate. The LSZ coefficient of the asymptotic field $\chi^{a}$ is $Z_{\chi}^{1/2}=M^{-1}$.

Under BRST transformations, the fields $\chi^{a}$, $\beta^{a}$, $\gamma^{a}$, and $\gammabar^{a}$ transform into each other as indicated in Eq.~\eqref{brste}: they form the elementary, scalar, massless BRST quartet of dynamically massive QCD. This result is expected, but nontrivial. It shows that 1. even in the fully nonperturbative setting in which gluons dynamically acquire a mass, the quartet structure of the elementary fields' gauge modes is preserved, and 2. one does not need to postulate the breaking of BRST invariance to account for the structure of the massive gluon propagator. As the next sections will confirm, at the level of propagators and of asymptotic states, BRST invariance and dynamical gluon mass generation are fully compatible with one another.

The assumptions under which these results were obtained are customary within the KO framework. In fact, they are overall weaker than the standard assumptions of \cite{KO79b,KO79c,NO90}. Let us discuss each of them in turn.

First, we assumed that the gluon and ghost propagators are diagonal in color space. This is expected in view of the preservation of color symmetry, confirmed by well-established lattice data \cite{SS96,Cuc97,LSWP98a,LSWP98b,BBLW00,BBLW01,BHLP04,SIMS05,BBMM06,CM07,CM08a,CM08b,BIMS09,ISI09,BMM10,ABBC12,BLLM12,OS12,BBCO15,CDMO18,FN03,FN06a,Ste06}, and is a standard assumption for QCD within the KO framework \cite{KO79b,KO79c,NO90}.

Second, we assumed that the massless discrete spectrum associated with the ghost propagator takes the form of a simple pole. With one exception which we will discuss in a moment, this is also a standard assumption within the KO framework \cite{KO79b,KO79c,NO90}. It avoids the introduction of higher-order multipole ghost fields which are not observed in any Standard Model theory \cite{KO79b,KO79c,NO90}, and whose existence is ruled out by well-established QCD lattice data \cite{SS96,Cuc97,SIMS05,BBMM06,CM08a,BIMS09,ABBC12,BLLM12,CDMO18}. While this assumption does not imply, per se, that the ghost dressing function $J(p^2)$ saturates at $p^{2}=0$, it raises the question of how can ghost have a discrete massless spectrum if the multiparticle continuum dominates over the single-particle spectrum at $p^{2}=0$. Specifically, it is unclear whether a non-saturating behavior for $J(p^{2})$ is compatible with the existence of ghost asymptotic fields, which are in turn essential to the quartet structure of unphysical degrees of freedom imposed by BRST symmetry. The saturating behavior displayed by the lattice data \cite{SS96,Cuc97,SIMS05,BBMM06,CM08a,BIMS09,ABBC12,BLLM12,CDMO18} is fully compatible with the existence of simple-pole ghost asymptotic fields, and poses no obstruction to the preservation of BRST symmetry. The exception we mentioned earlier is Kugo's \cite{Kug95}, which postulated an infrared-enhanced Landau-gauge $J(p^{2})$ in order to satisfy the criterion $1+u=0$ in the same gauge. As discussed in the Introduction and in Sec.~\ref{sec:koanal}, the logic of the criterion is flawed, and the criterion is formulated starting from an incorrect equation; therefore, we ignore \cite{Kug95}.

Third, we assumed standard behavior for renormalization and LSZ coefficients. Namely, we assumed that $\widetilde{Z}_{3}=J_{0}>0$, $Z_{3}>0$, the latter of which implies $M^{-2},M_{R}^{-2}\geq 0$ in view of the non-negativity of the Euclidean gluon propagator. While we did not exclude $M^{-2},M_{R}^{-2}=0$ a priori, a posteriori we found that $Z_{\chi}^{1/2}=M^{-1}$. $M^{-1}=0$ -- equivalently, $\Delta(0)=0$ -- thus implies non-standard behavior for the LSZ coefficient $Z_{\chi}^{1/2}$, favoring $\Delta(0)\neq0$, as is observed on the lattice \cite{Cuc97,LSWP98a,LSWP98b,BBLW00,BBLW01,BHLP04,SIMS05,BBMM06,CM07,CM08a,CM08b,BIMS09,ISI09,BMM10,ABBC12,BLLM12,OS12,BBCO15,FN03,FN06a,Ste06}. The analysis of dynamical gluon mass generation in the next section will provide further evidence against $\Delta(0)=0$.

Fourth, we assumed asymptotic completeness. As we saw in Sec.~\ref{sec:framework}, asymptotic completeness of the full Hilbert space with respect to \textit{some} set of asymptotic fields is a customary assumption in the KO framework. Our completeness postulate does not exclude the existence of asymptotic fields other than those associated with the elementary fields, as in \cite{KO79b,KO79c,NO90}. It only requires that any additional massless field with their quantum numbers (anti)commutes with $\chi_{\mu}^{a}$, $\beta^{a}$, $\gamma^{a}$, and $\gammabar^{a}$ -- in other words, that it represents massless degrees of freedom independent of them. The elementary channels of $A_{\mu}^{a}$, $B^{a}$, $c^{a}$, and $\cbar^{a}$ being saturated, any additional asymptotic field must arise from composite channels, hence it would represent a stable massless colored bound state. The existence of one such bound state has been proposed in the context of studies of the Schwinger mechanism in QCD \cite{JJ73,Smi74,AIMP12,IP13,ABP16,ABP17,Pap22,FP23,FP24,FP25}, and in the next section we will show that it is already associated with the asymptotic field $\chi^{a}$.

Fifth, we assumed that no massless bound states exist either in the gluon-ghost channel $f^{abc}A_{\mu}^{b}c^{c}$ or in the two-ghost channel $f^{abc}c^{b}c^{c}$. This assumption is customary as well \cite{KO79b,KO79c,NO90}. The non-existence of the former enforces the uniqueness of the BRST partner of $\chi^{a}$, identifying it with the ghost asymptotic field $\gamma^{a}$; the non-existence of the latter is then necessary to preserve the nilpotency of the BRST charge, which must annihilate $\gamma^{a}$.

Finally, at various stages, we discarded constants that would have violated spacelike commutativity, broken BRST invariance, or provided vacuum expectation values for the asymptotic fields.\\

What is the meaning of the structure we found? Does it have an analogoue in known theories? A complete answer to the first question will be provided in the coming sections. The answer to the second is yes: as we show in Appendices~\ref{app:stschw} and \ref{app:higgsmod}, this structure is found identical in Yang-Mills theories like the (Abelian) Stueckelberg model, the Schwinger model, and Higgs models in which a mass is generated equal for all gauge bosons\footnote{Theories in which gauge bosons acquire different masses do not behave in a qualitatively different way, but they violate the assumption of diagonality with respect to the color indices \cite{KO79d,KO79a}.} thanks to unphysical scalar Goldstone bosons. This fact is not surprising, given that these theories satisfy the assumptions listed above.

As discussed in the Introduction and in Sec.~\ref{sec:review}, modulo mixing with $\beta$, the first parents $\chi$ of the elementary BRST quartet of the Stueckelberg, Schwinger and Higgs models are, respectively, the Stueckelberg field, a fermion-antifermion state, and the unphysical components of the Higgs field. Establishing the identity of $\chi^{a}$ in dynamically massive QCD is the main objective of the next section.

\subsection{First-parent state dynamics, bound-state interpretation, and the Schwinger mechanism}\label{sec:goldbound}

In order to identify the first parent $\chi^{a}$ of the elementary BRST quartet of dynamically massive QCD, in this section we begin by studying the $\chi^{a}$-to-vacuum matrix element of the MYME,
\begin{align}\label{MYMbsa}
   &\bra{\Omega}\partial^{\nu}F_{\nu\mu}^{a}+gf^{abc}A^{b\,\nu}F^{c}_{\nu\mu}+igf^{abc}\partial_{\mu}\cbar^{b}c^{c}+\\
   \notag&\qquad\quad-\partial_{\mu}B^{a}\ket{\chi^{d}(p)}=0\ .
\end{align}
Roughly speaking, Eq.~\eqref{MYMbsa} can be interpreted as one of four momentum-space Schr\"{o}dinger equations describing the one-particle dynamics of $\chi^{a}$\footnote{Of the four equations in Eq.~\eqref{feqs}, only the MYME provide new dynamical information on $\ket{\chi^{a}(p)}$: the matrix elements of the ghost and antighost equations vanish by ghost number conservation, and the gauge-fixing condition just confirms that the momentum of the state satisfies $\hat{p}^{2}=0$.}.

Using Eq.~\eqref{1psme} and the identity
\begin{equation}\label{sumrp}
   \bra{\Omega}\partial_{\mu}B^{a}(x)\ket{\chi^{d}(p)}=M\bra{\Omega}\partial_{\mu}\beta^{a}(x)\ket{\chi^{d}(p)}\ ,
\end{equation}
we can rewrite Eq.~\eqref{MYMbsa} as
\begin{align}\label{sumr0}
   \notag&\bra{\Omega}\partial^{\nu}F_{\nu\mu}^{a}+gf^{abc}A^{b\,\nu}F^{c}_{\nu\mu}+igf^{abc}\partial_{\mu}\cbar^{b}c^{c}\ket{\chi^{d}(p)}=\\
   &\qquad=iZ_{3}^{-1/2}M_{R}\,\hat{p}_{\mu}\,\delta^{ad}\ e^{-i\hat{p}x}\ .
\end{align}
Here, as in the previous sections, $\hat{p}=(|{\bf p}|,{\bf p})$ is a massless on-shell momentum, and to go from Eq.~\eqref{sumrp} to Eq.~\eqref{sumr0} we used the known commutation relations between $\widetilde{\chi}^{a}$ and $\beta^{a}$. The field-strength tensor $F_{\mu\nu}^{a}=\partial_{\mu}A_{\nu}^{a}-\partial_{\nu}A_{\mu}^{a}+gf^{abc}A_{\mu}^{b}A_{\nu}^{c}$ in Eq.~\eqref{sumr0} contains two terms, $\partial_{\mu}A_{\nu}^{a}-\partial_{\nu}A_{\mu}^{a}$ and $gf^{abc}A_{\mu}^{b}A_{\nu}^{c}$, which are separately antisymmetric in their Lorentz indices. The matrix element of the first,
\begin{equation}
   \bra{\Omega}\partial_{\mu}A_{\nu}^{a}-\partial_{\nu}A_{\mu}^{a}\ket{\chi^{d}(p)}\ ,
\end{equation}
can be evaluated explicitly and vanishes because, asymptotically, $\chi^{a}$ enters $A_{\mu}^{a}$ as a gradient, $A_{\mu}^{a}\to M^{-1}\,\partial_{\mu}\chi^{a}+\cdots$. The matrix element of the second vanishes as well: in order for it to be nonzero, $f^{abc}A_{\mu}^{b}A_{\nu}^{c}$ would need to develop an asymptotic contribution from either $\chi^{a}$ or $\beta^{a}$, but this cannot happen in a way that yields an antisymmetric tensor \cite{KO79d,KO79a,NO90}. While $\bra{\Omega}\partial^{\nu}F_{\nu\mu}^{a}(x)\ket{\chi^{a}(p)}$ vanishes as a whole, for reasons that will become clear in a moment, we will nonetheless keep the contribution from $gf^{abc}A_{\mu}^{b}A_{\nu}^{c}$. Overall, we find
\begin{align}\label{sumr1}
   \notag&gf^{abc}\bra{\Omega}\partial^{\nu}(A_{\nu}^{b}A_{\mu}^{c})+A^{b\,\nu}(\partial_{\nu}A_{\mu}^{c}-\partial_{\mu}A_{\nu}^{c})\ket{\chi^{d}(p)}=\\
   &\ +g^{2}f^{abc}f^{cef}\bra{\Omega}A^{b\,\nu}A_{\nu}^{e}A_{\mu}^{f}\ket{\chi^{d}(p)}+\\
   \notag&\ +igf^{abc}\bra{\Omega}\partial_{\mu}\cbar^{b}c^{c}\ket{\chi^{d}(p)}=iM\,\hat{p}_{\mu}\,\delta^{ad}\ e^{-i\hat{p}x}\ .
\end{align}
Note that the LHS of Eq.~\eqref{sumr1} is but the matrix element $\braket{\Omega |K_{\mu}^{a}(x)|\chi^{d}(p)}$, where $K_{\mu}^{a}$ is the gluon polarization's kernel operator defined in Eq.~\eqref{kernelop}.

\begin{figure*}[t]
    \includegraphics[width=0.97\textwidth]{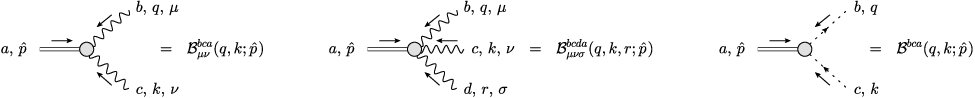}
    \caption{Bethe-Salpeter vertices (BSVs) of the first parent of the elementary BRST quartet $\chi^{a}$ of dynamically massive QCD. All momenta are ingoing. The scalar-three-gluon BSV $\mc{B}_{\mu\nu\sigma}^{abcd}$ is not one-particle irreducible.}
    \label{BSVdiag}
\end{figure*}

Eq.~\eqref{sumr1} is an unrenormalized sum rule between the Bethe-Salpeter amplitudes (BSA) \cite{BS51,Nak69,RW94,AvS01} of the first parent $\chi^{a}$ in the two-gluon, the three-gluon, and the ghost-antighost channels. To see this, let us define BSAs
\begin{align}\label{BSA}
   \notag B_{\mu\nu}^{abc}(x,y;\hat{p})&=\bra{\Omega}T\left\{A_{\mu}^{a}(x)A_{\nu}^{b}(y)\right\}\ket{\chi^{c}(p)}\ ,\\
   \notag B_{\mu\nu\sigma}^{abcd}(x,y,z;\hat{p})&=\bra{\Omega}T\{A_{\mu}^{a}(x)A_{\nu}^{b}(y)A_{\sigma}^{c}(z)\}\ket{\chi^{d}(p)}\ ,\\
   B^{abc}(x,y;\hat{p})&=\bra{\Omega}T\{c^{a}(x)\cbar^{b}(y)\}\ket{\chi^{c}(p)}\ ,
\end{align}
and -- after Fourier transforming and factorizing the external legs from the BSAs -- corresponding Bethe-Salpeter vertices (BSV),
\begin{align}\label{BSV}
   \notag&B^{abc}_{\mu\nu}(q,k;\hat{p})=\Delta_{\mu}^{aa^{\prime}\mu^{\prime}}(q)\Delta_{\nu}^{bb^{\prime}\nu^{\prime}}(k)\,\mc{B}^{a^{\prime}b^{\prime}c}_{\mu^{\prime}\nu^{\prime}}(-q,-k;\hat{p})\ ,\\
   &B^{abcd}_{\mu\nu\sigma}(q,k,r;\hat{p})=\\
   \notag&\ \ =\Delta_{\mu}^{aa^{\prime}\mu^{\prime}}(q)\Delta_{\nu}^{bb^{\prime}\nu^{\prime}}(k)\Delta_{\sigma}^{cc^{\prime}\sigma^{\prime}}(r)\,\mc{B}^{a^{\prime}b^{\prime}c^{\prime}d}_{\mu^{\prime}\nu^{\prime}\sigma^{\prime}}(-q,-k,-r;\hat{p})\ ,\\
   \notag &B^{abc}(q,k;\hat{p})=\mc{G}^{aa^{\prime}}(q)\mc{G}^{bb^{\prime}}(k)\,\mc{B}^{a^{\prime}b^{\prime}c}(-q,-k;\hat{p})\ .
\end{align}
The BSVs are displayed diagrammatically in Fig.~\ref{BSVdiag}. Note that, according to our definitions, the scalar-three-gluon BSV $\mc{B}^{abcd}_{\mu\nu\sigma}(q,k,r;\hat{p})$ is \textit{not} one-particle irreducible\footnote{In $B_{\mu\nu\sigma}^{abcd}(q,k,r;\hat{p})$, we also keep disconnected terms of the form $\bra{\Omega} A^{2}\ket{\Omega}\bra{\Omega}A\ket{\chi}$ implicit.}.

By adopting the customary definition
\begin{equation}\label{ope}
   \mc{O}_{1}(x)\,\mc{O}_{2}(x)=\lim_{y\to x}\ T\left\{\mc{O}_{1}(x)\mc{O}_{2}(y)\right\}\ ,
\end{equation}
in terms of the BSVs, Eq.~\eqref{sumr1} translates to
\begin{widetext}
   \begin{align}\label{sumr2}
      &\frac{g}{2}\ \int_{q,k}\ \Gamma^{\mu\nu\tau}_{(0)\,abc}(-\hat{p},q,k)\ \Delta_{\nu}^{bb^{\prime}\nu^{\prime}}(q)\,\Delta_{\tau}^{cc^{\prime}\tau^{\prime}}(k)\ \mc{B}_{\nu^{\prime}\tau^{\prime}}^{b^{\prime}c^{\prime}d}(-q,-k;\hat{p})\ \delta_{q+k,\hat{p}}+\\
      \notag&\ +\frac{g^{2}}{6}\ \int_{q,k,r}\ \Gamma^{\mu\nu\tau\lambda}_{(0)\,abcf}(-\hat{p},q,k,r)\ \Delta_{\nu}^{bb^{\prime}\nu^{\prime}}(q)\,\Delta_{\tau}^{cc^{\prime}\tau^{\prime}}(k)\,\Delta_{\lambda}^{ff^{\prime}\lambda^{\prime}}(r)\ \mc{B}_{\nu^{\prime}\tau^{\prime}\lambda^{\prime}}^{b^{\prime}c^{\prime}f^{\prime}d}(-q,-k,-r;\hat{p})\  \delta_{q+k+r,\hat{p}}+\\
      \notag&\qquad-g\int_{q,k}\ \Gamma_{(0)\,abc}^{\mu}(-\hat{p},q,k)\ \mc{G}^{bb^{\prime}}(q)\,\mc{G}^{cc^{\prime}}(k)\ \mc{B}^{b^{\prime}c^{\prime}d}(-q,k;\hat{p})\ \delta_{q-k,\hat{p}}=:gI^{\mu}(-\hat{p})\,\delta_{a}^{d}=-Z_{3}^{-1/2}M_{R}\,\hat{p}^{\mu}\,\delta_{a}^{d}\ ,
   \end{align}
\end{widetext}
where the $\Gamma_{(0)}$'s are the tree-level vertices of QCD -- see Eq.~\eqref{emvertices} and Fig.~\ref{QCDVertices} -- and we used the notation $\int_{q}=\int\frac{d^{4}q}{(2\pi)^{4}}$, $\delta_{q,k}=(2\pi)^{4}\,\delta^{(4)}(q-k)$. Retaining the vanishing matrix element $f^{abc}\bra{\Omega}\partial^{\nu}(A_{\nu}^{b}A_{\mu}^{c})\ket{\chi^{d}(p)}$ in the equation is useful to identify the tree-level three-gluon vertex $\Gamma_{(0)abc}^{\mu\nu\tau}$.

\begin{figure*}[t]
    \includegraphics[width=0.56\textwidth]{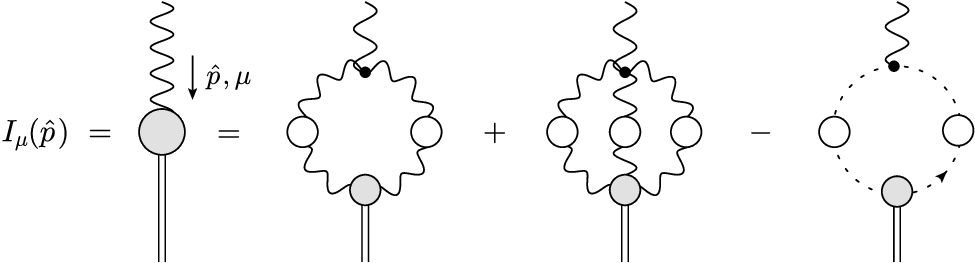}
    \caption{Gluon-scalar transition amplitude. On the RHS, solid dots represent tree-level vertices, grey blobs the BSVs of Fig.~\ref{BSVdiag}, and white blobs represent dressed propagators.}
    \label{GSTAdiag}
\end{figure*}

As anticipated above, since its RHS is different from zero, Eq.~\eqref{sumr2} implies that at least one of the BSAs $B_{\mu\nu}^{abc}$, $B_{\mu\nu\sigma}^{abcd}$ or $B^{abc}$ is different from zero, and that the three of them -- when integrated with tree-level vertices -- sum to a quantity which is proportional to the unrenormalized gluon mass. In the absence of symmetries protecting the corresponding channels, it is reasonable to assume all three amplitudes will be non-vanishing. In the presence of quarks, Eq.~\eqref{MYMbsa} also includes the quark current $g\sum_{f}\psibar_{f}\gamma^{\mu}T^{a}\psi_{f}$ -- $f$ denoting the flavor of the quark --, and the LHS of Eq.~\eqref{sumr2} contains a corresponding quark term. In full QCD, the scalar $\chi^{a}$ thus possesses a potentially -- and reasonably -- non-vanishing BSA in the quark-antiquark channel as well. For later reference, we observe that the non-vanishing of the BSAs occurs specifically in the channel of the Yang-Mills equations -- that is, in the channel of the gluon field $A_{\mu}^{a}$ and of the global gauge (color) current $J^{a}_{\mu}$.

The quantity $I_{\mu}(p)=I(p^{2})\,p_{\mu}$ whose on-shell $p^{2}\to0$ limit is defined by the LHS of Eq.~\eqref{sumr2},
\begin{align}\label{g2saker}
   &I_{\mu}(-\hat{p})\,\delta^{ab}\,e^{-i\hat{p}x}:=ig^{-1}\bra{\Omega}K_{\mu}^{a}(x)\ket{\chi^{b}(p)}\ ,
\end{align}
goes by the name of ``gluon-scalar transition amplitude''\footnote{See for instance Eq.~(7.23) of \cite{FP25}, in which only the two-gluon BSA is retained, or Eqs.~(7.16), (8.1), (8.7) and (8.11) of \cite{IP13}, which display its various contributions as per Fig.~5 therein. Note that our $\mc{B}^{abcd}_{\mu\nu\sigma}$, not being one-particle irreducible, includes the contributions of both Eqs.~(8.1) and (8.11). Our conventions are fixed so that $I_{\mu}(q)$ and the tree-level vertices coincide with the diagrammatic definitions of \cite{FP25}, whereas we retain the $-i$ in propagators, and do not add an extra $i$ in the BSVs.} -- see Fig.~\ref{GSTAdiag} -- and plays a central role in the dynamical generation of a gluon mass via the Schwinger mechanism \cite{JJ73,EF74,Smi74,AIMP12,IP13,FP25}. Its name reflects the fact that the index $\mu$ is carried by three-gluon, four-gluon and ghost-antighost-gluon tree-level vertices, all of which can be connected to an external gluon leg. The minus sign on the RHS of Eq.~\eqref{sumr2} is due to the momentum flow in our equation going from the scalar to the gluon, while the customary definition of $I_{\mu}(p)$ has momentum flowing in the opposite direction. By Eq.~\eqref{sumr2},
\begin{equation}\label{g2sa}
   I_{\mu}(\hat{p})=g^{-1}Z_{3}^{-1/2}\,M_{R}\,\hat{p}_{\mu}\ \Longleftrightarrow\ I(0)=g^{-1}Z_{3}^{-1/2}\,M_{R}\ .
\end{equation}

The fact that $\chi^{a}$ has non-vanishing BSAs in gluon, ghost and quark composite channels suggests that the first parent of the elementary BRST quartet of dynamically massive QCD is a bound-state superposition of the form\footnote{The following equation must not be understood as a literal relation between states. It only serves us to illustrate the valence content of $\chi^{a}$.}
\begin{equation}
   \ket{\chi}\sim \ket{AA}+\ket{AAA}+\ket{c\cbar}+\ket{\psi\psibar}+\cdots\ ,
\end{equation}
where the ellipsis denotes other potential channels. If this is the case, then $\chi^{a}$ is expected to generate massless poles in the appropriate amputated dressed vertices.

And indeed, this expectation is fulfilled. In Appendix~\ref{app:schwpoles}, with the aid of the one-particle projector $\mathbb{P}_{1G}$ of Eq.~\eqref{p1g}, we show that the three-gluon, the four-gluon, the ghost-antighost-gluon, and -- in the presence of quarks -- the quark-antiquark-gluon vertices, all develop massless longitudinal poles proportional to the BSVs of $\chi^{a}$ on each of their gluon legs. Specifically, employing for the dressed vertices $\Gamma$ the same conventions of the tree-level vertices in Eq.~\eqref{emvertices}, we find that, in the limit $p^{0}\to \hat{p}^{0}$,
\begin{align}\label{schwpoles}
   \notag\Gamma_{\mu\nu\sigma}^{abc}(p,q,k)&\to I_{\mu}(\hat{p})\,\frac{i}{p^{2}+i\epsilon}\,\mc{B}^{bca}_{\nu\sigma}(q,k;\hat{p})+\cdots\ ,\\
   \notag\Gamma_{\mu\nu\sigma\lambda}^{abcd}(p,q,k,r)&\to g^{-1}\,I_{\mu}(\hat{p})\,\frac{i}{p^{2}+i\epsilon}\,\mc{B}^{bcda}_{\nu\sigma\lambda}(q,k,r;\hat{p})+\\
   &\qquad\qquad+\cdots\ ,\\
   \notag\Gamma_{\mu}^{abc}(p,q,k)&\to I_{\mu}(\hat{p})\,\frac{i}{p^{2}+i\epsilon}\,\mc{B}^{cba}(-k,q;\hat{p})+\cdots\ ,
\end{align}
where the ellipses denote longitudinal poles produced by the NL field $B^{a}$ which do not contribute to the gluon polarization when summed together in its SDE. Analogous limits hold as $p^{0}\to -\hat{p}^{0}$, and when the momentum of any other external gluon leg is brought on the massless mass shell. In Eq.~\eqref{schwpoles}, like the scalar-3-gluon BSV on the RHS, the four-gluon vertex $\Gamma_{\mu\nu\sigma\lambda}^{abcd}(p,q,k,r)$ is not one-particle irreducible\footnote{Moreover, the singular limit of $\Gamma_{\mu\nu\sigma\tau}^{abcd}$ contains the same disconnected terms we left implicit in $B_{\mu\nu\sigma}^{abcd}$. These terms are thus not neglected, only visually omitted to avoid cluttering notation.}.

\begin{figure}[h]
    \includegraphics[width=0.38\textwidth]{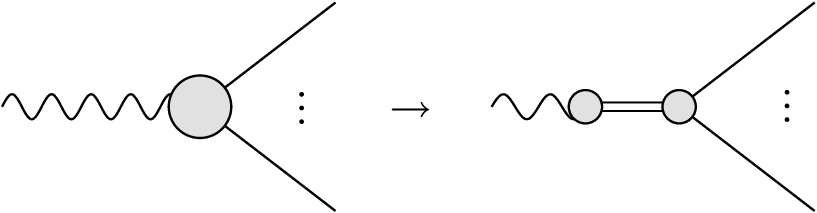}
    \caption{Formation of vertex Schwinger poles in dynamically massive QCD -- see Eq.~\eqref{schwpoles}. When a gluon leg is brought on the massless mass shell $p^{2}\to 0$, the vertex develops a pole $i/(p^{2}+i\epsilon)$, represented in the Figure as a double line, proportional to the BSV of $\chi^{a}$ in the corresponding channel -- see Fig.~\ref{BSVdiag} -- and to the gluon-scalar transition amplitude in Fig.~\ref{GSTAdiag}. In the same limit, the NL field $B^{a}$ produces vertex poles which do not contribute to dynamical gluon mass generation when summed together in the gluon polarization.}
    \label{VPole}
\end{figure}

The formation of the so-called ``vertex Schwinger poles'' in Eq.~\eqref{schwpoles} is depicted schematically in Fig.~\ref{VPole}. Their appearance in channels of composite operators, the non-vanishing of at least one -- reasonably, all -- of the associated Bethe-Salpeter amplitudes, and the absence from the theory of elementary scalar degrees of freedom other than $B^{a}$, $c^{a}$, and $\cbar^{a}$, makes the bound-state interpretation of the first parent $\chi^{a}$ of dynamically massive QCD essentially unavoidable. The relation of this interpretation to Bethe-Salpeter equations will be discussed below.

Note that the fact that $\chi^{a}$ enters the asymptotic limit of the elementary field $A_{\mu}^{a}$ is in no contradiction with the bound-state interpretation: as discussed in Sec.~\ref{sec:resgtgb}, this is a signature property of Higgs mechanisms, typical of Yang-Mills theories with gauge-boson mass generation, and independent of the identity of $\chi^{a}$. In the next section, we will confirm that $\chi^{a}$ is indeed the Goldstone boson associated with the spontaneous breaking of the global gauge charge $Q^{a}$, as is expected for the first parent of the elementary BRST quartet regardless of gauge-boson mass generation.\\

Eqs.~\eqref{sumr2} and \eqref{schwpoles} are the fundamental equations of dynamical gluon mass generation via the Schwinger mechanism. With these equations at hand, it is immediate to show that the mechanism is active. The proof is standard \cite{IP13,FP25}, and we report it here for completeness.

Let $\Pi_{\mu\nu}^{ab}(p)$ be the gluon polarization tensor defined as in Eq.~\eqref{glupoldef}. By Eq.~\eqref{spectral0}, $\Pi_{\mu\nu}^{ab}(p)$ is transverse,
\begin{equation}
   \Pi_{\mu\nu}^{ab}(p)=\Pi^{ab}(p^{2})\,(\eta_{\mu\nu}-p_{\mu}p_{\nu}/p^{2})\ ,
\end{equation}
with $\Pi^{ab}(p^{2})$ entering the form factor $\Delta^{ab}(p^2)$ as
\begin{equation}\label{deq350}
   \Delta^{-1}(p^{2})^{ab}=ip^{2}\,\delta^{ab}-\Pi^{ab}(p^{2})\ .
\end{equation}

\begin{figure*}[t]
    \includegraphics[width=0.95\textwidth]{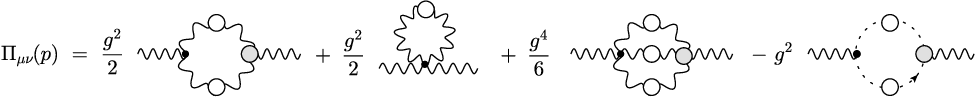}
    \caption{Schwinger-Dyson equation of the gluon polarization. Solid dots represent tree-level vertices, grey blobs represent dressed vertices, and white blobs represent dressed propagators. The four-gluon vertex is not one-particle irreducible.}
    \label{DSEdiag}
\end{figure*}

By Eq.~\eqref{gpsde}, using Eq.~\eqref{ope}, the SDE of $\Pi_{\mu\nu}^{ab}(p)$ reads
\begin{widetext}
   \begin{align}\label{sdeexpr}
      [\Pi^{\mu}_{\sigma}(p)]_{a}^{d}&=\frac{g^{2}}{2}\ \int_{q,k}\ \Gamma^{\mu\nu\tau}_{(0)\,abc}(-p,-q,-k)\ \Delta_{\nu}^{bb^{\prime}\nu^{\prime}}(q)\,\Delta_{\tau}^{cc^{\prime}\tau^{\prime}}(k)\ \Gamma_{\sigma\nu^{\prime}\tau^{\prime}}^{db^{\prime}c^{\prime}}(p,q,k)\ \delta_{q+k+p,0}+\\
      \notag&\ \ +\frac{g^{2}}{2}\ \int_{q}\ \Gamma^{\mu\nu\tau\ \ \ \ \ d}_{(0)\,\,\sigma\,abc}(-p,q,-q,p)\ \Delta_{\nu\tau}^{bc}(q)+\\
      \notag&\ \ +\frac{g^{4}}{6}\ \int_{q,k,r}\ \Gamma^{\mu\nu\tau\lambda}_{(0)\,abcf}(-p,-q,-k,-r)\ \Delta_{\nu}^{bb^{\prime}\nu^{\prime}}(q)\,\Delta_{\tau}^{cc^{\prime}\tau^{\prime}}(k)\,\Delta_{\lambda}^{ff^{\prime}\lambda^{\prime}}(r)\ \Gamma_{\sigma\nu^{\prime}\tau^{\prime}\lambda^{\prime}}^{db^{\prime}c^{\prime}f^{\prime}}(p,q,k,r)\  \delta_{q+k+r+p,0}+\\
      \notag&\ \ -g^{2}\int_{q,k}\ \Gamma_{(0)\,abc}^{\mu}(-p,q,k)\ \mc{G}^{bb^{\prime}}(q)\,\mc{G}^{cc^{\prime}}(k)\ \Gamma^{dc^{\prime}b^{\prime}}_{\sigma}(p,k,q)\ \delta_{-q+k+p,0}\ .
   \end{align}
   The SDE is depicted diagrammatically in Fig.~\ref{DSEdiag}. Following our conventions, we do not split the four-gluon vertex -- as well as the corresponding two-loop diagrams -- into one-particle-irreducible components\footnote{\label{fn:tadpole}Technically, the tadpole -- i.e. the second line in Eq.~\eqref{sdeexpr}, corresponding to the second diagram in Fig.~\ref{DSEdiag} -- is part of the nonreduced four-gluon term. Here, to comply with more ordinary notation, we keep the two terms separated.}.
   
   In the limit $p^{2}\to 0$, the dressed vertices in the equation develop Schwinger poles. In this limit, using Eqs.~\eqref{schwpoles} and \eqref{sumr2},
   \begin{align}\label{deq385}
      [\Pi^{\mu}_{\sigma}(p)]_{a}^{d}&\to\frac{i}{p^{2}}\, gI_{\sigma}(\hat{p})\bigg\{\frac{g}{2}\ \int_{q,k}\ \Gamma^{\mu\nu\tau}_{(0)\,abc}(-\hat{p},-q,-k)\ \Delta_{\nu}^{bb^{\prime}\nu^{\prime}}(q)\,\Delta_{\tau}^{cc^{\prime}\tau^{\prime}}(k)\ \mc{B}_{\nu^{\prime}\tau^{\prime}}^{b^{\prime}c^{\prime}d}(q,k;\hat{p})\ \delta_{q+k+\hat{p},0}+\\
      \notag&\qquad +\frac{g^{2}}{6}\ \int_{q,k,r}\ \Gamma^{\mu\nu\tau\lambda}_{(0)\,abcf}(-\hat{p},-q,-k,-r)\ \Delta_{\nu}^{bb^{\prime}\nu^{\prime}}(q)\,\Delta_{\tau}^{cc^{\prime}\tau^{\prime}}(k)\,\Delta_{\lambda}^{ff^{\prime}\lambda^{\prime}}(r)\ \mc{B}_{\nu^{\prime}\tau^{\prime}\lambda^{\prime}}^{b^{\prime}c^{\prime}f^{\prime}d}(q,k,r;\hat{p})\  \delta_{q+k+r+\hat{p},0}+\\
      \notag&\qquad-g\int_{q,k}\ \Gamma_{(0)\,abc}^{\mu}(-\hat{p},q,k)\ \mc{G}^{bb^{\prime}}(q)\,\mc{G}^{cc^{\prime}}(k)\ \mc{B}^{b^{\prime}c^{\prime}d}(-q,k;\hat{p})\ \delta_{-q+k+\hat{p},0}\bigg\}=\\
      \notag&=\frac{i}{p^{2}}\, g^{2}I_{\sigma}(\hat{p})I^{\mu}(-\hat{p})\,\delta_{a}^{d}=-iZ_{3}^{-1}M_{R}^{2}\ \frac{\hat{p}^{\mu}\hat{p}_{\sigma}}{p^{2}}\,\delta_{a}^{d}\ ,
   \end{align}
\end{widetext}
where we used the fact, proven in Appendix~\ref{app:schwpoles}, that the singular terms in the ellipses of Eq.~\eqref{schwpoles} sum to zero inside the polarization. At $p^{2}=0$, the non-pole part of $\Pi_{\mu\nu}^{ab}(p)$ attains the same value as its pole part by virtue of its transversality\footnote{In the literature on the Schwinger mechanism, transversality is explicitly shown to be realized by means of so-called ``Ward-identity displacements'' which are themselves induced by the presence of the Schwinger poles \cite{ABFP16,AFP22,ADF23,FP25}. The existence of such displacements was recently confirmed by lattice simulations \cite{ADF23}.}. From Eqs.~\eqref{deq350} and \eqref{deq385}, it then follows that
\begin{equation}\label{glueismassive}
   \Pi^{ab}(0)=iZ_{3}^{-1}M_{R}^{2}\,\delta^{ab}\quad\Longrightarrow\quad\Delta(0)=\frac{iZ_{3}}{M_{R}^{2}}\ .
\end{equation}

With the derivation of Eq.~\eqref{glueismassive}, we have come full circle to our starting point, Eq.~\eqref{gmass}. Note that at no stage in the previous or present sections we have \textit{assumed} that the Schwinger mechanism is active in dynamically massive QCD. What we have assumed is that the gluon propagator has no transverse massless pole. Then, within the KO framework, under customary assumptions, we have shown that the gluon spectrum contains a massless scalar excitation $\chi^{a}$ which plays the role of the first parent of the elementary BRST quartet. Using the algebraic relations that identify it as first parent, we have analyzed the dynamics of $\chi^{a}$, and we have determined that it can be interpreted as a bound state, that it generates Schwinger poles in the elementary QCD vertices, and that it triggers the Schwinger mechanism responsible for gluon mass generation. Under customary assumptions, BRST symmetry in the KO framework has selected a Schwinger mechanism triggered by bound states as the unique mechanism of dynamical gluon mass generation in the linear covariant gauges of QCD.

With the exception of singular terms in the elementary QCD vertices that do not contribute to gluon mass generation, the equations that govern the Schwinger mechanism in the KO framework are identical to those formulated by Aguilar, Binosi, Ferreira, and Papavassiliou in their ``massless bound-state formalism'' \cite{AP09,AIMP12,IP13,ABP16,ABP17,BP18,Pap22,FP23,FP24,FP25}. Since the additional singular terms vanish when summed into the polarization, in practical calculations that treat the Bethe-Salpeter vertices of the bound state as unknowns, these terms can be absorbed into the definition of the BSVs.

The formation of massless scalar bound states triggering the Schwinger mechanism in QCD has been extensively studied since the 1970s by formulating Bethe-Salpeter equations for the elementary QCD vertices and by solving them after employing various truncation ans\"{a}tze \cite{JJ73,EF74,Smi74,AIMP12,IP13,ABP16,BP18,FP24}. The state-of-the-art results are those obtained in the Landau gauge within the ABFP framework \cite{FP24}. Since the SDEs of QCD apply unchanged to the KO framework, these Bethe-Salpeter equations do as well. The analysis of the previous and present sections, by employing no truncation ansatz and under customary assumptions, demonstrates that, within the KO framework, in the presence of dynamical gluon mass generation, these BSEs always admit a solution in at least one -- if not all -- of the channels relevant to the Schwinger mechanism. Their solution is the first parent $\chi^{a}$ of the elementary BRST quartet, whose existence is guaranteed by BRST symmetry regardless of gluon mass generation, and which generates the required vertex Schwinger poles in its presence. Since $\chi^{a}$ has the same quantum numbers and Bethe-Salpeter amplitudes as the ABFP bound state, since it generates the same Schwinger poles, and since it triggers the Schwinger mechanism in the same formulation as that of ABFP, we conclude that $\chi^{a}$ is the ABFP bound state.\\

Before completing our discussion of mass generation, let us make a remark on the value of $\Delta(0)=i/M^{2}$. In this section we saw that, in the KO framework, in agreement with standard SM results, and modulo a factor of $g^{-1}$, $M$ is the on-shell value of the gluon-scalar transition amplitude $I(p^{2})$ -- see Eq.~\eqref{g2sa}. This value enters the numerator of Eq.~\eqref{schwpoles}, implying that $\Delta(0)=0$ can only be realized if at least one of the elementary vertices of QCD develops Schwinger poles with infinite, non-renormalizable residues. Together with the relation $Z_{\chi}^{1/2}=M^{-1}$, which we derived in the previous section, this observation gives us further reason to believe that, in agreement with the lattice data for QCD \cite{Cuc97,LSWP98a,LSWP98b,BBLW00,BBLW01,BHLP04,SIMS05,BBMM06,CM07,CM08a,CM08b,BIMS09,ISI09,BMM10,ABBC12,BLLM12,OS12,BBCO15,FN03,FN06a,Ste06}, $\Delta(0)=0$ should likely be excluded in the KO framework.\\

To conclude our discussion of the Schwinger mechanism, observe that, except for the identification of $\chi^{a}$, our derivation is general enough to apply not only to QCD, but to any Yang-Mills theory in which gauge bosons acquire a mass under customary KO assumptions. Indeed, Schwinger poles in the polarization and vertices are developed as a consequence of 1) sum rules for $\chi$ and $\beta$ that depend on the (in)equalities $\bra{\Omega}\partial_{\mu}B\ket{\chi}\propto \bra{\Omega}\partial_{\mu}\beta\ket{\chi}\neq 0$, $\bra{\Omega}\partial_{\mu}B\ket{\beta}\propto \bra{\Omega}\partial_{\mu}\beta\ket{\beta}=0$, and $\bra{\Omega}F_{\mu\nu}\ket{\chi},\bra{\Omega}F_{\mu\nu}\ket{\beta}=0$; 2) the form of the one-particle projector $\mathds{P}_{1G}$ in Eq.~\eqref{p1g}; and 3) the nonvanishing zero-momentum limit of the inverse gauge-boson form factor $\Delta^{-1}(0)$. Of these, 2) and 3) were used in Appendix~\ref{app:schwpoles} to prove the existence of vertex Schwinger poles in the elementary vertices of dynamically massive QCD, and all of them hold as a direct consequence of the customary assumptions we made in the previous section. Beyond QCD, if needed, the assumption of the diagonality of the gauge-boson and ghost propagators in color space can be relaxed, and the derivation can be carried out by first diagonalizing the symmetric mass matrix $(M^{-1})^{ab}=-i\Delta^{ab}(0)$, and then studying separately each diagonal massive direction.

While in the present section and in Appendix~\ref{app:schwpoles} we focused on dynamically massive QCD, its vertices, its SDEs, and the identity of its first parent $\chi^{a}$, it should be realized that the Schwinger mechanism ultimately originates from the fact that the kernel operator $K_{\mu}^{a}$ in the gluon propagator's SDE, Eq.~\eqref{gpsde}, is the very same operator that defines the sum rule encoded in the gluon-scalar transition amplitude $I_{\mu}(p)$, Eq.~\eqref{g2saker}. Once the necessary matter terms are included in $K_{\mu}^{a}$, under customary KO assumptions, the definition in Eq.~\eqref{g2saker} remains valid in any dynamically massive Yang-Mills theory, regardless of its matter content. The massless one-particle spectrum of the Green's function in Eq.~\eqref{gpsde} can then be studied like we did in Eqs.~\eqref{polschwpoleprel} and \eqref{polschwpole} of Appendix~\ref{app:schwpoles} to show that, in the presence of dynamical gauge-boson mass generation, the gauge-boson polarization tensor does indeed develop a Schwinger pole proportional to $I_{\mu}(-\hat{p})I_{\nu}(\hat{p})$, regardless of the vertices and matter content of the theory. Explicitly, as $p^{2}\to 0$,
\begin{align}
   \Pi_{\mu\nu}^{ab}(p)\to \frac{i\delta^{ab}}{p^{2}+i\epsilon}\ g^{2}I_{\mu}(-\hat{p})I_{\nu}(\hat{p})\ ,
\end{align}
implying
\begin{equation}
   \Pi(0)=ig^{2}I^{2}(0)\quad\Longrightarrow\quad i\Delta^{-1}(0)=g^{2}I^{2}(0)\ .
\end{equation}

In an arbitrary dynamically massive Yang-Mills theory quantized in the KO framework, the individual terms in $K_{\mu}^{a}$ can be written down explicitly like we did in the present section for QCD to yield corresponding terms in the sum rule $ig^{-1}\bra{\Omega}K_{\mu}^{a}\ket{\chi^{b}(p)}$ of the first parent of the elementary BRST quartet $\chi^{a}$. Just like in QCD, terms which are quadratic or higher-order in the fields are associated with cubic or higher-order integrated BSVs. Linear terms, on the other hand, contribute to the sum rule with one-particle matrix elements which are associated with no BSV. Typical examples of the latter -- see Appendices~\ref{app:stschw} and \ref{app:higgsmod} -- are exhibited by the Stueckelberg model, whose field equations are fully linear, $K_{\mu}=-m\partial_{\mu}\phi+m^{2}A_{\mu}$, and by Higgs models, whose current develops similar terms in the directions of the broken color symmetry as a consequence of the Higgs field acquiring a nonzero vacuum expectation value. To linear terms, obviously, are associated no vertex Schwinger poles. Nonlinear terms in $K_{\mu}^{a}$, on the other hand, produce Schwinger poles -- proportional to the BSVs -- in the vertices obtained by replacing the external scalar of the BSV with an external gauge-boson line.

As for the identity of the first parent $\chi$ of a generic dynamically massive Yang-Mills theory, needless to say, if $\chi$ lives in the channel of an elementary scalar field\footnote{Modulo, potentially, reparametrization of the fields in the Lagrangian, as it happens in Higgs models.}, then, in general, no bound-state interpretation of its nature needs to be advanced. Indeed, in the examples cited above, the first parent \textit{is} the elementary field that appears linearly in the equations alongside the gauge field -- namely, the Stueckelberg field in the Stueckelberg model, and the unphysical components of the Higgs field in Higgs models. From our discussion, it follows that the Stueckelberg mechanism and the Higgs mechanism of Higgs models are nothing but Schwinger mechanisms, that do not rely on the formation of bound states to generate a gauge-boson mass, and that produce Schwinger poles already at tree level.

More generally, the first parent of the elementary BRST quartet of any dynamically massive Yang-Mills theory that satisfies the assumptions of the previous section generates a Schwinger pole in the gauge-boson polarization, thus triggering the Schwinger mechanism. It follows that, in the KO framework, under customary assumptions, all dynamical gauge-boson mass generation is Schwinger mass generation.

\section{Spontaneous symmetry breaking and color confinement in dynamically massive QCD}\label{sec:symbreak}

In this section, we discuss spontaneous symmetry breaking and color confinement in dynamically massive QCD. In Sec.~\ref{sec:ssb} we start by rederiving the KO Goldstone equation under our working assumptions. We confirm that the global gauge (color) charge $Q^{a}$ is spontaneously broken, and we show that the ABFP bound state $\chi^{a}$ is the associated scalar Goldstone boson. We show that -- modulo unitary prefactors -- the matrix element of the global gauge (color) current $J_{\mu}^{a}$ between the vacuum and the one-Goldstone state is equal to the gluon-scalar transition amplitude evaluated in the PT-BFM, and that $1+u$ disappears from the equations, leaving behind the RGI quantity $\widehat{M}/g$. In Sec.~\ref{sec:chargestruct} we study the structure of the global gauge charge $Q^{a}$ in the presence of dynamical gluon mass generation. We show that $Q^{a}$ is dynamically rearranged so that, asymptotically, it does not generate mere color transformations, but rather global gauge transformations. The asymptotic color charge operator $\widetilde{Q}^{a}$ is uniquely obtained by isolating the former from the latter, and it is well-defined, BRST exact, and thus confining.

\subsection{Spontaneous symmetry breaking, the gluon-scalar transition amplitude, and the PT-BFM}\label{sec:ssb}

Let us begin this section by rederiving the KO Goldstone equation under our working hypotheses. While the derivation is essentially identical to that in Sec.~\ref{sec:koanal}, we believe it is important to go through it once more in order to make explicit the assumptions under which it is obtained in the context of the present study.

Let us once again express the global gauge current $J_{\mu}^{a}$,
\begin{equation}\label{deq396}
   J^{a}_{\mu}=f^{abc}\left(A^{b\nu}F_{\nu\mu}^{c}+A^{b}_{\mu}B^{c}+i\partial_{\mu}\cbar^{b}c^{c}-i\cbar^{b}D_{\mu}c^{c}\right)\ ,
\end{equation}
as
\begin{align}\label{KOcurr}
   gJ_{\mu}^{a}=\partial^{\nu}F_{\mu\nu}^{a}+\{Q_{B},D_{\mu}\cbar^{a}\}
\end{align}
using the MYME. The second term on the RHS of Eq.~\eqref{KOcurr} is conserved on its own,
\begin{align}
   \partial^{\mu}\{Q_{B},D_{\mu}\cbar^{a}\}=\{Q_{B},-\xi gf^{abc}B^{b}\cbar^{c}\}=0\ ,
\end{align}
hence it formally defines a Kugo-Ojima charge $N^{a}$ \cite{Oji78,KO79d,KO79a,NO90}
\begin{equation}
   N^{a}=\int d^{3}{\bf x}\ \mc{N}_{0}^{a}\ ,\ \mc{N}_{\mu}^{a}=\{Q_{B},D_{\mu}\cbar^{a}\}\ ,\ \partial^{\mu}\mc{N}_{\mu}^{a}=0\ .
\end{equation}
If the field-strength tensor $F_{\mu\nu}^{a}$ has no massless spectrum, denoting spatial indices with $i$, the integral
\begin{equation}
   \int d^{3}{\bf x}\ \partial^{\mu}F_{0\mu}=\int d^{3}{\bf x}\ \partial^{i}F_{0i}=\oint d^{2}{\bf x}_{i}\ F^{0i}
\end{equation}
vanishes \cite{KO79d,KO79a,NO90}, and -- modulo a factor of $g^{-1}$ -- the global gauge charge $Q^{a}=\int d^{3}{\bf x}\,J_{0}^{a}$ is given entirely by the Kugo-Ojima charge,
\begin{equation}
   Q^{a}=g^{-1}N^{a}\ .
\end{equation}

Consider now the one-particle state $\ket{\chi^{a}(p)}$ of Sec.~\ref{sec:mlessspect}. As discussed in Sec.~\ref{sec:goldbound}, $\bra{\Omega}F_{\mu\nu}^{a}\ket{\chi^{b}(p)}=0$. Therefore
\begin{align}
   \bra{\Omega}gJ_{\mu}^{a}\ket{\chi^{b}(p)}=\bra{\Omega}\mc{N}_{\mu}^{a}\ket{\chi^{b}(p)}\ .
\end{align}
Using the quartet relations we derived in Sec.~\ref{sec:asympt}, we evaluate
\begin{align}\label{deq942}
   &\bra{\Omega}gJ_{\mu}^{a}\ket{\chi^{b}(p)}=\\
   \notag&\ =(Z_{3}\widetilde{Z}_{3})^{-1/2}\,\bra{\Omega}D_{\mu}\cbar^{a}Q_{B,R}\ket{\chi^{b}(p)}=\\
   \notag&\ =-i(Z_{3}\widetilde{Z}_{3})^{-1/2}M_{R}\,\bra{\Omega}D_{\mu}\cbar^{a}\ket{\gamma^{b}(p)}\ .
\end{align}

The last line in Eq.~\eqref{deq942} contains the matrix element of the local operator $D_{\mu}\cbar^{a}$ between the vacuum and the one-particle asymptotic state $\ket{\gamma^{a}(p)}$. By Eq.~\eqref{1psme} and our assumption of asymptotic completeness, this quantity is different from zero if and only if the operator develops a non-vanishing asymptotic contribution of the form $\partial_{\mu}\gammabar^{a}$\footnote{Recall that, under the assumptions of Sec.~\ref{sec:mlessspect}, all asymptotic fields but $\gammabar^{a}$ (anti)commute with $\gamma^{a}$: other asymptotic fields, should they exist, do not contribute to the matrix element.}. Following Sec.~\ref{sec:koanal}, once we define a constant $u$ from the asymptotic limit
\begin{equation}\label{KOconst}
   gf^{abc}A_{\mu}^{b}\cbar^{c}\to \widetilde{Z}_{3}^{1/2}\,u^{ab}\,\partial_{\mu}\gammabar^{b}=\widetilde{Z}_{3}^{1/2}\,u\,\partial_{\mu}\gammabar^{a}\ ,
\end{equation}
where the assumption $u^{ab}=u\,\delta^{ab}$ is customary in QCD \cite{KO79d,KO79a,NO90}, we have
\begin{equation}\label{KOasym}
   D_{\mu}\cbar^{a}\to \widetilde{Z}_{3}^{1/2}(1+u)\,\partial_{\mu}\gammabar^{a}\ .
\end{equation}
Using
\begin{equation}
   \bra{\Omega}\partial_{\mu}\gammabar^{a}(x)\ket{\gamma^{b}(p)}=-\hat{p}_{\mu}\,\delta^{ab}\,e^{-i\hat{p}x}\ ,
\end{equation}
when plugged into Eq.~\eqref{deq942}, Eq.~\eqref{KOasym} then yields
\begin{align}\label{deq722}
   &\bra{\Omega}J_{\mu}^{a}(x)\ket{\chi^{b}(p)}=ig^{-1}M(1+u)\,\delta^{ab}\,\hat{p}_{\mu}\,e^{-i\hat{p}x}\ .
\end{align}

Observing that, by the same reasoning as above, $Q_{B,R}\ket{\beta^{b}(p)}=0$ implies $\bra{\Omega}J_{\mu}^{a}(x)\ket{\beta^{b}(p)}=0$, whence $\bra{\Omega}J_{\mu}^{a}(x)\beta^{b}(y)\ket{\Omega}=0$, it is easy to see that, when integrated over three-momentum with weight $\frac{1}{2\hat{p}^{0}}\,e^{i\hat{p}y}$, Eq.~\eqref{deq722} gives back Eq.~\eqref{KOGoldCorr} with $Z_{\chi}^{-1/2}=M$. On the other hand, by integrating the time component of Eq.~\eqref{deq722} in three-dimensional space, we obtain
\begin{align}\label{symbreak0}
   &\bra{\Omega}Q^{a}\ket{\chi^{b}(p)}=ig^{-1}M(1+u)\ \delta^{ab}\,\hat{p}^{0}\,(2\pi)^{3}\delta^{(3)}({\bf p})\ .
\end{align}
Note that the one-particle momentum-space measure of integration is $\int \frac{d^{3}{\bf p}}{(2\pi)^{3}}\frac{1}{2\hat{p}^{0}}$, so the RHS of the last equation does not vanish as a distribution even if ${\bf p}=0$ implies $\hat{p}^{0}=0$. And indeed,
\begin{align}
   \notag&\bra{\Omega}Q^{a}\widetilde{\chi}^{b}(x)\ket{\Omega}=\int \frac{d^{3}{\bf p}}{(2\pi)^{3}}\frac{1}{2\hat{p}^{0}}\ e^{-i\hat{p}x}\bra{\Omega}Q^{a}\ket{\chi^{b}(p)}=\\
   &\ =\frac{i}{2}\,g^{-1}M(1+u)\, \delta^{ab}\ .
\end{align}
Finally, observing that $\bra{\Omega}Q^{a}\beta^{b}(x)\ket{\Omega}=0$, after taking the commutator of $Q^{a}$ with $\chi^{b}(x)$\footnote{To avoid working directly with the charge $Q^{a}$, which, due to spontaneous symmetry breaking, is ill-defined, one can first evaluate the commutator using the current $J_{\mu}^{a}$, and then integrate over three-dimensional space. The result is the same as that in the next equation.}, we obtain the Goldstone equation
\begin{equation}\label{gstcomm0}
   \bra{\Omega}[Q^{a},\chi^{b}(x)]\ket{\Omega}=ig^{-1}M(1+u)\, \delta^{ab}\ .
\end{equation}

By the discussion in Sec.~\ref{sec:ssblcg}, there is no reason to believe that the RHSs of Eqs.~\eqref{deq722} and \eqref{gstcomm0} should be different from zero. Below, we will confirm it is. From Eqs.~\eqref{deq722} and \eqref{gstcomm0}, it then follows that, under assumptions which are customary in the KO framework, the global gauge charge $Q^{a}$ of dynamically massive QCD is spontaneously broken, $Q^{a}\ket{\Omega}\neq 0$, and that the ABFP bound state is the unphysical Goldstone boson associated with its breaking. As discussed in Sec.~\ref{sec:ssblcg}, this outcome is fully expected from the fact that $\chi^{a}$ is the first parent of the elementary BRST quartet. The precise relation between spontaneous symmetry breaking and residual gauge invariance will be established in the next section.

Since dynamical gluon mass generation is triggered by the eating of the scalar Goldstone boson associated with the spontaneous breaking of the global gauge charge $Q^{a}$, we see that the ABFP Schwinger mechanism is nothing but a fully nonperturbative Higgs mechanism acting in the absence of a Higgs field. In dynamically massive QCD, at variance with Higgs models, the Goldstone bosons of the Higgs mechanism are not the quanta of elementary scalar fields present in the Lagrangian, but rather bound states, generated by the dynamics of the strong interactions itself.\\

Is the ABFP bound state the unique Goldstone boson associated with the spontaneous breaking of $Q^{a}$? In the current $J_{\mu}^{a}$ of non-Abelian Yang-Mills theories, there are two potential sources of additional symmetry breaking which are seldom discussed in the literature, if at all. The first of these is in the field-strength tensor. Indeed, $F_{\mu\nu}^{a}=\partial_{\mu}A_{\nu}^{a}-\partial_{\nu}A_{\mu}^{a}+gf^{abc}A_{\mu}^{b}A_{\nu}^{c}$ may develop a massless asymptotic contribution independent of that of the elementary BRST quartet if a stable massless bound state is formed in the composite channel $f^{abc}A_{\mu}^{b}A_{\nu}^{c}$. Since the latter is antisymmetric in its Lorentz indices, the simplest such bound state would be created by an asymptotic massless vector field entering the expression as an antisymmetric derivative, just like the linear terms in $F_{\mu\nu}^{a}$ -- in other words, by the bound-state realization of a massless gluon. The second simplest possibility is a bound state associated with an asymptotic massless 2-form field $H^{a}_{\mu\nu}$. More complicated possibilities require the existence of stable massless bound states with spin higher than 1.

If these bound states existed in the channel of $f^{abc}A_{\mu}^{b}A_{\nu}^{c}$, the charge $\int d^{3}{\bf x}\ \partial^{i}F_{i0}^{a}$, present in $Q^{a}$ by Eq.~\eqref{KOcurr}, might be broken by corresponding Goldstone bosons, making it ill-defined and different from zero even in the presence of a gluon mass. It is, however, unlikely that the dynamics of QCD would provide gluons with a mass just to replace them with an identical massless bound-state realization; the same assessment can be made about the massless 2-form field, which would replace the field-strength tensor $F_{\mu\nu}^{a}$ itself. The possibility of stable massless bound states being generated in higher-spin channels is similarly unappealing.

The second potential source of additional symmetry breaking is an asymptotic contribution to the operator $D_{\mu}\cbar^{a}$ coming from a massless bound state in the composite channel $f^{abc}A_{\mu}^{b}\cbar^{c}$. The simplest such bound state would be an antighost scalar field entering the channel as a gradient -- in other words, a bound-state copy of the elementary asymptotic antighost field. If such a bound state existed, the first parent of its BRST quartet might break the Kugo-Ojima charge $N^{a}$, present in $Q^{a}$ by Eq.~\eqref{KOcurr}, independently of the elementary quartet. In Sec.~\ref{sec:asympt}, we have postulated the absence of massless bound states in the composite channel $f^{abc}A_{\mu}^{b}c^{c}$. In the Landau gauge ($\xi=0$), one has $\partial^{\mu}A_{\mu}=0$, and the identity $D^{\mu}\partial_{\mu}=\partial^{\mu}D_{\mu}$ holds, implying that the field equations of $c^{a}$ and $\cbar^{a}$ -- see Eq.~\eqref{feqs} -- are the same: the ghost and the antighost fields have the same dynamics. Therefore, in the Landau gauge at least, the absence of bound states in the $f^{abc}A_{\mu}^{b}\cbar^{c}$ channel follows from their absence in the channel $f^{abc}A_{\mu}^{b}c^{c}$. Beyond the Landau gauge, we deem the possibility that massless bound-state copies of $\gammabar^{a}$ exist unlikely.

For these reasons, we will henceforth make the customary -- albeit, in this context, usually implicit -- assumption that there are no massles bound states in the channels $f^{abc}A_{\mu}^{b}A_{\nu}^{c}$ and $f^{abc}A_{\mu}^{b}\cbar^{c}$. Under this assumption, the ABFP bound state is indeed the unique Goldstone boson associated with the spontaneous breaking of $Q^{a}$, consistent with its unique role of first parent of the elementary BRST quartet.\\

For what follows, it will be crucial to put $1+u$ in relation to known quantities. To do so, following \cite{KO79a,Kug95}, consider the asymptotic limit of the operator $D_{\mu}c^{a}$ -- by the third of Eq.~\eqref{quartetasym},
\begin{equation}
   D_{\mu}c^{a}\to \widetilde{Z}_{3}^{-1/2}\,\partial_{\mu}\gamma^{a}\ .
\end{equation}
Moreover, consider the unrenormalized ``Kugo-Ojima function'' \cite{Kug95,GHQ04,BQ13}
\begin{align}\label{kogf}
   \langle T\{D_{\mu}c^{a}(x)D_{\nu}\cbar^{b}(0)\}\rangle_{\text{F.T.}}=t_{\mu\nu}(p)\,u^{ab}(p^{2}) -\frac{p_{\mu}p_{\nu}}{p^{2}}\,\delta^{ab}\ .
\end{align}
Using the one-particle projector $\mathds{P}_{1G}$ in Eq.~\eqref{p1g}, it is easy to see that, since the LSZ coefficient of $D_{\mu}c^{a}$ is $\widetilde{Z}_{3}^{-1/2}$, and that of $D_{\mu}\cbar^{a}$ features its inverse $\widetilde{Z}_{3}^{1/2}$, $1+u$ is the coefficient of the massless pole generated by $\gamma^{a}$ and $\gammabar^{a}$ in the KO function,
\begin{align}\label{deq853}
   \langle T\{D_{\mu}c^{a}(x)D_{\nu}\cbar^{b}(0)\}\rangle_{\text{F.T.}}\to-(1+u)\,\frac{p_{\mu} p_{\nu}}{p^{2}}\,\delta^{ab}+\cdots\ ,
\end{align}
where the ellipsis denotes other potential one-particle contributions, and the multiparticle contributions. Since we assumed there are no massless bound states in the composite channels $f^{abc}A_{\mu}^{b}c^{c}$ and $f^{abc}A_{\mu}^{b}\cbar^{c}$, the asymptotic fields $\gamma^{a}$ and $\gammabar^{a}$ saturate the massless discrete spectrum of the KO function.

If the multiparticle continuum that contributes to the KO function is sufficiently separated from its massless one-particle spectrum, as is customary when asymptotic fields are well behaved, the zero-momentum limit $u^{ab}(0)$ is given entirely by the massless one-particle spectrum of the KO function, and we find that
\begin{equation}\label{BQIcoeff}
   1+u=1+u(0)\ ,
\end{equation}
where $u^{ab}(0)=u(0)\,\delta^{ab}$.\\

Before going on with our derivation, let us pause to make some remarks on the value of $1+u$ in the Landau gauge. For $\xi=0$, as we said before, the ghost and the antighost field have the same dynamics. Comparing the asymptotic limits
\begin{equation}\label{deq586}
   D_{\mu}c^{a}\to \widetilde{Z}_{3}^{-1/2}\,\partial_{\mu}\gamma^{a}\ ,\quad D_{\mu}\cbar^{a}\to\widetilde{Z}_{3}^{1/2}(1+u)\,\partial_{\mu}\gammabar^{a}\ ,
\end{equation}
and observing that $c^{a}\to \widetilde{Z}_{3}^{1/2}\,\gamma^{a}$, $\cbar^{a}\to \widetilde{Z}_{3}^{1/2}\,\gammabar^{a}$ are already given the same LSZ coefficients, we see that, in the Landau gauge, by ghost-antighost symmetry, one must have $\widetilde{Z}_{3}^{-1/2}=\widetilde{Z}_{3}^{1/2}(1+u)$. Therefore, for $\xi=0$,
\begin{equation}\label{1puz3}
   1+u=\widetilde{Z}_{3}^{-1}=J_{0}^{-1}\ .
\end{equation}
This simple relation was not derived in \cite{KO79d,KO79a,NO90}, and it was only obtained in \cite{Kug95} by a one-page calculation that involved discussions on the diagrammatic representation of ghost Green's functions. The reason why the simple proof reported above was not available to \cite{KO79d,KO79a,NO90,Kug95} is that -- as we saw in the previous sections -- those papers ignored the LSZ coefficients of the elementary BRST quartet. With no such coefficient at hand, the comparison in Eq.~\eqref{deq586} could not be made, and this simple proof could not be given.

We remark that -- by the discussion in Sec.~\ref{sec:asympt} -- $J_{0}$ is equal to $J(0)$, the zero-momentum limit of the form factor $J(p^{2})$ of the ghost propagator, only if $J(p^{2})$ saturates to a constant as $p^{2}\to 0$, and if the multiparticle continuum in the spectral representation of the ghost propagator $\mc{G}(p^{2})$ is sufficiently separated from its massless discrete spectrum for $p^{2}\mc{G}(p^{2})\to -J_{0}=-J(0)$ to hold. In particular, in the Landau gauge, if $J(p^2)$ is infrared enhanced as a result of the multiparticle continuum dominating on the single-particle spectrum, one can have both $J^{-1}(0)=0$ and $J_{0}^{-1}=1+u\neq 0$ at the same time: ghost enhancement alone does not guarantee the vanishing of $1+u$.

In the Landau gauge, $1+u=0$ can only be realized via an infinite, nonrenormalizable LSZ coefficient $\widetilde{Z}_{3}=J_{0}$ for the ghost and antighost fields. It is far from clear whether this notion makes any sense in the first place, and, if it does, whether ghost asymptotic fields are still allowed to exist in that case. As we noted in Sec.~\ref{sec:asympt}, and as is obvious, the existence of well-behaved ghost asymptotic fields is essential for preserving BRST symmetry.

Finally, we remark that the lattice data for Landau-gauge QCD \cite{SS96,Cuc97,SIMS05,BBMM06,CM08a,BIMS09,ABBC12,BLLM12,NF00,FN03,FN06a,Ste06,ABP09,ABFP24} display a saturating behavior for $J(p^{2})$, which is not affected by any of these issues, and is compatible with $J_{0}=J(0)=(1+u)^{-1}$. Moreover, we remark that whether $1+u=0$ or not is ultimately irrelevant, given that the KO criterion for confinement rests on inconsistent arguments and an incorrect equation -- see Sec.~\ref{sec:koanal}.\\

A direct connection between the Schwinger mechanism and spontaneous symmetry breaking in QCD can be established by observing that the RHS of the Goldstone equation Eq.~\eqref{deq722} is but the gluon-scalar transition amplitude $I_{\mu}(-\hat{p})$ -- Eq.~\eqref{g2saker} -- evaluated in the PT-BFM scheme \cite{Abb80,Cor81,CP89,Pil97,GHS99,GHS01,BP02b,BP02a,BP04,BP09,CPB11,BQ13}. Let us prove this statement.

By first writing $I_{\mu}(-\hat{p})$ as
\begin{align}
   &I_{\mu}(-\hat{p})\,\delta^{ab}\,e^{-i\hat{p}x}=\\
   \notag&=if^{acd}\bra{\Omega}\partial^{\nu}(A_{\nu}^{c}A_{\mu}^{d})+A^{c\,\nu}F^{d}_{\nu\mu}+i\partial_{\mu}\cbar^{c}c^{d}\ket{\chi^{b}(p)}\ ,
\end{align}
dropping the vanishing antisymmetric contribution coming from $\bra{\Omega}f^{acd}A_{\nu}^{c}A_{\mu}^{d}\ket{\chi^{b}(p)}$, and then using Eqs.~\eqref{deq396}, \eqref{KOconst} and \eqref{sumr2},
\begin{align}\label{deq683}
   &I_{\mu}(-\hat{p})\,\delta^{ab}\,e^{-i\hat{p}x}=\\
   \notag &\ =i\bra{\Omega}f^{acd}(A^{c\,\nu}F^{d}_{\nu\mu}+i\partial_{\mu}\cbar^{c}c^{d})\ket{\chi^{b}(p)}=\\
   \notag&\ =i\bra{\Omega}J^{a}_{\mu}\ket{\chi^{b}(p)}-i\bra{\Omega}\left\{Q_{B},f^{acd}A^{c}_{\mu}\cbar^{d}\right\}\ket{\chi^{b}(p)}=\\
   \notag&\ =i\bra{\Omega}J^{a}_{\mu}\ket{\chi^{b}(p)}-u\ Z_{3}^{-1/2}\,g^{-1}M_{R}\bra{\Omega}\partial_{\mu}\gammabar^{a}(x)\ket{\gamma^{b}(p)}=\\
   \notag&\ =i\bra{\Omega}J^{a}_{\mu}\ket{\chi^{b}(p)}+u\ Z_{3}^{-1/2}\,g^{-1}\,M_{R}\,\hat{p}_{\mu}\,\delta^{ab}\,e^{-i\hat{p}x}=\\
   \notag&\ =i\bra{\Omega}J^{a}_{\mu}\ket{\chi^{b}(p)}-u\,I_{\mu}(-\hat{p})\,\delta^{ab}\,e^{-i\hat{p}x}\ ,
\end{align}
we obtain
\begin{equation}\label{deq145}
   (1+u)\,I_{\mu}(-\hat{p})\,\delta^{ab}\,e^{-i\hat{p}x}=i\bra{\Omega}J^{a}_{\mu}(x)\ket{\chi^{b}(p)}\ .
\end{equation}
The constant $u$ has entered Eq.~\eqref{deq683} through the same operator of the original KO equation. This is obvious, given that they are the same equation.

In addition to appearing in the Kugo-Ojima criterion for confinement, the constant $u$ is known \cite{ABP09,BQ13,ABFP24,FP25} to realize the $p^{2}\to 0$ limit of a so-called ``Batalin-Vilkovisky (BV) function'' $G(p^{2})$,
\begin{equation}\label{uG0}
   u = G(0)\ ,
\end{equation}
which in all linear covariant gauges -- as it turns out, not by chance -- is defined exactly like the nontrivial form factor of the Kugo-Ojima function \cite{GHQ04,ABP09,BQ13,ABFP24,FP25}: suppressing color indices,
\begin{align}
   \langle T\{D_{\mu}c(x)D_{\nu}\cbar(0)\}\rangle_{\text{F.T.}}=t_{\mu\nu}(p)\,G(p^{2}) -\frac{p_{\mu}p_{\nu}}{p^{2}}\ .
\end{align}
The BV function $G(p^{2})$, in turn, appears in so-called ``Background-Quantum Identities'' (BQI) \cite{GHS01,BP02b,BP09,BQ13,IP13} that convert Green's functions from the linear covariant gauges to the background-field gauges (BFG) that define the PT-BFM. In particular, it converts the gluon-scalar transition amplitude $I_{\mu}(p)$ of LCGs into the gluon-scalar transition amplitude $\widetilde{I}_{\mu}(p)$ of BFGs via the relation \cite{IP13}
\begin{equation}\label{g2sabfg}
   \widetilde{I}_{\mu}(p)=[1+G(p^{2})]\,I_{\mu}(p)\ .
\end{equation}

From Eqs.~\eqref{deq145}, \eqref{uG0}, and \eqref{g2sabfg}, it follows that the matrix element of the global gauge (color) current $J_{\mu}^{a}$ between the vacuum and the one-Goldstone state $\ket{\chi^{b}(p)}$ is equal to the gluon-scalar transition amplitude evaluated in the PT-BFM:
\begin{align}\label{g2sacurr}
   \widetilde{I}_{\mu}(-\hat{p})\,\delta^{ab}\,e^{-i\hat{p}x}=i\bra{\Omega}J^{a}_{\mu}(x)\ket{\chi^{b}(p)}\ .
\end{align}

The reason why this relation is realized through quantities defined in the PT-BFM is easy to understand. In going from the second to the third line of Eq.~\eqref{deq683} to form the current $J_{\mu}^{a}$, we have added and subtracted the operator
\begin{align}\label{BVop}
   &\left\{Q_{B},f^{acd}A^{c}_{\mu}\cbar^{d}\right\}=f^{acd}\left(A^{c}_{\mu}B^{d}-i\cbar^{c}D_{\mu}c^{d}\right)=\\
   \notag&\qquad=-f^{acd}\left(\xi^{-1}\,A_{\mu}^{c}\partial^{\nu}A_{\nu}^{d}+i\cbar^{c}\partial_{\mu}c^{d}+igf^{def}\cbar^{c}A_{\mu}^{e}c^{f}\right)\ .
\end{align}
At the level of the gluon-scalar transition amplitude, written down analytically and represented diagrammatically as we did in Eq.~\eqref{sumr2} and in Fig.~\ref{GSTAdiag} --, the sole effect of adding this operator to the second line of Eq.~\eqref{deq683} is that of turning LCG tree-level vertices into corresponding BFG tree-level vertices with a single background gluon leg \cite{Abb80,Pil97,CPB11}. Specifically, denoting quantities defined in the BFG with a tilde, the first term on the second line of Eq.~\eqref{BVop} converts the three-gluon vertex,
\begin{align}
   &\Gamma^{\mu\nu\tau}_{(0)abc}(p,q,k)\to \widetilde{\Gamma}^{\mu\nu\tau}_{(0)abc}(p,q,k)=\\
   \notag&\qquad\quad =\Gamma^{\mu\nu\tau}_{(0)abc}(p,q,k)-f_{abc}\,\frac{1}{\xi}\,(\eta^{\mu\nu}k^{\tau}-\eta^{\mu\tau}q^{\nu})\ ,
\end{align}
the second one converts the ghost-antighost-gluon vertex
\begin{align}
   &\Gamma_{(0)abc}^{\mu}(p,q,k)\to \\
   \notag&\to\widetilde{\Gamma}_{(0)abc}^{\mu}(p,q,k)=\Gamma_{(0)abc}^{\mu}(p,q,k)+if_{abc}\,q^{\mu}\ ,
\end{align}
and the third one gives rise to a new tree-level quartic ghost-antighost-two-gluon vertex $\widetilde{\Gamma}_{(0)abcd}^{\mu\nu}$ which only exists in the BFG \cite{Abb80,Pil97,CPB11},
\begin{align}
   \widetilde{\Gamma}_{(0)abcd}^{\mu\nu}(p,q,k,r)=-f_{ace}f_{bde}\, \eta^{\mu\nu}\ .
\end{align}
As for the four-gluon vertex, no conversion is required in that case: at tree-level, the vertex that appears in the amplitude is the same in the two gauges \cite{CPB11,IP13}.

\begin{widetext}
   \begin{figure*}[t]
    \includegraphics[width=0.72\textwidth]{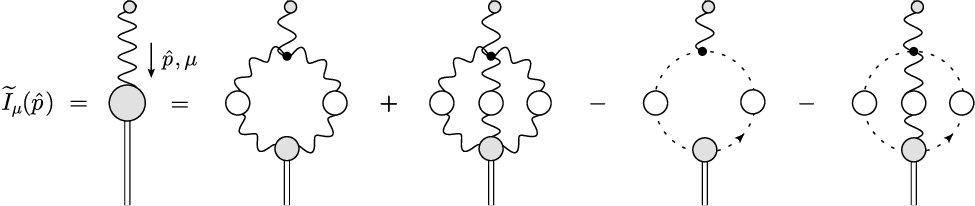}
    \caption{Gluon-scalar transition amplitude evaluated in the PT-BFM. On the RHS, solid dots represent BFG tree-level vertices, grey blobs the scalar's BSVs, and white blobs represent dressed propagators. Background PT-BFM objects are marked with a grey dot in their external gluon line. The three-gluon and the ghost-antighost-gluon BSVs are not one-particle irreducible.}
    \label{GSTAdiagPT}
\end{figure*}
\end{widetext}

In the analytic expression of Eq.~\eqref{g2sacurr}, the BSAs of $\chi^{a}$ in the two-gluon, three-gluon, ghost-antighost, and -- in the presence of quarks -- quark-antiquark channels are the same as in Eq.~\eqref{sumr2}. Additionally, a new BSA $B^{abcd}_{\mu}$,
\begin{equation}
   B^{abcd}_{\mu}(x,y,z;\hat{p})=\bra{\Omega}T\{c^{a}(x)\cbar^{b}(y)A_{\mu}^{c}(z)\}\ket{\chi^{d}(p)}\ ,
\end{equation}
appears in the ghost-antighost-gluon channel. The existence of Schwinger poles in the corresponding ghost-antighost-two-gluon vertex, proportional to its associated BSV, can be proven as we did in Appendix~\ref{app:schwpoles} for the other vertices. In Fig.~\ref{GSTAdiagPT}, we report the diagrammatic representation of the gluon-scalar transition amplitude $\widetilde{I}_{\mu}(\hat{p})$ evaluated in the PT-BFM -- see also \cite{IP13} --, derived by expanding the RHS of Eq.~\eqref{g2sacurr} as we did in Sec.~\ref{sec:goldbound}.

That achieving a conversion between the LCGs and the BFGs is possible at all at the operatorial level -- that is, without leaving the underlying Hilbert space of the theory -- is remarkable: in general, as far as the gauge sector is concerned, different gauges are associated with different solutions of the field equations, hence with different Hilbert spaces. This remarkable phenomenon nonetheless admits a simple explanation, suggested by the central result of \cite{BQ13}.

In \cite{BQ13}, it was shown that the BFGs are not gauges distinct from the linear covariant gauges, but rather a natural extension of the Batalin-Vilkoviski formalism of the linear covariant gauges from BRST symmetry alone, to both BRST symmetry and anti-BRST symmetry. Therein, the background gluon field $\widehat{A}_{\mu}^{a}$ employed to define the PT-BFM Green's functions was shown to coincide with the external current associated with the BRST-anti-BRST variation $s\overline{s}A_{\mu}^{a}\propto \mc{N}_{\mu}^{a}$.

Just like the external currents of ordinary functional formalisms, $\widehat{A}_{\mu}^{a}$ never appears in the final equations of the PT-BFM: once the BFG effective action is functionally differentiated with respect to $\widehat{A}_{\mu}^{a}$ to yield the PT-BFM Green's functions, the background field is set to zero. The net result of this operation is a set of Green's functions whose external topology is the same as that of their ordinary LCG counterpart -- that is, they are still propagators and vertices --, but whose internal structure is rearranged.

Once $\widehat{A}_{\mu}^{a}$ is set to zero, the dynamics of the LCGs is recovered, and the building blocks of the SDEs of the rearranged PT-BFM Green's functions become the propagators and vertices of the LCGs. Therefore, in every respect, quantities evaluated within the PT-BFM are, in fact, quantities defined in the LCGs. It is thus no surprise that these quantities can be obtained by operating on the Hilbert space of the LCGs.

What kind of rearrangement does the PT-BFM induce? The way by which we obtained $\widetilde{I}_{\mu}(p)$ from $I_{\mu}(p)$ suggests a new, nonperturbative answer to this question. Indeed, if we define a kernel operator $\widetilde{K}_{\mu}^{a}$ as
\begin{equation}
   \widetilde{K}_{\mu}^{a}=gf^{abc}\partial^{\nu}(A_{\nu}^{b}A_{\mu}^{c})+gJ_{\mu}^{a}\ ,
\end{equation}
we have
\begin{align}
   \widetilde{I}_{\mu}(-\hat{p})\,\delta^{ab}\,e^{-i\hat{p}x}=ig^{-1}\bra{\Omega}\widetilde{K}^{a}_{\mu}(x)\ket{\chi^{b}(p)}\ ,
\end{align}
in complete analogy with Eq.~\eqref{g2saker}. Moreover, since $\widetilde{K}_{\mu}^{a}$ contains the tree-level vertices of the BFG with a single background gluon leg, denoting the background-quantum PT-BFM gluon polarization tensor with $\widetilde{\Pi}_{\mu\nu}^{ab}(p)$, upon examination of the SDEs of the latter \cite{BP09,CPB11}, it is easy to see that
\begin{equation}\label{gpsdereppt}
   \widetilde{\Pi}_{\mu\nu}^{ab}(p)=i\,\left[\langle T\{\widetilde{K} A\}\rangle_{\text{F.T.}}(p)\cdot \Delta^{-1}(p)\right]_{\mu\nu}^{ab}\ ,
\end{equation}
in complete analogy with Eq.~\eqref{gpsde}.

It thus appears that one may be able to obtain PT-BFM Green's functions with one background external gluon leg by replacing the operator $j_{\mu}^{a}$,
\begin{equation}
   j_{\mu}^{a}=f^{abc}\left(A^{b\nu}F_{\nu\mu}^{c}+i\partial_{\mu}\cbar^{b}c^{c}\right)= J_{\mu}^{a}-\{Q_{B},f^{abc}A_{\mu}^{b}\cbar^{c}\}\ ,
\end{equation}
that appears in the kernel operator $K_{\mu}^{a}$ of the gluon SDEs Eq.~\eqref{kernelop}, with the global gauge current $J_{\mu}^{a}$. By doing so, one obtains Green's functions that dynamically couple directly to $J_{\mu}^{a}$, instead of $j_{\mu}^{a}$. At this stage, we have not established whether this construction remains valid beyond two-point Green's functions.

At variance with $K_{\mu}^{a}$, whose divergence is nonzero except in Abelian theories,
\begin{equation}
   \partial^{\mu}K_{\mu}^{a}=g\, \partial^{\mu}j_{\mu}^{a}=\{Q_{B},\partial^{2}\cbar^{a}\}=\partial^{2}B^{a}\ ,
\end{equation}
the divergence of $\widetilde{K}^{a}_{\mu}$ always vanishes,
\begin{equation}
   \partial^{\mu}\widetilde{K}_{\mu}^{a}=g\,\partial^{\mu}J_{\mu}^{a}=0\ ,
\end{equation}
suggesting an additional explanation for why the PT-BFM Green's functions have improved transversality properties compared with ordinary LCG Green's functions \cite{CPB11}.\\

As a corollary of the results presented up until this point, we are now in a position to answer a question that pervades contemporary studies of the BFGs, as they relate to QCD \cite{ABP09,BQ13,ABFP24,FP25}: since the Batalin-Vilkoviski function $G(p^{2})$ coincides with the Kugo-Ojima function, what is the deep connection between $G(p^{2})$ and the Kugo-Ojima criterion for confinement $u=-1$? The answer is that the constant $1+u$ has nothing to do with confinement. Modulo a missing LSZ coefficient equal to $M$, the Goldstone equation in which $u$ was originally introduced is nothing but the definition of the PT-BFM gluon-scalar transition amplitude. In it, the role of $1+u$ is precisely that of a BV function -- namely, that of transforming the amplitude from linear covariant gauges to background-field gauges. The Goldstone equation is realized in the BFGs because the matrix elements of the global gauge current $J_{\mu}^{a}$ between the vacuum and the one-Goldstone state can be expressed as an integral sum rule for the BSAs of the ABFP bound state, in which the LCG tree-level vertices are replaced by the tree-level vertices of the BFGs.\\

Eq.~\eqref{g2sacurr} cements the connection between dynamical gluon mass generation, the Schwinger mechanism, and spontaneous symmetry breaking in QCD. The mass scale $\widehat{M}=M(1+u)$ that appears in it,
\begin{equation}\label{CGold}
   \bra{\Omega}J_{\mu}^{a}(x)\ket{\chi^{b}(p)}=i\widehat{M}/g\ \hat{p}_{\mu}\,\delta^{ab}\,e^{-i\hat{p}x}\ ,
\end{equation}
and in its integrated counterpart,
\begin{equation}\label{gstcomm}
   \bra{\Omega}[Q^{a},\chi^{b}(x)]\ket{\Omega}=i\widehat{M}/g\ \delta^{ab}\ ,
\end{equation}
is the gluon mass defined from the transverse form factor $\widehat{\Delta}(p^{2})$ of the unrenormalized background-background PT-BFM gluon propagator \cite{BP02b,BP09,CPB11,BQ13,IP13,FP25}\footnote{Equivalently, $\widehat{M}$ appears in the zero-momentum limit $\widehat{M}\, M=i\widetilde{\Delta}^{-1}(0)$, where $\widetilde{\Delta}(p^{2})=-i/[p^{2}+i\widetilde{\Pi}(p^{2})]$ is the transverse form factor of the unrenormalized background-quantum PT-BFM gluon propagator \cite{CPB11,IP13}, obtained from the polarization tensor in Eq.~\eqref{gpsdereppt}. This relation can be derived by analizing the massless discrete spectrum of the Green's function on the RHS of Eq.~\eqref{gpsdereppt} like we did for the ordinary polarization in Eqs.~\eqref{polschwpoleprel} and \eqref{polschwpole} of Appendix~\ref{app:schwpoles}.},
\begin{equation}
   \widehat{\Delta}(0)=\frac{i}{\widehat{M}^{2}}\ .
\end{equation}
For massive gluons, it is hard to see how $\widehat{M}$ -- hence $\widehat{M}/g$ -- could be equal to zero \cite{Cor81}, confirming the Goldstone nature of $\chi^{a}$.

Once multiplied by $M$ to form $\widehat{M}$, the constant $1+u$ disappears completely from the Goldstone equation. $1+u$ thus plays no role in it, other than yielding $\widehat{M}$. Evidently, the fact that $u=0$ in Abelian theories agrees with the Abelian PT-BFM gauge-boson propagator being equal to its ordinary LCG counterpart.

The PT-BFM gluon propagator $\widehat{\Delta}(p^{2})$ is well-known to renormalize like the square of the inverse coupling constant $g$ \cite{Abb80,ABPR09,BQ13}: the product $g^{2}\,\widehat{\Delta}(p^{2})$ is finite and RG invariant. Therefore, the quantity $\widehat{M}/g$ on the RHS of Eq.~\eqref{CGold} is finite and RGI, as one would expect from a matrix element of the color current. En passing, we note that Kugo had in fact derived a corresponding renormalization relation in \cite{Kug95},
\begin{equation}\label{KOren}
   Z_{3}^{-1/2}g^{-1}(1+u)=g_{R}^{-1}\ ,
\end{equation}
where $g_{R}$ is a renormalized coupling constant. However, since the original KO Goldstone equation was missing a factor of $Z_{\chi}^{1/2}\propto Z_{3}^{1/2}$, Kugo did not realize that relation would make $1+u$ disappear from the equation, thus rendering the criterion $1+u=0$ meaningless.

Finally, given that the RHS of Eq.~\eqref{CGold} is finite and RGI, both the PT-BFM gluon-scalar transition amplitude $\widetilde{I}_{\mu}(\hat{p})$ and the sum rule it defines are too\footnote{This is at variance with $I_{\mu}(\hat{p})$ and the corresponding sum rule, which are neither RGI, nor finite.}. This strongly suggests that the complete valence content of $\chi^{a}$ as a bound state is, in fact,
\begin{equation}
   \ket{\chi}\sim \ket{AA}+\ket{AAA}+\ket{c\cbar}+\ket{c\cbar A}+\ket{\psi\psibar}\ ,
\end{equation}
with the channel $c\cbar A$ included.

\subsection{The structure of the global gauge charge, asymptotic residual gauge transformations, color symmetry, and color confinement}\label{sec:chargestruct}

In the previous section, we saw that, as expected for an interacting Yang-Mills theory quantized in the LCGs within the KO framework, the global gauge charge of dynamically massive QCD is spontaneously broken. The questions then arise of whether this symmetry breaking can be understood as discussed in Sec.~\ref{sec:resgtgb}, whether a well-defined color charge operator can still be uniquely identified and, if it can, whether the operator annihilates physical asymptotic states in the BRST cohomology, leading to color confinement in the original sense of Kugo and Ojima.

The aim of the present section is to give an answer to these questions. To do so, we start by analyzing the structure of the global gauge charge $Q^{a}$ in the light of asymptotic completeness. Gaining an understanding of the spontaneously broken transformations generated by $Q^{a}$ on the asymptotic fields of dynamically massive QCD will confirm our preliminary analysis of Sec.~\ref{sec:resgtgb}, and it will immediately point us to the definition of the unique asymptotic color charge operator.\\

Up until this point, we have not made use of the assumptions we formulated in Sec.~\ref{sec:resgtgb} concerning the symmetry behavior of dynamically massive QCD\footnote{While we have made use of the diagonality in color space of propagators and of the KO constant $u^{ab}$, we did not derive these properties from those assumptions.}. The reason is straightforward to understand: in Secs.~\ref{sec:mlessspect} and \ref{sec:ssb} we have primarily discussed dynamical gauge-boson mass generation, which can occur either in the presence or in the absence of the spontaneous breaking of the underlying symmetry of the theory -- in our case, color symmetry.

To appreciate this fact, observe that the Schwinger mechanism we described in Sec.~\ref{sec:goldbound} does not rest on any property specifically related to the violation of color symmetry. On the contrary, 1. at no point the proof of the existence of Schwinger poles in the elementary vertices of dynamically massive QCD required color symmetry to be violated, and 2. while the symmetry-breaking matrix element $\bra{\Omega}J_{\mu}^{a}(x)\ket{\chi^{b}(p)}$ is indeed required to be different from zero in order for a gluon mass to be generated, as we remarked many times, that same matrix element is different from zero in the LCGs of all Yang-Mills theories, including massless, symmetry-preserving theories like four-dimensional QED: $\bra{\Omega}J_{\mu}^{a}(x)\ket{\chi^{b}(p)}\neq 0$ on its own \textit{cannot} in any way be regarded as an indication that color symmetry is broken.

In Appendix~\ref{app:softtheorem}, to confirm this observation, we formulate a soft theorem for dynamically massive QCD. The theorem shows that, even in its strongest form, regardless of the spontaneous breaking of $Q^{a}$, whether color symmetry is preserved or not does not depend on the matrix elements of $\chi^{a}$, but rather on  the soft limit of those of $\beta^{a}$ -- equivalently, on the soft limit of the Slavnov-Taylor identities of the theory. As neither of these contributes to dynamical gluon mass generation, we must conclude that color symmetry breaking is not a necessary condition for gluons to acquire a mass.

Since there is no evidence that the Schwinger mechanism acting in dynamically massive QCD breaks color symmetry, one is free to assume -- as is customary -- that the mechanism preserves the symmetry, as long as the assumption is formulated in such a way as to account for the spontaneous breaking of the global gauge charge $Q^{a}$. As it turns out, and as we now show, the symmetry assumptions we introduced in Sec.~\ref{sec:resgtgb} can indeed account for such a breaking.

Let us therefore make use of our first symmetry assumption, Eq.~\eqref{symass}, to study the action of the global gauge charge $Q^{a}$ on the asymptotic fields of dynamically massive QCD. Recall that global gauge transformations act on the elementary fields of QCD as
\begin{align}
   &[Q^{a},A_{\mu}^{b}]=if^{abc}A_{\mu}^{c}\ ,\quad [Q^{a},B^{b}]=if^{abc}B^{c}\ ,\\
   \notag&\quad[Q^{a},c^{b}]=if^{abc}c^{c}\ ,\quad [Q^{a},\cbar^{b}]=if^{abc}\cbar^{c}\ .
\end{align}
By taking the asymptotic limit $t\to \pm \infty$ of both sides of the equations and using Eq.~\eqref{symass}, we obtain the transformations of the corresponding asymptotic fields,
\begin{align}\label{deq349}
   &[Q^{a},A_{\mu}^{b\,\text{as.}}]=if^{abc}A_{\mu}^{c\,\text{as.}}\ ,\quad [Q^{a},B^{b\,\text{as.}}]=if^{abc}B^{c\,\text{as.}}\ ,\\
   \notag&\quad[Q^{a},c^{b\,\text{as.}}]=if^{abc}c^{c\,\text{as.}}\ ,\quad [Q^{a},\cbar^{b\,\text{as.}}]=if^{abc}\cbar^{c\,\text{as.}}\ .
\end{align}
By separating asymptotic components of different masses, from Eq.~\eqref{deq349} we obtain the transformation laws of the elementary BRST quartet of dynamically massive QCD:
\begin{align}\label{gasymt}
   &[Q^{a},\partial_{\mu}\chi^{b}]=if^{abc}\partial_{\mu}\chi^{c}\ ,\quad [Q^{a},\beta^{b}]=if^{abc}\beta^{c}\ ,\\
   \notag&\quad[Q^{a},\gamma^{b}]=if^{abc}\gamma^{c}\ ,\quad [Q^{a},\gammabar^{b}]=if^{abc}\gammabar^{c}\ .
\end{align}
If massive asymptotic components are developed by the elementary fields, those are subject to analogous transformation laws. For instance, the massive gluon asymptotic field $U_{\mu}^{a}$ in Eq.~\eqref{glumasscomp} would transform as
\begin{equation}\label{Ucolort}
   [Q^{a},U_{\mu}^{b}]=if^{abc}U_{\mu}^{c}\ .
\end{equation}

The first of Eq.~\eqref{gasymt}, when integrated, yields the transformation laws of $\chi^{a}$. These contain constants, fixed by Eq.~\eqref{gstcomm}, that reflect its Goldstone nature:
\begin{equation}\label{gasymtchi}
   [Q^{a},\chi^{b}]=ig^{-1}\widehat{M}\,\delta^{ab}+if^{abc}\chi^{c}\ .
\end{equation}
The mechanism by which Eq.~\eqref{symass} does not contradict the spontaneous breaking of $Q^{a}$ is thus easily identified: being the first parent of the elementary BRST quartet, $\chi^{a}$ enters the asymptotic limit of the gluon field $A_{\mu}^{a}$ as a derivative, hence its transformation laws are still permitted to have a symmetry-breaking constant which attains the nonzero value determined by Eq.~\eqref{gstcomm}.

Armed with Eqs.~\eqref{gasymt} and \eqref{gasymtchi}, we now seek an expression of $Q^{a}$ in terms of the asymptotic fields. To this aim, we observe that, by the equal-time asymptotic (anti)commutation relations in Eq.~\eqref{ascommet}, modulo prefactors, $\dot{\chi}^{a}$, $\dot{\gamma}^{a}$, and $\dot{\gammabar}^{a}$ play the role of conjugate momenta for $\beta^{a}$, $\gammabar^{a}$, and $\gamma^{a}$, respectively. As for $\chi^{a}$, with it are formally associated two conjugate momenta: $\dot{\chi}^{a}$ and $\dot{\beta}^{a}$. An explicit calculation then easily shows that the asymptotic expression of $Q^{a}$ is
\begin{widetext}
   \begin{equation}\label{gcharge}
      Q^{a}=g^{-1}\widehat{M}\int d^{3}{\bf x}\ \dot{\beta}^{a}+f^{abc}\int d^{3}{\bf x}\ \left(\dot{\chi}^{b}\beta^{c}-\chi^{b}\dot{\beta}^{c}-\dot{\beta}^{b}\beta^{c}+i\,\dot{\gammabar}^{b}\gamma^{c}-i\,\gammabar^{b}\dot{\gamma^{c}}\right)+\cdots\ ,
   \end{equation}
\end{widetext}
where the ellipsis denotes contributions from asymptotic fields not belonging to the elementary BRST quartet. Note that, under our assumption of asymptotic completeness, since all other asymptotic fields (anti)commute with $\chi^{a}$, $\beta^{a}$, $\gamma^{a}$, and $\gammabar^{a}$, terms in the ellipsis do not influence the transformation laws of the elementary BRST quartet; specularly, the explicit terms in Eq.~\eqref{gcharge} do not influence the transformation laws of the other asymptotic fields.

Let us make some remarks on Eq.~\eqref{gcharge}. First of all, a straightforward calculation using Eq.~\eqref{chidip} and the definitions in Appendix~\ref{app:dipole} shows that the extra terms obtained by replacing the dipole field $\chi^{a}$ with the simple-pole field $\widetilde{\chi}^{a}$ in Eq.~\eqref{gcharge} drop out from the equation. It follows that the explicit terms in $Q^{a}$ can be expanded in the massless creation and annihilation operators of $\widetilde{\chi}^{a}$, $\beta^{a}$, $\gamma^{a}$, and $\gammabar^{a}$. The resulting expression needs no normal ordering thanks to the antisymmetry of the structure constants $f^{abc}$ and to the diagonality of the (anti)commutation relations in the color indices.

Second, as far as the elementary BRST quartet is concerned, the expression provided for $Q^{a}$ by Eq.~\eqref{gcharge} is unique: any additional term in the equation containing $\chi^{a}$, $\beta^{a}$, $\gamma^{a}$, or $\gammabar^{a}$ must commute with all the asymptotic fields, hence it must be a constant whose value is irrelevant, and can be set to zero.

Third, the explicit terms in $Q^{a}$ can be obtained by integrating the time component of the asymptotic current
\begin{widetext}
   \begin{align}
      \mc{J}^{a}_{\mu}=g^{-1}\widehat{M}\,\partial_{\mu}\beta^{a}+f^{abc}\,\left(\partial_{\mu}\chi^{b}\beta^{c}-\chi^{b}\partial_{\mu}\beta^{c}-\partial_{\mu}\beta^{b}\beta^{c}+i\,\partial_{\mu}\gammabar^{b}\gamma^{c}-i\,\gammabar^{b}\partial_{\mu}\gamma^{c}\right)\ .
   \end{align}
\end{widetext}
$\mc{J}^{a}_{\mu}$ is conserved, $\partial^{\mu}\mc{J}_{\mu}^{a}=0$, it is BRST exact,
\begin{equation}
   \mc{J}_{\mu}^{a}=M_{R}^{-1}\{Q_{B,R},\mc{R}_{\mu}^{a}\}\ ,
\end{equation}
where $\mc{R}_{\mu}^{a}$ is the asymptotic current defined as
\begin{align}
   &\mc{R}_{\mu}^{a}=g^{-1}\widehat{M}\,\partial_{\mu}\gammabar^{a}+f^{abc}\left(\partial_{\mu}\chi^{b}\gammabar^{c}-\chi^{b}\partial_{\mu}\gammabar^{c}-\partial_{\mu}\beta^{b}\gammabar^{c}\right)\ ,
\end{align}
and its linear term coincides with Eq.~\eqref{jaslim} after setting $Z_{\chi}^{-1/2}(1+u)=M(1+u)=\widehat{M}$.

Fourth, $Q^{a}$ is ill-defined as an operator acting on the Hilbert space. Indeed, the integral in the first term of Eq.~\eqref{gcharge},
\begin{equation}
   \int d^{3}{\bf x}\ \dot{\beta}^{a}=\frac{i}{2}\,\left[\sqrt{2\hat{p}^{0}}\,\left(\beta^{a\,\dagger}_{{\bf p}}-\beta^{a}_{{\bf p}}\right)\right]_{{\bf p}={\bf{0}}}\ ,
\end{equation}
is the difference between two operators, $\beta_{{\bf 0}}^{a\dagger}$ and its adjoint, that do not exist on the Hilbert space\footnote{See footnote~\ref{fn:symbreakilldef}.}. Nevertheless, $Q^{a}$ is perfectly meaningful as an operator acting on the asymptotic fields via commutation relations.

Fifth, the first term in Eq.~\eqref{gcharge} is the only term linear in the fields, and the only one responsible for spontaneous symmetry breaking. Indeed, as we saw in the previous section, under customary assumptions, $\chi^{a}$ is the unique Goldstone boson associated with $Q^{a}$. It follows that
\begin{equation}\label{deq500}
   \bra{\Omega}[Q^{a},\phi_{I}^{\text{as.}}(x)]\ket{\Omega}=0
\end{equation}
for every asymptotic field $\phi_{I}^{\text{as.}}(x)$ different from $\chi^{a}(x)$, hence that the ellipsis can contain no linear terms.

Finally, by Eqs.~\eqref{symass} and \eqref{deq500} -- see also \cite{NO78} --, terms in the ellipsis generate transformations linear in the remaining asymptotic fields, hence they are quadratic in the asymptotic fields\footnote{Modulo potential irrelevant constants, which we assume to be zero.}. The corresponding spatial integral is thus well-defined, and it can be obtained from an asymptotic current which is conserved, since $J_{\mu}^{a}$ and $\mc{J}_{\mu}^{a}$ both are.
\\

Because of the linear term in Eq.~\eqref{gcharge} -- equivalently, because of the constants in Eq.~\eqref{gasymtchi} --, it is clear that $Q^{a}$ does not merely generate color transformations. For instance, ignoring questions of well-definedness, it is easy to check that the $Q^{a}$'s do not satisfy the SU(3) color algebra\footnote{Terms in the ellipsis of Eq.~\eqref{gcharge} commute with the explicit terms, so the former cannot fix the violation of the color algebra already caused by the latter.},
\begin{equation}
   [Q^{a},Q^{b}]\neq if^{abc}Q^{c}\ ,
\end{equation}
as announced in Sec.~\ref{sec:pymt}.

In order to understand the transformations generated by $Q^{a}$, in what follows we study the properties of a closely related Hermitian operator $\mc{Q}[\Lambda,\lambda]$, defined as

\begin{widetext}
   \begin{align}\label{gchargegen}
      \mathcal{Q}[\Lambda,\lambda]=g^{-1}\widehat{M}\int d^{3}{\bf x}\ \left(\Lambda^{a}\dot{\beta}^{a}-\dot{\Lambda}^{a}\beta^{a}\right)+\lambda^{a}f^{abc}\int d^{3}{\bf x}\ \left(\dot{\chi}^{b}\beta^{c}-\chi^{b}\dot{\beta}^{c}-\dot{\beta}^{b}\beta^{c}+i\,\dot{\gammabar}^{b}\gamma^{c}-i\,\gammabar^{b}\dot{\gamma^{c}}\right)+\lambda^{a}(\cdots)\ ,
   \end{align}
\end{widetext}
where the $\Lambda^{a}(x)$'s are real functions such that $\partial^{2}\Lambda^{a}=0$, the $\lambda^{a}$'s are real constants, and the ellipsis corresponds to that in Eq.~\eqref{gcharge}. It is easy to see that, due to the individual conservation of all currents involved in the equation, $\frac{d}{dt}\mc{Q}[\Lambda,\lambda]=0$. Moreover, for all non-constant $\Lambda^{a}$'s, the charge $\mc{Q}[\Lambda,\lambda]$ is well-defined as an operator acting on the Hilbert space: we have
\begin{equation}\label{deq792}
   \mc{Q}[\Lambda,\lambda]=ig^{-1}\widehat{M}\ \int\frac{d^{3}{\bf p}}{(2\pi)^{3}}\ \left(\Lambda_{{\bf p}}^{a}\beta_{{\bf p}}^{a\,\dagger}-\Lambda_{{\bf p}}^{a\,\star}\beta_{{\bf p}}^{a}\right)+\cdots\ ,
\end{equation}
where $\Lambda_{{\bf p}}^{a}$ is the Fourier transform of $\Lambda^{a}(x)$,
\begin{equation}
   \Lambda^{a}(x)=\int \frac{d^{3}{\bf p}}{(2\pi)^{3}}\,\frac{1}{\sqrt{2\hat{p}^{0}}}\ \left(\Lambda_{{\bf p}}^{a}\,e^{-i\hat{p}x}+\Lambda_{{\bf p}}^{a\,\star}\,e^{i\hat{p}x}\right)\ .
\end{equation}
The ill-defined global gauge charge $Q^{a}$ is recovered from $\mc{Q}[\Lambda,\lambda]$ by setting $\Lambda^{a}(x)=\lambda^{a}$,
\begin{equation}
   \mc{Q}[\lambda,\lambda]=\lambda^{a}Q^{a}\ .
\end{equation}

Let us compute the action of $\mc{Q}[\Lambda,\lambda]$ on the asymptotic fields. Since the first integral in Eq.~\eqref{gchargegen} only contains $\beta^{a}$, $\mc{Q}[\Lambda,\lambda]$ acts like the global gauge charge $Q^{a}$ on all asymptotic fields $X$ except for $\chi^{a}$:
\begin{equation}\label{astransX}
   [\mathcal{Q}[\Lambda,\lambda],X(x)]=\lambda^{a}[Q^{a},X(x)]\qquad(X\,\neq \chi^{a})\ .
\end{equation}
As for $\chi^{a}$, an explicit calculation yields
\begin{equation}\label{astranschi}
   [\mathcal{Q}[\Lambda,\lambda],\chi^{a}(x)]=ig^{-1}\widehat{M}\,\Lambda^{a}(x)+if^{bac}\lambda^{b}\chi^{c}(x)\ .
\end{equation}
For $\Lambda^{a}(x)=\lambda^{a}$, we recover Eq.~\eqref{gasymtchi}.

The transformations generated by $\mc{Q}[\Lambda,\lambda]$ are symmetries of the asymptotic field equations of the elementary BRST quartet. Indeed, going back to Eq.~\eqref{asfieldeq},
\begin{align}\label{asfieldeq2}
   &\partial^{2}\chi^{a}+\xi_{R}M_{R}^{2}\,\beta^{a}=0\ ,\quad \partial^{2}\beta^{a}=0\ ,\\
   \notag&\qquad \partial^{2}\gamma^{a}=0\ ,\quad \partial^{2}\gammabar^{a}=0\ ,
\end{align}
we see that these equations are still satisfied after operating the transformations
\begin{align}
   \notag\chi^{a}(x)&\to \chi^{a}(x)+g^{-1}\widehat{M}\,\Lambda^{a}(x)+f^{bac}\lambda^{b}\chi^{c}(x)\ ,\\
   \beta^{a}(x)&\to \beta^{a}(x)+f^{bac}\lambda^{b}\beta^{c}(x)\ ,\\
   \notag\gamma^{a}(x)&\to \gamma^{a}(x)+f^{bac}\lambda^{b}\gamma^{c}(x)\ ,\\
   \notag\gammabar^{a}(x)&\to \gammabar^{a}(x)+f^{bac}\lambda^{b}\gammabar^{c}(x)\ .
\end{align}
The fact that these transformations depend on two distinct sets of parameters $\Lambda^{a}$ and $\lambda^{a}$ -- the first being functions that satisfy the wave equation, the second being constants -- is not arbitrary. To see this, let us define separately well-defined charges $C[\Lambda]$ and $\widetilde{Q}[\lambda]$ as
\begin{align}\label{asymsplit}
   &C[\Lambda]=\mc{Q}[\Lambda,0]\ ,\qquad \widetilde{Q}[\lambda]=\mc{Q}[0,\lambda]\ ,\\
   \notag&\qquad\quad\mc{Q}[\Lambda,\lambda]=C[\Lambda]+\widetilde{Q}[\lambda]\ .
\end{align}
The charge $C[\Lambda]$ only acts on the ABFP Goldstone boson,
\begin{equation}
   \chi^{a}(x)\to \chi^{a}(x)+g^{-1}\widehat{M}\,\Lambda^{a}(x)\ ,
\end{equation}
whereas, on the fields $\varphi^{a}$ of the elementary quartet, the charge $\widetilde{Q}[\lambda]$ acts as
\begin{equation}\label{deq497}
   \varphi^{a}(x)\to \varphi^{a}(x)+f^{bac}\lambda^{b}\varphi^{c}(x)\ .
\end{equation}
The structure of the parameters of $\mc{Q}[\Lambda,\lambda]$ is then explained by observing that 1. the asymptotic field equations in Eq.~\eqref{asfieldeq2} are separately invariant under the transformations generated by $C[\Lambda]$ and $\widetilde{Q}[\lambda]$, and 2. they are not invariant under transformations such that $\partial^{2}\Lambda^{a}\neq 0$, nor under transformations with nonconstant $\lambda^{a}$'s.

The transformations generated by $C[\Lambda]$ are symmetries of the asymptotic field equations, but not of the vacuum: under them, the ABFP Goldstone boson acquires a (coordinate-dependent) vacuum expectation value,
\begin{equation}\label{deq629}
   \bra{\Omega}\chi^{a}(x)\ket{\Omega}\to \bra{\Omega}\chi^{a}(x)\ket{\Omega}+g^{-1}\widehat{M}\,\Lambda^{a}(x)\neq 0\ ,
\end{equation}
making these symmetries spontaneously broken, $C[\Lambda]\ket{\Omega}\neq 0$. The spontaneous breaking of $\mc{Q}[\Lambda,\lambda]$ and of the global gauge charge $Q^{a}$ are in one-to-one correspondence with that of $C[\Lambda]$.

On the other hand, the charge $\widetilde{Q}[\lambda]$ generates unbroken transformations: due to its structure quadratic in the asymptotic fields, $\widetilde{Q}[\lambda]$ is well-defined as an operator acting on the Hilbert space, it satisfies
\begin{equation}\label{ubcharge}
   \bra{\Omega}[\widetilde{Q}[\lambda],\phi_{I}^{\text{as.}}(x)]\ket{\Omega}= 0
\end{equation}
for every asymptotic field $\phi_{I}^{\text{as.}}(x)$, and it annihilates the vacuum, $\widetilde{Q}[\lambda]\ket{\Omega}=0$.

In order to establish the connection anticipated in Sec.~\ref{sec:resgtgb} between the spontaneously broken transformations generated by $Q^{a}$ and residual gauge transformations, we follow Kugo \cite{Kug95} and define a renormalized coupling constant $g_{R}$ as in Eq.~\eqref{KOren}. Like in the previous sections, we define $Z_{3}$ from the zero-momentum limit $\Delta(0)=iZ_{3}/M_{R}^{2}$ of the form factor of the massive gluon propagator, and use it to renormalize $A_{\mu}^{a\,\text{as.}}$. Let us apply the transformation $\mathcal{Q}[\Lambda,\lambda]$ to the asymptotic gauge field $A_{\mu}^{a\,\text{as.}}$. Observing that $C[\Lambda]$ does not transform a potential massive asymptotic component $U_{\mu}^{a}$ contained in $A_{\mu}^{a\,\text{as.}}$, by Eqs.~\eqref{Ucolort} and \eqref{astranschi}, we find that
\begin{align}\label{asymgauge}
   [\mathcal{Q}[\Lambda,\lambda],A_{\mu}^{a\,\text{as.}}]&=i\left(g_{R}^{-1}\,\partial_{\mu}\Lambda^{a}+f^{abc}A_{\mu}^{b\,\text{as.}}\lambda^{c}\right)\ .
\end{align}
The above transformation is nothing more than a gauge transformation in which the parameter functions $\Lambda^{a}$ of the ``Abelian'' derivative term have become independent of the parameters $\lambda^{a}$ of the non-Abelian term, while the latter have become constants. As discussed above, this configuration of parameters is not arbitrary: it reflects the actual symmetries of the asymptotic field equations of dynamically massive QCD.

The transformations in Eq.~\eqref{asymgauge} act on the asymptotic fields of the theory. Moreover, they are symmetries of the asymptotic field equations, and they preserve the asymptotic limit of the gauge-fixing condition\footnote{Recall that the gauge-fixing condition constrains the massive asymptotic component of $B^{a\,\text{as.}}$ to vanish, see footnote \ref{fn:bmassc}.}:
\begin{equation}
   [\mathcal{Q}[\Lambda,\lambda],\partial^{\mu}A_{\mu}^{a\,\text{as.}}+\xi_{R}\,B^{a\,\text{as.}}]=ig_{R}^{-1}\partial^{2}\Lambda^{a}=0\ .
\end{equation}
Therefore, they are asymptotic residual gauge transformations. Given that, under these transformations, the asymptotic field $A_{\mu}^{a\,\text{as.}}$ acquires a vacuum expectation value,
\begin{equation}
   \bra{\Omega}A_{\mu}^{a\,\text{as.}}(x)\ket{\Omega}\to\bra{\Omega}A_{\mu}^{a\,\text{as.}}(x)\ket{\Omega}+g_{R}^{-1}\,\partial_{\mu}\Lambda^{a}(x)\ ,
\end{equation}
their generator must necessarily be spontaneously broken.

When $\Lambda^{a}(x)=\lambda^{a}$, the global gauge charge $Q^{a}$ is recovered from $\mc{Q}[\Lambda,\lambda]$, and the derivative seems to disappear from Eq.~\eqref{asymgauge}, yielding pure color transformations of the gauge field. This is, however, an illusion: the \textit{generator} of the derivative is \textit{still present} in $Q^{a}$. In complete analogy with the Abelian case discussed in Sec.~\ref{sec:resgtgb}, the derivative survives as a constant in the transformation laws of the first parent of the elementary BRST quartet $\chi^{a}$, making it the Goldstone boson associated with the spontaneous breaking of asymptotic residual (global) gauge symmetry.\\

Let us make some remarks on $\mc{Q}[\Lambda,\lambda]$ and $Q^{a}$, and on the transformations they generate. First of all, as is apparent from the fact the transformations preserve the gauge-fixing condition, the structure of symmetry breaking exhibited by these charges is specific to the linear covariant gauges.

Second, the spontaneously broken but well-defined charge $C[\Lambda]$,
\begin{equation}
   C[\Lambda]=g^{-1}\widehat{M}\int d^{3}{\bf x}\ \left(\Lambda^{a}\dot{\beta}^{a}-\dot{\Lambda}^{a}\beta^{a}\right)\ ,
\end{equation}
is BRST exact, $C[\Lambda]=M_{R}^{-1}\{Q_{B,R},\mc{Y}[\Lambda]\}$, where
\begin{equation}
   \mc{Y}[\Lambda]=g^{-1}\widehat{M}\int d^{3}{\bf x}\ \left(\Lambda^{a}\dot{\gammabar}^{a}-\dot{\Lambda}^{a}\gammabar^{a}\right)\ .
\end{equation}
Just like $C[\Lambda]$, $\mc{Y}[\Lambda]$ generates spontaneously broken symmetries $[\theta \mc{Y}[\Lambda],\gamma^{a}(x)]$ of the asymptotic field equations -- specifically, of the asymptotic ghost field,
\begin{equation}
   \gamma^{a}(x)\to \gamma^{a}(x)+g^{-1}\widehat{M}\,\theta\Lambda^{a}(x)\ ,
\end{equation}
where $\theta$ is a Grassmann number\footnote{The operator $\overline{\mc{Y}}[\Lambda]=g_{R}^{-1}M_{R}\int d^{3}{\bf x}\ \left(\Lambda^{a}\dot{\gamma}^{a}-\dot{\Lambda}^{a}\gamma^{a}\right)$ that generates the spontaneously broken asymptotic antighost symmetries $\gammabar^{a}(x)\to \gammabar^{a}(x)+g_{R}^{-1}M_{R}\theta\Lambda^{a}(x)$ can then be expected to satisfy $C[\Lambda]=M_{R}^{-1}\{\overline{Q}_{B,R},\overline{\mc{Y}}[\Lambda]\}$, where $\overline{Q}_{B,R}$ is a renormalized anti-BRST operator. On the contrary, the transformation $\beta^{a}(x)\to \beta^{a}(x)+g_{R}^{-1}M_{R}\Lambda^{a}(x)$ is not a symmetry of the asymptotic field equations unless $\xi=\xi_{R}=0$, so no charge is expected to generate it in the general case.}.

Third, like the Abelian theories in Sec.~\ref{sec:resgtgb}, with asymptotic residual gauge symmetry spontaneously broken by structure, $\mc{Q}[\Lambda,\lambda]\ket{\Omega}\neq 0$, QCD must have a way to express the invariance of physical asymptotic states with respect to these transformations. In this respect, going back to Eq.~\eqref{qbasym}, we see that -- modulo prefactors -- the explicit term in the redefined BRST charge $Q_{B,R}$ is nothing more than the broken charge $C[\Lambda]$, in which the parameter functions $\Lambda^{a}$ have been replaced with the asymptotic ghost fields $\gamma^{a}$:
\begin{equation}
   Q_{B,R}=-g_{R}\,C[\gamma]+\cdots\ .
\end{equation}
With $\Lambda^{a}$ replaced by $\gamma^{a}$, the non-vanishing vacuum expectation value of the former is replaced by that of the latter,
\begin{align}
   0&=\bra{\Omega}A_{\mu}^{a\,\text{as.}}(x)\ket{\Omega}\to\\
   \notag&\to\bra{\Omega}A_{\mu}^{a\,\text{as.}}(x)\ket{\Omega}-g_{R}^{-1}\,\theta\bra{\Omega}\partial_{\mu}\gamma^{a}(x)\ket{\Omega}=0\ ,
\end{align}
yielding an unbroken and well-defined charge. Therefore, limited to the elementary BRST quartet, we see that the BRST charge $Q_{B,R}$ is but the unbroken realization of the broken charge $C[\Lambda]$: the invariance of physical states under $C[\Lambda]$ can be expressed as their invariance under $Q_{B,R}$. As for the operator $\widetilde{Q}[\lambda]$, the invariance of physical states under this charge is, essentially, synonymous with color confinement. We postpone this discussion until later.\\

From the above analysis, it is apparent that the generator of asymptotic color transformations is the operator $\widetilde{Q}[\lambda]$ defined in Eq.~\eqref{asymsplit}. Factoring out the parameter $\lambda^{a}$, $\widetilde{Q}^{a}$ can be obtained by subtracting the generator $C[\Lambda=\lambda]$ of the derivative term of asymptotic residual gauge transformations from the global gauge charge $Q^{a}$,
\begin{equation}\label{colorcharge}
   (\widetilde{Q}^{a})^{\txtin/\txtout}=Q^{a}-g^{-1}\widehat{M}\int d^{3}{\bf x}\ (\dot{\beta}^{a})^{\txtin/\txtout}\ .
\end{equation}
Since $\beta^{a}$ commutes with all asymptotic fields except for $\chi^{a}$, the subtraction has the only effect of removing the well-understood symmetry-breaking constant from the transformation laws of $\chi^{a}$, yielding
\begin{equation}\label{deq394}
   [\widetilde{Q}^{a},\chi^{b}(x)]=if^{abc}\chi^{c}(x)\ ,
\end{equation}
as reported in Eq.~\eqref{deq497}. On all other asymptotic fields $X(x)$, as well as on the asymptotic gauge field $A_{\mu}^{a\,\text{as.}}(x)$, $\widetilde{Q}^{a}$ acts like the global gauge charge $Q^{a}$,
\begin{equation}\label{deq956}
   [\widetilde{Q}^{a},X(x)]=[Q^{a},X(x)]\qquad(X\neq \chi^{a})\ ,
\end{equation}
\begin{equation}\label{deq859}
   [\widetilde{Q}^{a},A_{\mu}^{b\,\text{as.}}(x)]=if^{abc}A_{\mu}^{c\,\text{as.}}(x)\ .
\end{equation}
Moreover, since $\beta^{a}$ commutes with the BRST and ghost charges $Q_{B}$ and $Q_{C}$\footnote{In terms of the asymptotic fields, $Q_{C}$ can be expressed as $Q_{C}=i\int d^{3}{\bf x}\ \left(\gammabar^{a}\dot{\gamma}^{a}-\dot{\gammabar}^{a}\gamma^{a}\right)+\cdots$, where the ellipsis contains potential contributions due to asymptotic fields not belonging to the elementary BRST quartet \cite{KO78b,KO78a,KO79b,KO79c}.}, $\widetilde{Q}^{a}$ has the same commutation relations as $Q^{a}$ with the BRST algebra -- see Eq.~\eqref{chargealgebra} --,
\begin{equation}
   [\widetilde{Q}^{a},Q_{B}]=0\ ,\qquad [\widetilde{Q}^{a},Q_{C}]=0\ .
\end{equation}

Let us confirm that, under the symmetry assumptions of Sec.~\ref{sec:resgtgb}, as can be expected from Eqs.~\eqref{deq394} and \eqref{deq956}, $\widetilde{Q}^{a}$ does indeed generate color transformations. First of all, note that, by Eq.~\eqref{symass}, the global gauge charge $Q^{a}$ generates color transformations of the asymptotic limit of all local fields. If no local field develops a non-derivative asymptotic contribution from $\chi^{a}$, then $\widetilde{Q}^{a}$ generates the same transformations as $Q^{a}$, as displayed e.g. by Eq.~\eqref{deq859}. To show that no such contribution exists, let $\phi_{I}(x)$ be a local field transforming nontrivially and irreducibly under the local action of $Q^{a}$. Suppose that $\phi_{I}(x)$ develops a non-derivative asymptotic contribution from $\chi^{a}$,
\begin{equation}
   \phi_{I}(x)\to Z_{Ia}^{1/2}\,\chi^{a}(x)+\cdots\ ,
\end{equation}
where the $Z_{Ia}^{1/2}$'s are LSZ coefficients, and the ellipsis denotes other potential asymptotic contributions. Then, in the asymptotic limit,
\begin{equation}
   [Q^{a},\phi_{I}(x)]\to iZ_{Ia}^{1/2}\,g^{-1}\widehat{M}+\cdots\ ,
\end{equation}
where the ellipsis are terms linear in the asymptotic fields. From Eq.~\eqref{symass}, and from the fact that the vacuum expectation value of asymptotic fields is zero, it follows that the vacuum expectation value of the RHS of the above equation must vanish, hence that $Z_{Ia}^{1/2}=0$. If $\phi_{I}(x)$ transforms trivially under $Q^{a}$, then $Z_{Ia}^{1/2}=0$ because the transformation must yield a vanishing RHS. Since $Z_{Ia}^{1/2}=0$, $\chi^{a}$ can only enter the asymptotic limit of local fields through derivatives, as is the case for $A_{\mu}^{a\,\text{as.}}$.

From this property, and from the fact that $\widetilde{Q}^{a}$ acts on derivatives of $\chi^{a}$ like $Q^{a}$, it follows that $\widetilde{Q}^{a}$ generates color transformations of the asymptotic limit of all the local fields in the theory. Now, since every asymptotic field $\phi_{I}^{\text{as.}}(x)$ is obtained from one of these limits, in which it enters via a differential operator $D$ (potentially of order zero), we have that
\begin{equation}
   [\widetilde{Q}^{a},D\phi_{I}^{\text{as.}}(x)]=T^{a}_{IJ}\,D\phi_{J}^{\text{as.}}(x)\ ,
\end{equation}
where the $T^{a}_{IJ}$'s realize representations of the color algebra. By integrating the differential operator, we then obtain
\begin{equation}\label{Qcolorasym}
   [\widetilde{Q}^{a},\phi_{I}^{\text{as.}}(x)]=T^{a}_{IJ}\,\phi_{J}^{\text{as.}}(x)\ ,
\end{equation}
where the integration constants vanish by virtue of Eq.~\eqref{ubcharge}. Since this relation holds for every asymptotic field, the first of Eq.~\eqref{ccond} is proven, and $\widetilde{Q}^{a}$ is seen to generate the color algebra,
\begin{equation}
   [\widetilde{Q}^{a},\widetilde{Q}^{b}]=if^{abc}\widetilde{Q}^{c}\ ,
\end{equation}
as a consequence of the $T_{IJ}^{a}$'s realizing representations of the algebra.

Going ahead with our discussion, observe that the color charge $\widetilde{Q}^{a}$ is an asymptotic charge: while $Q^{a}$ is the spatial integral of the time component of the conserved local current $J_{\mu}^{a}$, $\widetilde{Q}^{a}$ is obtained by first determining the action of $Q^{a}$ on the in/out fields of the theory, and then isolating the color part of this action. While the resulting operator $(\widetilde{Q}^{a})^{\txtin/\txtout}$ can be expressed as a well-defined spatial integral quadratic in the asymptotic fields, the action of $(\widetilde{Q}^{a})^{\txtin}$ (resp. $(\widetilde{Q}^{a})^{\txtout}$) is only known on the in (resp. out) asymptotic fields and states of the theory. In particular, with no additional assumptions, one cannot conclude that $(\widetilde{Q}^{a})^{\txtin}$ and $(\widetilde{Q}^{a})^{\txtout}$ are the same operator.

Let us show that the pair $(\widetilde{Q}^{a})^{\txtin/\txtout}$ nevertheless satisfies the second of Eq.~\eqref{ccond}. To this aim, let $\phi_{I}^{\txtin}(x)$ be an in field, let $\phi_{I}^{\txtout}(x)$ be the corresponding out field, and let $S$ be the $S$-matrix. By Eqs.~\eqref{Sop} and \eqref{Qcolorasym},
\begin{align}
   &[S(\widetilde{Q}^{a})^{\txtout}S^{-1},\phi_{I}^{\txtin}]=S[(\widetilde{Q}^{a})^{\txtout},S^{-1}\phi_{I}^{\txtin}S]S^{-1}=\\
   \notag&\ =S[(\widetilde{Q}^{a})^{\txtout},\phi_{I}^{\txtout}]S^{-1}=T_{IJ}^{a}\,S\phi_{J}^{\txtout}S^{-1}=\\
   \notag&\ =T_{IJ}^{a}\,\phi_{J}^{\txtin}=[(\widetilde{Q}^{a})^{\txtin},\phi_{I}^{\txtin}]\ ,
\end{align}
hence
\begin{equation}
   [S(\widetilde{Q}^{a})^{\txtout}S^{-1}-(\widetilde{Q}^{a})^{\txtin},\phi_{I}^{\txtin}]=0\ .
\end{equation}
Since this equation holds for every in field, by asymptotic completeness, the left-hand operand of the commutator must be proportional to a constant, which is fixed to zero by virtue of $S^{-1}\ket{\Omega}=\ket{\Omega}$ and $(\widetilde{Q}^{a})^{\txtin/\txtout}\ket{\Omega}=0$. Therefore,
\begin{equation}\label{deq041}
   S(\widetilde{Q}^{a})^{\txtout}=(\widetilde{Q}^{a})^{\txtin}S\ .
\end{equation}

By Eq.~\eqref{deq041}, the $S$-matrix defines an isomorphism between the color structures of the in and of the out Fock spaces of the theory. Note that this is a statement about the organization of the spectrum; it \textit{does not} imply that color is conserved in scattering processes, \textit{nor} that the Green's functions have the correct transformation properties under color symmetry. As we show in Appendix~\ref{app:softtheorem}, in order for these properties to be satisfied, the asymptotic field $\beta^{a}$ must satisfy an additional property -- namely, its matrix elements must vanish in the soft limit $p\to 0$; if this happens, then color symmetry is preserved in its strongest form in dynamically massive QCD. For the purpose of this paper, whose main concern is dynamical gluon mass generation and color confinement, it will not be necessary to assume that $\beta^{a}$ satisfies this additional property.\\

Let us now turn to color confinement. In the presence of dynamical gluon mass generation, if no massless bound states are formed in the channel $f^{abc}A_{\mu}^{b}A_{\nu}^{c}$, the field-strength operator $F_{\mu\nu}^{a}$ has no massless discrete spectrum, the chromoelectric flux at spatial infinity vanishes,
\begin{equation}
   \oint_{R\to \infty}dS\ \hat{n}\cdot\vec{\mc{E}}^{a}=0\ ,
\end{equation}
and the global gauge charge $Q^{a}$ is given entirely by the Kugo-Ojima charge $N^{a}$ \cite{KO79d,KO79a,NO90},
\begin{equation}\label{deq834}
   Q^{a}=g^{-1}N^{a}=g^{-1}\int d^{3}{\bf x}\ \left\{Q_{B},D_{0}c^{a}\right\} .
\end{equation}
It follows that $Q^{a}$ is formally BRST-exact, albeit ill-defined as an operator.

The ill-definedness of $Q^{a}$ has no effect on its BRST-exactness. Indeed, as we saw above, the terms in $Q^{a}$ which belong to the sector responsible for symmetry breaking -- namely, to the elementary BRST quartet -- can be computed explicitly, they are expressly BRST-exact, and they are nevertheless well-defined in their action on the asymptotic fields. Since potential obstructions to BRST-exactness can only originate from symmetry breaking, and since the symmetry-breaking sector evidently causes no such issue, we conclude that there is no obstruction to the BRST-exactness of $Q^{a}$.

The asymptotic color charge operator $\widetilde{Q}^{a}$ can be obtained by subtracting the BRST-exact symmetry-breaking term from $Q^{a}$. It follows that $\widetilde{Q}^{a}$ is a BRST-exact operator\footnote{We remark that the combination $1+u$ disappears from the equation once all LSZ coefficients are written down explicitly.},
\begin{equation}\label{cchargeexact}
   \widetilde{Q}^{a}=\left\{Q_{B},g^{-1}\int d^{3}{\bf x}\, \left[D_{0}\cbar^{a}-\widetilde{Z}_{3}^{1/2}\,(1+u)\,\partial_{0}\gammabar^{a}\right]\right\}\ .
\end{equation}
Note that this expression only makes sense because, by virtue of our assumptions, we know that $\widetilde{Q}^{a}$ is a well-defined operator generating color transformations of the asymptotic fields and states.

As for the action of $\widetilde{Q}^{a}$ on physical states, recall that, in the full Hilbert space $\mc{H}$ of the theory, every state in the BRST cohomology is represented by a BRST-closed state $\ket{\text{phys}}$, $Q_{B}\ket{\text{phys}}=0$. By operating on $\ket{\text{phys}}$ with the BRST-exact asymptotic color charge $(\widetilde{Q}^{a})^{\txtin/\txtout}=\{Q_{B},(R^{a})^{\text{in}/\text{out}}\}$, one obtains
\begin{equation}
   (\widetilde{Q}^{a})^{\txtin/\txtout}\ket{\text{phys}}=Q_{B}\ket{\psi^{a}}
\end{equation}
for some state $\ket{\psi^{a}}$. It follows that $(\widetilde{Q}^{a})^{\txtin/\txtout}$ descend to well-defined and vanishing operators on the BRST cohomology:
\begin{equation}\label{confinement}
   (\widetilde{Q}^{a})^{\txtin/\txtout}\,\mc{H}_{\text{phys}}=0\ .
\end{equation}
Eq.~\eqref{confinement} expresses the confinement of color in QCD, in the sense of the absence of colored asymptotic states from the BRST cohomology. Because of this equation, all colored asymptotic states in $\mc{H}$ are organized into BRST quartets, that decouple from the physical dynamics of the theory by virtue of the quartet mechanism.

Explicitly, suppose that $\ket{\psi}$ is a simultaneous eigenstate of a Cartan subalgebra of $\{\widetilde{Q}^{a}\}_{a=1}^{8}$ with eigenvalues not all equal to zero. Without loss of generality, let the Cartan subalgebra be indexed by a subset of $a=1,\dots,8$. Then there exists an index $a$ such that, with $q^{a}\neq 0$,
\begin{equation}
   \ket{\psi}=(q^{a})^{-1}\,\widetilde{Q}^{a}\ket{\psi}=(q^{a})^{-1}\,\{Q_{B},R^{a}\}\ket{\psi}\ .
\end{equation}
Thus either $Q_{B}\ket{\psi}\neq 0$, in which case $\ket{\psi}$ is one of two parent states of a BRST quartet, or
\begin{equation}
   \ket{\psi}=Q_{B}\left((q^{a})^{-1}\,R^{a}\right)\,\ket{\psi}\ ,
\end{equation}
in which case it is one of two daughter states.

Despite the algebraic nature of equations such as~\eqref{deq834} and \eqref{cchargeexact}, the mechanism yielding Eq.~\eqref{confinement} is all but kinematic. When $g=0$, the dynamics of QCD in LCGs reduces to that of free gluons and ghosts; the operator $Q^{a}$ can be evaluated exactly, and one finds that $Q^{a}$ generates ordinary color transformations of all gluonic and ghost degrees of freedom, including the massless physical transverse gluon polarizations\footnote{For the non-interacting theory, this is most easily seen in the Feynman gauge $\xi=1$, in which $B^{a}=-\partial^{\mu}A_{\mu}^{a}$, the gluon field satisfies the wave equation $\partial^{2}A_{\mu}^{a}=0$, and one can avoid dealing with the dipolar nature of $A_{\mu}^{a}$.}. In the non-interacting theory, $Q^{a}$ is unbroken and it is not BRST exact; the physical transverse gluon polarizations transform nontrivially under $Q^{a}$, hence color is not confined.

When interactions are turned on and gluons acquire a mass, the charge $Q^{a}$ is dynamically rearranged so as to generate asymptotic residual global gauge transformations, instead of mere color transformations. The rearrangement is induced by the Maxwell-Yang-Mills field equations $\partial^{\nu}F_{\nu\mu}^{a}-\{Q_{B},D_{\mu}\cbar^{a}\}+gJ_{\mu}^{a}=0$, which enforce a relation between the global gauge current $J_{\mu}^{a}$, the field-strength tensor $F_{\mu\nu}^{a}$, and the Kugo-Ojima current $\mc{N}_{\mu}^{a}$. Dynamics renders $Q^{a}$ BRST-exact and spontaneously broken, as is expected for the generator of a gauge transformation.

Despite dynamical gluon mass generation and the spontaneous breaking of $Q^{a}$, under customary assumptions, the color symmetry of the spectrum of dynamically massive QCD is preserved. This is possible because the ABFP Schwinger mechanism does not rely on a Higgs field acquiring a nonzero vacuum expectation value to generate its Goldstone bosons. The Goldstone bosons of the ABFP Schwinger mechanism are instead generated by forming scalar bound states out of the color-symmetric dynamics of the theory itself. As a consequence, a well-defined asymptotic color charge operator $\widetilde{Q}^{a}$ still exists on the Hilbert space of the theory. Because of the vanishing of the chromoelectric flux at infinity induced by the gluon mass, $\widetilde{Q}^{a}$ vanishes on physical asymptotic states, leading to color confinement in the sense of the absence of colored asymptotic states from the BRST cohomology.

In the KO framework, the mechanism by which color is confined in the linear covariant gauges of QCD is the dynamical mechanism that generates a mass for gluons without breaking the color symmetry of the spectrum. As we saw in the previous sections, under customary assumptions, this mechanism is unique within the framework, and it is the Schwinger mechanism \cite{Sch62a,JJ73,Smi74,Cor81} in the formulation of Aguilar, Binosi, Ferreira, and Papavassiliou \cite{AP06,ABP08,AP09,AIMP12,IP13,ABP16,ABFP16,ABP17,BP18,Pap22,FP23,FP24,FP25}.

\section{Discussion}\label{sec:discussion}

In this paper, we have studied dynamical gluon mass generation, spontaneous symmetry breaking, and color confinement in the framework of the canonical BRST quantization of the linear covariant gauges of QCD, as formulated by T. Kugo and I. Ojima in \cite{KO78b,Oji78,KO78a,KO79b,KO79c,KO79d,KO79a,Kug78}. Our main objective has been to provide an answer to several open problems concerning the status of symmetries and the mechanism by which gluons acquire a mass within the framework, and to reassess the issue of confinement in light of those answers. In this section, we summarize our main results and we present our conclusions.\\

The first question we addressed in this paper is whether, in the KO framework, dynamical gluon mass generation actually implies the spontaneous breaking of color symmetry unless the KO criterion $1+u=0$ is satisfied. The traditional answer to this question has been in the positive: the so-called ``converse of the Higgs theorem'', proven by Kugo and Ojima in \cite{KO79d,KO79a}, purported to show that, if $1+u\neq 0$, massive gluons imply the spontaneous breaking of color symmetry.

While the core result of KO's theorem is correct -- in the KO framework, when gauge bosons acquire a mass, the global gauge charge $Q^{a}$ is indeed spontaneously broken --, in this paper we have shown that the theorem has largely been overinterpreted. In fact, as discussed in Sec.~\ref{sec:ssblcg}, in the LCGs of interacting Yang-Mills theories, the global gauge charge $Q^{a}$ is \textit{always} spontaneously broken, regardless of whether or not gauge bosons acquire a mass, and regardless of whether or not the underlying symmetry of the theory -- e.g., U(1) symmetry in QED, or SU(3) symmetry in QCD -- is preserved. It follows that the spontaneous breaking of $Q^{a}$, taken on its own, \textit{cannot} be regarded as evidence that the underlying symmetry of the theory -- in our case, color symmetry -- is violated: the ``converse of the Higgs theorem'' does not actually provide a converse to the Higgs theorem.

As for the condition $1+u=0$ which in the KO confinement scenario would have purportedly avoided the spontaneous breaking of the global gauge charge $Q^{a}$ in dynamically massive QCD, in Secs.~\ref{sec:framework} and \ref{sec:koanal} we showed that, in addition to resting on flawed arguments, the condition was derived from an incorrect equation. Indeed, throughout their works \cite{KO79d,KO79a,NO90,Kug95}, Kugo and Ojima systematically overlooked the importance of explicitly writing down the LSZ coefficients of the elementary BRST quartet -- see footnote \ref{fn:koerror} -- when evaluating the matrix element $\bra{\Omega}J_{\mu}^{a}(x)\ket{\chi^{b}(p)}$, where $J_{\mu}^{a}(x)$ is the global gauge (color) current, $\ket{\Omega}$ is the vacuum state of the theory, and $\ket{\chi^{a}(p)}$ is the one-particle state of the first parent $\chi^{a}$ of the elementary BRST quartet. In QCD, under customary assumptions -- see Sec.~\ref{sec:ssb} --, this matrix element is responsible for the spontaneous breaking and for the ill-definedness of the global gauge charge $Q^{a}$. As we showed in Secs.~\ref{sec:koanal}, \ref{sec:asympt} and \ref{sec:symbreak}, once the LSZ coefficients of the elementary BRST quartet are written down explicitly and the corresponding asymptotic states are normalized canonically, the asymptotic limit of the color current of dynamically massive QCD reads
\begin{equation}\label{deq357}
   J_{\mu}^{a}\to M(1+u)/g\,\partial_{\mu}\beta^{a}\ .
\end{equation}
In the above equation, $\beta^{a}$ is the massless asymptotic field associated with the Nakanishi-Lautrup field $B^{a}$, $g$ is the bare strong coupling constant, $M=Z_{\chi}^{-1/2}$ is both the inverse of the LSZ coefficient of $\chi^{a}$,
\begin{equation}\label{deq358}
   A_{\mu}^{a}\to Z_{\chi}^{1/2}\,\partial_{\mu}\chi^{a}+\cdots\ ,
\end{equation}
and the gluon mass as defined by the zero-momentum value of the transverse form factor $\Delta(p^{2})$ of the \textit{unrenormalized} gluon propagator,
\begin{equation}
   \Delta(0)=\frac{i}{M^{2}}\ ,\quad \Delta_{\mu\nu}(p)=\Delta(p^{2})\,t_{\mu\nu}(p)+\frac{-i\xi}{p^{2}}\,\ell_{\mu\nu}(p)\ ,
\end{equation}
and $1+u$ is both a ratio between LSZ coefficients \cite{KO79d,KO79a,Kug95},
\begin{equation}
   D_{\mu}\cbar^{a}\to \widetilde{Z}_{3}^{1/2}(1+u)\,\partial_{\mu}\gammabar^{a}\ ,
\end{equation}
$\widetilde{Z}_{3}^{1/2}$ being the LSZ coefficient of the antighost asymptotic field $\gammabar^{a}$, and, under customary KO assumptions, the residue $1+u=1+u(0)$ of the \textit{unrenormalized} KO/BV function \cite{Kug95,GHQ04,ABP09,BQ13,ABFP24,FP25}
\begin{equation}
   \left\langle T\left\{D_{\mu}c\,D_{\nu}\cbar\right\}\right\rangle_{\text{F.T.}}(p)=u(p^{2})\,t_{\mu\nu}(p)-\frac{p_{\mu}p_{\nu}}{p^{2}}
\end{equation}
at its massless pole $p^{2}=0$, multiplied by $-1$.

As is well known, and as we have confirmed independently in the present paper within the KO framework -- see Sec.~\ref{sec:ssb} -- the quantity $\widehat{M}=M(1+u)$ that appears on the RHS of Eq.~\eqref{deq357} is the gluon mass as defined by the zero-momentum value of the transverse form factor $\widehat{\Delta}(p^{2})$ of the background-background PT-BFM gluon propagator \cite{BP02b,BP09,CPB11,BQ13,IP13,FP25},
\begin{equation}
   \widehat{\Delta}(0)=\frac{i}{\widehat{M}^{2}}\ .
\end{equation}
It follows that, for massive gluons, neither $\widehat{M}=M(1+u)$ nor $M(1+u)/g$ are equal to zero: like in every Yang-Mills theory, in the LCGs of dynamically massive QCD as well the global gauge charge $Q^{a}$ is spontaneously broken,
\begin{equation}
   \bra{\Omega}J_{\mu}^{a}(x)\ket{\chi^{b}(p)}=i\widehat{M}/g\,\hat{p}_{\mu}\,\delta^{ab}\,e^{-i\hat{p}x}\neq 0\ ,
\end{equation}
where $\hat{p}=(|{\bf p}|,{\bf p})$.

Once multiplied by $M$ to form the nonvanishing PT-BFM gluon mass $\widehat{M}$, the constant $1+u$ disappears completely from the equations. Therefore, in dynamically massive QCD, $1+u=0$ is a sufficient condition neither for the absence of spontaneous symmetry breaking, nor for the well-definedness of the global gauge charge $Q^{a}$, nor for color confinement. The original KO equations \cite{KO79d,KO79a,NO90,Kug95} were, quite simply, missing a factor of $Z_{\chi}^{-1/2}=M=[-i\Delta(0)]^{-1/2}$ on their RHSs.

As for the meaning of the spontaneous breaking of the global gauge charge $Q^{a}$ in the LCGs of interacting Yang-Mills theories, in Secs.~\ref{sec:resgtgb} and \ref{sec:chargestruct} we showed that, far from being a symptom of the spontaneous breaking of the underlying symmetry of the theory -- e.g., U(1) in QED or SU(3) in QCD --, this phenomenon is a structural property of the LCGs that originates from the existence of a residual gauge symmetry. The existence of this symmetry is well known in the context of Abelian theories, where it already acts at the level of the local fields and of the local field equations. In non-Abelian theories, on the other hand, an analysis of the massless asymptotic component of the elementary fields -- see Sec.~\ref{sec:asympt} -- is required to show that the symmetry exists at the level of the asymptotic fields and of the asymptotic field equations.

The spontaneous breaking of the global gauge charge can be understood as follows. First of all, as we said -- see Secs.~\ref{sec:resgtgb} and \ref{sec:chargestruct} -- a residual gauge symmetry, generated by the operator
\begin{equation}\label{deq356}
   C[\Lambda]\propto \int d^{3}{\bf x}\ \left(\Lambda^{a}\dot{\beta}^{a}-\dot{\Lambda}^{a}\beta^{a}\right)\ ,
\end{equation}
where $\Lambda^{a}$ is a function such that $\partial^{2}\Lambda^{a}=0$, exists in the theory at the level of the asymptotic fields. The transformation acts on the first parent of the elementary BRST quartet $\chi^{a}$ as
\begin{equation}
   \delta\chi^{a}(x)\propto\Lambda^{a}(x)\ ,
\end{equation}
and, by Eq.~\eqref{deq358}, on the asymptotic gauge field $A_{\mu}^{a\,\text{as.}}$ as
\begin{equation}\label{deq739}
   \delta A_{\mu}^{a\,\text{as.}}(x)\propto\partial_{\mu}\Lambda^{a}(x)\ .
\end{equation}
Since it realizes a derivative transformation, the symmetry modifies the vacuum expectation value of $A_{\mu}^{a\,\text{as.}}(x)$,
\begin{equation}
   \langle \delta A_{\mu}^{a\,\text{as.}}\rangle \propto \partial_{\mu}\Lambda^{a}(x)\neq 0\ .
\end{equation}
It follows that the generator $C[\Lambda]$ is necessarily spontaneously broken, $C[\Lambda]\ket{\Omega}\neq 0$. The spontaneous breaking of $C[\Lambda]$ is, quite simply, a structural property of the transformation it generates.

The first parent of the elementary BRST quartet $\chi^{a}$ is the Goldstone boson associated with the spontaneous breaking of $C[\Lambda]$:
\begin{equation}
   \bra{\Omega}[C[\Lambda],\chi^{a}(x)]\ket{\Omega}\propto i\Lambda^{a}(x)\neq 0\ .
\end{equation}
When $\Lambda^{a}(x)$ tends to a constant, $\Lambda^{a}(x)\to \lambda^{a}$, the transformation of $A_{\mu}^{a\,\text{as.}}$ becomes trivial, but that of $\chi^{a}$ does not:
\begin{equation}\label{deq740}
   [C[\lambda],\chi^{a}]\propto i\lambda^{a}\ .
\end{equation}
It follows that $\chi^{a}$ is a Goldstone boson also for the constant-parameter limit of the transformation.

Second, in the LCGs of interacting Yang-Mills theories, the global gauge current $J_{\mu}^{a}$ always receives a contribution from $C[\lambda]$. Indeed, assuming that the matter fields are minimally coupled to the gauge field, and using the Maxwell-Yang-Mills equations, the global gauge current $J_{\mu}^{a}$ can always be expressed as \cite{Oji78,KO79a,NO90}
\begin{align}\label{deq346}
   gJ_{\mu}^{a}=\partial^{\nu}F_{\mu\nu}^{a}+\{Q_{B},D_{\mu}\cbar^{a}\}\ .
\end{align}
The two terms in the above equation separately contribute to the asymptotic limit of $J_{\mu}^{a}$ with a term proportional to $\partial_{\mu}\beta^{a}$,
\begin{equation}
   \partial^{\nu}F_{\mu\nu}^{a}\to -w\,\partial_{\mu}\beta^{a}\ ,\quad \{Q_{B},D_{\mu}\cbar^{a}\}\to Z_{\chi}^{-1/2}(1+u)\,\partial_{\mu}\beta^{a}\ ,
\end{equation}
where $w=0$ if gauge bosons acquire a mass -- see Secs.~\ref{sec:ssblcg} and \ref{sec:goldbound}. In no known theory these contributions sum to zero.

The operator $\partial_{\mu}\beta^{a}$ is but the (asymptotic) current corresponding to the charge $C[\lambda]$ that generates the constant-parameter transformations in Eq.~\eqref{deq740}:
\begin{equation}
   C[\lambda]\propto \int d^{3}{\bf x}\ \lambda^{a}\dot{\beta}^{a}\ .
\end{equation}
Therefore, the global gauge charge $Q^{a}=\int d^{3}{\bf x}\ J_{0}^{a}$ acts on the first parent of the elementary BRST quartet as\footnote{If the underlying symmetry of the theory is actually broken, the RHS of the next equation might not be diagonal in the algebra indices.}
\begin{equation}
   \bra{\Omega}[Q^{a},\chi^{b}(x)]\ket{\Omega}\propto i\delta^{ab}\neq 0\ .
\end{equation}
In particular, in the LCGs of all interacting Yang-Mills theories, the global gauge charge $Q^{a}$ is spontaneously broken, and the first parent of the elementary BRST quartet is its associated Goldstone boson.

Being a structural property of the LCGs of interacting Yang-Mills theories, originating from a residual gauge symmetry, the spontaneous breaking of $Q^{a}$, on its own, does not imply the spontaneous breaking of the underlying symmetry of the theory. Nonetheless, it does signal that, in the LCGs, the global gauge charge that appears in the MYME does not merely generate the expected symmetries of the asymptotic fields -- e.g., U(1) transformations in QED, or SU(3) transformations in QCD: at the very least, $Q^{a}$ also acts by a constant shift of the first parent of the elementary BRST quartet.

Since the global gauge charge of all interacting Yang-Mills theories quantized in the LCGs within the KO framework is spontaneously broken, the actual spontaneous breaking of the underlying symmetry of the theory, if any, must be characterized in terms of some other property. To provide such a characterization, in Appendix~\ref{app:softtheorem}, we formulated a soft theorem for dynamically massive QCD that relates the vacuum expectation value of the color variation of an arbitrary time-ordered product of local operators to the soft limit of the matrix element of the same product between the one-particle state $\ket{\beta^{a}(p)}$ and the vacuum,
\begin{equation}
   \bra{\Omega}[Q^{a},T\{\cdots\}]\ket{\Omega}=-i\widehat{M}/g\,\lim_{{\bf p}\to {\bf 0}}\, \bra{\beta^{a}(p)}T\{\cdots\}\ket{\Omega}\ .
\end{equation}
In its strongest form, color symmetry is preserved if and only if all the Green's functions of the theory transform covariantly under color transformations -- that is, if and only if the LHS of the above equation vanishes for every time ordered product of local operators. It follows that color symmetry is preserved in its strongest form if and only if the soft limit of the matrix elements of $\ket{\beta^{a}(p)}$ vanishes,
\begin{equation}\label{deq742}
   \lim_{{\bf p}\to {\bf 0}}\, \bra{\beta^{a}(p)}T\{\cdots\}\ket{\Omega}=0\ .
\end{equation}
In Appendix~\ref{app:softtheorem}, we showed that Eq.~\eqref{deq742} is equivalent to the condition
\begin{equation}\label{deq743}
   \lim_{p\to 0}\,p^{\mu}\Gamma_{\mu}^{a}(p,\dots)=0\ ,
\end{equation}
where $\Gamma_{\mu}^{a}(p,\dots)$ is an arbitrary amputated Green's function with an external gluon leg $A_{\mu}^{a}$ of momentum $p$. Eq.~\eqref{deq743} thus provides a characterization of the preservation of color symmetry in terms of the vanishing of the soft limit of the Slavnov-Taylor identities of the theory.

In light of the findings on dynamical gluon mass generation which we are going to discuss shortly, in Appendix~\ref{app:softtheorem} we observed that the violation of color symmetry expressed by a potential failure of Eqs.~\eqref{deq742} and \eqref{deq743} is not a necessary condition for gluons to acquire a mass: neither the matrix elements of $\beta^{a}$ nor the Slavnov-Taylor identities of the theory actively participate in the mechanism responsible for gluon mass generation. It follows that, even in its strongest form, the preservation of color symmetry is in no contradiction with the massiveness of gluons.

In summary, our analysis of spontaneous symmetry breaking within the KO quantization framework led us to two main conclusions. First of all, the value of the constant $1+u$ is irrelevant to whether or not either the global gauge charge or the underlying symmetry of the theory are spontaneously broken. Second, there is no evidence that, in the linear covariant gauges of QCD, dynamical gluon mass generation implies the spontaneous breaking of color symmetry. These conclusions are in sharp contrast with the traditional result provided by KO's so-called ``converse of the Higgs theorem'' -- or rather, with the way the theorem has been traditionally interpreted.\\

The second question we addressed concerns the mechanism by which gluons acquire a mass in the KO framework. Remarkably, despite the central role played by the gluon mass in the traditional formulation of the KO confinement scenario, this question has remained unanswered since the framework's inception, nearly fifty years ago. It is quite likely that the misinterpretation of the spontaneous breaking of the global gauge charge, together with the formulation of the invalid criterion $1+u=0$, hindered any progress on this front, shifting the focus from understanding gluon mass generation to avoiding spontaneous symmetry breaking.

As discussed in Sec.~\ref{sec:resgtgb}, in the KO framework, scalar degrees of freedom must exist in the spectrum of the theory in order for gauge bosons to acquire a mass. In fact, when gauge bosons become massive, their longitudinal polarization either becomes massive or it disappears from the spectrum together with the transverse polarizations, leaving a gap in the elementary BRST quartet which must be filled in by a scalar. For the Stueckelberg, Schwinger, and Higgs models, all of which admit a formulation in the KO framework, these degrees of freedom are provided, respectively, by the Stueckelberg field, by a fermion-antifermion state arising from the gauge-invariant point splitting of the electric current, and by the unphysical degrees of freedom of the Higgs field -- see Appendices~\ref{app:stschw} and \ref{app:higgsmod}.

In Sec.~\ref{sec:asympt}, by an analysis of the massless component of the asymptotic spectrum of the elementary Yang-Mills fields carried out under customary KO assumptions, we confirmed that unphysical massless scalars $\chi^{a}$ exist in the spectrum of dynamically massive QCD as well. As expected for a massive theory \cite{Nak75}, the scalars are dipole states in all linear covariant gauges except for the Landau gauge ($\xi=0$), and they form the unphysical elementary BRST quartet together with the massless asymptotic fields $\beta^{a}$, $\gamma^{a}$, and $\gammabar^{a}$ of the Nakanishi-Lautrup and Faddeev-Popov ghost and antighost fields $B^{a}$, $c^{a}$, and $\cbar^{a}$. Within the quartet, $\chi^{a}$ is its first parent.

In Sec.~\ref{sec:goldbound}, we sought to the determine the identity of $\chi^{a}$ in dynamically massive QCD. To this aim, we studied the dynamics of $\chi^{a}$ as encoded in the matrix element of the Maxwell-Yang-Mills equations between the vacuum and the one-particle state $\ket{\chi^{a}(p)}$, and we established that the MYME enforce a non-vanishing (unrenormalized) sum rule between the Bethe-Salpeter amplitudes of $\chi^{a}$ in the two-gluon, three-gluon, ghost-antighost, and -- in the presence of quarks -- quark-antiquark channels. Modulo momenta and unitary prefactors, the RHS of the sum rule, compactly expressed as
\begin{equation}\label{deq444}
   \bra{\Omega}K_{\mu}^{a}(x)\ket{\chi^{b}(p)}=iM\hat{p}_{\mu}\,\delta^{ab}\,e^{-i\hat{p}x}\ ,
\end{equation}
where $K_{\mu}^{a}$ is the kernel operator that appears in the SDE of the gluon polarization,
\begin{equation}
   \Pi_{\mu\nu}^{ab}(p)=i\left[\langle T\{KA\}\rangle_{\text{F.T.}}(p)\cdot \Delta^{-1}(p)\right]_{\mu\nu}^{ab}\ ,
\end{equation}
is none other than the gluon mass $M$.

The quantity $I_{\mu}(p)$ that, modulo prefactors and sign conventions, appears on the LHS of Eq.~\eqref{deq444},
\begin{equation}\label{deq445}
   I_{\mu}(-\hat{p})\,\delta^{ab}\,e^{-i\hat{p}x}=ig^{-1}\bra{\Omega}K_{\mu}^{a}(x)\ket{\chi^{b}(p)}\ ,
\end{equation}
is a well-known object in the literature on the Schwinger mechanism acting in QCD, where it goes by the name of gluon-scalar transition amplitude \cite{JJ73,EF74,Smi74,AIMP12,IP13,FP25}. Its nonvanishing, guaranteed by Eq.~\eqref{deq444}, is one of two necessary conditions for the Schwinger mechanism to be active.

The second condition for the Schwinger mechanism to be active in QCD is that, in the limit in which an external gluon leg goes on the massless shell, the elementary dressed QCD vertices develop massless longitudinal poles -- so-called vertex Schwinger poles -- proportional both to the gluon-scalar transition amplitude and to the Bethe-Salpeter vertices of the scalar \cite{AP06,ABP08,AP09,AIMP12,IP13,ABP16,ABFP16,ABP17,BP18,Pap22,FP23,FP24,FP25}. In Appendix~\ref{app:schwpoles}, an analysis of the massless one-particle spectrum of the elementary QCD vertices showed that the first parent $\chi^{a}$ of the elementary BRST quartet of dynamically massive QCD does indeed generate these poles, with the required residues. By plugging the singular $p^{2}\to 0$ limit of the elementary QCD vertices into the Schwinger-Dyson equation of the gluon polarization, we were then able to reproduce the classic proof \cite{IP13,FP25} that vertex Schwinger poles, in our case generated by $\chi^{a}$, trigger the dynamical generation of a gluon mass via the Schwinger mechanism.

With the exception of vertex singularities originating from the STIs of the theory which do not contribute to the gluon polarization\footnote{The existence of these singularities does not contradict Eqs.~\eqref{deq742} and \eqref{deq743}. It just reflects the on-shell relation $\lim_{p^{2}\to 0}\,p^{\mu}\Gamma_{\mu}^{a}(p,\dots)\propto\bra{\beta^{a}(p)}T\{\cdots\}\ket{\Omega}$, which in turn implies the equivalence of the vanishing of the two soft limits.} -- hence to the gluon mass --, and which in practical calculations can be absorbed into a redefinition of the Bethe-Salpeter vertices of $\chi^{a}$, the equations that govern the Schwinger mechanism in the KO framework are identical to those formulated by Aguilar, Binosi, Ferreira and Papavassiliou within their massless bound-state formalism \cite{AP09,AIMP12,IP13,ABP16,ABP17,BP18,Pap22,FP23,FP24,FP25}. Specifically, in the KO framework, the first parent $\chi^{a}$ of the elementary BRST quartet of dynamically massive QCD is a massless colored scalar with nonvanishing Bethe-Salpeter amplitudes in at least one -- reasonably, all -- of the channels of the ABFP bound state, that generates the same vertex Schwinger poles as the ABFP bound state, and that triggers the Schwinger mechanism in the ABFP formulation just like the ABFP bound state.

Once fully coupled, untruncated Bethe-Salpeter equations are written down for the vertices relevant to the Schwinger mechanism -- namely, the three-gluon, four-gluon, ghost-antighost-gluon, and quark-antiquark-gluon vertices --, the one-particle state of the first parent $\chi^{a}$ of the elementary BRST quartet of dynamically massive QCD provides a solution to those equations. We thus conclude that, in the KO framework, the first parent $\chi^{a}$ of the elementary BRST quartet of dynamically massive QCD is the ABFP bound state, and that the mechanism by which gluons acquire a mass within the framework is the ABFP Schwinger mechanism.

We remark that, all throughout the present work, we \textit{assumed} that gluons dynamically acquire a mass. Therefore, our results \textit{do not} establish that gluons are massive in the KO framework. What they establish is that, within the framework, under customary assumptions, the mechanism of dynamical gluon mass generation is unique, and it coincides with the ABFP Schwinger mechanism.

As a corollary of our results, in Sec.~\ref{sec:goldbound} we observed that, once the necessary matter terms are added to the kernel operator $K_{\mu}^{a}$ on the RHS of Eq.~\eqref{deq445}, the derivations we provided in the context of dynamically massive QCD remain valid in the LCGs of all Yang-Mills theories in which gauge bosons acquire a mass under customary KO assumptions. These include the Stueckelberg model and the Higgs models, which we reviewed in Appendices~\ref{app:stschw} and \ref{app:higgsmod}. Specifically, in these theories, the first parent of the elementary BRST quartet generates at the very least a Schwinger pole proportional to the product of two gauge-boson-scalar transition amplitudes in the gauge-boson polarization,
\begin{align}
   \Pi_{\mu\nu}^{ab}(p)\to \frac{i\delta^{ab}}{p^{2}+i\epsilon}\ g^{2}I_{\mu}(-\hat{p})I_{\nu}(\hat{p})\qquad(p^{2}\to 0)\ ,
\end{align}
thus triggering the Schwinger mechanism responsible for the gauge boson's mass. It follows that, within the KO framework, under customary assumptions, all gauge-boson mass generation is Schwinger mass generation, triggered by the first parent of the elementary BRST quartet. This observation makes it clear that a unified theory of gauge-boson mass generation in the LCGs of Yang-Mills theory can be formulated in the KO framework.\\

The third question we addressed concerns the Goldstone nature of the ABFP bound state. In the literature on the Schwinger mechanism, it is often reported that the ABFP bound state acts like a dynamical composite Goldstone scalar, in that it displays all its typical characteristics, albeit not being associated with the spontaneous breaking of any global symmetry \cite{Cor81,Pap11,AIMP12,Cor15,FA16}. In this paper, we showed that the last statement is incorrect: the ABFP bound state is the \textit{literal} Goldstone boson associated with the spontaneous breaking of the global gauge (color) charge $Q^{a}$. To see this, it is sufficient to realize that the gluon-scalar transition amplitude $\widetilde{I}_{\mu}(\hat{p})$ evaluated in the PT-BFM scheme \cite{IP13} is but the matrix element of the global gauge (color) current $J_{\mu}^{a}$ between the vacuum and the one-particle state $\ket{\chi^{a}(p)}$ of the ABFP bound state -- see Sec.~\ref{sec:ssb} --,
\begin{align}\label{deq446}
   \widetilde{I}_{\mu}(-\hat{p})\,\delta^{ab}=-\widehat{M}/g\,\hat{p}_{\mu}\,\delta^{ab}=i\bra{\Omega}J^{a}_{\mu}(0)\ket{\chi^{b}(p)}\ .
\end{align}
Since $\widetilde{I}_{\mu}(p)\neq 0$ in order for the Schwinger mechanism to be active, the RHS of Eq.~\eqref{deq446} is different from zero, and the Goldstone nature of the ABFP bound state follows. Note that Eq.~\eqref{deq446} is framework-independent: when written in the equivalent form
\begin{align}
   \widetilde{I}_{\mu}(-\hat{p})\,\delta^{ab}=ig^{-1}\bra{\Omega}\widetilde{K}^{a}_{\mu}(0)\ket{\chi^{b}(p)}\ ,
\end{align}
where $\widetilde{K}_{\mu}^{a}$ is the kernel operator that appears in the Schwinger-Dyson equation of the background-quantum PT-BFM gluon polarization,
\begin{equation}\label{deq447}
   \widetilde{\Pi}_{\mu\nu}^{ab}(p)=i\left[\langle T\{\widetilde{K}A\}\rangle_{\text{F.T.}}(p)\cdot \Delta^{-1}(p)\right]_{\mu\nu}^{ab}\ ,
\end{equation}
that the equation holds can be checked without any reference to the KO framework, by employing standard operatorial definitions for the Bethe-Salpeter amplitudes and vertices of the bound state -- see Sec.~\ref{sec:goldbound}.

Still, it is true that the ABFP bound state is not the Goldstone boson associated with the spontaneous breaking of \textit{color} symmetry. Indeed, as discussed above, the spontaneous breaking of the global gauge charge $Q^{a}$ does not imply that color symmetry is broken, and there is no evidence that the Schwinger mechanism breaks it in QCD. In the present work, we did not go any further in the formulation of the mechanism than to show that the first parent of the elementary BRST quartet $\chi^{a}$ generates the gluon mass in the KO framework, but the topic of symmetry preservation is, of course, not at all new to studies of the Schwinger mechanism. In the context of QCD \cite{Smi74,Cor81,AP06,ABP08,AP09,AIMP12,IP13,ABP16,ABFP16,ABP17,BP18,Pap22,FP23,FP24,FP25}, the Schwinger mechanism has been traditionally formulated in such a way as to explicitly preserve color symmetry in its strongest form -- namely, in the form of the covariance of all Green's functions under color transformations, and there is even some evidence that conditions similar to those in Eqs.~\eqref{deq742} and \eqref{deq743} are already being implemented for the declared purpose of preserving color symmetry\footnote{See e.g. the discussion below and around Eq.~(5.23) in \cite{FP25}. There, vertices having ``no poles in $q=0$'' is equivalent to amplitudes vanishing in the $q\to 0$ limit, which is in fact the behavior displayed by Eq.~(5.23) of \cite{FP25}, and a condition similar to the vanishing of the soft limit in our Eq.~\eqref{deq742}.}.

How color symmetry can be preserved in the presence of the spontaneous breaking of the global gauge charge $Q^{a}$ is explained by the simple observation -- see Sec.~\ref{sec:ssblcg} and the discussion above -- that, in the LCGs of interacting Yang-Mills theories, the global gauge charge always acts by a shift of the first parent of the elementary BRST quartet, regardless of whether or not a gauge-boson mass is generated, and regardless of whether or not the underlying symmetry of the theory is preserved. This already happens in theories like four-dimensional, massless QED, in which the preservation of U(1) symmetry is uncontroversial, yet the global gauge (electric) charge $Q$ is spontaneously broken, and it acts by a shift of the longitudinal photon polarization. The ABFP bound state, in its role of first parent of the elementary BRST quartet of dynamically massive QCD, is the Goldstone boson associated with the spontaneous breaking of \textit{this} shift symmetry. As we saw above, and as we will discuss in more detail below, this symmetry, in turn, is the expression of the derivative term of a residual gauge symmetry, acting at the level of the asymptotic fields.

We remark that the ABFP bound state, as member of a BRST quartet, is an \textit{unphysical} Goldstone boson: it belongs to the unphysical sector of the spectrum of the linear covariant gauges of dynamically massive QCD. In this respect, its status is the exact same as that of the quanta of the unphysical components of the Higgs field in Higgs theories, which are not observable as physical particles.

As a matter of fact, being activated by the eating -- see Sec.~\ref{sec:resgtgb} -- of the scalar Goldstone bosons associated with the spontaneous breaking of the global gauge charge, the Schwinger mechanism of dynamically massive QCD is but a Higgs mechanism in which the Goldstone bosons are generated nonperturbatively, rather than being the quanta of elementary fields present in the Lagrangian. It is understood that the Higgs-Schwinger mechanism acting in dynamically massive QCD does not rely on colored operators acquiring a nonzero vacuum expectation value, hence on color symmetry to be spontaneously broken.

Eqs.~\eqref{deq446} - \eqref{deq447}, together with the well known relation \cite{IP13}
\begin{equation}\label{deq448}
   \widetilde{I}_{\mu}(\hat{p})=(1+u)\,I_{\mu}(\hat{p})\ ,
\end{equation}
which we have rederived independently in Sec.~\ref{sec:ssb} from the original KO Goldstone equation, reveal the true meaning of the KO constant $1+u$. Far from having anything to do with spontaneous symmetry breaking or color confinement, the role of $1+u$ both in the KO framework and in the ABFP Schwinger mechanism is that of converting the gluon-scalar transition amplitude from the linear covariant gauges to the background-field gauges that define the PT-BFM\footnote{In Sec.~\ref{sec:ssb} we have explained how objects defined in the PT-BFM's background-field gauges can exist in the Hilbert space of the LCGs.}. The connection between $1+u$ and the background-field gauges was not unknown to Kugo, who in \cite{Kug95} reported to have first discovered the renormalization relation
\begin{equation}\label{deq824}
   Z_{3}^{-1/2}g^{-1}(1+u)=g_{R}^{-1}\ ,
\end{equation}
where $Z_{3}^{1/2}\propto M^{-1}$ is a renormalization constant for the gluon propagator and $g_{R}$ is a renormalized coupling, using background-field methods. Since however the factor of $M=Z_{\chi}^{-1/2}$ was missing from the original KO Goldstone equation, it could never be realized that, in \cite{KO79d,KO79a}, $1+u$ is simply realizing the most elementary Background-Quantum Identity \cite{GHS01,BP02b,BP09,BQ13,IP13} of QCD.

The PT-BFM gluon-scalar transition amplitude $\widetilde{I}_{\mu}(\hat{p})$ sets the RGI scale $\widehat{M}/g$ for dynamical gluon mass generation. By Eq.~\eqref{deq446}, this scale plays the same role in dynamically massive QCD as the pion decay constant in low-energy strong interactions and as the vacuum expectation value of the Higgs field -- equivalently, as the Fermi constant -- in Higgs models. While for arbitrary values of momentum the background-background PT-BFM gluon propagator $\widehat{\Delta}(p^{2})$ is well-known to be gauge-parameter-dependent \cite{CPB11}, it would be interesting to explore whether symmetry protects the zero-momentum limit $\widehat{\Delta}(0)$ in such a way that the combination $\widehat{M}/g$ is gauge-independent. If this turned out to be the case, then $\widehat{M}/g$ would be the obvious candidate for the fundamental nonperturbative scale of the strong interactions, whose definition -- at variance with $\Lambda_{{\text{QCD}}}$ -- relies neither on the perturbative running of the strong coupling constant nor on a choice of renormalization scheme.

Eqs.~\eqref{deq446} - \eqref{deq448} cement the connection between spontaneous symmetry breaking, the Schwinger mechanism, the PT-BFM, and dynamical gluon mass generation in the linear covariant gauges of QCD. Their derivation, which involves completing the color current $J_{\mu}^{a}$ in the kernel operator $K_{\mu}^{a}$ to form $\widetilde{K}_{\mu}^{a}$ -- see Sec.~\ref{sec:ssb} -- suggests that the peculiar properties of the PT-BFM Green's functions may be explained by the fact that these functions dynamically couple directly to the color current. While we have established this is the case for the gluon-scalar transition amplitude $\widetilde{I}_{\mu}(\hat{p})$ and for the background-quantum gluon polarization $\widetilde{\Pi}_{\mu\nu}^{ab}(p)$, whether this remains true for higher-point Green's functions is yet to be determined.\\

The last question we addressed in this paper is whether a well-defined color charge operator can be uniquely identified in dynamically massive QCD quantized in the KO framework and, if it can, whether the operator realizes color confinement in the sense of the absence of colored asymptotic states from the BRST cohomology. Since the Schwinger mechanism does not rely on the spontaneous breaking of color symmetry to generate a mass for gluons, in Sec.~\ref{sec:chargestruct} -- see also Sec.~\ref{sec:resgtgb} -- we assumed that color symmetry is preserved in a form which is suitable to answer this question. Namely, we assumed that the global gauge charge $Q^{a}$ acts on the asymptotic limit of the local fields of the theory by color transformations,
\begin{equation}\label{deq402}
   [Q^{a},\phi_{I}^{\text{as.}}]=T_{IJ}^{a}\,\phi_{J}^{\text{as.}}\ ,
\end{equation}
where, asymptotically, if the $T_{IJ}^{a}$'s realize nontrivial representations of the color algebra, $\phi_{I}\to Z_{\phi}^{1/2}\phi_{I}^{\text{as.}}$ for some local operator $\phi_{I}$ transforming irreducibly under $Q^{a}$,
\begin{equation}
   [Q^{a},\phi_{I}]=T_{IJ}^{a}\,\phi_{J}\ .
\end{equation}
This assumption is not satisfied in Higgs models, in which the nonvanishing vacuum expectation value of the Higgs field breaks color symmetry, thus preventing the asymptotic fields from having definite transformation properties under the global gauge charge. On the other hand, it would be automatically satisfied in color-symmetry-preserving QCD if $Q^{a}$ were not spontaneously broken.

Crucially, the assumption in Eq.~\eqref{deq402} does not contradict the spontaneous breaking of $Q^{a}$. In fact, if Eq.~\eqref{deq402} holds, then $Q^{a}$ generates color transformations of the asymptotic gluon field $A_{\mu}^{a\,\text{as.}}$,
\begin{equation}
   [Q^{a},A_{\mu}^{b\,\text{as.}}]=if^{abc}A_{\mu}^{c\,\text{as.}}\ ;
\end{equation}
since in dynamically massive QCD the ABFP bound state enters the asymptotic limit of $A_{\mu}^{a}$ as a derivative, $A_{\mu}^{a}\to Z_{\chi}^{1/2}\partial_{\mu}\chi^{a}+\cdots$, $Q^{a}$ generates the transformation
\begin{equation}
   [Q^{a},\partial_{\mu}\chi^{b}]=if^{abc}\,\partial_{\mu}\chi^{c}\ ,
\end{equation}
which, when integrated, yields
\begin{equation}\label{deq065}
   [Q^{a},\chi^{b}]=i\widehat{M}/g\,\delta^{ab}+if^{abc}\,\chi^{c}\ .
\end{equation}
The constant on the RHS of Eq.~\eqref{deq065} reflects the spontaneous breaking of $Q^{a}$.

In Sec.~\ref{sec:chargestruct}, we wrote down the unique expression\footnote{Under the assumption of asymptotic completeness, unicity is a trivial consequence of the fact that $Q^{a}$ generates specific -- and, in principle, well-known -- transformations of the asymptotic fields.} of $Q^{a}$ in terms of the asymptotic fields of dynamically massive QCD,
\begin{widetext}
   \begin{equation}\label{deq066}
      Q^{a}=g^{-1}\widehat{M}\int d^{3}{\bf x}\ \dot{\beta}^{a}+f^{abc}\int d^{3}{\bf x}\ \left(\dot{\chi}^{b}\beta^{c}-\chi^{b}\dot{\beta}^{c}-\dot{\beta}^{b}\beta^{c}+i\,\dot{\gammabar}^{b}\gamma^{c}-i\,\gammabar^{b}\dot{\gamma^{c}}\right)+\cdots\ ,
   \end{equation}
\end{widetext}
where the ellipsis denotes contributions from asymptotic fields that do not belong to the elementary BRST quartet, and that -- under customary assumptions -- do not participate in spontaneous symmetry breaking. Due to the first term, linear in $\dot{\beta}^{a}$ and responsible for the shift of $\chi^{a}$, $Q^{a}$ is spontaneously broken and ill-defined as an operator acting on the Hilbert space of the theory. Nevertheless, it is well-defined as an operator acting on the asymptotic fields via commutators. The symmetry-breaking term in Eq.~\eqref{deq066} is BRST-exact, $\dot{\beta}^{a}\propto \{Q_{B},\dot{\gammabar}^{a}\}$, in agreement with the fact that Eqs.~\eqref{deq742} and \eqref{deq743} are always satisfied for BRST-invariant products of local operators -- see Appendix~\ref{app:softtheorem}.

We then noted that $Q^{a}$ can be obtained from a two-parameter operator $\mc{Q}[\Lambda,\lambda]$ which is well-defined on the Hilbert space of the theory. $\mc{Q}[\Lambda,\lambda]$ generates the asymptotic residual gauge transformations
\begin{equation}
   [\mc{Q}[\Lambda,\lambda],A_{\mu}^{a\,\text{as.}}]=i\left(g_{R}^{-1}\,\partial_{\mu}\Lambda^{a}+f^{abc}A_{\mu}^{b\,\text{as.}}\lambda^{c}\right)\ ,
\end{equation}
where the functions $\Lambda^{a}$ are such that $\partial^{2}\Lambda^{a}=0$, the $\lambda^{a}$'s are constants, and $g_{R}$ is the coupling defined in Eq.~\eqref{deq824}. $Q^{a}$ is recovered from $\mc{Q}[\Lambda,\lambda]$ as $\mc{Q}[\lambda,\lambda]=\lambda^{a}Q^{a}$.

The parameters of $\mc{Q}[\Lambda,\lambda]$ reflect the actual symmetries of the asymptotic field equations: $\mc{Q}[\Lambda,\lambda]$ can be split as
\begin{equation}\label{deq067}
   \mc{Q}[\Lambda,\lambda]=C[\Lambda]+\widetilde{Q}[\lambda]\ ,
\end{equation}
where $C[\Lambda]$ is the operator in Eq.~\eqref{deq356}, and the asymptotic field equations are separately symmetric under the transformations generated by $C[\Lambda]$ and by $\widetilde{Q}[\lambda]$. As we said, the vacuum is not invariant under $C[\Lambda]$ due to the fact that $C[\Lambda]$ modifies the vacuum expectation value of $\chi^{a}$ and, for nonconstant $\Lambda^{a}$'s, of $A_{\mu}^{a\,\text{as.}}$.

As far as the elementary BRST quartet is concerned, the BRST charge $Q_{B}$ is but the unbroken realization of the operator $C[\Lambda]$ in Eq.~\eqref{deq067}. Indeed, as we saw in Sec.~\ref{sec:chargestruct},
\begin{equation}
   Q_{B}\propto -C[\gamma]+\cdots\ ,
\end{equation}
where $\gamma$ is the asymptotic ghost field, and the ellipsis denotes contributions due to potential asymptotic fields not belonging to the elementary BRST quartet. It follows that BRST symmetry quite literally replaces the spontaneously broken derivative symmetry generated by $C[\Lambda]$ by providing an unbroken implementation of the same transformation.

The charge $\widetilde{Q}^{a}$ such that $\widetilde{Q}[\lambda]=\lambda^{a}\widetilde{Q}^{a}$ in Eq.~\eqref{deq067} is not spontaneously broken,
\begin{equation}
   \bra{\Omega}[\widetilde{Q}^{a},\phi_{I}^{\text{as.}}]\ket{\Omega}=0\qquad \forall \phi_{I}^{\text{as.}}\ ,
\end{equation}
it is well-defined, it annihilates the vacuum,
\begin{equation}
   \widetilde{Q}^{a}\ket{\Omega}=0\ ,
\end{equation}
it generates the color transformations of all the asymptotic fields of the theory,
\begin{equation}
   [\widetilde{Q}^{a},\phi_{I}^{\text{as.}}]=T_{IJ}^{a}\,\phi_{J}^{\text{as.}}\ ,
\end{equation}
hence it generates the color algebra,
\begin{equation}
   [\widetilde{Q}^{a},\widetilde{Q}^{b}]=if^{abc}\,\widetilde{Q}^{c}\ ,
\end{equation}
and it commutes with the BRST algebra,
\begin{equation}
   [\widetilde{Q}^{a},Q_{B}]=0\ ,\qquad [\widetilde{Q}^{a},Q_{C}]=0\ .
\end{equation}
Therefore, the operator $\widetilde{Q}^{a}$ given by
\begin{widetext}
   \begin{equation}\label{deq068}
      \widetilde{Q}^{a}=f^{abc}\int d^{3}{\bf x}\ \left(\dot{\chi}^{b}\beta^{c}-\chi^{b}\dot{\beta}^{c}-\dot{\beta}^{b}\beta^{c}+i\,\dot{\gammabar}^{b}\gamma^{c}-i\,\gammabar^{b}\dot{\gamma^{c}}\right)+\cdots\ ,
   \end{equation}
\end{widetext}
where the ellipsis is uniquely determined by the color transformations of potential asymptotic fields not belonging to the elementary BRST quartet, is the sought-after color charge operator. As shown in Sec.~\ref{sec:chargestruct}, the explicit terms in Eq.~\eqref{deq068} are BRST-exact. 

Being built from the transformation laws of the asymptotic fields, $\widetilde{Q}^{a}$ is an \textit{asymptotic} color charge operator: the action of $(\widetilde{Q}^{a})^{\text{in}/\text{out}}$ -- which may be distinct operators -- is only known on the in/out states. Under the assumption of asymptotic completeness, this is in principle sufficient to know the action of $(\widetilde{Q}^{a})^{\text{in}/\text{out}}$ on the whole Hilbert space of the theory\footnote{In practice, one needs to know how each state in the Hilbert space is expressed in terms of the in/out asymptotic states.}. In Sec.~\ref{sec:chargestruct}, we observed that the mere existence of $\widetilde{Q}^{a}$ does not imply that color is conserved in scattering processes, nor that color symmetry is preserved at the level of Green's functions. For that to be true, the conditions provided by the soft theorem in Appendix~\ref{app:softtheorem} must still be satisfied. Rather, the existence of $\widetilde{Q}^{a}$ reflects the organization of the asymptotic (in/out) spectrum of dynamically massive QCD into representations of color SU(3), to which the Schwinger mechanism poses no obstruction.

Since the chromoelectric flux at spatial infinity vanishes in dynamically massive QCD quantized in the KO framework \cite{KO79d,KO79a,NO90},
\begin{equation}
   \oint_{R\to{\infty}}dS\ \hat{n}\cdot\vec{\mathcal{E}}^{a}=0\ ,
\end{equation}
by Eq.~\eqref{deq346}, the asymptotic color charge $\widetilde{Q}^{a}$ can be expressed in the BRST-exact form
\begin{equation}
   \widetilde{Q}^{a}=\left\{Q_{B},g^{-1}\int d^{3}{\bf x}\, \left[D_{0}\cbar^{a}-\widetilde{Z}_{3}^{1/2}\,(1+u)\,\partial_{0}\gammabar^{a}\right]\right\}\ .
\end{equation}
It follows that the asymptotic charge operators $(\widetilde{Q}^{a})^{\text{in}/\text{out}}$ descend to well-defined and vanishing operators on the BRST cohomology,
\begin{equation}
   (\widetilde{Q}^{a})^{\txtin/\txtout}\,\mc{H}_{\text{phys}}=0\ ,
\end{equation}
thus realizing color confinement in the sense of the absence of colored asymptotic states from the physical Hilbert space $\mc{H}_{\text{phys}}$ of QCD. Because of the BRST-exactness of $\widetilde{Q}^{a}$, any colored asymptotic state in the full Hilbert space of the theory belongs to an unphysical BRST quartet.

In conclusion, in the KO framework, under symmetry assumptions which are customary in QCD and which do not conflict either with the spontaneous breaking of the global gauge charge $Q^{a}$ or with the massiveness of gluons induced by the action of the Schwinger mechanism, a well-defined, unique asymptotic color charge operator $\widetilde{Q}^{a}$ exists in dynamically massive QCD, and it realizes color confinement in the sense of the absence of colored asymptotic states from the BRST cohomology.\\

Let us enumerate the central assumptions under which the results of the present work were obtained. In addition to gluons acquiring a mass, as defined by the finiteness of the transverse form factor of the gluon propagator at zero momentum, $|\Delta(0)|<\infty$ modulo renormalization, in Sec.~\ref{sec:asympt} we assumed that:
\begin{itemize}
   \item the gluon and the ghost propagator are diagonal in color space
   \item the massless discrete spectrum of the ghost propagator consists of a simple pole, instead of higher-order poles
   \item the LSZ coefficient $\widetilde{Z}_{3}$ of the ghost and antighost asymptotic fields and the renormalization factor $Z_{3}$ of the transverse form factor of the gluon propagator are positive, $\widetilde{Z}_{3},Z_{3}>0$; moreover, the transverse form factor of the Euclidean gluon propagator is non-negative (at least at zero momentum), $\Delta_{E}(0)\geq 0$
   \item the indefinite-metric Hilbert space of the theory is asymptotically complete with respect to a set of asymptotic fields which includes the massless asymptotic component of the elementary Yang-Mills fields $A_{\mu}^{a}$, $B^{a}$, $c^{a}$, and $\cbar^{a}$, and other asymptotic fields which (anti)commute with said component
   \item no massless bound states exist in the gluon-ghost channel $f^{abc}A_{\mu}^{b}c^{c}$ and in the two-ghost channel $f^{abc}c^{b}c^{c}$
\end{itemize}
Moreover, in Sec.~\ref{sec:ssb} we assumed that:
\begin{itemize}
   \item the KO constant $u^{ab}$ is diagonal in color space, $u^{ab}=u\,\delta^{ab}$
   \item no massless bound states exist in the antisymmetric two-gluon channel $f^{abc}A_{\mu}^{b}A_{\nu}^{c}$ and in the gluon-antighost channel $f^{abc}A_{\mu}\cbar^{c}$
   \item the spectrum of the KO/BV Green's function $\langle T\{D_{\mu}c\, D_{\nu}\cbar\}\rangle_{\text{F.T.}}(p)=u(p^{2})t_{\mu\nu}(p)-p_{\mu}p_{\nu}/p^{2}$ is sufficiently well-behaved for the relation $u=u(0)$ to hold
\end{itemize}
Finally, in Sec.~\ref{sec:chargestruct} we assumed that:
\begin{itemize}
   \item the global gauge charge $Q^{a}$ acts by color transformations of the asymptotic limit of the local fields of the theory
\end{itemize}
This assumption was formulated and discussed in Sec.~\ref{sec:resgtgb}.

While we did not assume that the transverse form factor of the gluon propagator is different from zero at zero momentum, $\Delta(0)\neq 0$, or that the ghost dressing function saturates in the same limit, $J(0)=J_{0}<\infty$ modulo renormalization, at various stages throughout this paper we observed that it is unclear whether $\Delta(0)=0$ and $J(0)=\infty$ can be actually realized under the listed assumptions. At the very least, within the KO framework, these values appear unnatural.

Of the assumptions listed above, those which do not concern color symmetry are customary in Standard Model theories quantized in the KO framework \cite{KO78b,Oji78,KO78a,KO79b,KO79c,KO79d,KO79a,Kug78}; those which do are customary in QCD. That the global gauge charge $Q^{a}$ acts by color transformations of the asymptotic limit of local fields would be an immediate consequence of the unbrokenness of $Q^{a}$; however, since in the KO framework the global gauge charge of the LCGs of interacting Yang-Mills theories is always spontaneously broken, one must instead introduce the property as a standalone symmetry-preserving assumption\footnote{While it is possible that this property can be derived from other symmetry assumptions, we did not attempt to do so in the present paper.}.\\

Let us now discuss the unitarity of the linear covariant gauges of QCD quantized in the KO framework, in the presence of a gluon mass.

While approaches that predict a saturating zero-momentum value for $\Delta(p^{2})$ and $J(p^{2})$ have long been believed to break either BRST symmetry or color symmetry \cite{AvS01,FMP09} mainly on the basis of the misinterpretation of KO's ``converse of the Higgs theorem'', in this paper we have shown this not to be the case. Quite the opposite, as discussed above, there is no evidence that the Schwinger mechanism\footnote{Technically, the Schwinger mechanism, \textit{per se}, does not predict the saturation of $\Delta(p^{2})$ and $J(p^{2})$. Still, the values $\Delta(0)=0$ and $J(0)=\infty$ are just as unnatural in its ordinary formulation as they are in the KO framework. As it turns out, this is not a coincidence.} breaks color symmetry and, both in its ordinary formulation \cite{Smi74,Cor81,AP06,ABP08,AP09,AIMP12,IP13,ABP16,ABFP16,ABP17,BP18,Pap22,FP23,FP24,FP25} and in the one we derived here within the KO framework, the mechanism has its very foundation in the preservation of BRST symmetry.

In the KO framework, the role of the ABFP bound state is that of providing a nonperturbative completion to the elementary BRST quartet of dynamically massive QCD, which loses its massless longitudinal polarization when gluons acquire a mass. Completing the quartet is in turn required to preserve the unitarity of the theory, realized via the quartet mechanism \cite{KO78b,KO78a,KO79a,Kug78}.

As we saw in Sec.~\ref{sec:spectr}, in the KO framework, the unitarity of the physical $S$-matrix defined on the BRST cohomology, in the sense of $S_{\text{phys}}^{\dagger}S_{\text{phys}}=S_{\text{phys}}S_{\text{phys}}^{\dagger}=1$, and the decoupling of the unphysical degrees of freedom from BRST-invariant observables, are immediate consequences of the unitarity of the full $S$-matrix of the theory and of BRST symmetry: a sufficient condition for a gauge theory to be unitary and for its unphysical degrees of freedom to decouple from physical space is for the theory to admit a formulation within the framework \cite{Nak79}. In the present paper, we have shown not only that the ABFP Schwinger mechanism admits such a formulation, but also that, under customary assumptions, it is unique within the framework as the mechanism of dynamical gluon mass generation. It follows that the physical $S$-matrix of the linear covariant gauges of QCD quantized in the KO framework, in which gluons acquire a mass via the ABFP Schwinger mechanism, is unitary: their unphysical degrees of freedom do not contribute either to the correlators of BRST-invariant operators, or to the scattering amplitudes of physical asymptotic states belonging to the BRST cohomology.

For what concerns colored asymptotic states, when the Schwinger mechanism is active in the KO framework, as discussed above, these are organized into BRST quartets that do not contribute to physical observables. For instance, just like in ordinary perturbative QCD, the unphysical thresholds produced by ghosts in physical observables will be cancelled by corresponding thresholds produced by the quanta of the Nakanishi-Lautrup field $\beta^{a}$ and by the ABFP bound state $\chi^{a}$, which are part of the same massless quartet.

Similarly, it has long been known that, if the gluon and quark spectra have massive discrete components, then the corresponding asymptotic fields must be part of massive BRST quartets triggering analogous cancellations \cite{Kug78,AA10,AA11}. For instance, if it existed, the transverse massive asymptotic vector field $U_{\mu}^{a}$ we introduced in Secs.~\ref{sec:asympt} and \ref{sec:chargestruct} would be part of a massive vector BRST quartet together with massive vector bound states in the channels $f^{abc}A_{\mu}^{b}c^{c}$ (first daughter), $f^{abc}A_{\mu}^{b}\cbar^{c}$ (second parent), and $\{Q_{B},f^{abc}A_{\mu}^{b}\cbar^{c}\}=f^{abc}A_{\mu}^{b}B^{c}-if^{abc}D_{\mu}c^{b}\cbar^{c}$ (second daughter) \cite{Kug78,AA10,AA11}.

Another interesting possibility is that, in the LCGs, the unphysical sector of the spectrum of dynamically massive QCD may contain no colored asymptotic states other than those of the elementary BRST quartet. This would be the case if the gluon and quark propagators developed no poles at $p^{2}=m^{2}>0$, and if no bound states other than the ABFP scalars were formed in colored channels. If this scenario were to be realized, then the explicit term in Eq.~\eqref{deq068} would provide the complete expression of the asymptotic color charge in dynamically massive QCD.

In addition to colored quartets, color-neutral BRST quartets can also exist in the unphysical sector of the spectrum. In the presence of quarks, one well-known candidate \cite{Kug79,KO79a} is the quartet featuring the unphysical Goldstone boson associated with the anomalous U(1)$_{A}$ axial symmetry \cite{Adl69,BW69,Wei75}, which is spontaneously broken in the chiral limit\footnote{It is well-known that the Adler-Bell-Jackiw anomaly does not solve the problem of the existence of the Goldstone boson; it merely shifts it to a different, gauge-dependent current \cite{Bar74,KO79a,Kug79,Stro2021}.}. This quartet does indeed exist in the two-dimensional Schwinger model, but there it coincides with the elementary BRST quartet owing to the linearity of the anomaly with respect to the gauge fields \cite{Kug79,KO79a}. In four dimensional QCD, the anomaly is quadratic in the field-strength tensor $F_{\mu\nu}^{a}$, the elementary BRST quartet is colored, and one expects the U(1)$_{A}$ quartet, if any, to be distinct from the elementary quartet.

Finally, let us make a remark on the positivity of the inner product of the BRST cohomology. Since the present paper is concerned with the behavior of the unphysical degrees of freedom of the linear covariant gauges of QCD, determining the norm of the asymptotic states belonging to the BRST cohomology goes well beyond its scope. Nevertheless, it should be realized that, once the unphysical degrees of freedom of the theory decouple from the BRST cohomology thanks to the quartet mechanism, whether or not the inner product of $\mc{H}_{\text{phys}}$ is positive definite is a matter of what stable particles exist in its spectrum. For instance, if we assumed, as is customary, that the only stable particles in $\mc{H}_{\text{phys}}$ are the known stable hadrons\footnote{Given that all the elementary fields in QCD are colored, such stable particles must in any case be BRST-invariant color-neutral bound states.}, then the positivity of the inner product of $\mc{H}_{\text{phys}}$ would immediately follow from the well-known positivity of the norm of the corresponding one-particle states, and from asymptotic completeness -- see Sec.~\ref{sec:framework}. Therefore, while in the present paper we did not show that the inner product of the BRST cohomology of the linear covariant gauges of dynamically massive QCD is positive definite, it should be understood that this property does indeed hold under additional customary assumptions on the content of $\mc{H}_{\text{phys}}$.\\

To conclude this paper, let us make some remarks on the experimental evidence in favor of the Schwinger mechanism within the boundaries set by the KO framework.

In the linear covariant gauges of QCD, only one of two possibilities can be realized: either the transverse form factor $\Delta(p^{2})$ of the gluon propagator has a pole at $p^{2}=0$, or it does not, the latter possibility corresponding to gluons having dynamically acquired a mass. Even setting aside the overwhelming wealth of results accumulated over the past 30 years from lattice calculations concerning the finiteness of $\Delta(0)$ in the LCGs \cite{Cuc97,LSWP98a,LSWP98b,BBLW00,BBLW01,BHLP04,SIMS05,BBMM06,CM07,CM08a,CM08b,BIMS09,ISI09,BMM10,ABBC12,BLLM12,OS12,BBCO15,FN03,FN06a,Ste06}, if one assumes that the KO framework is the correct way to formalize Yang-Mills theories, and that the customary assumptions listed in this section hold both in the presence and in the absence of dynamical gluon mass generation, then the experimental evidence confirms that the Schwinger mechanism is active in QCD. Indeed, as we noted in Sec.~\ref{sec:spectr}, within the KO framework, under those same assumptions, a massless pole in the transverse form factor of the gluon propagator necessarily implies the existence of massless asymptotic gluon states in the \textit{physical} sector of the spectrum of QCD \cite{KO79c}. These states are, of course, not observed experimentally, and their existence is in open conflict with the confinement of gluons. Furthermore, massless gluons provide neither a clear mechanism for the confinement of colored asymptotic states nor an explanation for the absence of long-range strong interactions. Within the KO framework, under customary assumptions, there is thus no doubt as to which of the two possibilities above is realized in Nature.

Preventing the gauge bosons of the strong interactions from appearing in the physical spectrum, at a time when the concept of a gluon had not been invented yet, was the main motivation that led Schwinger to formulate the mechanism that bears his name \cite{Sch62a}. That the mechanism could also explain confinement was first proposed by A. Casher, J. B. Kogut, and L. Susskind in \cite{KS74,CKS74}. In \cite{KO79d,KO79a}, framing the problem in the language of the BRST cohomology and of asymptotic states, Kugo and Ojima predicted that dynamical gluon mass generation without the spontaneous breaking of color symmetry would be a sufficient condition for color confinement. While based on an incorrect interpretation of symmetry breaking and on a missing LSZ coefficient they ruled out gluon mass generation mechanisms for which $1+u\neq 0$, like the Schwinger mechanism, in this paper we have shown that $1+u$ plays no role either in spontaneous symmetry breaking or in color confinement. Within the KO framework, the unique mechanism of dynamical gluon mass generation is indeed the Schwinger mechanism, in the modern formulation of Aguilar, Binosi, Ferreira, and Papavassiliou \cite{AP06,ABP08,AP09,AIMP12,IP13,ABP16,ABFP16,ABP17,BP18,Pap22,FP23,FP24,FP25}.

Given the strong case for a gluon mass stemming from fundamental properties of QCD such as confinement, we believe it is of utmost importance that significant effort be devoted to pursuing an analytical proof of dynamical gluon mass generation. Our hope is that this work will prove helpful in achieving that goal.

\acknowledgments

The author wishes to thank A. Borys, V. Branchina, M. N. Ferreira, C. Gigliuto, G. La Rosa, M. Ruggieri, and F. Siringo for useful discussions. He is grateful to D. Binosi and to J. Papavassiliou for carefully reading the manuscript, and to O. Sipala for granting the leave necessary to complete this work. This work received no financial support.

\appendix

\section{Dipole functions and states}\label{app:dipole}

In the present Appendix, we review some useful properties of dipole functions and of dipole states. More information on this subject can be found in Appendix A.5 of \cite{NO90} and in \cite{Nak72}.\\

Let $f(x)$ be a function that satisfies the massless wave equation $\partial^{2}f=0$, where $\partial^{2}=\partial_{0}^{2}-\boldsymbol{\nabla}^{2}$ is the wave operator, and $\boldsymbol{\nabla}^{2}=-\partial^{i}\partial_{i}$ -- $i$ being spatial indices -- is the Laplacian. Define an integro-differential operator $\mc{E}^{(1/2)}_{x}$ as \cite{Lau67,KO79b,KO79c,NO90}
\begin{equation}\label{dipopapp}
   \mc{E}^{(1/2)}_{x}=\frac{1}{2\boldsymbol{\nabla}^{2}}\,(x^{0}\partial_{0}-1/2)\ .
\end{equation}
$\mc{E}^{(1/2)}_{x}$ is non-local and not Poincar\'{e}-covariant. It is easy to see that
\begin{equation}
   \partial^{2}\big(\mc{E}^{(1/2)}f\big)=f\ ,
\end{equation}
so that, limited to functions that satisfy the wave equation, $\mc{E}^{(1/2)}$ can be used as a noncovariant inverse of the wave operator. $\mc{E}^{(1/2)}$ commutes with the spatial gradient,
\begin{equation}
   [\partial_{i},\mc{E}^{(1/2)}]=0\ ,
\end{equation}
but not with the time derivative:
\begin{equation}
   [\partial_{0},\mc{E}^{(1/2)}]=\frac{1}{2\boldsymbol{\nabla}^{2}}\,\partial_{0}\ .
\end{equation}

Let $D^{(\cdot)}(x;m^{2})$ be a simple-pole function of mass $m$, like the retarded/advanced Green's functions in Eq.~\eqref{ffretadv}, the commutator $D$-function in Eq.~\eqref{ffcommut}, the Feynman (time-ordered) Green's function
\begin{equation}
   D_{F}(x;m^{2})=\int\frac{d^{4}p}{(2\pi)^{4}}\ e^{-ipx}\ \frac{i}{p^{2}-m^{2}+i\epsilon}\ ,
\end{equation}
and the positive/negative-frequency Green's functions
\begin{equation}
   D^{(\pm)}(x;m^{2})=\pm\int\frac{d^{4}p}{(2\pi)^{3}}\ e^{-ipx}\ \theta(\pm p_{0})\ \delta(p^{2}-m^{2})\ .
\end{equation}
The corresponding massless dipole function $E^{(\cdot)}(x)$ is defined as
\begin{equation}
   E^{(\cdot)}(x)=-\frac{\partial}{\partial m^{2}}\bigg|_{m^{2}=0}D(x;m^{2})\ .
\end{equation}
With $D^{(\cdot)}(x)=D^{(\cdot)}(x;m^{2}=0)$, one can prove the following relation \cite{NO90}:
\begin{equation}
   E^{(\cdot)}(x-y)=\left(\mc{E}^{(1/2)}_{x}+\mc{E}^{(1/2)}_{y}\right)D^{(\cdot)}(x-y)\ .
\end{equation}

In order to compute the equal-time commutation relations of free simple-pole and dipole fields, one needs the relations \cite{Nak72,NO90}:
\begin{equation}
   D(x;m^{2})\Big|_{x^{0}=0}=0\ ,\quad \frac{\partial}{\partial x^{0}}\bigg|_{x^{0}=0}D(x;m^{2})=-\delta^{(3)}({\bf x})\ ,
\end{equation}
\begin{equation}
   \frac{\partial^{n}}{\partial (x^{0})^{n}}\bigg|_{x^{0}=0}E(x)=0\ ,\quad \frac{\partial^{3}}{\partial (x^{0})^{3}}\bigg|_{x^{0}=0}E(x)=-\delta^{(3)}({\bf x})\ ,
\end{equation}
for $n=0,1,2$.\\

Let $\chi(x)$ and $\beta(x)$ be scalar quantum fields,
\begin{equation}\label{poincareapp}
   [P_{\mu},\chi]=-i\partial_{\mu}\chi\ ,\qquad [M_{\mu\nu},\chi]=i(x_{\mu}\partial_{\nu}-x_{\nu}\partial_{\mu})\chi\ ,
\end{equation}
and similarly for $\beta(x)$, where $P_{\mu}$ and $M_{\mu\nu}$ are the generators of translations and of Lorentz transformations, respectively. Suppose that
\begin{equation}
   \partial^{2}\chi+\kappa\beta=0\ ,\qquad \partial^{2}\beta=0\ ,
\end{equation}
where $\kappa$ is a constant. If $\kappa\neq 0$, then $\chi(x)$ is a massless dipole field,
\begin{equation}
   (\partial^{2})^{2}\chi=0\ ,\qquad \partial^{2}\chi\neq 0\ .
\end{equation}

Let $\widetilde{\chi}(x)$ be the field defined as
\begin{equation}\label{dipdefapp}
   \widetilde{\chi}(x)=\chi(x)+\kappa\,\mc{E}^{(1/2)}_{x}\beta(x)\ .
\end{equation}
$\widetilde{\chi}(x)$ satisfies the wave equation $\partial^{2}\widetilde{\chi}=0$, but it is neither local, nor Poincar\'{e} covariant. Nevertheless, one can define its associated creation an annihilation operators as
\begin{equation}\label{creatapp}
   \sqrt{2\hat{p}^{0}}\chi_{{\bf p}}^{\dagger}=-i\int d^{3}{\bf x}\ e^{-i\hat{p}x}\,\overleftrightarrow{\partial_{0}}\widetilde{\chi}(x)\ ,
\end{equation}
where $f\overleftrightarrow{\partial_{0}}g=f(\partial_{0}g)-(\partial_{0}f)g$ and $\hat{p}=(|{\bf p}|,{\bf p})$. An analogous equation holds for the creation operators $\beta_{{\bf p}}^{\dagger}$ of $\beta(x)$.

Since $\widetilde{\chi}(x)$ does not satisfy Eq.~\eqref{poincareapp}, the one-particle state $\ket{\chi(p)}=\sqrt{2\hat{p}^{0}}\chi_{{\bf p}}^{\dagger}\ket{\Omega}$ -- $\ket{\Omega}$ being the vacuum state of the theory -- is neither an eigenstate of the momentum operator $P_{\mu}$, nor it is Poincar\'{e} covariant. Instead, using
\begin{equation}
   [P_{\mu},\widetilde{\chi}(x)]=-i\partial_{\mu}\widetilde{\chi}(x)+\frac{i\kappa}{2}\,\delta_{\mu}^{0}\ \boldsymbol{\nabla}^{-2}\,\partial_{0}\beta(x)\ ,
\end{equation}
which can be easily derived from Eqs.~\eqref{poincareapp} and \eqref{dipdefapp}, one finds that
\begin{equation}\label{deq496}
   \left(P_{\mu}-\hat{p}_{\mu}\right)\ket{\chi(p)}=\frac{\kappa}{2\hat{p}^{0}}\,\delta_{\mu}^{0}\,\ket{\beta(p)}\ .
\end{equation}
The last equation implies that $\ket{\chi(p)}$ is an eigenstate of the spatial momentum operator $P_{i}$ only. Moreover, since $(P_{\mu}-\hat{p}_{\mu})\ket{\beta(p)}=0$ by the analogues of Eqs.~\eqref{poincareapp} and \eqref{creatapp} rewritten for $\beta(x)$, so that
\begin{equation}\label{deq615}
   (P_{0}-\hat{p}_{0})\ket{\beta(p)}=0\quad\Longrightarrow\quad (P_{0}-\hat{p}_{0})^{2}\ket{\chi(p)}=0\ ,
\end{equation}
Eq.~\eqref{deq496} implies that $\ket{\chi(p)}$ is a dipole state of the Hamiltonian $H=P_{0}$. Finally, $\ket{\chi(p)}$ satisfies the equation
\begin{equation}
   P^{2}\ket{\chi(p)}=\kappa\ket{\beta(p)}\neq0\ ,\qquad (P^{2})^{2}\ket{\chi(p)}=0\ ,
\end{equation}
hence it is a massless dipole of the mass operator $P^{2}$.

Let $\Lambda$ be a matrix and $a$ be a vector defining the Poincar\'{e} transformation $x^{\mu}\to\Lambda^{\mu}_{\nu}x^{\nu}+a^{\mu}$ of the spacetime coordinates. Unlike ordinary momentum eigenstates, the corresponding Poincar\'{e} transformation $U(\Lambda,a)$ on the Hilbert space of the theory does not leave the subspace spanned by the one-particle states $\ket{\chi(p)}$ invariant. Instead, under the transformation, given that $\widetilde{\chi}(x)$ is the sum of the covariant field $\chi(x)$ and of the noncovariant field $\mc{E}^{(1/2)}_{x}\beta(x)$, the state $\ket{\chi(p)}$ acquires an extra term containing a linear combination of $\ket{\beta(p)}$'s,
\begin{align}\label{deq395}
   &U(\Lambda,a)\ket{\chi(p)}=\\
   \notag&=\ket{U(\Lambda,a)\chi(p)}+\int\frac{d^{3}{\bf q}}{(2\pi)^{3}}\ \mc{K}({\bf q},{\bf p};\Lambda,a)\,\ket{\beta(q)}\ ,
\end{align}
where the first term takes the form of an ordinary Poincar\'{e} transformation, and $\mc{K}({\bf q},{\bf p};\Lambda,a)$ is a kernel expressing the deviation from the former.

Let us work out explicitly the action of the translation $U(a)=e^{-iPa}$ on the state $\ket{\chi(p)}$. By Eqs.~\eqref{deq496} and \eqref{deq615}, we have
\begin{align}
   U(a)\ket{\chi(p)}&=e^{-iPa}\ket{\chi(p)}=\\
   \notag&=e^{-i\hat{p}a}e^{-i(P^{0}-\hat{p}^{0})a^{0}}\ket{\chi(p)}=\\
   \notag&=e^{-i\hat{p}a}[1-i(P^{0}-\hat{p}^{0})a^{0}]\ket{\chi(p)}\ .
\end{align}
Therefore,
\begin{equation}\label{dipoletransl}
   U(a)\ket{\chi(p)}=e^{-i\hat{p}a}\left(\ket{\chi(p)}-\frac{i\kappa a^{0}}{2\hat{p}_{0}}\ket{\beta(p)}\right)\ .
\end{equation}
For translations, we then find
\begin{equation}
   \mc{K}({\bf q},{\bf p};1,a)=-\frac{i\kappa a^{0}}{2\hat{p}^{0}}\,e^{-i\hat{p}a}\ (2\pi)^{3}\,\delta^{(3)}({\bf q}-{\bf p})\ .
\end{equation}

To end this Appendix, we observe that a sufficient condition for a matrix element $\mc{M}(p)$ of the form
\begin{equation}
   \mc{M}(p)=\bra{\psi}\mc{O}\ket{\chi(p)}\ ,
\end{equation}
where $\ket{\psi}$ is a state and $\mc{O}$ is an operator, to be Poincar\'{e} covariant, is that $\ket{\psi}$ and $\mc{O}$ be both covariant, and that
\begin{equation}
   \bra{\psi}\mc{O}\ket{\beta(q)}=0\qquad\forall\, {\bf q}\ .
\end{equation}
Indeed, if these properties hold, then the extra term on the second line of Eq.~\eqref{deq395} drops out from the transformation, yielding a matrix element with ordinary transformation properties.

\section{Asymptotic structure of the photon field in QED with no photon mass generation}\label{app:photon}

The asymptotic analysis of QED with no photon mass generation, quantized in the LCGs within the KO framework, was first carried out by Kugo and Ojima in \cite{KO79c} in the wider context of pure Yang-Mills theories\footnote{While \cite{KO79c,KO79a,NO90} assume the theory is symmetric under a simple compact gauge group, this assumption plays no role in the asymptotic analysis. The latter only rests on the (massless) pole structure of the elementary two-point functions, and of Slavnov-Taylor identities which also hold in the Abelian case.}. In this Appendix, we report the asymptotic structure of the photon field as presented in \cite{KO79a,NO90}, slightly adapting the notation to the one used in this paper.\\

In the asymptotic limit $t=x^{0}\to \pm \infty$, the massless photon field $A_{\mu}(x)$ and the NL field $B(x)$ take the form\footnote{$B(x)$ is a massless free field, so it is equal to its asymptotic field, modulo normalization.}
\begin{equation}\label{deq335}
   A_{\mu}(x)\to Z_{3}^{1/2}\,A_{\mu}^{\text{as.}}(x)\ ,\qquad B(x)= Z_{3}^{-1/2}\,B^{\text{as.}}(x)\ ,
\end{equation}
where
\begin{equation}
   Z_{3}=\frac{1}{1+i\widehat{\Pi}(0)}\ ,\qquad \Delta(p^{2})=\frac{-i}{p^{2}[1+i\widehat{\Pi}(p^{2})]}\ ,
\end{equation}
$\Delta(p^{2})$ being the transverse form factor of the unrenormalized photon propagator $\Delta_{\mu\nu}(p)$,
\begin{equation}
   \Delta_{\mu\nu}(p)=\Delta(p^{2})\,t_{\mu\nu}(p)+\frac{-i\xi}{p^{2}}\,\ell_{\mu\nu}(p)\ .
\end{equation}
The asymptotic fields $A_{\mu}^{\text{as.}}(x)$ and $B^{\text{as.}}(x)$ in Eq.~\eqref{deq335} read
\begin{align}\label{deq131}
   \notag&A_{\mu}^{\text{as.}}(x)=\int \frac{d^{3}{\bf p}}{(2\pi)^{3}}\ \Big\{\sum_{\sigma=1,2}\, A_{{\bf p},\sigma}\,f_{{\bf p},\mu}^{\sigma}(x)+A_{{\bf p}}^{L}\,\partial_{\mu}g_{{\bf p}}(x)+\\
   &\quad +B_{{\bf p}}\,\left[f_{{\bf p},S,\mu}(x)+(1-\xi_{R})\,\partial_{\mu}h_{{\bf p}}(x)\right]\Big\}+\text{h.c.}\ ,\\
   \notag&B^{\text{as.}}(x)=\int \frac{d^{3}{\bf p}}{(2\pi)^{3}}\ B_{{\bf p}}\,g_{{\bf p}}(x)+\text{h.c.}\ ,
\end{align}
where $\xi_{R}=Z_{3}^{-1}\xi$ is a renormalized gauge parameter, whereas $g_{{\bf p}}(x)$, $f_{{\bf p},\mu}^{\sigma=1,2}(x)$, $f_{{\bf p},S,\mu}(x)$, and $h_{{\bf p}}(x)$ are positive-frequency wavepacket systems, defined in Appendix A.5 of \cite{NO90}.

The functions $g_{{\bf p}}(x)$, $f_{{\bf p},\mu}^{\sigma=1,2}(x)$, and $f_{{\bf p},S,\mu}(x)$ satisfy the massless wave equation,
\begin{equation}\label{deq132}
   \partial^{2}g_{{\bf p}}(x)=0\ ,\quad \partial^{2}f_{{\bf p},\mu}^{\sigma=1,2}(x)=0\ ,\quad \partial^{2}f_{{\bf p},S,\mu}(x)=0\ .
\end{equation}
The functions $f_{{\bf p},\mu}^{\sigma=1,2}(x)$ are transverse,
\begin{equation}\label{deq133}
   \partial^{\mu}f_{{\bf p},\mu}^{\sigma=1,2}(x)=0\ ,
\end{equation}
whereas $f_{{\bf p},S,\mu}(x)$ satisfies
\begin{equation}\label{deq134}
   \partial^{\mu}f_{{\bf p},S,\mu}(x)=-g_{{\bf p}}(x)\ .
\end{equation}
The function $h_{{\bf p}}(x)$, defined as
\begin{equation}
   h_{{\bf p}}(x)=\mc{E}^{(1/2)}_{x}g_{{\bf p}}(x)\ ,   
\end{equation}
with $\mc{E}^{(1/2)}_{x}$ given in Eq.~\eqref{dipopapp}, is a dipole wavepacket,
\begin{equation}
   \partial^{2}h_{{\bf p}}(x)=g_{{\bf p}}(x)\neq 0\ ,\qquad (\partial^{2})^{2}h_{{\bf p}}(x)=0\ .
\end{equation}
It follows that, asymptotically, the photon field is a dipole field in all but the renormalized Feynman gauge $\xi_{R}=1$.

The operators $A_{{\bf p},\sigma=1,2}$, $A_{{\bf p}}^{L}$, and $A_{{\bf p}}^{S}=B_{{\bf p}}$ are annihilation operators for the transverse, the longitudinal, and the scalar photon polarizations, respectively. Their nonvanishing commutation relations are
\begin{align}
   \notag[A_{{\bf p},\sigma},A_{{\bf q},\sigma^{\prime}}^{\dagger}]&=\delta_{\sigma\sigma^{\prime}}\,(2\pi)^{3}\,\delta^{(3)}({\bf p}-{\bf q})\ ,\\
   [A_{{\bf p}}^{L},A_{{\bf q}}^{L\,\dagger}]&=-K\,(2\pi)^{3}\,\delta^{(3)}({\bf p}-{\bf q})\ ,\\
   \notag[A_{{\bf p}}^{L},B_{{\bf q}}^{\dagger}]&=-\,(2\pi)^{3}\,\delta^{(3)}({\bf p}-{\bf q})\ ,
\end{align}
where $K$ is a constant obtained from the photon propagator\footnote{Note that, since $K\neq 1$, neither $A_{{\bf p}}^{L}$ nor $B_{{\bf p}}$ are canonically normalized. This is irrelevant for our purposes.} -- see \cite{KO79c,KO79a,NO90}.

Once (free) ghost fields $c(x)$, $\cbar(x)$ are introduced in the theory, one finds that the transverse polarizations belong to the BRST cohomology,
\begin{equation}
   [Q_{B,R},A_{{\bf p},\sigma=1,2}]=0\ ,
\end{equation}
whereas $A_{{\bf p}}^{L}$, and $A_{{\bf p}}^{S}=B_{{\bf p}}$ form the elementary BRST quartet together with ghosts,
\begin{align}
   &[Q_{B,R},A_{{\bf p}}^{L}]=-ic_{{\bf p}}\ ,\qquad \{Q_{B,R},\cbar_{{\bf p}}\}=B_{{\bf p}}\ ,\\
   \notag &\qquad\{Q_{B,R},c_{{\bf p}}\}=0\ ,\qquad [Q_{B,R},B_{\bf p}]=0\ ,
\end{align}
where $Q_{B,R}=Z_{3}^{1/2}Q_{B}$ ($\widetilde{Z}_{3}=1$).
\\

Define a field $\chi(x)$ as
\begin{align}
   \notag\chi(x)&=Z_{3}^{1/2}\int \frac{d^{3}{\bf p}}{(2\pi)^{3}}\ \Big\{A_{{\bf p}}^{L}\ g_{{\bf p}}(x)+(1-\xi_{R})\,B_{{\bf p}}\,h_{{\bf p}}(x)\Big\}+\\
   &\qquad+\text{h.c.}\ .
\end{align}
Then, in the asymptotic limit,
\begin{equation}
   A_{\mu}(x)\to \partial_{\mu}\chi(x)+\cdots\ ,
\end{equation}
where the ellipsis commutes with the NL field $B$. Using the completeness relations \cite{NO90}
\begin{equation}
   \int\frac{d^{3}{\bf p}}{(2\pi)^{3}}\ g_{{\bf p}}(x)\,g^{\star}_{{\bf p}}(y)=D^{(+)}(x-y)\ ,
\end{equation}
it is easy to evaluate the commutator
\begin{equation}\label{deq055}
   [\chi(x),B(y)]=-iD(x-y)\ .
\end{equation}
Observe that only the longitudinal photon polarization -- that is, the first parent of the elementary BRST quartet of the theory -- contributes to the commutator. In Sec.~\ref{sec:resgtgb}, using Eq.~\eqref{deq055}, we show that, in the interacting theory, even in the absence of photon mass generation, the longitudinal photon polarization is the unphysical Goldstone boson associated with the spontaneous breaking of the global gauge (electric) charge.\\

The asymptotic limit of the electromagnetic field-strength tensor $F_{\mu\nu}=\partial_{\mu}A_{\nu}-\partial_{\nu}A_{\mu}$ reads
\begin{equation}
   F_{\mu\nu}\to Z_{3}^{1/2}\,F_{\mu\nu}^{\text{as.}}\ ,
\end{equation}
where, using Eq.~\eqref{deq131},
\begin{align}
   \notag&F_{\mu\nu}^{\text{as.}}(x)=\int \frac{d^{3}{\bf p}}{(2\pi)^{3}}\ \Big\{\sum_{\sigma=1,2}\, A_{{\bf p},\sigma}\,[\partial_{\mu}f_{{\bf p},\nu}^{\sigma}(x)-\partial_{\nu}f_{{\bf p},\mu}^{\sigma}(x)]+\\
   &\quad +B_{{\bf p}}\,\left[\partial_{\mu}f_{{\bf p},S,\nu}(x)-\partial_{\nu}f_{{\bf p},S,\mu}(x)\right]\Big\}+\text{h.c.}\ .
\end{align}
Using Eqs.~\eqref{deq131} - \eqref{deq134}, one then computes
\begin{equation}
   \partial^{\mu}F_{\mu\nu}^{\text{as.}}=\partial_{\nu}B^{\text{as.}}=Z_{3}^{1/2}\,\partial_{\nu}B\ .
\end{equation}
The last equation implies that, in the asymptotic limit,
\begin{equation}
   \partial^{\mu}F_{\mu\nu}\to Z_{3}\,\partial_{\nu}B\ .
\end{equation}

\section{The Stueckelberg model and the Schwinger model}\label{app:stschw}

In this Appendix, we review the (Abelian) Stueckelberg model quantized in linear covariant gauges and the Schwinger model solved exactly in the KO framework.\\

The Stueckelberg model \cite{Stu38a,Stu38b,DTT88,RRA04} in the LCGs is defined by the Lagrangian
\begin{equation}
   \mc{L}=-\frac{1}{4}\,F^{2}+\frac{\xi}{2}\,B^{2}+B\,\partial^{\mu}A_{\mu}+\frac{1}{2}\,\left(\partial\phi-mA\right)^{2}\ ,
\end{equation}
where $\phi$ is a scalar field -- the Stueckelberg field -- and $m$ is a mass parameter.

The field equations of the model are
\begin{align}\label{stuckfeq}
   &\partial^{\nu}F_{\nu\mu}-\partial_{\mu}B-m\partial_{\mu}\phi+m^{2}A_{\mu}=0\ ,\\
   \notag&\partial^{\mu}A_{\mu}+\xi B=0\ ,\qquad \partial^{2}\phi-m\partial^{\mu}A_{\mu}=0\ .
\end{align}
Its solutions are given in terms of the triplet of fields $(U_{\mu},\phi,B)$, where $U_{\mu}$ is a transverse vector field of mass $m$,
\begin{equation}
   (\partial^{2}+m^{2})U_{\mu}=0\ ,\qquad\partial^{\mu}U_{\mu}=0\ ,
\end{equation}
$\phi$ and $B$ satisfy
\begin{equation}
   \partial^{2}\phi+\xi mB=0\ ,\qquad \partial^{2}B=0\ ,
\end{equation}
the nonvanishing commutators between the fields are
\begin{align}
   \notag[U_{\mu}(x),U_{\nu}(y)]&=-i\left(\eta_{\mu\nu}+m^{-2}\partial_{\mu}\partial_{\nu}\right)D(x-y;m^{2})\ ,\\
   [\phi(x),\phi(y)]&=iD(x-y)+i\xi m^{2}E(x-y)\ ,\\
   \notag[\phi(x),B(y)]&=-im D(x-y)\ ,
\end{align}
and the photon field $A_{\mu}$ is given by
\begin{equation}
   A_{\mu}=U_{\mu}+m^{-1}\partial_{\mu}\phi+m^{-2}\partial_{\mu}B\ .
\end{equation}
Note that the Stueckelberg field is a dipole field in all linear covariant gauges except for the Landau gauge ($\xi=0$).

By evaluating
\begin{align}
   \notag&[A_{\mu}(x),A_{\nu}(y)]=-i\left(\eta_{\mu\nu}+m^{-2}\partial_{\mu}\partial_{\nu}\right)D(x-y;m^{2})+\\
   &\qquad+im^{-2}\partial_{\mu}\partial_{\nu}D(x-y)-i\xi\partial_{\mu}\partial_{\nu}E(x-y)\ ,
\end{align}
replacing commutator functions with the corresponding Feynman (time-ordered) Green's functions,
\begin{align}
   \notag&\left\langle T\left\{A_{\mu}(x)A_{\nu}(y)\right\}\right\rangle=-\left(\eta_{\mu\nu}+m^{-2}\partial_{\mu}\partial_{\nu}\right)D_{F}(x-y;m^{2})+\\
   &\qquad+m^{-2}\partial_{\mu}\partial_{\nu}D_{F}(x-y)-\xi\partial_{\mu}\partial_{\nu}E_{F}(x-y)\ ,\\
   \notag&\left\langle T\left\{\phi(x)\phi(y)\right\}\right\rangle=D_{F}(x-y)+\xi m^{2}E_{F}(x-y)\ ,
\end{align}
and Fourier-transforming, one obtains the propagators in Eq.~\eqref{stueckdipoles}. Note that the massless poles in the photon propagator are produced by the fields $\phi(x)$ and $B(x)$.

Define fields $\chi(x)$ and $\beta(x)$ as
\begin{equation}
   \chi(x)=\phi(x)+m^{-1}B(x)\ ,\qquad\beta(x)=m^{-1}B(x)\ .
\end{equation}
Then
\begin{equation}\label{deq953}
   A_{\mu}=U_{\mu}+m^{-1}\partial_{\mu}\chi\ ,
\end{equation}
and
\begin{align}
   \notag[A_{\mu}(x),\chi(y)]&=-im^{-1}\partial_{\mu}D(x-y)+i\xi m\partial_{\mu}E(x-y)\ ,\\
   [A_{\mu}(x),\beta(y)]&=-im^{-1}\partial_{\mu}D(x-y)\ ,\\
   \notag[\chi(x),\chi(y)]&=-iD(x-y)+i\xi m^{2}E(x-y)\ ,\\
   \notag[\chi(x),\beta(y)]&=-iD(x-y)\ .
\end{align}
In Sec.~\ref{sec:asympt} we will encounter the same algebra in the context of dynamically massive QCD. Note that the field $U_{\mu}(x)$ does not contribute to these equations, as it commutes with both $\chi(x)$ and $\beta(x)$.

Once (free) ghost fields $c(x)$ and $\cbar(x)$ are introduced in the model, its nonvanishing BRST transformations read
\begin{align}
   &[iQ_{B},A_{\mu}(x)]=\partial_{\mu}c\ ,\qquad[iQ_{B},\phi(x)]=mc(x)\ ,\\
   \notag&\qquad\qquad\qquad \{iQ_{B},\cbar(x)\}=iB(x)\ .
\end{align}
In terms of $(U_{\mu},\chi,\beta,c,\cbar)$, these translate to
\begin{align}
   &\qquad\qquad\qquad[iQ_{B},U_{\mu}(x)]=0\ ,\\
   \notag &\ \ [iQ_{B},\chi(x)]=mc(x)\ ,\qquad \{iQ_{B},c(x)\}=0\ ,\\
   \notag&\{iQ_{B},\cbar(x)\}=im\beta(x)\ ,\qquad[iQ_{B},\beta(x)]=0\ .
\end{align}
Evidently, $U_{\mu}(x)$ belongs to the BRST cohomology, hence it creates/annihilates physical quanta, whereas $\chi(x)$, $\beta(x)$, $c(x)$, and $\cbar(x)$ form the unphysical elementary BRST quartet of the Stueckelberg model. Note that, modulo mixing with $\beta$, $\chi$ is nothing but the Stueckelberg field $\phi$. Since $\beta$ is BRST exact, mixing does not modify the BRST transformations of $\phi$.

The Fock space of the BRST-covariant model is constructed by defining creation and annihilation operators for the elementary BRST quartet as in Sec.~\ref{sec:asympt}, and for the field $U_{\mu}(x)$ as
\begin{equation}
   \sqrt{2\hat{p}^{0}}\,a_{{\bf p},\sigma}^{\dagger}=i\varepsilon^{\mu\star}_{\sigma}(p)\int d^{3}{\bf x}\ e^{-i\hat{p}x}\overleftrightarrow{\partial_{0}}U_{\mu}(x)\ ,
\end{equation}
where $\hat{p}=(\sqrt{m^{2}+{\bf p}^{2}},{\bf p})$, and $\varepsilon_{\sigma=0,\pm 1}^{\mu}(p)$ are transverse polarization vectors,
\begin{align}
   \notag&\hat{p}_{\mu}\,\varepsilon_{\sigma}^{\mu}(p)=0\ ,\qquad\sum_{\sigma}\varepsilon_{\sigma}^{\mu}(p)\,\varepsilon_{\sigma}^{\nu\star}(p)=-\eta^{\mu\nu}+\frac{\hat{p}^{\mu}\hat{p}^{\nu}}{m^{2}}\ ,\\
   &\qquad\qquad\qquad\varepsilon_{\sigma}^{\mu}(p)\,\varepsilon_{\sigma^{\prime}\mu}^{\star}(p)=-\delta_{\sigma\sigma^{\prime}}\ .
\end{align}
Observe that, since the Stueckelberg model is essentially a free field theory, its BRST-covariant Fock-space solutions automatically satisfy the postulate of asymptotic completeness, hence they are automatically formulated in the KO framework.

The current $J_{\mu}$ of the Stueckelberg model can be read off from the field equations, or it can be derived from the Lagrangian as the Noether current associated with the constant-parameter limit of gauge transformations
\begin{align}
   A_{\mu}(x)&\to A_{\mu}(x)+m^{-1}\partial_{\mu}\Lambda(x)\ ,\\
   \notag\phi(x)&\to \phi(x)+\Lambda(x)\ .
\end{align}
It is given by
\begin{equation}
   J_{\mu}=-\partial_{\mu}\phi+mA_{\mu}=m^{-1}\partial_{\mu}B+mU_{\mu}\ ,\quad \partial^{\mu}J_{\mu}=0\ .
\end{equation}
A simple calculation using
\begin{equation}
   [\partial_{0}B(y),\phi(x)]_{y^{0}=x^{0}}=im\,\delta^{(3)}({\bf y}-{\bf x})
\end{equation}
shows that
\begin{equation}
   [Q,\phi(x)]=[Q,\chi(x)]=i\ ,
\end{equation}
where $Q=\int d^{3}{\bf x}\ J_{0}$ is the global gauge charge. Therefore
\begin{equation}
   \bra{\Omega}[Q,\phi(x)]\ket{\Omega}=i\ .
\end{equation}
It follows that the global gauge charge $Q$ of the Stueckelberg model is spontaneously broken, and that the quanta of the Stueckelberg field are its associated unphysical Goldstone bosons. As discussed in Sec.~\ref{sec:resgtgb}, these are generic features of Yang-Mills theory quantized in linear covariant gauges; in particular, the Goldstone nature of the Stueckelberg field is expected on the grounds that it is the first parent of the elementary BRST quartet.\\

The Schwinger model \cite{Sch62a,Sch62b,LS71,KS74,CKS74} quantized in the linear covariant gauges \cite{Ito75,Nak75,Nak76a,Nak77a,Nak77b,Nak77c} is defined in spacetime dimension $d=2$ by the Lagrangian
\begin{equation}
   \mc{L}=-\frac{1}{4}\,F^{2}+\frac{\xi}{2}\,B^{2}+B\,\partial^{\mu}A_{\mu}+\psibar(i\slashed{\partial}+e\slashed{A})\psi\ ,
\end{equation}
where $\psi$ is a fermion field. The field equations of the model are
\begin{align}
   &\qquad\ \partial^{\nu}F_{\nu\mu}-\partial_{\mu}B+e J_{\mu}=0\ ,\\
   \notag&\partial^{\mu}A_{\mu}+\xi B=0\ ,\qquad (i\slashed{\partial}+e\slashed{A})\psi=0\ ,
\end{align}
where $J_{\mu}$ is the electric current,
\begin{equation}
   J_{\mu}=\psibar\gamma_{\mu}\psi\ .
\end{equation}

By gauge-invariantly point splitting the current, $J_{\mu}$ can be expressed as \cite{Sch62b,LS71,Nak75}
\begin{equation}\label{schwcurr}
   eJ_{\mu}=-m\partial_{\mu}X+m^{2}A_{\mu}\ ,   
\end{equation}
where
\begin{equation}
   m=\frac{e}{\sqrt{\pi}}
\end{equation}
is a mass parameter\footnote{Note that, in two spacetime dimensions, the electromagnetic coupling constant $e$ has mass dimension 1.} and $X$ is a field in the fermion-antifermion channel. Because of current conservation, $X$ satisfies the equation
\begin{equation}
   \partial^{2}X-m\partial^{\mu}A_{\mu}=0\ .
\end{equation}

When expressed in terms of the field $X$, the field equations of the Schwinger model,
\begin{align}\label{deq532}
   \notag&\partial^{\nu}F_{\nu\mu}-\partial_{\mu}B-m\partial_{\mu}X+m^{2}A_{\mu}=0\ ,\\
   &\partial^{\mu}A_{\mu}+\xi B=0\ ,\quad (i\slashed{\partial}+e\slashed{A})\psi=0\ ,\\
   \notag&\qquad\qquad\partial^{2}X-m\partial^{\mu}A_{\mu}=0\ ,
\end{align}
are identical to those of the two-dimensional Stueckelberg model, with the Stueckelberg field $\phi$ replaced by $X$, plus an additional fermion equation. The solutions we reviewed above for the Stueckelberg model thus also solve the Schwinger model in the $(A_{\mu},X,B)$ sector. In \cite{Nak75}, building on these solutions, Nakanishi provided an expression for the field $\psi(x)$ in terms of the free fields $(U_{\mu},X,B)$ that solves the fermion equation in Eq.~\eqref{deq532} as well. Nakanishi's solutions of the Schwinger model are thus asymptotically complete with respect to the fields $(U_{\mu},X,B)$. Since the fermion field has no associated asymptotic field, and its propagator has no pole \cite{LS71,CKS74,Nak75}, in these solutions, fermions are confined in the sense of the absence of asymptotic fermion states from the spectrum of the theory.

Given that the (two-dimensional) Stueckelberg and Schwinger models have the same solutions in the $(A_{\mu},\phi/X,B)$ sector, everything we said for the former also holds for the latter. In particular, once (free) ghost fields are added to the Schwinger model, Nakanishi's solutions are BRST covariant and asymptotically complete with respect to the fields $(U_{\mu},X,B,c,\cbar)$, with the first parent of the elementary BRST quartet being the fermion-antifermion state $X$. These solutions thus satisfy the assumptions of the KO framework. Since the fermion field has no associated asymptotic field, it has no associated quanta; hence it is neither in the BRST cohomology, nor it is part of an unphysical BRST quartet.

With the electric current expressed as
\begin{equation}
   J_{\mu}=e^{-1}\left(-m\partial_{\mu}X+m^{2}A_{\mu}\right)=e^{-1}\left(\partial_{\mu}B+m^{2} U_{\mu}\right)\ ,
\end{equation}
and the field $\chi(x)$ such that $A_{\mu}=U_{\mu}+m^{-1}\partial_{\mu}\chi$ defined as
\begin{equation}
   \chi(x)=X(x)+m^{-1}B(x)\ ,
\end{equation}
it is easy to see that
\begin{equation}
   [Q,X(x)]=[Q,\chi(x)]=ie^{-1}m\ ,
\end{equation}
Therefore, as expected, in the Schwinger model quantized within the KO framework, the fermion-antifermion first-parent state is the unphysical Goldstone boson associated with the spontaneous breaking of the global gauge (electric) charge.

\section{Higgs models}\label{app:higgsmod}

In the present Appendix, we review the SU(2) Higgs-Kibble model quantized in linear covariant gauges as an example of a Yang-Mills theory in which gauge bosons acquire a mass via the spontaneous breaking of the underlying symmetry of the theory -- in this case, SU(2) color symmetry.\\

The SU(2) Higgs-Kibble model quantized in the LCGs is defined by the Lagrangian \cite{Kib67,tHoo71,KO79b}
\begin{align}
   \notag\mc{L}&=-\frac{1}{4}\,F^{i}_{\mu\nu}F^{i\mu\nu}+\frac{\xi}{2}\,B^{i}B^{i}+B^{i}\partial^{\mu}A_{\mu}^{i}-i\partial^{\mu}\cbar^{i}D_{\mu}c^{i}+\\
   &\qquad+(D^{\mu}\Phi)^{\dagger}(D_{\mu}\Phi)-V(\Phi)\ ,
\end{align}
where $i=1,2,3$ are SU(2) color indices, $\Phi$ is a scalar field -- the Higgs field -- in the fundamental representation of SU(2), $D_{\mu}\Phi$ is its covariant derivative,
\begin{equation}
   D_{\mu}\Phi=(\partial_{\mu}-igT^{i}A_{\mu}^{i})\Phi\ ,\qquad T^{i}=\frac{\sigma^{i}}{2}\ ,
\end{equation}
$\sigma^{i}$ being the Pauli matrices, and $V(\Phi)$ is the Higgs potential,
\begin{equation}
   V(\Phi)=\frac{\lambda}{2}\left(|\Phi|^{2}-\frac{v^{2}_{0}}{2}\right)^{2}\ ,\qquad \lambda>0\ ,
\end{equation}
where $\lambda$ and $v_{0}$ are real constants. The field equations of the model read
\begin{align}\label{hkmfeqs}
   &\partial^{\nu}F_{\nu\mu}^{i}-\partial_{\mu}B^{i}+\\
   \notag&\ \, +g[\epsilon^{ijk}A^{j\,\nu}F^{k}_{\nu\mu}+i\epsilon^{ijk}\partial_{\mu}\cbar^{j}c^{k}+(J_{\mu}^{i})_{\text{matter}}]=0\ ,\\
   \notag&\ \partial^{\mu}D_{\mu}c^{i}=0\ ,\quad D^{\mu}\partial_{\mu}\cbar^{i}=0\ ,\quad\partial^{\mu}A_{\mu}^{i}+\xi B^{i}=0\ ,\\
   \notag &\qquad\qquad\qquad\quad D^{\mu}D_{\mu}\Phi+V^{\prime}(\Phi)=0\ ,
\end{align}
where $\epsilon^{ijk}$ are the SU(2) structure constants, $V^{\prime}(\Phi)=\partial V/\partial\Phi^{\dagger}$, and $(J_{\mu}^{i})_{\text{matter}}$ is the matter current
\begin{equation}
   (J_{\mu}^{i})_{\text{matter}}=i\Phi^{\dagger}T^{i}D_{\mu}\Phi-i(D_{\mu}\Phi)^{\dagger}T^{i}\Phi\ .
\end{equation}

Together with the gauge current $(J_{\mu}^{i})_{\text{gauge}}$ in Eq.~\eqref{colorcurr},
\begin{equation}\label{hkmcolorcurr}
   (J^{i}_{\mu})_{\text{gauge}}=\epsilon^{ijk}\left(A^{j\nu}F_{\nu\mu}^{k}+A^{j}_{\mu}B^{k}+i\partial_{\mu}\cbar^{j}c^{k}-i\cbar^{j}D_{\mu}c^{k}\right)\ ,
\end{equation}
the matter current $(J_{\mu}^{i})_{\text{matter}}$ forms the conserved SU(2) color current $J_{\mu}^{i}$,
\begin{equation}
   J_{\mu}^{i}=(J_{\mu}^{i})_{\text{gauge}}+(J_{\mu}^{i})_{\text{matter}}\ ,\qquad \partial^{\mu}J_{\mu}^{i}=0\ .
\end{equation}
The corresponding SU(2) global gauge (color) charge $Q^{i}=\int d^{3}{\bf x}\,J_{0}^{i}$ acts on the gauge fields as in Eq.~\eqref{color}, and on the Higgs field as
\begin{equation}\label{higgscolor}
   [Q^{i},\Phi]=-T^{i}\Phi\ .
\end{equation}
In terms of $J_{\mu}^{i}$, the MYME of the model take the form \cite{Oji78,KO79a}
\begin{equation}\label{hkmmyme2}
   \partial^{\nu}F_{\nu\mu}^{i}-\{Q_{B},D_{\mu}\cbar^{i}\}+gJ_{\mu}^{i}=0\ ,
\end{equation}
where $Q_{B}$ is the BRST charge
\begin{align}
   \notag Q_{B}&=\int d^{3}{\bf x}\, \bigg\{B^{i}D_{0}c^{i}-\partial_{0}B^{i}c^{i}+\frac{ig}{2}\,\epsilon^{ijk}\partial_{0}\cbar^{i}c^{j}c^{k}+\\
   &\qquad\qquad\qquad-g c^{i}(J_{0}^{i})_{\text{matter}}\bigg\}\ .
\end{align}
$Q_{B}$ acts on the gauge fields as in Eq.~\eqref{brst}, and on the Higgs field as
\begin{equation}\label{higgsbrst}
   [Q_{B},\Phi]=gc^{i}T^{i}\Phi\ .
\end{equation}

Since the classical minimum of the Higgs potential lies at a nonvanishing value of the Higgs field, $|\Phi|^{2}=v_{0}^{2}/2$, one assumes that the exact solutions of the model are such that the vacuum expectation value $\avg{\Phi}$ is different from zero,
\begin{equation}\label{higgsavg}
   \avg{\Phi}=\frac{v}{\sqrt{2}}\,\varphi_{0}\ ,\qquad v\neq 0\ ,\qquad \varphi^{\dagger}_{0}\varphi_{0}=1\ ,
\end{equation}
where $\varphi_{0}$ is a unitary vector in color space. Conventionally, one then reparametrizes the Higgs field as
\begin{equation}\label{higgsrepar}
   \Phi(x)=\frac{1}{\sqrt{2}}\left[v+h(x)+i\pi^{i}(x)\,\sigma^{i}\right]\varphi_{0}\ ,
\end{equation}
where $h(x)$ and $\pi^{i}(x)$ -- respectively, the Higgs boson and Goldstone boson fields -- are real scalar fields such that $\avg{h(x)},\avg{\pi^{i}(x)}=0$.

Since the Higgs field acquires a nonzero vacuum expectation value despite transforming nontrivially under color transformations, the SU(2) color symmetry of the model is spontaneously broken. It follows that the asymptotic fields are not organized into representations of color SU(2): the discrete spectrum of symmetry-violating two-point time-ordered correlators -- hence the corresponding asymptotic fields -- does not have definite transformation properties under color transformations. In particular, in Higgs models, an equation like Eq.~\eqref{symass} has no meaning.

The transformation laws of the fields $h$ and $\pi^{i}$ under the global gauge symmetry can be easily derived from Eqs.~\eqref{higgscolor} and \eqref{higgsrepar}: they read
\begin{align}\label{hkmcol}
   [Q^{i},h]=-\frac{i}{2}\,\pi^{i}\ ,\quad[Q^{i},\pi^{j}]=\frac{i}{2}\left[(v+h)\delta^{ij}+\epsilon^{ijk}\pi^{k}\right]\ .
\end{align}
By taking the vacuum expectation value of the second of these, one obtains
\begin{equation}\label{hkmgold0}
   \bra{\Omega}[Q^{i},\pi^{j}]\ket{\Omega}=\frac{iv}{2}\,\delta^{ij}\ .
\end{equation}
Therefore, the quanta of $\pi^{i}$ are the scalar Goldstone bosons associated with the spontaneous breaking of the global gauge charge $Q^{i}$.

Similarly, from Eqs.~\eqref{higgsbrst} and \eqref{higgsrepar}, one derives the following BRST transformations of the fields $h$ and $\pi^{i}$:
\begin{align}\label{hkmbrst}
   [Q_{B},h]&=\frac{ig}{2}\,c^{i}\pi^{i}\ ,\\
   \notag[Q_{B},\pi^{i}]&=-\frac{ig}{2}c^{j}\left[(v+h)\delta^{ij}+\epsilon^{jik}\pi^{k}\right]\ .
\end{align}
By introducing asymptotic fields for $h$ and $\pi^{i}$ as
\begin{equation}\label{higgsas0}
   h\to Z_{h}^{1/2}h^{\text{as.}}\ ,\qquad\pi^{i}\to Z_{\pi}^{1/2}\pi^{i\,\text{as.}}\ ,
\end{equation}
under the customary assumptions that BRST symmetry is unbroken and that there are no bound states in the channels $c^{i}\pi^{i}$, $c^{i}h$, and $\epsilon^{ijk}c^{j}\pi^{k}$ \cite{KO79b}, Eqs.~\eqref{asymsymm} and \eqref{hkmbrst} yield
\begin{equation}
   [Q_{B},h^{\text{as.}}]=0\ ,\qquad [Q_{B},\pi^{i\,\text{as.}}]\propto-i\gamma^{i}\neq 0\ ,
\end{equation}
where $\gamma^{i}$ is the asymptotic field associated with the ghost field $c^{i}$. It follows that the Higgs boson belongs to the BRST cohomology -- hence to the physical sector of the spectrum --, whereas Goldstone bosons do not -- hence they are unphysical particles. Since $\pi^{i\,\text{as.}}$ is the Goldstone boson associated with the spontaneous breaking of the global gauge charge $Q^{i}$, and since it transforms into $\gamma^{i}$ under BRST transformations, by the discussion in Sec.~\ref{sec:resgtgb} and by the first of Eq.~\eqref{brste}, one should expect that it is the first parent of the elementary BRST quartet of the model. As we will see below, this is in fact the case.

When expressed in terms of the fields $h$ and $\pi^{i}$, the matter current $(J_{\mu}^{i})_{\text{matter}}$ reads
\begin{widetext}
\begin{align}
   (J_{\mu}^{i})_{\text{matter}}=\frac{1}{2}\left\{\pi^{i}\left(\partial_{\mu}h+\frac{g}{2}\,A_{\mu}^{j}\pi^{j}\right)-\left[(v+h)\delta^{ij}+\epsilon^{ijk}\pi^{k}\right]\left[\partial_{\mu}\pi^{j}-\frac{g}{2}(v+h)A_{\mu}^{j}+\frac{g}{2}\,\epsilon^{jlm}A_{\mu}^{l}\pi^{m}\right]\right\}\ ,
\end{align}
\end{widetext}
and it contains a linear term $(J_{\mu}^{i})_{\text{linear}}$,
\begin{equation}\label{hkmlcurr}
   g(J_{\mu}^{i})_{\text{linear}}=-m\partial_{\mu}\pi^{i}+m^{2}A_{\mu}^{i}\ ,\qquad m=\frac{gv}{2}\ ,
\end{equation}
which has the same form as the current of the Stueckelberg model and as the point-split current of the Schwinger model -- see Eqs.~\eqref{stuckfeq} and \eqref{schwcurr}. The second term on the RHS of the first of Eq.~\eqref{hkmlcurr} is a mass term for the gauge bosons.

When rewritten in terms of $h$ and $\pi^{i}$, the field equations of the model are symmetric under the infinitesimal SU(2) transformations
\begin{align}\label{higgscustodial}
   \notag&A_{\mu}^{i}\to A_{\mu}^{i}+\epsilon^{ijk}\lambda^{j}A_{\mu}^{k}\ ,\qquad B^{i}\to B^{i}+\epsilon^{ijk}\lambda^{j}B^{k}\ ,\\
   &\quad\ c^{i}\to c^{i}+\epsilon^{ijk}\lambda^{j}c^{k}\ ,\qquad \cbar^{i}\to \cbar^{i}+\epsilon^{ijk}\lambda^{j}\cbar^{k}\ ,\\
   \notag&\qquad\qquad\qquad\ h\to h\ ,\qquad \pi^{i}\to \pi^{i}+\epsilon^{ijk}\lambda^{j}\pi^{k}\ ,
\end{align}
$\lambda^{i}$ being a real parameter. These are not the global SU(2) gauge transformations: under them, $\pi^{i}$ transforms in the adjoint and $h$ is a scalar, instead of transforming into each other as in Eq.~\eqref{hkmcol}. Rather, the transformations in Eq.~\eqref{higgscustodial} are \textit{custodial} SU(2) transformations, independent of the gauge symmetry, and specific to the SU(2) Higgs-Kibble model. Under perturbative assumptions -- see below -- the asymptotic states of the model are organized into representations of the custodial SU(2) symmetry.\\

The asymptotic analysis of the SU(2) Higgs-Kibble model quantized in linear covariant gauges was carried out by Kugo and Ojima in \cite{KO79b} under the assumption that the exact pole structure of its propagators is the same as the one obtained by expanding perturbatively around the Higgs vacuum -- that is, as the one encoded in the linear terms of the field equations expressed in terms of $h$ and $\pi^{i}$. Under this assumption, like in the Stueckelberg and Schwinger models, the transverse form factor $\Delta(p^{2})$ of the gauge-boson propagator $\Delta_{\mu\nu}(p)$,
\begin{equation}\label{deq045}
   \Delta_{\mu\nu}(p)=\Delta(p^{2})\,t_{\mu\nu}(p)+\frac{-i\xi}{p^{2}}\,\ell_{\mu\nu}(p)\ ,
\end{equation}
develops a pole at $p^{2}=m_{U}^{2}$, where $m_{U}^{2}$ is a finite, renormalized, and RGI mass,
\begin{equation}
   \Delta(p^{2})\to \frac{-i Z_{3}}{p^{2}-m_{U}^{2}}\qquad(p^{2}\to m_{U}^{2})\ ,
\end{equation}
implying that the gauge boson propagator has acquired a mass, and that a massive particle of spin $1$ and mass $m_{U}$ has appeared in the spectrum. Moreover, the ghost propagator $\mc{G}(p^{2})$ has a simple pole at $p^{2}=0$,
\begin{equation}
   \mc{G}(p^{2})\to -\frac{\widetilde{Z}_{3}}{p^{2}}\qquad(p^{2}\to0)\ ,
\end{equation}
the Higgs boson propagator $\Delta_{hh}(p^{2})$ has a pole at the finite nonvanishing RGI mass $p^{2}=m_{h}^{2}\neq 0$,
\begin{equation}
   \Delta_{hh}(p^{2})\to \frac{iZ_{h}}{p^{2}-m_{h}^{2}}\qquad(p^{2}\to m_{h}^{2})\ ,
\end{equation}
and the Goldstone propagator $\Delta_{\pi\pi}(p^{2})$ has a simple pole and a dipole at $p^{2}=0$ \cite{KO79b,KO79a},
\begin{equation}
   \Delta_{\pi\pi}(p^{2})\to\frac{iZ_{\pi}(1-2\xi\mc{N})}{p^{2}}-\frac{iZ_{\pi}\xi M^{2}}{(p^{2})^{2}}\quad(p^{2}\to 0)\ .
\end{equation}
In the last equation, the constant $\mc{N}$ is proportional to the residue of the massless simple pole of the Green's function $\langle T\{A_{\mu}^{i}\pi^{j}\}\rangle$ \cite{KO79b,KO79a}, and $M$ is the zero-momentum value of the gauge-boson propagator,
\begin{equation}\label{deq046}
   \Delta(0)=\frac{i}{M^{2}}\ .
\end{equation}
The constants $\mc{N}$ and $M$ appear in the Goldstone propagator in fixed combinations as a consequence of the Slavnov-Taylor identities of the model -- i.e., of BRST symmetry.

For ease of comparison, we note down the following relations between the constants $L$, $K$, $N$, $\alpha$ defined in \cite{KO79b,KO79a}, and $M$, $\mc{N}$, $\xi$ used in this Appendix\footnote{In \cite{KO79b,KO79a}, the fields $h$ and $\pi^{i}$ are denoted with $\psi$ and $\chi^{a}$, respectively.}:
\begin{align}\label{hkmkoconv}
   &\quad L=M^{-2}\ ,\quad K=Z_{3}^{-1}M^{-2}\ ,\\
   \notag&\mc{N}=Z_{3}^{-1}K^{-1/2}N\ ,\qquad \alpha=Z_{3}^{-1}\xi\ .
\end{align}
We remark that the propagators in Eqs.~\eqref{deq045} - \eqref{deq046} are not renormalized, and that we have suppressed diagonal indices in the equations.

From the pole structure described above, under the assumption that the SU(2) Higgs-Kibble model is asymptotically complete with respect to the asymptotic fields of $A_{\mu}^{i}$, $B^{i}$, $c^{i}$, $\cbar^{i}$, $h$, and $\pi^{i}$, one deduces the following asymptotic limits \cite{KO79b}:
\begin{align}
   \notag&A_{\mu}^{i}\to Z_{3}^{1/2}\,U_{\mu}^{i}+M^{-1}\partial_{\mu}\chi^{i}\ ,\qquad B^{i}\to M\beta^{i}\ ,\\
   &\qquad\qquad c^{i}\to \widetilde{Z}_{3}^{1/2}\gamma^{i}\ ,\qquad \cbar^{i}\to \widetilde{Z}_{3}^{1/2}\gammabar^{i}\ ,\\
   \notag&\qquad\quad h\to Z_{h}^{1/2}\,h^{\text{as.}}\ ,\qquad \pi^{i}\to Z_{\pi}^{1/2}\,\pi^{i\,\text{as.}}\ .
\end{align}
Here, modulo mixing with $\beta^{i}$, the asymptotic field $\chi^{i}$ is provided by $\pi^{i\,\text{as.}}$ itself,
\begin{equation}
   \chi^{i}=\pi^{i\,\text{as.}}+(1-\xi\mc{N})\beta^{i}\ ,
\end{equation}
and the asymptotic fields satisfy the equations
\begin{align}
   \notag &(\partial^{2}+m_{U}^{2})U_{\mu}^{i}=0\ ,\quad \partial^{\mu}U_{\mu}^{i}=0\ ,\quad (\partial^{2}+m_{h}^{2})h=0\ ,\\
   &\qquad\partial^{2}\chi^{i}=\partial^{2}\pi^{i\,\text{as.}}=-\xi M^{2}\beta^{i}\ ,\qquad \partial^{2}\beta^{i}=0\ ,\\
   \notag &\qquad\qquad\quad\ \partial^{2}\gamma^{i}=0\ ,\qquad \partial^{2}\gammabar^{i}=0\ .
\end{align}
Beyond the Landau gauge ($\xi=0$), $\chi^{i}$ and $\pi^{i\,\text{as.}}$ are dipole fields.

The commutation relations of $\chi^{i}$ are not reported explicitly in \cite{KO79b}. They can however be derived easily from those of $\pi^{i\,\text{as.}}$: suppressing diagonal indices,
\begin{align}
   \notag[\pi^{\text{as.}}(x),\pi^{\text{as.}}(y)]&=(1-2\xi\mc{N})iD(x-y)+i\xi M^{2}E(x-y)\ ,\\
   [\pi^{\text{as.}}(x),\beta(y)]&=-iD(x-y)\ .
\end{align}
The nonvanishing (anti)commutation relations between the asymptotic fields $U_{\mu}^{i}$, $h^{\text{as.}}$, $\chi^{i}$, $\beta^{i}$, $\gamma^{i}$, and $\gammabar^{i}$ are then found to be
\begin{align}\label{deq088}
   \notag[U_{\mu}(x),U_{\nu}(y)]&=-i\left(\eta_{\mu\nu}+m^{-2}_{U}\partial_{\mu}\partial_{\nu}\right)D(x-y;m^{2}_{U})\ ,\\
   [h^{\text{as.}}(x),h^{\text{as.}}(y)]&=iD(x-y;m^{2}_{h})\ ,\\
   \notag[\chi(x),\chi(y)]&=-iD(x-y)+i\xi M^{2}E(x-y)\ ,\\
   \notag[\chi(x),\beta(y)]&=-iD(x-y)\ ,\\
   \notag\{\gamma(x),\gammabar(y)\}&=-D(x-y)\ .
\end{align}
Since $U_{\mu}^{i}$ commutes with all massless asymptotic fields, one also has
\begin{align}\label{deq089}
   \notag[A_{\mu}(x),\chi(y)]&=-iM^{-1}\partial_{\mu}D(x-y)+i\xi M\partial_{\mu}E(x-y)\ ,\\
   [A_{\mu}(x),\beta(y)]&=-iM^{-1}\partial_{\mu}D(x-y)\ .
\end{align}

With the exception of the Higgs boson relations, that in Eqs.~\eqref{deq088} and \eqref{deq089} is the same algebra we reported in Appendix~\ref{app:stschw} for the Stueckelberg and Schwinger models. We remark, as we did in Appendix~\ref{app:stschw}, that the field $U_{\mu}^{i}$ does not contribute to the massless component of this algebra.

Under BRST transformations, the asymptotic fields $U_{\mu}^{i}$, $h^{\text{as.}}$, $\chi^{i}$, $\beta^{i}$, $\gamma^{i}$, and $\gammabar^{i}$ transform as
\begin{align}\label{hkmbrstfull}
   &\qquad [iQ_{B,R},U_{\mu}^{i}]=0\ ,\qquad [iQ_{B,R},h^{\text{as.}}]=0\ ,\\
   \notag&[iQ_{B,R},\chi^{i}]=[iQ_{B,R},\pi^{i\,\text{as.}}]=\gamma^{i}\ ,\qquad \{iQ_{B,R},\gamma^{i}\}=0\ ,\\
   \notag &\qquad \{i Q_{B,R},\gammabar^{i}\}=i\beta^{i}\ ,\qquad [i Q_{B,R},\beta^{i}]=0\ ,
\end{align}
where $Q_{B,R}$ is the redefined BRST charge
\begin{equation}
   Q_{B,R}=\widetilde{Z}_{3}^{1/2}M^{-1}\ Q_{B}\ .
\end{equation}
By the assumption of asymptotic completeness, Eq.~\eqref{hkmbrstfull} shows that the BRST cohomology is spanned by the multiparticle in/out states of the physical massive gauge boson $U_{\mu}^{i}$ and of the physical Higgs boson $h^{\text{as.}}$. Moreover, it shows that the unphysical elementary BRST quartet is formed by $\chi^{i}$ -- equivalently, $\pi^{i\,\text{as.}}$ --, $\beta^{i}$, $\gamma^{i}$, and $\gammabar^{i}$. As anticipated, the unphysical components of the Higgs field -- i.e., the Goldstone bosons associated with the spontaneous breaking of the global gauge charge $Q^{i}$ -- are the first parent of the elementary BRST quartet.

Since the model's MYME can be rewritten as in Eq.~\eqref{hkmmyme2}, as a consequence of Eq.~\eqref{hkmbrstfull}, the KO Goldstone equation in its corrected form -- see Sec.~\ref{sec:ssb} --,
\begin{align}
   &\bra{\Omega}J_{\mu}^{i}(x)\ket{\chi^{j}(p)}=\bra{\Omega}J_{\mu}^{i}(x)\ket{\pi^{j}(p)}=\\
   \notag&\quad=iM(1+u)/g\,\delta^{ij}\,\hat{p}_{\mu}\,e^{-i\hat{p}x}\ ,
\end{align}
where $\hat{p}=(|{\bf p}|,{\bf p})$, holds. In \cite{KO79a}, limited to the SU(2) Higgs-Kibble model, the factor of $M^{-1}$ was indeed reported in the definition of the asymptotic field $\chi^{i}$, albeit in the obscure form $Z_{3}^{1/2}K^{1/2}$ -- see Eq.~\eqref{hkmkoconv} and Eq.~(6.26) of \cite{KO79a}. However, the factor of $1+u$ had been previously absorbed into a redefinition of $\chi^{i}$ -- see the Remark below Eq.~(6.22) of \cite{KO79a} --, so the combination $M(1+u)$ never appeared explicitly in the original analysis of Kugo and Ojima.

\section{Vertex Schwinger poles and massless longitudinal singularities in dynamically massive QCD}\label{app:schwpoles}

In this Appendix, we study the massless pole structure of the elementary vertices of dynamically massive QCD. Here we show that, under the customary assumptions of Sec.~\ref{sec:mlessspect}, the first parent $\chi^{a}$ of the elementary BRST quartet of dynamically massive QCD produces Schwinger poles in the three-gluon, four-gluon, ghost-antighost-gluon, and -- in the presence of quarks -- quark-antiquark-gluon vertices. Moreover, we show that the vertices contain additional massless longitudinal singularities produced by the NL field $B^{a}$ which, when summed together in the polarization, do not contribute to dynamical gluon mass generation.\\

\begin{widetext}

Consider the three-point Green's function $G_{\mu\nu\sigma}^{abc}(x,y,z)$ defined as
\begin{equation}
   G_{\mu\nu\sigma}^{abc}(x,y,z)=\bra{\Omega}T\{A_{\mu}^{a}(x)A_{\nu}^{b}(y)A_{\sigma}^{c}(z)\}\ket{\Omega}\ .
\end{equation}
An elementary perturbative calculation carried out to fix conventions shows that, in order to yield the first of Eq.~\eqref{emvertices} at tree level, modulo an overall momentum-conserving delta distribution, the dressed three-gluon vertex $\Gamma_{\mu\nu\sigma}^{abc}(p,q,k)$ must be computed from $G_{\mu\nu\sigma}^{abc}(x,y,z)$ as
\begin{align}\label{amputdef}
   &\Gamma_{\mu\nu\sigma}^{abc}(p,q,k)=g^{-1}\Delta_{\mu}^{-1\,\mu^{\prime}}(p)\Delta_{\nu}^{-1\,\nu^{\prime}}(q)\Delta_{\sigma}^{-1\,\sigma^{\prime}}(k)\int_{x,y,z}e^{-i(px+qy+kz)}\ G_{\mu^{\prime}\nu^{\prime}\sigma^{\prime}}^{abc}(x,y,z)\ .
\end{align}

As our starting point, we observe that the time-ordered product in $G_{\mu\nu\sigma}^{abc}(x,y,z)$ contains terms
\begin{align}\label{deq654}
   \notag&G_{\mu\nu\sigma}^{abc}(x,y,z)=\bra{\Omega}A_{\nu}^{b}(y)A_{\sigma}^{c}(z)A_{\mu}^{a}(x)\ket{\Omega}\,\theta(y^{0}-z^{0})\theta(z^{0}-x^{0})+\bra{\Omega}A_{\sigma}^{c}(z)A_{\nu}^{b}(y)A_{\mu}^{a}(x)\ket{\Omega}\,\theta(z^{0}-y^{0})\theta(y^{0}-x^{0})+\\
   &\ +\bra{\Omega}A_{\mu}^{a}(x)A_{\nu}^{b}(y)A_{\sigma}^{c}(z)\ket{\Omega}\,\theta(x^{0}-y^{0})\theta(y^{0}-z^{0})+\bra{\Omega}A_{\mu}^{a}(x)A_{\sigma}^{c}(z)A_{\nu}^{b}(y)\ket{\Omega}\,\theta(x^{0}-z^{0})\theta(z^{0}-y^{0})+\cdots
\end{align}
in which the gluon field $A_{\mu}^{a}(x)$ acts on the vacuum $\ket{\Omega}$ either from the left or from the right. The one-particle projector $\mathbb{P}_{1G}$ defined in Eq.~\eqref{p1g} can be used to extract the massless discrete contribution associated with this action:
\begin{align}\label{deq549}
   &G_{\mu\nu\sigma}^{abc}(x,y,z)=\\
   \notag&\ =\bra{\Omega}A_{\nu}^{b}(y)A_{\sigma}^{c}(z)\,\mathbb{P}_{1G}\,A_{\mu}^{a}(x)\ket{\Omega}\,\theta(y^{0}-z^{0})\theta(z^{0}-x^{0})+\bra{\Omega}A_{\sigma}^{c}(z)A_{\nu}^{b}(y)\,\mathbb{P}_{1G}\,A_{\mu}^{a}(x)\ket{\Omega}\,\theta(z^{0}-y^{0})\theta(y^{0}-x^{0})+\\
   \notag&\ +\bra{\Omega}A_{\mu}^{a}(x)\mathbb{P}_{1G}A_{\nu}^{b}(y)A_{\sigma}^{c}(z)\ket{\Omega}\,\theta(x^{0}-y^{0})\theta(y^{0}-z^{0})+\bra{\Omega}A_{\mu}^{a}(x)\mathbb{P}_{1G}A_{\sigma}^{c}(z)A_{\nu}^{b}(y)\ket{\Omega}\,\theta(x^{0}-z^{0})\theta(z^{0}-y^{0})+\cdots\ ,
\end{align}
where the dots now also include multiparticle contributions to the matrix elements in Eq.~\eqref{deq654}, and potential massive discrete contributions. By our assumption of asymptotic completeness, since $A_{\mu}^{a}\to M^{-1}\partial_{\mu}\chi^{a}$, and since all asymptotic fields but $\chi^{a}$ and $\beta^{a}$ have vanishing commutation relations with $\chi^{a}$, the four explicit terms in Eq.~\eqref{deq549} saturate the discrete spectrum associated with the external leg $A_{\mu}^{a}(x)$ going on the massless mass shell, $p^{2}\to 0$.

Let us focus on the first term in Eq.~\eqref{deq549}. The matrix element contained in it reads
\begin{align}\label{schwmem0}
   &\bra{\Omega}A_{\nu}^{b}(y)A_{\sigma}^{c}(z)\,\mathds{P}_{1G}\,A_{\mu}^{a}(x)\ket{\Omega}=-M^{-1}\int\frac{d^{3}{\bf p}}{(2\pi)^{3}}\frac{1}{2\hat{p}^{0}}\,\big[\bra{\Omega}A_{\nu}^{b}(y)A_{\sigma}^{c}(z)\ket{\chi^{d}(p)}\bra{\beta^{d}(p)}\partial_{\mu}\chi^{a}(x)\ket{\Omega}+\\
   \notag&\qquad\qquad\qquad +\bra{\Omega}A_{\nu}^{b}(y)A_{\sigma}^{c}(z)\ket{\beta^{d}(p)}\big(\bra{\chi^{d}(p)}-\bra{\beta^{d}(p)}\big)\,\partial_{\mu}\chi^{a}(x)\ket{\Omega}\big]\ ,
\end{align}
where $\hat{p}=(|{\bf p}|,{\bf p})$. We see that the matrix element $\bra{\Omega}A_{\nu}^{b}(y)A_{\sigma}^{c}(z)\ket{\chi^{d}(p)}$ that defines the BSA of $\chi^{a}$ in the two-gluon channel -- see Eq.~\eqref{BSA} -- has appeared in the equation together with a matrix element $\bra{\Omega}A_{\nu}^{b}(y)A_{\sigma}^{c}(z)\ket{\beta^{d}(p)}$ that defines an analogous amplitude for $\beta^{a}$. By repeating the analysis of Sec.~\ref{sec:goldbound} for the state $\ket{\beta^{a}(p)}$, it is easy to see that the latter has a vanishing sum rule with its three-gluon and ghost-antighost counterparts, $ig^{-1}\bra{\Omega}K_{\mu}^{a}(x)\ket{\beta^{b}(p)}=0$, as a consequence of $\bra{\Omega}B^{a}(x)\ket{\beta^{b}(p)}=0$.

In Eq.~\eqref{schwmem0}, using Eq.~\eqref{dipopapp}, the matrix element that multiplies $\bra{\Omega}A_{\nu}^{b}(y)A_{\sigma}^{c}(z)\ket{\beta^{d}(p)}$ can be evaluated as follows:
\begin{align}
   &\big(\bra{\chi^{d}(p)}-\bra{\beta^{d}(p)}\big)\,\partial_{\mu}\chi^{a}\ket{\Omega}=\big(\bra{\chi^{d}(p)}-\bra{\beta^{d}(p)}\big)\,\partial_{\mu}(\widetilde{\chi}^{a}-\xi M^{2}\mc{E}^{(1/2)}\beta^{a})\ket{\Omega}=\\
   \notag &=-\xi M^{2}\big(\bra{\chi^{d}(p)}-\bra{\beta^{d}(p)}\big)\,\partial_{\mu}(\mc{E}^{(1/2)}\beta^{a})\ket{\Omega}=-\xi M^{2}\,\partial_{\mu}(\mc{E}^{(1/2)}\bra{\chi^{d}(p)}\,\beta^{a}\ket{\Omega})=\\
   \notag &=\xi M^{2}\,\delta^{ad}\,\partial_{\mu}(\mc{E}^{(1/2)}e^{i\hat{p}x})=\xi M^{2}\,\delta^{ad}\,\partial_{\mu}\left[-\frac{1}{2\hat{p}_{0}^{2}}(i\hat{p}^{0}x^{0}-1/2)\right]e^{i\hat{p}x}=\\
   \notag&=-i\xi M^{2}\,\delta^{ad}\left[\frac{\hat{p}_{\mu}}{2\hat{p}_{0}^{2}}(i\hat{p}^{0}x^{0}-1/2)+\frac{\delta_{\mu}^{0}}{2\hat{p}_{0}}\right]e^{i\hat{p}x}=-i\xi M^{2}\,\delta^{ad}\left[\frac{\hat{p}_{\mu}}{2\hat{p}_{0}^{2}}(\hat{p}^{0}\partial_{\hat{p}^{0}}-1/2)+\frac{\delta_{\mu}^{0}}{2\hat{p}_{0}}\right]e^{i\hat{p}x}=\\
   \notag&=-i\xi M^{2}\,\delta^{ad}\,L_{\mu}(\hat{p})\,e^{i\hat{p}x}\ .
\end{align}
In the above equation, to go from the first to the second line, we used the fact that $\widetilde{\chi}^{a}$ has the same commutation relations with itself and with $\beta^{a}$. Moreover, on the last line, we defined a differential operator $L_{\mu}(\hat{p})$. It is understood that the derivative $\partial_{\hat{p}^{0}}$ in $L_{\mu}(\hat{p})$ acts by first differentiating its operand with respect to $\hat{p}^{0}$ treated as an independent variable, and then setting $\hat{p}^{0}$ equal to $|{\bf p}|$ in the resulting expression. As for the matrix element that multiplies $\bra{\Omega}A_{\nu}^{b}(y)A_{\sigma}^{c}(z)\ket{\chi^{d}(p)}$ in Eq.~\eqref{schwmem0}, it reads
\begin{align}
   \bra{\beta^{d}(p)}\partial_{\mu}\chi^{a}(x)\ket{\Omega}&=\bra{\beta^{d}(p)}\partial_{\mu}\widetilde{\chi}^{a}(x)\ket{\Omega}=-i\hat{p}_{\mu}\,e^{i\hat{p}x}\,\delta^{ad}\ .
\end{align}
We thus obtain
\begin{align}\label{schwmem1}
   &\bra{\Omega}A_{\nu}^{b}(y)A_{\sigma}^{c}(z)\,\mathds{P}_{1G}\,A_{\mu}^{a}(x)\ket{\Omega}=\\
   \notag&\qquad =iM^{-1}\int\frac{d^{3}{\bf p}}{(2\pi)^{3}}\frac{1}{2\hat{p}^{0}}\,\big[\bra{\Omega}A_{\nu}^{b}(y)A_{\sigma}^{c}(z)\ket{\chi^{a}(p)}\,\hat{p}_{\mu}+\xi M^{2}\bra{\Omega}A_{\nu}^{b}(y)A_{\sigma}^{c}(z)\ket{\beta^{a}(p)}\,L_{\mu}(\hat{p})\big]e^{i\hat{p}x}\ .
\end{align}

Let us Fourier-transform $G_{\mu\nu\sigma}^{abc}(x,y,z)$ with respect to the variable $x$. We have
\begin{align}
   &\int d^{4}x\ e^{-irx}\ \theta(z^{0}-x^{0})\,e^{i\hat{p}x}=(2\pi)^{3}\,\delta^{(3)}({\bf r}-{\bf p})\,\frac{ie^{-i(r^{0}-\hat{p}^{0})z^{0}}}{r^{0}-\hat{p^{0}}+i\epsilon}\ .
\end{align}
Moreover, using $(r^{0}-\hat{p}^{0})/(r^{0}-\hat{p}^{0}+i\epsilon)=1$,
\begin{align}
   L_{\mu}(\hat{p})\,\frac{ie^{-i(r^{0}-\hat{p}^{0})z^{0}}}{r^{0}-\hat{p^{0}}+i\epsilon}&=\frac{ie^{-i(r^{0}-\hat{p}^{0})z^{0}}}{(r^{0}-\hat{p^{0}}+i\epsilon)^{2}}\frac{\hat{p}_{\mu}}{2\hat{p}_{0}}+\frac{ie^{-i(r^{0}-\hat{p}^{0})z^{0}}}{r^{0}-\hat{p^{0}}+i\epsilon}\left[\frac{\hat{p}_{\mu}}{2\hat{p}_{0}^{2}}(i\hat{p}^{0}z^{0}-1/2)+\frac{\delta_{\mu}^{0}}{2\hat{p}_{0}}\right]=\\
   \notag&=\frac{ie^{-i(r^{0}-\hat{p}^{0})z^{0}}}{(r^{0}-\hat{p^{0}}+i\epsilon)^{2}}\frac{\hat{p}_{\mu}+\delta_{\mu}^{0}(r^{0}-\hat{p}^{0})}{2\hat{p}_{0}}+\frac{ie^{-i(r^{0}-\hat{p}^{0})z^{0}}}{r^{0}-\hat{p^{0}}+i\epsilon}\frac{\hat{p}_{\mu}}{2\hat{p}_{0}^{2}}(i\hat{p}^{0}z^{0}-1/2)\ .
\end{align}
Therefore
   \begin{align}
     \notag&\int d^{4}x\ e^{-ipx}\, \bra{\Omega}A_{\nu}^{b}(y)A_{\sigma}^{c}(z)\,\mathds{P}_{1G}\,A_{\mu}^{a}(x)\ket{\Omega}\, \theta(y^{0}-z^{0})\,\theta(z^{0}-x^{0})=iM^{-1}\,\frac{e^{-i(p^{0}-\hat{p}^{0})z^{0}}}{2\hat{p}^{0}}\,\Bigg\{\frac{i\hat{p}_{\mu}\,\bra{\Omega}A_{\nu}^{b}(y)A_{\sigma}^{c}(z)\ket{\chi^{a}(p)}}{p^{0}-\hat{p^{0}}+i\epsilon}+\\
      &\qquad+\xi M^{2}\,\bra{\Omega}A_{\nu}^{b}(y)A_{\sigma}^{c}(z)\ket{\beta^{a}(p)}\left[\frac{i}{(p^{0}-\hat{p^{0}}+i\epsilon)^{2}}\frac{p_{\mu}}{2\hat{p}_{0}}+\frac{i}{p^{0}-\hat{p^{0}}+i\epsilon}\frac{\hat{p}_{\mu}}{2\hat{p}_{0}^{2}}(i\hat{p}^{0}z^{0}-1/2)\right]\Bigg\}\ \theta(y^{0}-z^{0})\ .
   \end{align}
In the limit $p^{0}\to \hat{p}^{0}$, which implies $p^{2}\to0$, we see that the terms on the RHS of the above equation develop simple and double poles. On the other hand, they remain finite as $p^{0}\to -\hat{p}^{0}$. The same goes for the Fourier transform of the second term in Eq.~\eqref{deq549}, which is obtained from the last equation by interchanging $(y,\nu,b)$ and $(z,\sigma,c)$. Specularly, the third and fourth terms in Eq.~\eqref{deq549} develop poles as $p^{0}\to -\hat{p}^{0}$, while they remain finite as $p^{0}\to \hat{p}^{0}$. In the limit $p^{0}\to \hat{p}^{0}$, using $(p^{0}+\hat{p}^{0})/(2\hat{p}^{0})\to 1$, we find
\begin{align}\label{deq380}
     \notag&\int d^{4}x\ e^{-ipx}\, \bra{\Omega}A_{\nu}^{b}(y)A_{\sigma}^{c}(z)\,\mathds{P}_{1G}\,A_{\mu}^{a}(x)\ket{\Omega}\, \theta(y^{0}-z^{0})\,\theta(z^{0}-x^{0})\to iM^{-1}\,\bigg\{\frac{ip_{\mu}\,\bra{\Omega}A_{\nu}^{b}(y)A_{\sigma}^{c}(z)\ket{\chi^{a}(p)}}{p^{2}+i\epsilon}+\\
      &\qquad+\xi M^{2}\,\bra{\Omega}A_{\nu}^{b}(y)A_{\sigma}^{c}(z)\ket{\beta^{a}(p)}\left[\frac{ip_{\mu}}{(p^{2}+i\epsilon)^{2}}+\frac{ip_{\mu}}{p^{2}+i\epsilon}\frac{i\hat{p}^{0}z^{0}-1/2}{2\hat{p}_{0}^{2}}\right]\bigg\}\ \theta(y^{0}-z^{0})\ .
\end{align}

Except for the Landau gauge ($\xi=0$), the first and the third term on the RHS of Eq.~\eqref{deq380} are not individually Poincar\'{e} covariant: the third term contains the noncovariant quantities $\hat{p}_{0}^{2}$ and $\hat{p}^{0}z^{0}$, whereas the first one contains the matrix element $\bra{\Omega}A_{\nu}^{b}(y)A_{\sigma}^{c}(z)\ket{\chi^{a}(p)}$, which is not covariant for $\xi\neq0$ due to the dipolar nature of $\ket{\chi^{a}(p)}$ -- see the discussion in Appendix~\ref{app:dipole}. Nevertheless, since the Green's function as a whole is Poincar\'{e} covariant, and since the four terms in Eq.~\eqref{deq549} yield its massless $p^{2}\to 0$ pole structure in full, we know that these terms, when expressed as in Eq.~\eqref{deq380}, will eventually sum to a quantity which is Poincar\'{e} covariant modulo higher-order, non-pole corrections even for $\xi\neq 0$.

Let us check this is the case explicitly for translational covariance. Indeed, since the LHS of Eq.~\eqref{deq380} is covariant under spacetime translations, it follows that the RHS must be as well. Using Eq.~\eqref{dipoletransl} with $\kappa=\xi M^{2}$ by the third of Eq.~\eqref{chistatedip}, we have that
\begin{align}
   \bra{\Omega}A_{\nu}^{b}(y)A_{\sigma}^{c}(z)\ket{\chi^{a}(p)}&=\bra{\Omega}A_{\nu}^{b}(y-z)A_{\sigma}^{c}(0)e^{-iPz}\ket{\chi^{a}(p)}=e^{-i\hat{p}z}\bra{\Omega}A_{\nu}^{b}(y-z)A_{\sigma}^{c}(0)\left(\ket{\chi^{a}(p)}-\frac{i\xi M^{2} z^{0}}{2\hat{p}_{0}}\ket{\beta^{a}(p)}\right)\ .
\end{align}
The second, noncovariant term in the above transformation is cancelled by an equal and opposite term on the RHS of Eq.~\eqref{deq380}, so the sum of the first and of the third term on the RHS of Eq.~\eqref{deq380} indeed has the correct transformation properties under spacetime translations.

In what follows, we will ignore noncovariant splitting of terms. Instead, we will treat all quantities as if, even beyond the Landau gauge, they appeared in the equations in suitable Poincar\'{e}-covariant combinations. As we will discuss below Eq.~\eqref{polschwpole}, in Poincar\'{e}-covariant quantities like the gluon-scalar transition amplitude, these combinations are in any case interchangeable with the original noncovariant quantities. We thus set
\begin{align}
     &\int d^{4}x\ e^{-ipx}\, \bra{\Omega}A_{\nu}^{b}(y)A_{\sigma}^{c}(z)\,\mathds{P}_{1G}\,A_{\mu}^{a}(x)\ket{\Omega}\, \theta(y^{0}-z^{0})\,\theta(z^{0}-x^{0})\to\\
     \notag&\qquad\to \bigg\{-\frac{M^{-1}p_{\mu}}{p^{2}+i\epsilon}\,\bra{\Omega}A_{\nu}^{b}(y)A_{\sigma}^{c}(z)\ket{\chi^{a}(p)}-\frac{\xi M p_{\mu}}{(p^{2}+i\epsilon)^{2}}\bra{\Omega}A_{\nu}^{b}(y)A_{\sigma}^{c}(z)\ket{\beta^{a}(p)}\bigg\}\ \theta(y^{0}-z^{0})
\end{align}
as $p^{0}\to \hat{p}^{0}$, it being understood that the matrix element of $\ket{\chi^{a}(p)}$ must be treated as a Poincar\'{e}-covariant quantity even when $\xi \neq 0$. With this understanding, we find that, in the limit $p^{0}\to \hat{p}^{0}$,
   \begin{align}\label{deq756}
      &G_{\mu\nu\sigma}^{abc}(p,y,z)=\int d^{4}x\ e^{-ipx}\ G_{\mu\nu\sigma}^{abc}(x,y,z)=\\
      \notag&\quad=-\frac{M^{-1}p_{\mu}}{p^{2}+i\epsilon}\ \bra{\Omega}T\{A_{\nu}^{b}(y)A_{\sigma}^{c}(z)\}\ket{\chi^{a}(p)}-\frac{\xi M p_{\mu}}{(p^{2}+i\epsilon)^{2}}\ \bra{\Omega}T\{A_{\nu}^{b}(y)A_{\sigma}^{c}(z)\}\ket{\beta^{a}(p)}+\cdots=\\
      \notag&\quad=-\frac{M^{-1}p_{\mu}}{p^{2}+i\epsilon}\ B^{bca}_{\nu\sigma}(y,z;\hat{p})-\frac{\xi M p_{\mu}}{(p^{2}+i\epsilon)^{2}}\ B^{bca}_{(\beta)\nu\sigma}(y,z;\hat{p})+\cdots\ ,
   \end{align}
where $B^{bca}_{\nu\sigma}(y,z;\hat{p})$ is the (Poincar\'{e}-covariant counterpart of the) BSA defined in Eq.~\eqref{BSA}, $B^{bca}_{(\beta)\nu\sigma}(y,z;\hat{p})$ is the corresponding amplitude of $\beta^{a}$, and the ellipsis denotes terms which are nonsingular as $p^{0}\to \hat{p}^{0}$.

In the limit $p^{0}\to -\hat{p}^{0}$, the poles produced by the third and fourth term in Eq.~\eqref{deq549} take the same form as the last line of Eq.~\eqref{deq756}, with the amplitudes on its second line replaced by $\bra{\chi^{a}(-p)}T\{A_{\nu}^{b}(y)A_{\sigma}^{c}(z)\}\ket{\Omega}$ and $\bra{\beta^{a}(-p)}T\{A_{\nu}^{b}(y)A_{\sigma}^{c}(z)\}\ket{\Omega}$. Therefore, overall, modulo a momentum-conserving delta distribution, we find that
\begin{align}\label{greenspoles}
      G_{\mu\nu\sigma}^{abc}(p,q,k)&=\int_{x,y,z}e^{-i(px+qy+kz)}\ G_{\mu\nu\sigma}^{abc}(x,y,z)=\\
      \notag&=R_{\mu\nu\sigma}^{abc}(p,q,k)-\frac{M^{-1}p_{\mu}}{p^{2}+i\epsilon}\ B^{bca}_{\nu\sigma}(-q,-k,p)-\frac{\xi M p_{\mu}}{(p^{2}+i\epsilon)^{2}}\ B^{bca}_{(\beta)\nu\sigma}(-q,-k,p)\ ,
   \end{align}
where $R_{\mu\nu\sigma}^{abc}(p,q,k)$ is a tensor which is regular as $p^{2}\to 0$, whereas, in the same limit, $B^{bca}_{\nu\sigma}(-q,-k,p)$ and $B^{bca}_{(\beta)\nu\sigma}(-q,-k,p)$ approach the Fourier transforms of the respective covariant two-gluon amplitudes\footnote{Here the minus signs reflect the definition of the Fourier transform of $\chi^{a}$'s BSAs, whose momenta have the opposite sign -- see Eq.~\eqref{BSV} and below.}.

As an intermediate step in our derivation, consider the Green's function $G_{B\nu\sigma}^{abc}(x,y,z)$ defined as
\begin{equation}
   G_{B\nu\sigma}^{abc}(x,y,z)=\bra{\Omega}T\{B^{a}(x)A_{\nu}^{b}(y)A_{\sigma}^{c}(z)\}\ket{\Omega}=-\xi^{-1}\partial^{\mu}_{(x)}\bra{\Omega}T\{A_{\mu}^{a}(x)A_{\nu}^{b}(y)A_{\sigma}^{c}(z)\}\ket{\Omega}\ ;
\end{equation}
equivalently, again modulo a momentum-conserving delta distribution,
\begin{equation}\label{gbnb}
   G_{B\nu\sigma}^{abc}(p,q,k)=\int_{x,y,z}e^{-i(px+qy+kz)}\ G_{B\nu\sigma}^{abc}(x,y,z)=-i\xi^{-1}\,p^{\mu}G_{\mu\nu\sigma}^{abc}(p,q,k)\ .
\end{equation}
Since, by our assumption of asymptotic completeness, the only nonvanishing matrix element of $B^{a}(x)$ between the vacuum and a one-particle state is $\bra{\chi^{b}(p)}B^{a}(x)\ket{\Omega}=Me^{i\hat{p}x}$, a calculation identical to the one we carried out for $G_{\mu\nu\sigma}^{abc}(p,q,k)$ yields
\begin{equation}\label{deq471}
   G_{B\nu\sigma}^{abc}(p,q,k)=\frac{iM}{p^{2}+i\epsilon}\,B_{(\beta)\nu\sigma}^{bca}(-q,-k,p)+\text{subdominant terms}\ .
\end{equation}
Moreover, using Eqs.~\eqref{greenspoles} and \eqref{gbnb}, dropping $i\epsilon$'s, we find that
\begin{equation}\label{rdiv}
   p^{\mu}R_{\mu\nu\sigma}^{abc}(p,q,k)=i\xi\,G_{B\nu\sigma}^{abc}(p,q,k)+M^{-1}\, B^{bca}_{\nu\sigma}(-q,-k,p)+\frac{\xi M}{p^{2}}\, B^{bca}_{(\beta)\nu\sigma}(-q,-k,p)\ .
\end{equation}

Let us now turn to the amputation of the Green's function. Suppressing all indices and momenta except for $\mu$ and $p$ to simplify notation, and focusing on the amputation of the external leg $(p,a,\mu)$ alone, with the inverse gluon propagator $\Delta^{-1}_{\mu\nu}(p)$ expressed as
\begin{equation}
   \Delta_{\mu\nu}^{-1}(p)=\Delta^{-1}(p^{2})\,t_{\mu\nu}(p)+i\xi^{-1}p_{\mu}p_{\nu}\ ,
\end{equation}
using Eqs.~\eqref{greenspoles} and \eqref{gbnb}, we find that the partially amputated vertex $\Gamma_{\mu}^{\text{part.}}(p)$ reads
\begin{equation}\label{gammaimpl}
   g\Gamma_{\mu}^{\text{part.}}(p)=\Delta^{-1}(p)_{\mu}^{\nu}G_{\nu}(p)=\Delta^{-1}(p^{2})\,t_{\mu}^{\nu}(p)R_{\nu}(p)-p_{\mu}\,G_{B}(p)\ .
\end{equation}
Observe that $p^{\mu}\Gamma_{\mu}^{\text{part.}}(p)=-g^{-1}p^{2}G_{B}(p)$: the tensor $-g^{-1}p^{2}G_{B}(p)$ determines the RHS of the Slavnov-Taylor identities of the vertex.

Going ahead, plugging Eq.~\eqref{rdiv} into \eqref{gammaimpl}, we obtain
\begin{equation}\label{fullvertex}
   g\Gamma_{\mu}^{\text{part.}}(p)=\Delta^{-1}(p^{2})\,R_{\mu}(p)-\Delta^{-1}(p^{2})\,\frac{p_{\mu}}{p^{2}}\left\{M^{-1}\, B(p)+\frac{\xi M}{p^{2}}\left[\frac{ip^{2}}{M}\,G_{B}(p)+B_{(\beta)}(p)\right]\right\}-p_{\mu}\,G_{B}(p)\ .
\end{equation}
From this equation, it is easy to isolate the singular part of the partially amputated vertex $\Gamma_{\mu}^{\text{part.}}(p)$. With $\Delta^{-1}(0)=-iM^{2}$, using Eq.~\eqref{deq471}, we find that
\begin{equation}\label{singularvertex}
   [\Gamma_{\mu}^{\text{part.}}(p)]_{\text{singular}}=\frac{ig^{-1}Mp_{\mu}}{p^{2}}\left[B(p)-B_{(\beta)}(p)+\xi\,V_{B}(p)\right]\ ,
\end{equation}
where
\begin{equation}
   V_{B}(p)=\lim_{p^{2}\to 0}\,iM\left[G_{B}(p)-\frac{iM}{p^{2}}\,B_{(\beta)}(p)\right]\ .
\end{equation}
Note that the third term in Eq.~\eqref{singularvertex} vanishes in the Landau gauge ($\xi=0$); $V_{B}(p)$ thus only contributes to the singular part of the vertex for $\xi \neq 0$.

Reinstating indices and momenta, and carrying out the amputations of the remaining external legs, we find that the singular part of the vertex $\Gamma_{\mu\nu\sigma}^{abc}(p,q,k)$ in Eq.~\eqref{amputdef} reads
\begin{equation}\label{singularvertexexp}
   [\Gamma_{\mu\nu\sigma}^{abc}(p,q,k)]_{\text{singular}}=I_{\mu}(p)\,\frac{i}{p^{2}+i\epsilon}\left\{\mc{B}_{\nu\sigma}^{bca}(q,k,p)-\mc{B}_{(\beta)\nu\sigma}^{bca}(q,k,p)+\xi \,\mc{V}_{B\nu\sigma}^{abc}(p,q,k)\right\}\ .
\end{equation}
In the above equation, $\mc{B}_{\nu\sigma}^{bca}(q,k,p)$ approaches the (Poincar\'{e}-covariant counterpart of the) BSV of $\chi^{a}$ defined in Sec.~\ref{sec:goldbound} as $p^{2}\to 0$, $\mc{B}_{(\beta)\nu\sigma}^{bca}(q,k,p)$ approaches the corresponding vertex of $\beta^{a}$, $\mc{V}_{B\nu\sigma}^{abc}(p,q,k)$ is amputated like the other quantities in the equation, and we have introduced the gluon-scalar transition amplitude $I_{\mu}(p)=g^{-1}Mp_{\mu}$ -- see Eq.~\eqref{sumr2}.\\

The derivation of the singular $p^{2}\to 0$ limits of the four-gluon, ghost-antighost-gluon and quark-antiquark-gluon vertices, where $p$ is an external gluon leg, proceeds along the same lines as that of the three-gluon vertex. With the tree-level quark-antiquark-gluon vertex $\Gamma_{(0)\mu}^{a}(p,q,k)$ defined as
\begin{equation}
   \Gamma_{(0)\mu}^{a}(p,q,k)=i\gamma_{\mu}T^{a}\ ,
\end{equation}
and the remaining tree-level vertices as in Eq.~\eqref{emvertices}, modulo overall momentum-space conserving delta distributions, the definitions of the dressed vertices read
\begin{align}
   \notag\Gamma_{\mu\nu\sigma\tau}^{abcd}(p,q,k,r)&=g^{-2}\Delta_{\mu}^{-1\,\mu^{\prime}}(p)\Delta_{\nu}^{-1\,\nu^{\prime}}(q)\Delta_{\sigma}^{-1\,\sigma^{\prime}}(k)\Delta_{\tau}^{-1\,\tau^{\prime}}(r)\times\\
   &\qquad\qquad\times\int_{x,y,z,w}e^{-i(px+qy+kz+rw)}\ \left\langle T\left\{A_{\mu^{\prime}}^{a}(x)A_{\nu^{\prime}}^{b}(y)A_{\sigma^{\prime}}^{c}(z)A_{\tau^{\prime}}^{d}(w)\right\}\right\rangle\ ,\\
   \notag\Gamma_{\mu}^{abc}(p,q,k)&=g^{-1}\Delta_{\mu}^{-1\,\mu^{\prime}}(p)\mc{G}^{-1}(q)\mc{G}^{-1}(k)\int_{x,y,z}e^{-i(px-ky+qz)}\ \left\langle T\left\{A_{\mu^{\prime}}^{a}(x)c^{c}(y)\overline{c}^{b}(z)\right\}\right\rangle\ ,\\
   \notag\Gamma_{\mu}^{a}(p,q,k)&=g^{-1}\Delta_{\mu}^{-1\,\mu^{\prime}}(p)S^{-1}(q)S^{-1}(k)\int_{x,y,z}e^{-i(px-qy+kz)}\ \left\langle T\left\{A_{\mu^{\prime}}^{a}(x)\psi(y)\psibar(z)\right\}\right\rangle\ ,
\end{align}
where $S(q)$ is the dressed quark propagator. By isolating the one-particle contributions of $\chi^{a}$ and $\beta^{a}$, Fourier transforming, and amputating the Green's functions using the obvious analogues of Eq.~\eqref{rdiv}, one finds that
\begin{align}\label{singularverticesexp}
   [\Gamma_{\mu\nu\sigma\tau}^{abcd}(p,q,k,r)]_{\text{singular}}&=g^{-1}I_{\mu}(p)\,\frac{i}{p^{2}+i\epsilon}\left\{\mc{B}_{\nu\sigma\tau}^{bcda}(q,k,r,p)-\mc{B}_{(\beta)\nu\sigma\tau}^{bcda}(q,k,r,p)+\xi \,\mc{V}_{B\nu\sigma\tau}^{abcd}(p,q,k,r)\right\}\ ,\\
   \notag [\Gamma_{\mu}^{abc}(p,q,k)]_{\text{singular}}&=I_{\mu}(p)\,\frac{i}{p^{2}+i\epsilon}\left\{\mc{B}^{cba}(-k,q,p)-\mc{B}_{(\beta)}^{cba}(-k,q,p)+\xi \,\mc{V}_{B}^{abc}(p,-k,q)\right\}\ ,\\
   \notag[\Gamma_{\mu}^{a}(p,q,k)]_{\text{singular}}&=I_{\mu}(p)\,\frac{i}{p^{2}+i\epsilon}\left\{\mc{B}^{a}(-q,k,p)-\mc{B}_{(\beta)}^{a}(-q,k,p)+\xi \,\mc{V}_{B}^{a}(p,-q,k)\right\}\ ,
\end{align}
where the quark-antiquark BSA and BSV are defined as
\begin{equation}
   B^{a}(x,y;\hat{p})=\bra{\Omega}T\{\psi(x)\psibar(y)\}\ket{\chi^{a}(p)}\ ,\qquad B^{a}(q,k;\hat{p})=S(q)S(-k)\mc{B}^{a}(-q,-k;\hat{p})\ ,
\end{equation}
whereas the other quantities are defined as in Sec.~\ref{sec:goldbound} and as we did in this Appendix for the three-gluon vertex. It is understood that analogous limits can be derived for any other gluon leg going on the massless mass shell.

The singular limit of the four-gluon vertex in the first of Eq.~\eqref{singularverticesexp} contains an extra factor of $g^{-1}$ due to a mismatch between the definition of the vertex, which is multiplied by $g^{-2}$, and that of the gluon-scalar transition amplitude, which is multiplied by $g^{-1}$. Moreover, neither the LHS nor the RHS are connected and one-particle irreducible, having been computed straight from a Green's function.\\

When summed together inside the gluon polarization, the vertex singularities produced by $\beta^{a}$ and by the STI Green's functions $G_{B}(p)$ do not contribute to the zero-momentum limit of the propagator -- i.e., to the gluon mass. Let us provide both a simple, indirect proof of this statement, and a lengthier, direct proof.

First of all, consider the SDE of the gluon propagator, Eq.~\eqref{gpsde},
\begin{equation}
   \Pi_{\mu\nu}^{ab}(p)=i\,\left[\langle T\{K A\}\rangle_{\text{F.T.}}(p)\cdot \Delta^{-1}(p)\right]_{\mu\nu}^{ab}\ .
\end{equation}
Since the polarization tensor $\Pi_{\mu\nu}^{ab}(p)$ is transverse and the inverse gluon propagator $\Delta^{-1ab}_{\mu\nu}(p)$ is not, we know that the Green's function $\langle T\{K_{\mu}^{a}A_{\nu}^{b}\}\rangle$ is transverse. Therefore, in the above equation, we can suppress the longitudinal component of $\Delta^{-1ab}_{\mu\nu}(p)$, and rewrite the equation as
\begin{equation}
   \Pi_{\mu\nu}^{ab}(p)=i\,\langle T\{K_{\mu}^{a} A_{\nu}^{b}\}\rangle_{\text{F.T.}}(p)\ \Delta^{-1}(p^{2})\ .
\end{equation}
By inserting the one-particle projector $\mathds{P}_{1G}$ of Eq.~\eqref{p1g} into the Green's function, observing -- as we did earlier in this Appendix -- that $\bra{\Omega}K_{\mu}^{a}(x)\ket{\beta^{b}(p)}=0$, we find that its dominant $p^{2}\to 0$ contribution is given by
\begin{equation}\label{polschwpoleprel}
   \bra{\Omega}K_{\mu}^{a}(x)A_{\nu}^{b}(0)\ket{\Omega}\to-\int\frac{d^{3}{\bf p}}{(2\pi)^{3}}\,\frac{1}{2\hat{p}^{0}}\bra{\Omega}K_{\mu}^{a}(x)\ket{\chi^{c}(p)}\bra{\beta^{c}(p)}A_{\nu}^{b}(0)\ket{\Omega}=\int\frac{d^{3}{\bf p}}{(2\pi)^{3}}\,\frac{1}{2\hat{p}^{0}}\ gM^{-1}I_{\mu}(-\hat{p})\,\hat{p}_{\nu}\ \delta^{ab}\ e^{-i\hat{p}x}\ ,
\end{equation}
where $I_{\mu}(-\hat{p})\,\delta^{ab}=ig^{-1}\bra{\Omega}K_{\mu}^{a}(0)\ket{\chi^{b}(p)}=-g^{-1}M\hat{p}_{\mu}\,\delta^{ab}$ is the gluon-scalar transition amplitude -- see Eqs.~\eqref{sumr2} and \eqref{g2saker}. It follows that, in the $p^{2}\to 0$ limit,
\begin{equation}\label{polschwpole}
   i\langle T\{K_{\mu}^{a} A_{\nu}^{b}\}\rangle_{\text{F.T.}}(p)\ \Delta^{-1}(p^{2})\to\frac{i}{p^{2}}\ g\,I_{\mu}(-\hat{p})\,M\hat{p}_{\nu}\,\delta^{ab}=\frac{i}{p^{2}}\,g^{2}I_{\mu}(-\hat{p})I_{\nu}(\hat{p})\,\delta^{ab}=-\Pi(0)\,\frac{\hat{p}_{\mu}\hat{p}_{\nu}}{p^{2}}\ \delta^{ab}\ ,
\end{equation}
where $\Pi_{\mu\nu}^{ab}(p)=\Pi(p^{2})\,\delta^{ab}t_{\mu\nu}(p)$. By inspection of the analytic expression of the gluon-scalar transition amplitude $I_{\mu}(-\hat{p})$, Eq.~\eqref{sumr2}, we see that the only quantities that appear in it, besides the tree-level vertices and the dressed propagators, are the BSVs of $\chi^{a}$. Therefore, only these BSVs contribute to the zero-momentum limit of the gluon propagator: the vertex singularities produced by $\beta^{a}$ and by the STI Green's functions $G_{B}(p)$ do not. Note that, as a consequence of $\bra{\Omega}K_{\mu}^{a}(x)\ket{\beta^{b}(p)}=0$, the gluon-scalar transition amplitude is Poincar\'{e}-covariant also beyond the Landau gauge ($\xi=0$) -- see the discussion at the end of Appendix~\ref{app:dipole}. Therefore, the BSVs that appear in the analytic expression of the amplitude can be interpreted either as noncovariant or as covariant quantities: any noncovariant term will cancel in the sum in any case.

The cancellation of the singular $\beta^{a}$ and $G_{B}(p)$ vertex contributions inside the gluon polarization can be regarded as a consequence of the transversality of the latter -- equivalently, of the transversality of the Green's function $\langle T\{K_{\mu}^{a}A_{\nu}^{b}\}\rangle$,
\begin{equation}\label{deq205}
   0=p^{\nu}\Pi_{\mu\nu}^{ab}(p)=ip^{\nu}\left[\langle T\{K A\}\rangle_{\text{F.T.}}(p)\cdot \Delta^{-1}(p)\right]_{\mu\nu}^{ab}=-\xi^{-1}p^{2}\,p^{\nu}\,\langle T\{K_{\mu}^{a} A_{\nu}^{b}\}\rangle_{\text{F.T.}}(p)\ .
\end{equation}
Indeed, as we will show in a moment, the transversality of the polarization is in one to one correspondence with the vanishing of the Green's function $\langle T\{K_{\mu}^{a} B^{b}\}\rangle$, which in turn can be used to prove that the singular $\beta^{a}$ and $G_{B}(p)$ vertex terms do not contribute to the polarization when summed together.

To prove the cancellation, let us first check that the equal-time vacuum expectation value $\langle [K_{\mu}^{a}({\bf x}),A_{0}^{b}({\bf y})]\rangle$ vanishes, so that, by Eq.~\eqref{deq205},
\begin{equation}
   p^{\nu}\Pi_{\mu\nu}^{ab}(p)=ip^{2}\,\langle T\{K_{\mu}^{a} B^{b}\}\rangle_{\text{F.T.}}(p)\ .
\end{equation}
Using Eq.~\eqref{kernelop}, together with $[\partial_{0}\cbar^{a}({\bf x}),A_{0}^{b}({\bf y})]=-i[\Pi^{a}({\bf x}),A_{0}^{b}({\bf y})]=0$, $[F_{i0}^{a}({\bf x}),A_{0}^{b}({\bf y})]=[\Pi_{i}^{a}({\bf x}),A_{0}^{b}({\bf y})]=0$ -- see Eqs.~\eqref{conjmom} - \eqref{cancomm} --, we have
\begin{align}
   [K_{\mu}^{a}({\bf x}),A_{0}^{b}({\bf y})]&=gf^{acd}[\partial^{\nu}(A_{\nu}^{c}A_{\mu}^{d})({\bf x}),A_{0}^{b}({\bf y})]=gf^{acd}[-\xi B^{c}({\bf x})A_{\mu}^{d}({\bf x})+A_{\nu}^{c}({\bf x})\partial^{\nu}A_{\mu}^{d}({\bf x}),A_{0}^{b}({\bf y})]=\\
   \notag&=gf^{acd}\left\{i\xi\delta^{bc}A_{\mu}^{d}({\bf x})\,\delta^{(3)}({\bf x}-{\bf y})+A_{0}^{c}({\bf x})[\partial_{0}A_{\mu}^{d}({\bf x}),A_{0}^{b}({\bf y})]\right\}\ .
\end{align}
Then, using $\partial_{0}A_{i}^{a}=-\Pi_{i}^{a}+\partial_{i}A_{0}^{a}+gf^{abc}A^{b}_{i}A^{c}_{0}$, $\partial_{0}A_{0}^{a}=-\xi B^{a}-\partial^{i}A_{i}^{a}$, and $\langle A_{\mu}^{a}\rangle=0$,
\begin{align}
   \left\langle[K_{\mu}^{a}({\bf x}),A_{0}^{b}({\bf y})]\right\rangle&=gf^{acd}\left\langle i\xi\delta^{bc}A_{\mu}^{d}({\bf x})\,\delta^{(3)}({\bf x}-{\bf y})+i\xi\delta_{\mu}^{0}\delta^{bd} A_{0}^{c}({\bf x})\,\delta^{(3)}({\bf x}-{\bf y})\right\rangle=0\ ,
\end{align}
and the commutator vanishes. In particular,
\begin{equation}
   p^{\nu}\Pi_{\mu\nu}^{ab}(p)=0\qquad\Longleftrightarrow\qquad \langle T\{K_{\mu}^{a} B^{b}\}\rangle=0\ .
\end{equation}

Let us now write down the analytic expression of the polarization's SDE as in Sec.~\ref{sec:goldbound},
\begin{align}\label{deq002}
      [\Pi^{\mu}_{\sigma}(p)]_{a}^{d}&=\frac{g^{2}}{2}\ \int_{q,k}\ \Gamma^{\mu\nu\tau}_{(0)\,abc}(-p,-q,-k)\ \Delta_{\nu}^{bb^{\prime}\nu^{\prime}}(q)\,\Delta_{\tau}^{cc^{\prime}\tau^{\prime}}(k)\ \Gamma_{\sigma\nu^{\prime}\tau^{\prime}}^{db^{\prime}c^{\prime}}(p,q,k)\ \delta_{q+k+p,0}+\\
      \notag&\ \ +\frac{g^{4}}{6}\ \int_{q,k,r}\ \Gamma^{\mu\nu\tau\lambda}_{(0)\,abcf}(-p,-q,-k,-r)\ \Delta_{\nu}^{bb^{\prime}\nu^{\prime}}(q)\,\Delta_{\tau}^{cc^{\prime}\tau^{\prime}}(k)\,\Delta_{\lambda}^{ff^{\prime}\lambda^{\prime}}(r)\ \Gamma_{\sigma\nu^{\prime}\tau^{\prime}\lambda^{\prime}}^{db^{\prime}c^{\prime}f^{\prime}}(p,q,k,r)\  \delta_{q+k+r+p,0}+\\
      \notag&\ \ -g^{2}\int_{q,k}\ \Gamma_{(0)\,abc}^{\mu}(-p,q,k)\ \mc{G}^{bb^{\prime}}(q)\,\mc{G}^{cc^{\prime}}(k)\ \Gamma^{dc^{\prime}b^{\prime}}_{\sigma}(p,k,q)\ \delta_{-q+k+p,0}\ .
\end{align}
Here, for the purposes of this Appendix, we have absorbed the gluon tadpole into the nonreduced four-gluon vertex $\Gamma_{\sigma\nu^{\prime}\tau^{\prime}\lambda^{\prime}}^{db^{\prime}c^{\prime}f^{\prime}}(p,q,k,r)$ -- see footnote~\ref{fn:tadpole}. Observing that the products of propagators and dressed vertices in Eq.~\eqref{deq002} are but the partially amputated vertices of Eq.~\eqref{fullvertex} and of its four-gluon and ghost-antighost-gluon analogues, we see that the STI Green's functions $G_{B}(p)$ contained in the elementary dressed QCD vertices by Eq.~\eqref{fullvertex} contribute a polarization term $\delta\Pi_{\mu\nu}^{ab}(p)$ whose expression reads
\begin{align}\label{deq555}
      [\delta\Pi^{\mu}_{\sigma}(p)]_{a}^{d}&=f(p^{2})\,p_{\sigma}\,\Bigg\{\frac{g}{2}\ \int_{q,k}\ \Gamma^{\mu\nu\tau}_{(0)\,abc}(-p,-q,-k)\ G_{B\nu\tau}^{dbc}(p,q,k)\ \delta_{q+k+p,0}+\\
      \notag&\qquad\qquad+\frac{g^{2}}{6}\ \int_{q,k,r}\ \Gamma^{\mu\nu\tau\lambda}_{(0)\,abcf}(-p,-q,-k,-r)\ G_{B\nu\tau\lambda}^{dbcf}(p,q,k,r)\  \delta_{q+k+r+p,0}+\\
      \notag&\qquad\qquad\ -g\int_{q,k}\ \Gamma_{(0)\,abc}^{\mu}(-p,q,k)\ G^{dcb}_{B}(p,-q,k)\ \delta_{-q+k+p,0}\Bigg\}\ ,
\end{align}
where $f(p^{2})$ is a function of $p^{2}$ alone, whereas the four-gluon and ghost-antighost-gluon STI Green's functions $G_{B}(p)$, modulo momentum-conserving delta distributions, are given by
\begin{align}
   G_{B\nu\tau\lambda}^{dbcf}(p,q,k,r)&=\int_{x,y,z,w}e^{-i(px+qy+kz+rw)}\ \left\langle T\left\{B^{d}(x)A_{\nu}^{b}(y)A_{\tau}^{c}(z)A_{\lambda}^{f}(w)\right\}\right\rangle\ ,\\
   \notag G_{B}^{dcb}(p,q,k)&=\int_{x,y,z}e^{-i(px+qy+kz)}\ \left\langle T\left\{B^{d}(x)c^{b}(y)\overline{c}^{c}(z)\right\}\right\rangle\ .
\end{align}
By carrying out the integrations, it is easy to see that $\delta\Pi_{\mu\nu}^{ab}(p)$ in Eq.~\eqref{deq555} is proportional to the Fourier transform of the vanishing Green's function $\langle T\{K_{\mu}^{a} B^{b}\}\rangle$:
\begin{equation}
   \delta\Pi_{\mu\nu}^{ab}(p)=if(p^{2})\,p_{\nu}\ \langle T\{K^{a}_{\mu} B^{b}\}\rangle_{\text{F.T.}}(p)=0\ .
\end{equation}
It follows that the STI Green's functions $G_{B}(p)$ that enter the elementary dressed QCD vertices by Eq.~\eqref{fullvertex} do not contribute to the gluon polarization when all diagrams are summed together. In particular, the singularities they produce in the QCD vertices do not contribute to the zero-momentum limit of the propagator, hence to the gluon mass.

As for the vertex singularities produced by the amplitudes of $\beta^{a}$, the vanishing of their contribution to the polarization can be regarded either 1. by Eq.~\eqref{deq471} and its four-gluon and ghost-antighost-gluon analogues, as a direct consequence of the vanishing of the STI contributions, or 2. as a consequence of the vanishing of the sum rule of $\beta^{a}$, $ig^{-1}\bra{\Omega}K_{\mu}^{a}(x)\ket{\beta^{b}(p)}=0$. Indeed, in reference to the latter, it is sufficient to note that, modulo a $p^{2}$-dependent prefactor, the vertices $\mc{B}_{(\beta)}(p)$ of $\beta^{a}$ enter Eqs.~\eqref{singularvertexexp} and \eqref{singularverticesexp} exactly like the BSVs $\mc{B}(p)$ of $\chi^{a}$. Since the BSVs $\mc{B}(p)$ of $\chi^{a}$ enter the zero-momentum limit of the polarization through the gluon-scalar transition amplitude $ig^{-1}\bra{\Omega}K_{\mu}^{a}(x)\ket{\chi^{b}(p)}\neq 0$, the corresponding vertices of $\beta^{a}$ enter the polarization through a term proportional to the vanishing amplitude $ig^{-1}\bra{\Omega}K_{\mu}^{a}(x)\ket{\beta^{b}(p)}$.
\end{widetext}

\phantom{}
\newpage
\phantom{}
\newpage

\section{Soft theorem for dynamically massive QCD and color symmetry preservation}\label{app:softtheorem}

In this Appendix, we derive a necessary and sufficient condition for color symmetry to be preserved in its strongest form -- namely, in the form of the covariance of Green's functions under color transformations. The condition is obtained from a soft theorem which has very well-known analogues in QED and in gravity, where it enforces the conservation, respectively, of electric charge and of energy-momentum \cite{Low58,Wei64,Wei65}. Here we formulate the corresponding theorem for dynamically massive QCD quantized in the LCGs within the KO framework. Our derivation is fully nonperturbative and makes no reference to gluon asymptotic states other than those of the members of the elementary BRST quartet, which always exist in the (unphysical sector of the) spectrum -- see Sec.~\ref{sec:framework}. As a result, the theorem remains valid in a confining setting, even in the event that no gluon asymptotic states (other than those of the members of the quartet) exist in the spectrum of the theory.\\

Consider the time-ordered correlator
\begin{align}\label{deq933}
   \langle T\{J_{\mu}^{a}(x)\prod_{k}\mc{O}_{I_{k}}(x_{k})\}\rangle\ ,
\end{align}
where $J_{\mu}^{a}(x)$ is the color current, $k=1,\dots, N$ is an integer index, and $\mc{O}_{I_{k}}(x_{k})$ are local field operators, indexed by $I_{k}$, whose color transformations read
\begin{equation}\label{deq932}
   [Q^{a},\mc{O}_{I_{k}}(x)]=T_{I_{k}J_{k}}^{a}\,\mc{O}_{J_{k}}(x)\ ,
\end{equation}
$T_{I_{k}J_{k}}^{a}$ being coefficients that realize a representation of the color algebra. Since the color current is conserved, $\partial^{\mu}J_{\mu}^{a}=0$, by taking the divergence of the correlator in Eq.~\eqref{deq933}, we find that
\begin{align}
   &\partial^{\mu}_{(x)}\langle T\{J_{\mu}^{a}(x)\prod_{k}\mc{O}_{I_{k}}(x_{k})\}\rangle=\\
   \notag&=\sum_{k}\ \langle T\{[J_{0}^{a}(x),\mc{O}_{I_{k}}(x_{k})]\prod_{k^{\prime}\neq k}\mc{O}_{I_{k^{\prime}}}(x_{k^{\prime}})\}\rangle\ \delta(x^{0}-x^{0}_{k})\ ,
\end{align}
where for simplicity of notation we disregard signs due to the potential anticommuting nature of some of the fields in the average. By Fourier transforming the above equation with respect to the variable $x$ and by taking its zero-momentum limit, we obtain the equation
\begin{align}\label{deq934}
   &\lim_{p\to 0}\int d^{4}x\ e^{ipx}\ \partial^{\mu}_{(x)}\langle T\{J_{\mu}^{a}(x)\prod_{k}\mc{O}_{I_{k}}(x_{k})\}\rangle=\\
   \notag&=\sum_{k}\ T_{I_{k}J_{k}}^{a}\langle T\{\mc{O}_{J_{k}}(x_{k})\prod_{k^{\prime}\neq k}\mc{O}_{I_{k^{\prime}}}(x_{k^{\prime}})\}\rangle\ ,
\end{align}
where the RHS is but the vacuum expectation value of the commutator $[Q^{a},T\{\prod_{k}\mc{O}_{I_{k}}\}]$. In order for color symmetry to be preserved by the Green's function $\langle T\{\prod_{k}\mc{O}_{I_{k}}\}\rangle$, this vacuum expectation value must vanish.

As we saw in Secs.~\ref{sec:ssblcg} and \ref{sec:symbreak}, $J_{\mu}^{a}$ generates spontaneously broken transformations. It follows that the Fourier transform of the correlator in Eq.~\eqref{deq933} contains a massless pole proportional to the matrix element $\bra{\Omega} J_{\mu}^{a}(x)\ket{\chi^{b}(p)}$ responsible for spontaneous symmetry breaking. Let us compute this pole explicitly. By inserting the one-particle projector $\mathds{P}_{1G}$ of Eq.~\eqref{p1g} into the correlator, we find that
\begin{widetext}
   \begin{align}
      \langle T\{J_{\mu}^{a}(x)\prod_{k}\mc{O}_{I_{k}}(x_{k})\}\rangle&=\bra{\Omega} J_{\mu}^{a}(x)\,T\{\prod_{k}\mc{O}_{I_{k}}(x_{k})\}\ket{\Omega}\,\theta(x^{0}-\max_{k^{\prime}}\{x_{k^{\prime}}^{0}\})+\cdots=\\
      \notag &=-\int\frac{d^{3}{\bf p}}{(2\pi)^{3}}\frac{1}{2\hat{p}^{0}}\bra{\Omega} J_{\mu}^{a}(x)\ket{\chi^{b}(p)}\bra{\beta^{b}(p)}T\{\prod_{k}\mc{O}_{I_{k}}(x_{k})\}\ket{\Omega}\,\theta(x^{0}-\max_{k^{\prime}}\{x_{k^{\prime}}^{0}\})+\cdots=\\
      \notag &=-i\widehat{M}/g\int\frac{d^{3}{\bf p}}{(2\pi)^{3}}\frac{\hat{p}_{\mu}}{2\hat{p}^{0}}\ e^{-i\hat{p}x}\ \bra{\beta^{a}(p)}T\{\prod_{k}\mc{O}_{I_{k}}(x_{k})\}\ket{\Omega}\,\theta(x^{0}-\max_{k^{\prime}}\{x_{k^{\prime}}^{0}\})+\cdots\ ,
   \end{align}
where $\hat{p}=(|{\bf p}|,{\bf p})$, the ellipsis denotes terms which, after Fourier transforming, are subdominant in the $p^{0}\to\hat{p}^{0}$ limit, and on the last line we have used Eq.~\eqref{CGold}. By Fourier transforming with respect to $x$ as we did in Appendix~\ref{app:schwpoles}, we then obtain
\begin{align}
   &\langle T\{J_{\mu}^{a}(x)\prod_{k}\mc{O}_{I_{k}}(x_{k})\}\rangle_{\text{F.T.}}(p)=\widehat{M}/g\ \frac{p_{\mu}}{p^{2}}\ \bra{\beta^{a}(p)}T\{\prod_{k}\mc{O}_{I_{k}}(x_{k})\}\ket{\Omega}+\cdots\ .
\end{align}
By multiplying the previous equation by $-ip^{\mu}$, taking the $p\to 0$ limit, and using Eq.~\eqref{deq934}, we finally find
   \begin{equation}\label{softtheorem}
      \sum_{k}T_{I_{k}J_{k}}^{a}\langle T\{\mc{O}_{J_{k}}(x_{k})\prod_{k^{\prime}\neq k}\mc{O}_{I_{k^{\prime}}}(x_{k^{\prime}})\}\rangle=-i\widehat{M}/g\ \lim_{{\bf p}\to {\bf 0}}\ \bra{\beta^{a}(p)}T\{\prod_{k}\mc{O}_{I_{k}}(x_{k})\}\ket{\Omega}\ .
   \end{equation}
\end{widetext}

Eq.~\eqref{softtheorem} expresses the equality between the vacuum expectation value of the color variation of the operator $T\{\prod_{k}\mc{O}_{I_{k}}(x_{k})\}$ and the zero-momentum limit of the matrix element of the same operator between the vacuum and the state $\ket{\beta^{a}(p)}$.The equality is verified regardless of whether color symmetry is violated or not, and it rests on the fact that, in Yang-Mills theories quantized in the LCGs within the KO framework, the global gauge charge $Q^{a}$ is always spontaneously broken. The reason why $\beta^{a}$ appears on the RHS of the equation, instead of the Goldstone boson $\chi^{a}$, is that $\chi^{a}$ and $\beta^{a}$ are coupled non-diagonally in the projector $\mathds{P}_{1G}$. For analogous reasons, the symmetry-breaking term in $Q^{a}$ features the operator $\beta^{a}$, and not $\chi^{a}$ -- see e.g. Sec.~\ref{sec:chargestruct}.

If asymptotic fields are associated with the operators $\mc{O}_{I_{k}}(x)$, by interpreting $\ket{\beta^{a}(p)}$ in Eq.~\eqref{softtheorem} as an out state\footnote{In principle, in our derivation, $\ket{\beta^{a}(p)}$ can also be an in state.}, and by applying the LSZ reduction formula identically to both of its members, from Eq.~\eqref{softtheorem} we obtain the $S$-matrix relation
\begin{equation}\label{softtheoremsmat}
   \lim_{{\bf p}\to {\bf 0}}\ \mc{M}(A\to B+\beta^{a}(p))=ig/\widehat{M}\ T^{a}\mc{M}(A\to B)\ ,
\end{equation}
where $A$ and $B$ are, respectively, the initial and the final state of a scattering process, $\mc{M}$ denote $S$-matrix elements -- i.e. scattering amplitudes --, and $T^{a}$ is a color operator acting on each of the particles in the initial and final states in turn\footnote{In Eq.~\eqref{softtheoremsmat}, $T^{a}$ must not be interpreted as a Hilbert-space operator. $T^{a}\mc{M}(A\to B)$ is just a compact expression for the the color transformation of the amplitude.}. We thus see that Eq.~\eqref{softtheorem} truly yields a soft theorem \cite{Low58,Wei64,Wei65}, in which $\beta^{a}$ plays the role of the final-state external gauge boson whose polarization vector is equal to its momentum $p_{\mu}$.

As discussed in Secs~\ref{sec:ssblcg} and \ref{sec:symbreak}, the spontaneous breaking of $Q^{a}$, \textit{per se}, does not imply the spontaneous breaking of color symmetry. Rather, it reflects the existence of a residual gauge symmetry, which by its very nature is spontaneously broken. The soft theorem in Eq.~\eqref{softtheorem} then provides a way to concretely distinguish between a theory in which color symmetry is preserved and one in which it is not.

Indeed, a necessary and sufficient condition for color symmetry to be preserved in its strongest form is that all the Green's functions -- i.e., the vacuum correlators -- transform covariantly under color transformations. This condition, in turn, is equivalent to the vanishing of the LHS of Eq.~\eqref{softtheorem}; therefore, by the soft theorem in Eq.~\eqref{softtheorem}, regardless of the spontaneous breaking of $Q^{a}$, color symmetry is preserved in its strongest form if and only if
\begin{equation}\label{softcond}
   \lim_{{\bf p}\to {\bf 0}}\ \bra{\beta^{a}(p)}T\{\prod_{k}\mc{O}_{I_{k}}(x_{k})\}\ket{\Omega}=0
\end{equation}
for every time-ordered product $T\{\prod_{k}\mc{O}_{I_{k}}(x_{k})\}$. 

Observe that the condition in Eq.~\eqref{softcond} is automatically satisfied for a BRST-invariant product of operators. Indeed, if the product is BRST invariant,
\begin{equation}
   \Big[Q_{B},T\{\prod_{k}\mc{O}_{I_{k}}(x_{k})\}\Big]=0\ ,
\end{equation}
then
\begin{align}
   &\bra{\beta^{a}(p)}T\{\prod_{k}\mc{O}_{I_{k}}(x_{k})\}\ket{\Omega}\propto\\
   \notag&\qquad \propto\bra{\gammabar^{a}(p)}Q_{B}T\{\prod_{k}\mc{O}_{I_{k}}(x_{k})\}\ket{\Omega}=0\ ,
\end{align}
where to obtain the last zero we made $Q_{B}$ commute past the time-ordered product. On the other hand, it is not automatically satisfied for operators that transform nontrivially under $Q_{B}$, and it fails in theories like Higgs models, in which color symmetry is violated.

Let us provide an example of the failure of Eq.~\eqref{softcond} caused by the violation of color symmetry. Since the massless algebra of the SU(2) Higgs-Kibble model quantized in the LCGs -- see Appendix~\ref{app:higgsmod} -- is the same as the one of dynamically massive QCD\footnote{Additionally, quantities like the gauge-boson propagator and $u^{ij}=u\,\delta^{ij}$ are diagonal in their indices because of the custodial SU(2) symmetry \cite{KO79b,KO79a}.}, and since the model, in its classic KO treatment, satisfies the assumptions under which we obtained the results in Secs.~\ref{sec:mlessspect} and \ref{sec:ssb}, it follows that Eq.~\eqref{softtheorem} holds for the SU(2) Higgs-Kibble model as well, with $\widehat{M}$ being the gauge-boson mass defined from the zero-momentum limit of the corresponding PT-BFM propagator. Plugging the Higgs field $\Phi(x)$ into Eq.~\eqref{softtheorem}, and using Eq.~\eqref{higgscolor}, we obtain
\begin{equation}
   -T^{i}\langle \Phi(x)\rangle=-i\widehat{M}/g\,\lim_{{\bf p}\to {\bf 0}}\ \bra{\beta^{i}(p)}\Phi(x)\ket{\Omega}\ .
\end{equation}
Then, using Eqs.~\eqref{higgsavg}, \eqref{higgsrepar} and \eqref{higgsas0}, and the fact that the Goldstone bosons $\pi^{i}$ are the first parent of the elementary BRST quartet of the model, so that
\begin{equation}
   \bra{\beta^{i}(p)}\pi^{j\,\text{as.}}(x)\ket{\Omega}=-\delta^{ij}\,e^{i\hat{p}x}\ ,
\end{equation}
we find that Eq.~\eqref{softtheorem} is equivalent to
\begin{equation}
   -T^{i}\,\frac{v}{\sqrt{2}}\,\varphi_{0}=-\widehat{M}/g\,Z_{\pi}^{1/2}\,\frac{1}{\sqrt{2}}\,\sigma^{i}\,\varphi_{0}\ ,
\end{equation}
hence to
\begin{equation}
   \frac{v}{2}=\frac{Z_{\pi}^{1/2}\widehat{M}}{g}\ .
\end{equation}
This relation is a nonperturbative, exact generalization of the well-known tree-level relation $m=gv/2$ -- see Eq.~\eqref{hkmlcurr} --, and it is enforced by the nonvanishing of both members of Eq.~\eqref{softtheorem}. The latter, in turn, is caused by the violation of SU(2) color symmetry triggered by $\langle\Phi\rangle\neq 0$.

By contrast, consider the only elementary field of dynamically massive QCD whose matrix element between $\ket{\beta^{a}(p)}$ and the vacuum does not vanish -- namely, the gluon field $A_{\mu}^{a}$. Plugging the latter into Eq.~\eqref{softtheorem}, using Eq.~\eqref{color}, we obtain
\begin{align}
   \notag&if^{abc}\langle A_{\mu}^{c}(x)\rangle=-i\widehat{M}/g\ \lim_{{\bf p}\to {\bf 0}}\ \bra{\beta^{a}(p)}A_{\mu}^{b}(x)\ket{\Omega}=\\
   &\quad=-i(1+u)/g\ \lim_{{\bf p}\to {\bf 0}}\ \bra{\beta^{a}(p)}\partial_{\mu}\chi^{b}(x)\ket{\Omega}=\\
   \notag&\quad=-(1+u)/g\ \lim_{{\bf p}\to {\bf 0}}\ \hat{p}_{\mu}\,\delta^{ab}\,e^{i\hat{p}x}=0\ ,
\end{align}
consistent with the vanishing of $\langle A_{\mu}^{a}(x)\rangle$, which is in any case enforced by Lorentz covariance. The vanishing of $\langle X^{a}(x)\rangle$, where $X^{a}=B^{a},c^{a},\cbar^{a}$, similarly corresponds to the vanishing of $\bra{\beta^{a}(p)}X^{b}(x)\ket{\Omega}$.

Observe that, if Eq.~\eqref{softcond} is satisfied, then color is conserved in scattering processes\footnote{In the light of color confinement, this must be understood as a statement about the unphysical sector of the theory.}. Indeed, in that case, by Eq.~\eqref{softtheoremsmat}, we have
\begin{equation}
   T^{a}\mc{M}(A\to B)=0
\end{equation}
for every scattering amplitude, implying that
\begin{equation}
   T^{a}_{A}\,\mc{M}(A\to B)=(T^{a}_{B})^{\star}\,\mc{M}(A\to B)\ ,
\end{equation}
where $T^{a}_{A}$ and $(T^{a}_{B})^{\star}$ are color operators acting on each of the particles in the initial state $A$ and in the final state $B$\footnote{By Eq.~\eqref{deq932}, if $\mc{O}_{I_{k}}$ transforms with coefficients $T_{I_{k}J_{k}}^{a}$ and creates particles in the initial state, then $\mc{O}_{I_{k}}^{\dagger}$ creates particles in the final state and it transforms with coefficients $-T_{I_{k}J_{k}}^{a\,\star}$.}.

What is the meaning of the necessary and sufficient condition in Eq.~\eqref{softcond}? To answer this question, note that we already encountered matrix elements of the form $\bra{\beta^{a}(p)}T\{\cdots\}\ket{\Omega}$ in Appendix~\ref{app:schwpoles} in the process of amputating the elementary QCD vertices. There we saw that Green's functions of the form
\begin{equation}
   \bra{\Omega}T\{B^{a}(x)\cdots\}\ket{\Omega}\ ,
\end{equation}
when Fourier transformed with respect to the variable $x$ and multiplied by $p^{2}$, reduce to these matrix elements in the limit $p^{2}\to 0$:
\begin{align}\label{deq069}
   &\bra{\beta^{a}(p)}T\{\cdots\}\ket{\Omega}=\\
   \notag&\quad=\lim_{p^{2}\to 0}\ \int d^{4}x\ e^{ipx}\ \frac{p^{2}}{iM}\ \bra{\Omega}T\{B^{a}(x)\cdots\}\ket{\Omega}\ .
\end{align}
The above equation is nothing but the LSZ reduction formula applied to $B^{a}$.

Now, operating on Eq.~\eqref{deq069} as we did in Appendix~\ref{app:schwpoles}, we find that
\begin{align}\label{deq070}
   &\bra{\beta^{a}(p)}T\{\cdots\}\ket{\Omega}=\\
   \notag&\ =\lim_{p^{2}\to 0}\ \int d^{4}x\ e^{ipx}\ \frac{-\xi^{-1}p^{2}}{iM}\, \partial^{\mu}\bra{\Omega}T\{A_{\mu}^{a}(x)\cdots\}\ket{\Omega}=\\
   \notag&\ =\lim_{p^{2}\to 0}\ \int d^{4}x\ e^{ipx}\ \frac{i\xi^{-1}p^{2}p^{\mu}}{iM}\, \bra{\Omega}T\{A_{\mu}^{a}(x)\cdots\}\ket{\Omega}=\\
   \notag&\ =\lim_{p^{2}\to 0}\ \frac{i\xi^{-1}p^{2}p^{\mu}}{iM}\ G_{\mu}^{a}(p,\dots)\ ,
\end{align}
where the derivatives of the time-ordering operator on the second line of Eq.~\eqref{deq070} vanish in the limit $p^{2}\to0$, and we have introduced the Green's function $G_{\mu}^{a}(x,\cdots)$ defined as
\begin{equation}
   G_{\mu}^{a}(x,\dots)=\bra{\Omega}T\{A_{\mu}^{a}(x)\cdots\}\ket{\Omega}\ .
\end{equation}
Passing to the amputated vertex $\Gamma_{\mu}^{a}(p,\dots)$ corresponding to $G_{\mu}^{a}(p,\dots)$, defined so that $G_{\mu}^{a}(p,\dots)=\Delta_{\mu}^{\nu}(p)^{ab}\,\Gamma_{\nu}^{b}(p,\dots)$, we then obtain
\begin{align}\label{deq071}
   &\bra{\beta^{a}(p)}T\{\cdots\}\ket{\Omega}=(iM)^{-1}\lim_{p^{2}\to 0}\ p^{\mu}\Gamma_{\mu}^{a}(p,\dots)\ .
\end{align}
Finally, since $p\to 0$ implies $p^{2}\to 0$, we find that the condition in Eq.~\eqref{softcond} is equivalent to
\begin{equation}\label{softcond2}
   \lim_{p\to 0}\ p^{\mu}\Gamma_{\mu}^{a}(p,\dots)=0
\end{equation}
for every amputated vertex $\Gamma_{\mu}^{a}(p,\dots)$ with an external gluon leg of momentum $p$.

Eq.~\eqref{softcond2} is a statement about the soft limit of the Slavnov-Taylor identities of the theory. Since in it $\Gamma_{\mu}^{a}(p,\dots)$ is contracted with $p^{\mu}$, it is not hard to see that the equation will be generally satisfied, except for the special case in which the vertex develops specific singularities. If these singularities are present, then the Green's functions violate color symmetry.

Insofar as dynamical gluon mass generation is concerned, it is crucial to realize that the singularities that may cause the failure of Eqs.~\eqref{softcond} and \eqref{softcond2} \textit{are not the singularities responsible for dynamical gluon mass generation}. Indeed, as we saw in Sec.~\ref{sec:goldbound} and in Appendix~\ref{app:schwpoles}, the singularities responsible for the gluon mass -- namely, the Schwinger poles, proportional to the BSVs of $\chi^{a}$ --, albeit being longitudinally coupled, are found in the \textit{transverse part} of amputated vertices -- see Eqs.~\eqref{gammaimpl} - \eqref{singularvertex}. The singularities that may cause the failure of Eqs.~\eqref{softcond} and \eqref{softcond2}, on the other hand, are found in the \textit{longitudinal part} of amputated vertices -- namely, in the part that survives the contraction $p^{\mu}\Gamma_{\mu}^{a}(p,\dots)$.

Similarly, the matrix elements of $\beta^{a}$ do not contribute to the zero-momentum limit of the gluon propagator. Therefore, whether or not Eq.~\eqref{softcond} is satisfied is irrelevant to dynamical gluon mass generation: the violation of Eq.~\eqref{softcond} is not a necessary condition for gluons to acquire a mass.

In the present paper, we do not assume explicitly that Eqs.~\eqref{softcond} and \eqref{softcond2} are satisfied -- i.e. that color symmetry is preserved in its strongest form in dynamically massive QCD. Other, more specific assumptions are better suited to our purposes. Nevertheless, by the above discussion, we conclude that there is no contradiction in assuming that a gluon mass is dynamically generated \textit{and} that color symmetry is preserved. The spontaneous breaking of the global gauge charge $Q^{a}$, the preservation of color symmetry, and the dynamical generation of a gluon mass are distinct features of the theory that can coexist peacefully.

\bibliography{QCD}

\end{document}